\documentclass[prd,10pt,twocolumn,showpacs,nofootinbib,aps,superscriptaddress,preprintnumbers]{revtex4-1}

\usepackage{graphicx}
\usepackage{epsfig}
\usepackage{dcolumn}
\usepackage{bm}
\usepackage{epstopdf}
\usepackage{url}
\usepackage{siunitx}
\usepackage{hyperref}
\usepackage{float}
\usepackage{lineno}
\usepackage{upgreek}
\usepackage{multirow}
\usepackage{tabularx} 
\usepackage{eurosym} %
\usepackage[T1]{fontenc}
\usepackage{xspace}
\usepackage{units}
\usepackage{siunitx}
\usepackage{amstext}
\usepackage{graphicx}
\usepackage{float}
\usepackage{color}
\usepackage{rotating}
\usepackage[dvipsnames,table]{xcolor}
\usepackage{soul}
\hypersetup{urlcolor=RedViolet,
	    citecolor=ForestGreen ,
	    linkcolor=RoyalBlue, 
	    colorlinks=true}
\definecolor{lightgray}{rgb}{0.9,0.9,0.9}	    
\definecolor{green}{rgb}{0,0.5,0}
\definecolor{red}{rgb}{1,0,0}
\definecolor{blue}{rgb}{0,0,0.5}
\usepackage{booktabs}
\usepackage[utf8]{inputenc}
\usepackage{pifont}

\usepackage{todonotes}

\DeclareRobustCommand{\rchi}{{\mathpalette\irchi\relax}}
\newcommand{\irchi}[2]{\raisebox{\depth}{$#1\chi$}} 

\newcommand{\kms}{km\,s\ensuremath{^{-1}}}
\newcommand{\cevns}{CE$\nu$NS\xspace}
\newcommand{\elec}{e\ensuremath{^-}}
\newcommand{\mupic}{$\upmu$-PIC\xspace}

\newcommand{\gevcc}{GeV$/c^{2}$\xspace}
\newcommand{\kevr}{keV$_{\rm r}$\xspace}
\newcommand{\kevee}{keV$_{\rm ee}$\xspace}
\newcommand{\cff}{CF$_{4}$}

\newcommand{\sfsix}{SF$_{6}$}
\newcommand{\hesfsix}{He:SF$_{6}$\xspace}
\newcommand{\hecffour}{He:CF$_{4}$\xspace}
\newcommand{\ie}{{\em i.e.}}
\newcommand{\eg}{{\em e.g.}}

\newcommand{\nocontentsline}[3]{}
\newcommand{\tocless}[2]{\bgroup\let\addcontentsline=\nocontentsline#1{#2}\egroup}
\newcommand{\tick}{\ding{51}}

\newcommand{\Cygnus}{\textsc{Cygnus}\xspace}

\def\lsim{\mathrel{\rlap{\lower4pt\hbox{\hskip1pt$\sim$}}
    \raise1pt\hbox{$<$}}}         
\def\gsim{\mathrel{\rlap{\lower4pt\hbox{\hskip1pt$\sim$}}
    \raise1pt\hbox{$>$}}}         

\newcommand{\EASPUB}{EAS Publ. Ser.}

\newcommand{\ASTROPART}{Astropart. Phys.}

\newcommand{\IJMPA}{Int. J. Mod. Phys.}

\newcommand{\JCAP}{J. Cosmol. Astropart. Phys.}

\newcommand{\JINST}{J. Instrum.}

\newcommand{\NIM}{Nucl. Instrum. Methods}

\newcommand{\PHYSLETT}{Phys. Lett.}
\newcommand{\PHYSDARKU}{Phys. Dark Univ.}

\newcommand{\PHYSREV}{Phys. Rev.}
\newcommand{\PRL}{Phys. Rev. Lett.}
\newcommand{\REVSCIINSTRUM}{Rev. Sci. Instrum.}

\begin{document}

\date{\today}

\title{\textsc{Cygnus}: Feasibility of a nuclear recoil observatory with directional sensitivity to dark matter and neutrinos}

\author{S.~E.~Vahsen}
\affiliation{Department of Physics and Astronomy, University of Hawaii, Honolulu, Hawaii 96822, USA}

\author{C.~A.~J.~O'Hare}
\affiliation{The University of Sydney, School of Physics, NSW 2006, Australia}

\author{W.~A.~Lynch}
\affiliation{Department of Physics and Astronomy, University of Sheffield, S3 7RH, Sheffield, United Kingdom}

\author{N.~J.~C.~Spooner}
\affiliation{Department of Physics and Astronomy, University of Sheffield, S3 7RH, Sheffield, United Kingdom}

\author{E.~Baracchini}
\affiliation{Istituto Nazionale di Fisica Nucleare, Laboratori Nazionali di Frascati, I-00040, Italy}
\affiliation{Istituto Nazionale di Fisica Nucleare, Sezione di Roma, I-00185, Italy}
\affiliation{Department of Astroparticle Physics, Gran Sasso Science Institute, L'Aquila, I-67100, Italy}

\author{P.~Barbeau}
\affiliation{Department of Physics, Duke University, Durham, NC 27708 USA}

\author{J.~B.~R.~Battat}
\affiliation{Department of Physics, Wellesley College, Wellesley, Massachusetts 02481, USA}

\author{B.~Crow}
\affiliation{Department of Physics and Astronomy, University of Hawaii, Honolulu, Hawaii 96822, USA}

\author{C.~Deaconu}
\affiliation{Department of Physics, Enrico Fermi Inst., Kavli Inst. for Cosmological Physics, Univ. of Chicago , Chicago, IL 60637, USA}

\author{C.~Eldridge}
\affiliation{Department of Physics and Astronomy, University of Sheffield, S3 7RH, Sheffield, United Kingdom}

\author{A.~C.~Ezeribe}
\affiliation{Department of Physics and Astronomy, University of Sheffield, S3 7RH, Sheffield, United Kingdom}

\author{M.~Ghrear}
\affiliation{Department of Physics and Astronomy, University of Hawaii, Honolulu, Hawaii 96822, USA}

\author{D.~Loomba}
\affiliation{Department of Physics and Astronomy, University of New Mexico, NM 87131, USA}

\author{K.~J.~Mack}
\affiliation{Department of Physics, North Carolina State University, Raleigh, NC 27695, USA}

\author{K.~Miuchi} 
\affiliation{Department of Physics, Kobe University, Rokkodaicho, Nada-ku, Hyogo 657-8501, Japan}

\author{F.~M.~Mouton}
\affiliation{Department of Physics and Astronomy, University of Sheffield, S3 7RH, Sheffield, United Kingdom}

\author{N.~S.~Phan}
\affiliation{Los Alamos National Laboratory, P.O. Box 1663, Los Alamos, NM 87545, USA}
 
\author{K.~Scholberg}
\affiliation{Department of Physics, Duke University, Durham, NC 27708 USA}

\author{T.~N.~Thorpe}
\affiliation{Department of Physics and Astronomy, University of Hawaii, Honolulu, Hawaii 96822, USA}
\affiliation{Department of Astroparticle Physics, Gran Sasso Science Institute, L'Aquila, I-67100, Italy}

\begin{abstract}
Now that conventional weakly interacting massive particle (WIMP) dark matter searches are approaching the neutrino floor, there has been a resurgence of interest in detectors with sensitivity to nuclear recoil directions. A large-scale directional detector is attractive in that it would have sensitivity below the neutrino floor, be capable of unambiguously establishing the galactic origin of a purported dark matter signal, and could serve a dual purpose as a neutrino observatory. We present the first detailed analysis of a 1000~m$^3$-scale detector capable of measuring a directional nuclear recoil signal at low energies. We propose a modular and multi-site observatory consisting of time projection chambers (TPCs) filled with helium and SF$_6$ at atmospheric pressure. By comparing several available readout technologies, we identify high-resolution strip readout TPCs as the optimal tradeoff between performance and cost. We estimate that suitable angular resolution and head-tail recognition is achievable down to helium recoil energies of 6~\kevr. Depending on the readout technology, an average of only 4--5 detected 100~\gevcc WIMP-fluorine recoils above 50 \kevr are sufficient to rule out an isotropic recoil distribution at 90\% CL. An average of 10--20 helium recoils above 6~\kevr or only 3--4 helium recoils above 20~\kevr would suffice to distinguish a 10~\gevcc WIMP signal from the solar neutrino background. High-resolution TPC charge readout also enables powerful electron background rejection capabilities well below 10~keV. We detail background and site requirements at the 1000~m$^3$-scale, and identify materials that require improved radiopurity. The final experiment, which we name \Cygnus-1000, will be able to observe 10--40 neutrinos from the Sun, depending on the final energy threshold. With the same exposure, the sensitivity to spin independent cross sections will extend into presently unexplored sub-10~\gevcc parameter space. For spin dependent interactions, already a 10~m$^3$-scale experiment could compete with upcoming generation-two detectors, but \Cygnus-1000 would improve upon this considerably.  Larger volumes would bring sensitivity to neutrinos from an even wider range of sources, including galactic supernovae, nuclear reactors, and geological processes.
\end{abstract}

\maketitle








\section{Introduction}
\label{sec:introduction}
A wide range of astrophysical observations across galactic and cosmological scales indicate that dark matter (DM) dominates the mass budget of the Universe. While the gravitational evidence for DM is now overwhelmingly strong (for a review, see Ref.~\cite{Bertone:2016nfn}), its particle identity remains unknown. A definitive detection of DM is expected to be a gateway to physics beyond the Standard Model. Being of such fundamental importance, the quest to uncover the nature of DM remains one of the most important experimental challenges in contemporary physics~\cite{Battaglieri:2017aum}. 

Experimental efforts in direct DM detection have largely focused on the possibility that it consists of weakly interacting massive particles (WIMPs) whose scattering cross sections with ordinary matter are small but nonzero, detectable in low-background nuclear recoil experiments. After years of incremental improvements, large swaths of WIMP parameter space are now ruled out, including those in which detections were previously claimed~\cite{Angloher:2015ewa,Agnese:2017jvy,Hehn:2016nll,Amole:2015lsj,Amole:2015pla,Agnes:2015ftt,Tan:2016zwf,Akerib:2016vxi,Aprile:2017iyp,Adhikari:2019off}. As detectors become sensitive to lower masses and weaker cross sections, the previously negligible neutrino background will become important, potentially swamping any WIMP signal. Models with weak signals that are hidden beneath the neutrino background are said to live below the ``neutrino floor'', a theoretical boundary beyond which conventional DM detectors cannot efficiently probe. 

DM detectors with \textit{directional sensitivity}---the ability to reconstruct the directions of WIMP-induced nuclear recoil events---have the potential to address two of the main challenges to current detection techniques: (1) the difficulty of positively identifying detected interactions as being caused by the DM making up the galactic halo, and (2) the potential for the saturation of the signal by the neutrino background. Directionality addresses both challenges by identifying the direction from which interacting particles originate, thus distinguishing between events coming from the Milky Way DM halo and those originating elsewhere. While this is a powerful background discrimination technique in principle, in practice it remains challenging to build a directional detector that is large enough to compete with non-directional experiments.

This article lays out the science case and goals for a large directional nuclear recoil observatory, and details the challenges that need to be faced in order to achieve those goals with existing technology. The motivation for this work draws on results summarized in three extensive reviews: the first summarized the status of existing directional detection projects~\cite{Ahlen:2009ev}, the second detailed the full discovery potential for directional detectors~\cite{Mayet:2016zxu}; and the third presented a broad summary of available direction-sensitive readout technologies~\cite{Battat:2016pap}. Here, we combine all present knowledge in the field and develop a single strategy for a directional detector named \Cygnus. This work bridges the gap between, on the one hand, theoretical literature on directional detection which often involves many idealized assumptions about detector capabilities; and on the other, purely experimental studies which are relatively far removed from a potential future utility in a competitive DM search.

The primary goal of \Cygnus will be to discover DM even if its cross section lies beyond the neutrino floor. WIMP sensitivity at this level also brings a secondary motivation: the study of the neutrino background itself. There is substantial physics motivation for the detection of natural or human-made neutrinos, via either coherent neutrino-nucleus scattering (\cevns) or neutrino-electron scattering, both of which will be measurable in \Cygnus. In particular, directional measurements of these processes would be a highly-novel additional signal for many terrestrial and astrophysical sources of neutrino. \Cygnus, therefore, retains a strong science case, even in the absence of a positive DM signal.

\begin{figure*}[hbt]
\includegraphics[width=0.99\textwidth,trim={0.7cm 0cm 0cm 0cm},clip]{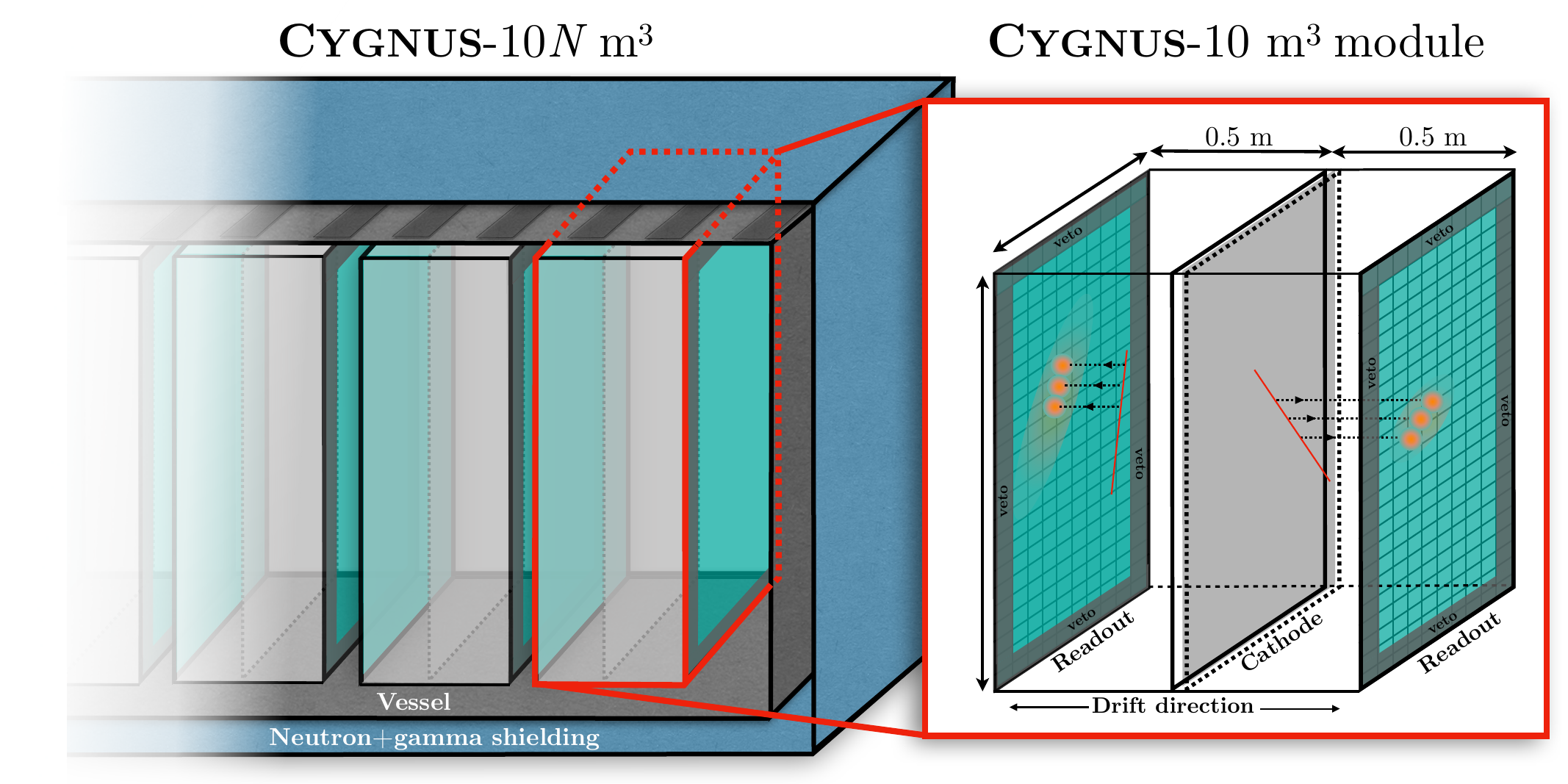}
\caption{Schematic of a modular $N\times10$~m$^3$ directional detector made up of $N$ back-to-back TPC modules. Each module would have two readout planes and a central cathode to ensure a short maximum drift distance of 50~cm.}\label{fig:TPCdiagram}
\end{figure*}

\begin{figure}[hbt]
\includegraphics[width=0.49\textwidth]{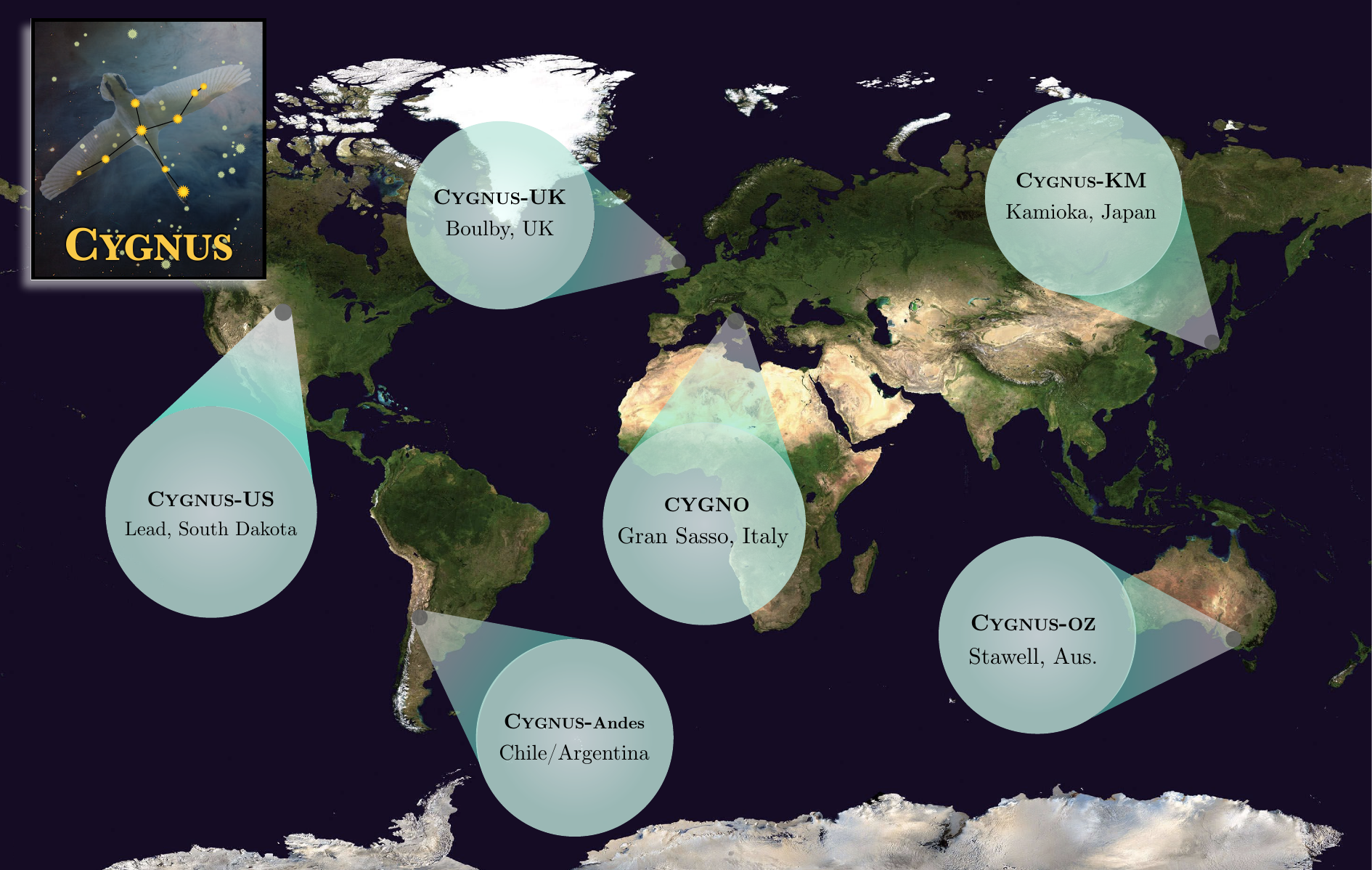}
\caption{Proposed sites for a network of \Cygnus detectors. Site considerations for a large-scale TPC are discussed in Sec.~\ref{sec:sites}.}\label{fig:map}
\end{figure}

Observing the neutrino background and WIMP cross sections lying below the neutrino floor requires experiments with very large total masses. Hence the final goal for \Cygnus will be a network of modular and multi-site gas experiments each individually based on the strategies set out here. Organizing the experiment in a modular and multi-site manner alleviates some of the issues regarding the very large volumes needed for a gas target to reach the required mass. A schematic for a style of modularity is shown in Fig.~\ref{fig:TPCdiagram}, which shows how several smaller ``back-to-back'' gas time projection chambers (TPCs) can be placed within the same background shielding. Our study focuses on the design of a single module that can be scaled up in this fashion. In Fig.~\ref{fig:map} we show the locations of some of the candidate sites under consideration. A globally distributed experiment is also advantageous as it provides an additional method to deal with seasonal or location-dependent backgrounds as well as help to overcome possible site-restrictions on the overall size of a detector.

The organization of this article is as follows: in Sec.~\ref{sec:science_case} we describe the advantages of directionality for the robust discovery and characterization of a DM signal, and to contribute to studies of solar, supernova, atmospheric, and geological neutrinos. Section~\ref{sec:technology_choices} reviews the existing and proposed detector technologies that may enable directional detection, including the gas TPC, which forms the basis of \Cygnus. In Sec.~\ref{sec:technology_comparison}, we describe how different TPC gases and readout technologies compare in their detection capabilities while keeping in mind practical considerations for construction. Section~\ref{sec:backgrounds} discusses the feasibility of achieving zero-background in large-scale TPCs, while Sec.~\ref{sec:sites} describes the requirements for detector location and installation. Finally in Sec.~\ref{sec:conclusion}, we conclude the technological considerations described in the above sections and outline a specific direction-sensitive detector that is both feasible and cost-effective.

\section{Science case for a large nuclear recoil observatory}
\label{sec:science_case}
Past and ongoing directional detection experiments and R\&D efforts such as DRIFT~\cite{Battat:2014oqa}, MIMAC~\cite{mimac}, D$^3$~\cite{Vahsen:2011qx}, DMTPC~\cite{DMTPC_ucladm14}, NEWAGE~\cite{Miuchi:2010hn}, and CYGNO~\cite{Baracchini:2020btb} have demonstrated the low-density gas TPC concept, and have made many impressive technological advances over the years. 
However, current stringent WIMP limits from non-directional experiments dictate a significant scaling-up of these technologies, which will pose new experimental challenges.
This section provides the motivation for facing those challenges by laying out the strong science case for a direction-sensitive ``recoil observatory.'' 

\subsection{Motivation}
The most popular technique to directly detect WIMP DM is to search for nuclear recoils with energies $\mathcal{O}(1~-~100)$~keV~\cite{Undagoitia:2015gya,Schumann:2019eaa}. 
Experimental strategies using a wide range of targets have been developed, exploiting charge, light, or heat signals (for a review see \eg{}~Ref.~\cite{Undagoitia:2015gya}). After decades of progress, the tightest limits now challenge favored models from supersymmetry~\cite{Arcadi:2017kky}. Several experiments have in the past reported detections that are consistent with a DM interpretation, but were in tension with limits from other experiments. The only outstanding WIMP detection claim remaining is from the DAMA/LIBRA collaboration who report a 9.3$\sigma$ annual modulation of the event rate in their NaI(Tl) crystal scintillators~\cite{Bernabei:2013xsa,Bernabei:2018yyw}. While the annual modulation they observe is ostensibly DM-like, such an interpretation is without support from any other experiment. The annual modulation of galactic DM was often touted as a clear signature of DM, however; the DAMA/LIBRA result has demonstrated this approach can be entirely frustrated by systematics. So persistent is this disagreement that there are now several experiments around the world designed to test the DAMA result in the most direct way possible: by replicating the experiment (DM-Ice~\cite{deSouza:2016fxg}, KIMS~\cite{Kim:2018wcl}, SABRE~\cite{Froborg:2016ova}, COSINE-100~\cite{Adhikari:2018ljm,Adhikari:2019off}, ANAIS~\cite{Amare:2019jul,Amare:2018sxx}) and COSINUS~\cite{Angloher:2017sft,Angloher:2016ooq}. Of the currently released results from ANAIS and COSINE-100, both are still consistent with a modulation amplitude of zero. However they will need several more years of exposure to begin to definitively rule out the presence of a modulation at the level of DAMA. 

Though experience has uncovered the limitations of the annual modulation signature -- the distribution of recoil directions promises more robust and convincing evidence for DM.
The directional DM signature arises from the motion of the Earth (and solar system) through the galactic halo. The flux of DM particles incident on Earth is strongly anisotropic. This so-called DM ``wind'' peaks close to the star Deneb in the constellation Cygnus~\cite{Spergel:1987kx}. For DM-induced recoil directions, this leads to two prominent phenomena. Firstly, in galactic coordinates the angular distribution of recoils should be a dipole pointing back towards this well-understood direction. Secondly, in lab-centric coordinates, the anisotropy will cycle across the sky due to the Earth's rotation, with a period of one \emph{sidereal} day. By definition, these effects are observable in \emph{direction-sensitive} experiments, but not otherwise. The measurement of recoil directions is therefore critical to establishing unequivocally that the source of some signal excess in events is due to the same particle that makes up the DM in the Milky Way. Such a detector would also have a strongly-enhanced ability to remove backgrounds. In an idealized situation, directional capability would make a WIMP search maximally reliable and robust by eliminating all possible backgrounds~\cite{Grothaus:2014hja,O'Hare:2015mda,OHare:2017rag} and demonstrating the galactic origin of the detected particle~\cite{Billard:2011zj}.

As anticipated long before the advent of ton-scale detectors, natural sources of neutrinos are the ultimate irreducible background to WIMP searches. Indeed, the upcoming generation of direct detection experiments are anticipated to be so large and so sensitive that the coherent scattering between neutrinos and the target will already be one of the most important backgrounds. Neutrinos are unshieldable so without a strategy to discriminate against them, the discovery of some well-motivated low-mass WIMP models~\cite{Arcadi:2017wqi} would require prohibitively large exposures. However, even in the eventuality that no WIMP signal is detected, the neutrino background itself constitutes an interesting signal. A measurement of the angular distribution of neutrinos is of great interest in the context of supernovae~\cite{Linzer:2019swe}, geoneutrinos~\cite{geoneutrinos}, and atmospheric neutrinos~\cite{OHare:2020lva}. Finally, we note that in the event of a positive WIMP detection, a directional experiment would be able to probe some more exotic dark particle physics as well as  to map the currently unknown three-dimensional velocity structure of the DM distribution around the solar system~\cite{Morgan:2004ys,Billard:2009mf,Lee:2012pf,O'Hare:2014oxa,Mayet:2016zxu,Kavanagh:2016xfi}. This latter achievement would greatly improve our understanding of the cosmological accretion and merger history of our galaxy, making such an experiment an instrument of astrophysics as well as particle physics.

\subsection{Dark matter}
\label{sec:DM}

\subsubsection{WIMP scattering review}\label{sec:WIMPscattering}

\begin{figure*}[hbt]
\includegraphics[width=0.49\textwidth] {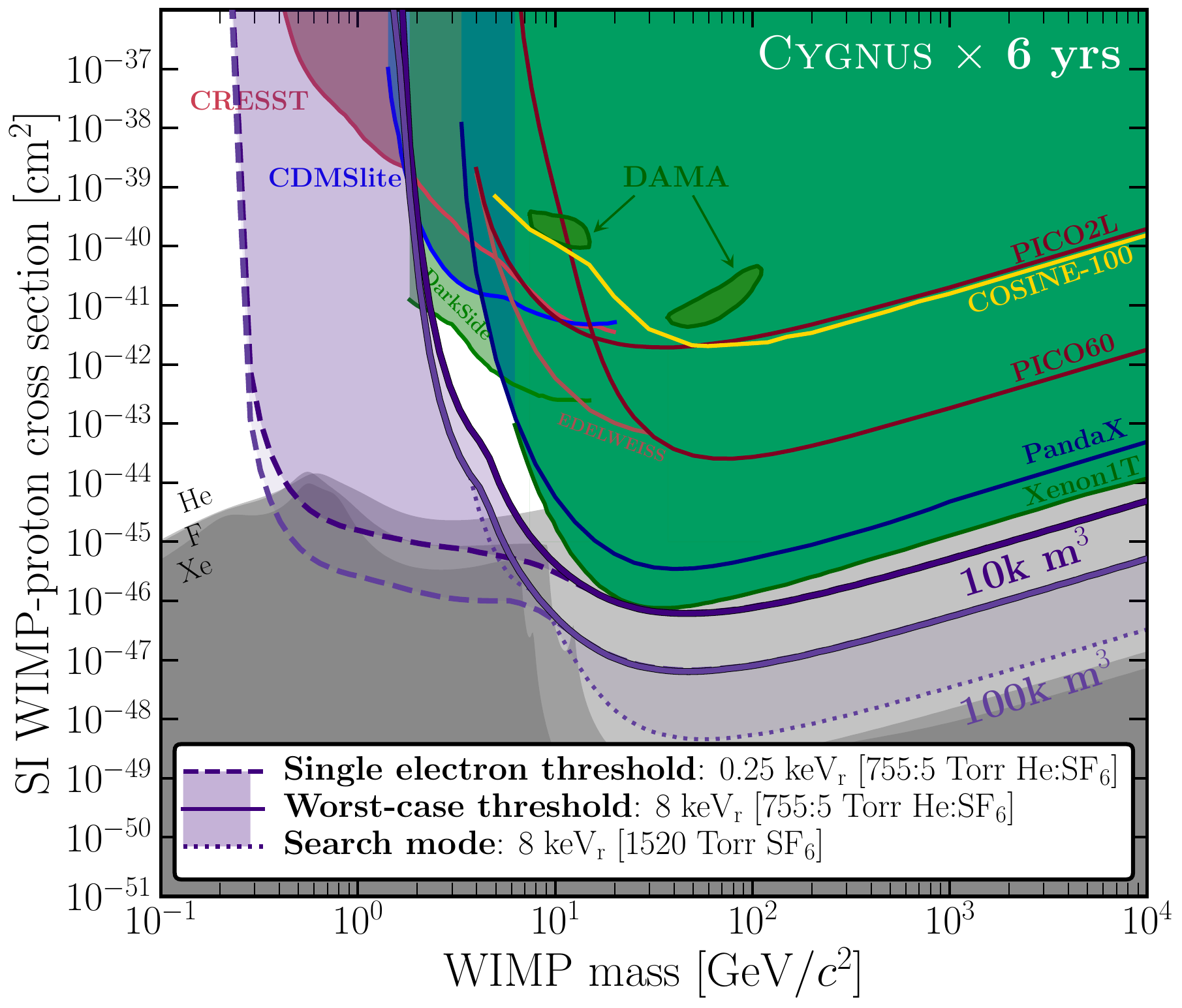}
\includegraphics[width=0.49\textwidth] {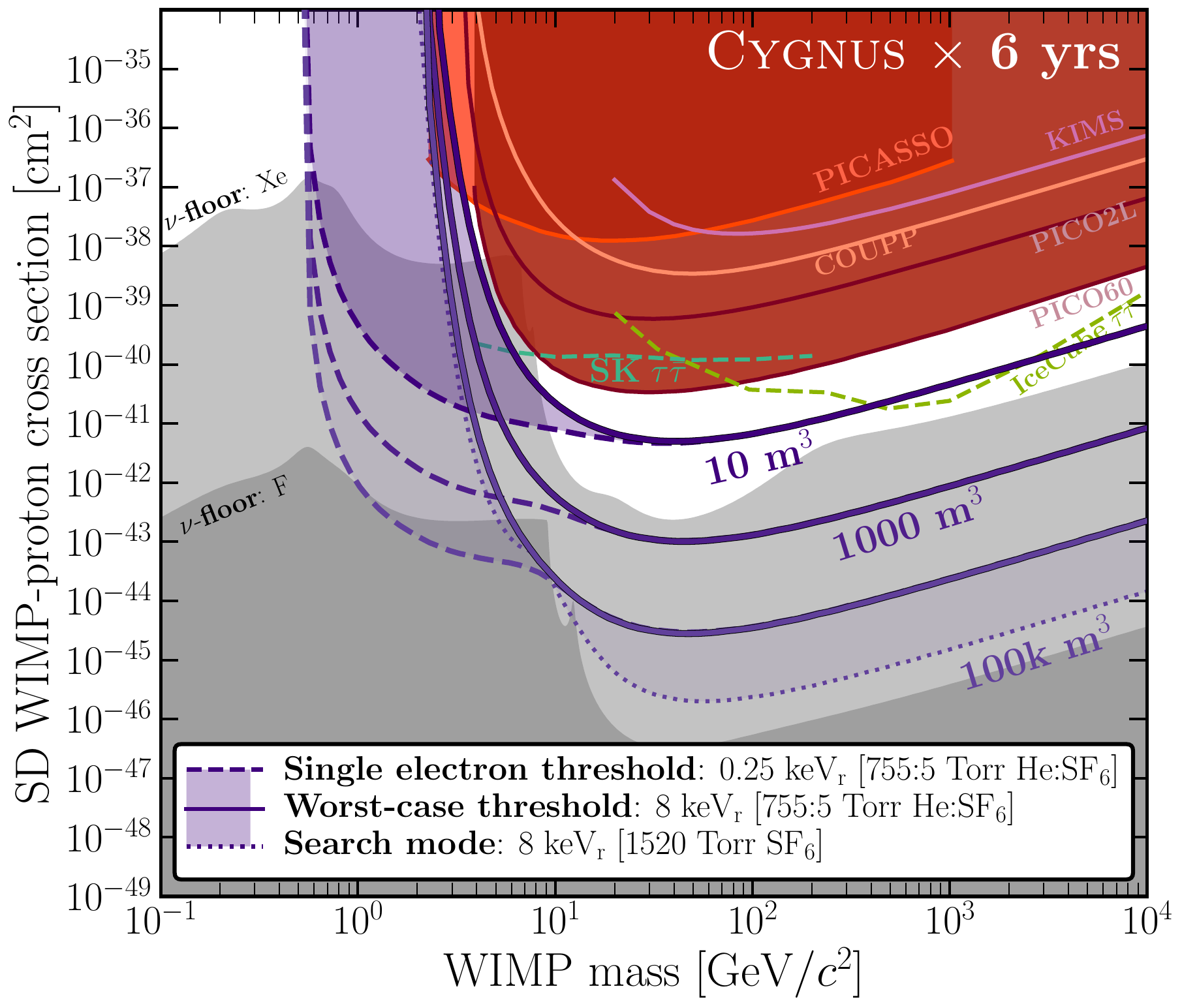}
\caption{Constraints on the spin-independent WIMP-nucleon (left) and spin-dependent WIMP-proton (right) cross sections. We show the existing constraints and detections from various experiments as labeled (see text for the associated references). In purple solid and dashed lines we show our projected 90\% CL exclusion limits for the \Cygnus experiment operating for six years with 10 m$^3$ up to 100,000 m$^3$ of \hesfsix gas at 755:5 Torr (where 6 years$\times$1000 m$^3$ corresponds to a $\sim$1 ton-year exposure). For each volume we show limits for two possible thresholds, ranging from the worst-case threshold for electron-background-free operation of 8 keV$_{\rm r}$ to a very best-case minimum energy threshold corresponding to a single electron, 0.25 keV$_{\rm r}$. We emphasize however that we anticipate excellent electron discrimination well below 8~\kevr. For the 100k m$^3$ limits we add a third dotted line which corresponds to a mode with purely SF$_6$ gas at 1520 Torr (two atmospheres). This experiment would have a significantly higher total mass but would come at the cost of any directional sensitivity. This `search mode' could be used to extend the high mass sensitivity to just within reach of the neutrino floor. For the SI panel, we shade in gray the neutrino floor for helium, fluorine, and xenon targets (top to bottom), and for SD we show only fluorine and xenon. The neutrino floor is derived following a procedure described in the text. It roughly corresponds to the cross sections where the rate of sensitivity improvement with increasing detector size is slowed the most by the various sources of neutrino background.}\label{fig:limits}
\end{figure*}

The event rate for WIMP-induced nuclear recoils is derived by integrating the product of the incoming flux of DM and the cross section $\sigma$ for the relevant WIMP-nucleus interaction. This is usually written in terms of the differential event rate $R$ per unit detector mass, as a function of recoil energy $E_r$ and time $t$,
\begin{equation}
\frac{\textrm{d}R}{\textrm{d}E_r}(E_r,t) = \frac{\rho_0}{m_\chi m_A} \int_{v > v_\textrm{min}}{v f({\bf v},t)} \frac{\textrm{d} \sigma}{\textrm{d}E_r}(E_r,v) \textrm{d}^3 v ,
\end{equation}
where $\rho_0$ is the local DM mass density, $m_\chi$ is the WIMP mass, $m_A$ is the nucleus mass and $\mathbf{v}$ is the DM velocity in the detector rest frame. The integral is performed for WIMP speeds $v$ larger than $v_{\rm min}(E_r) =  \sqrt{m_A E_r/2 \mu^2_{\rchi A}}$  which is the minimum speed kinematically permitted to produce a recoil with energy $E_r$. The factor $\mu_{\rchi A}$ is the WIMP-nucleus reduced mass. The integral over velocities is weighted by the flux of DM particles $v f({\bf v},t)$ where $f$ is the galactic distribution of DM velocities. For DM in the form of WIMPs, over the timescales of observation, the velocity distribution is constant in time. In this formula however, it picks up a time dependence after a boost into the laboratory rest frame by $\mathbf{v}_{\rm lab}$, the velocity with which we are moving through the galactic halo. The differential cross section is proportional to the squared matrix element for a particular WIMP-nucleus interaction, so is therefore specifically model-dependent. However one can work with general formulae. The most common approach is to divide the interaction into two possible channels, both of which may contribute to the total rate,
\begin{equation}
\frac{\textrm{d}\sigma}{\textrm{d}E_r} = \frac{m_A}{2 \mu_{\rchi A}^2 v^2} \left( \sigma_0^{\rm SI} F_{\rm SI}^2(E_r) + \sigma_0^{\rm SD} F_{\rm SD}^2(E_r) \right) \, .
\end{equation}
Here, the first term describes spin-independent (SI) interactions such as those arising from scalar or vector WIMP-quark couplings, whereas the second includes the spin-dependent (SD) contributions from, for example, axial-vector couplings. The cross sections $\sigma^{\rm SI,\, SD}_0$ are defined at zero momentum transfer, and nuclear form factors $F^2_{\rm SI,\, SD}$ (where $F(0)\equiv1$) are invoked to describe the nuclear structure and its response to WIMP scattering events at different energies: in particular the suppression of the rate due to the loss in coherence over the nucleus at high momentum transfers. Note that the form factors depend entirely on nuclear physics, and all WIMP model dependence is contained in $\sigma^{\rm SI, \, SD}_0$. Keep in mind that this is only the baseline set of WIMP interactions which are most commonly used to compare experimental results. The list of all the physically-allowed non-relativistic WIMP-nucleus interaction operators is much longer than the two mentioned so far, and accounting for them reveals a much richer array of possible signals (as will be discussed in Sec.~\ref{sec:models}). But for simply setting and comparing null results, the convention has been to focus on the SI and SD cross sections. Simplifying things even further, it is also commonplace to write all nuclear cross sections as proportional to those on the proton $\sigma_p^{\rm SI, SD}$,
\begin{eqnarray}
\sigma_0^{\rm SI} &=& \big| Z + (f_n/f_p) (A-Z)\big|^2 \sigma_p^{\rm SI} \, ,\\
\sigma_0^{\rm SD} &=& \frac{4}{3} \frac{J+1}{J} \big| \langle S_p \rangle + (a_n/a_p) \langle S_n \rangle \big|^2 \sigma_p^{\rm SD} \, ,
\end{eqnarray}
where $A$ is the nucleus mass number, $Z$ is the nucleus atomic number, $\langle S_{p,n} \rangle$ are the average of the proton and neutron spins in the nucleus, and $J$ is the total nuclear spin. The ratios of the couplings to the proton and neutron are written as $f_p/f_n$ and $a_p/a_n$ for SI and SD interactions respectively. In the SI case, the total nuclear cross section is enhanced by the number of nucleons squared (assuming, as we do here, isospin conserving DM where $f_n/f_p = 1$), meaning that large target nuclei tend to have the potential to set the most stringent limits. In the spin-dependent case, the interaction probability is not amplified by mass number, but depends instead on the spin content of the target nuclei, hence; constraints tend to be weaker.

In the case of directional detection we want to know how the differential event rate depends on a recoil's direction as well as its energy. We can derive this by considering the kinematic relationship between the incoming WIMP velocity, $\mathbf{v}$, with the outgoing recoil direction $\hat{\mathbf{r}}$,
\begin{equation}\label{eq:scatteringangle}
\textbf{v} \cdot \hat{\textbf{r}} = \sqrt{\frac{m_A E_r}{2 \mu_{\rchi A}^2}} \equiv v_{\rm min}\, ,
\end{equation}
which can be enforced in the differential cross section with a delta function,
\begin{equation}
 \frac{\textrm{d}^2\sigma}{\textrm{d}E_r\,\textrm{d}\Omega} = \frac{\textrm{d}\sigma}{\textrm{d}E_r} \frac{1}{2\pi} v \, \delta \left(\textbf{v} \cdot \hat{\textbf{r}} - v_{\rm min}\right) \, ,
\end{equation} 
where $\textrm{d}\Omega$ is the solid angle element around $\hat{\textbf{r}}$. The event rate then has the structure,
\begin{eqnarray}\label{eq:finaleventrate-directional}
 \frac{\textrm{d}^2R}{\textrm{d}E_r\textrm{d}\,\Omega}(E_r,\hat{\mathbf{r}},t) &=& \frac{\rho_0}{4\pi\mu_{\chi p}^2 m_\chi} (\sigma^{\rm SI}_0 F^2_{\rm SI}(E) + \sigma^{\rm SD}_0 F^2_{\rm SD}(E))  \, \nonumber \\
 & \times & \hat{f}(v_{\rm min},\hat{\textbf{r}},t) \, ,
\end{eqnarray}
where the velocity distribution now enters in the form of its Radon transform~\cite{Gondolo:2002np,Radon},
\begin{equation}
 \hat{f}(v_{\rm min},\hat{\textbf{r}},t) = \int \delta\left(\textbf{v} \cdot \hat{\textbf{r}} - v_{\rm min}\right) f(\textbf{v},t)\, \textrm{d}^3 \textbf{v}\, .
\end{equation}

The characteristic angular structure of the DM flux on Earth is the reason why directional detection could be such a powerful means to discover DM. The primary signal is a dipole anisotropy towards the direction $\hat{\bf{r}} = -\hat{{\bf v}}_{\rm lab}$, leading to an $\mathcal{O}(10)$ forward-backward asymmetry in the number of recoil events. The strength of this dipole means that in ideal circumstances (\ie{}~perfect recoil direction reconstruction) an isotropic null hypothesis for the recoil direction distribution can be rejected at 90\% confidence with only $\mathcal{O}(10)$ events~\cite{Morgan:2004ys,Green:2010zm}. With $\mathcal{O}(30)$ recoil directions it becomes possible to point back towards Cygnus and confirm the galactic origin of the signal~\cite{Billard:2009mf}. Secondary signals such as a ring feature at low energies~\cite{Bozorgnia:2011vc}, and the aberration of recoil directions over time~\cite{Bozorgnia:2012eg}, may also aid in the confirmation of a DM signal, as well as for characterizing astrophysical properties of the DM halo.

\subsubsection*{Existing limits and projections for \Cygnus}
\begin{figure*}[hbt]
\includegraphics[width=0.49\textwidth]{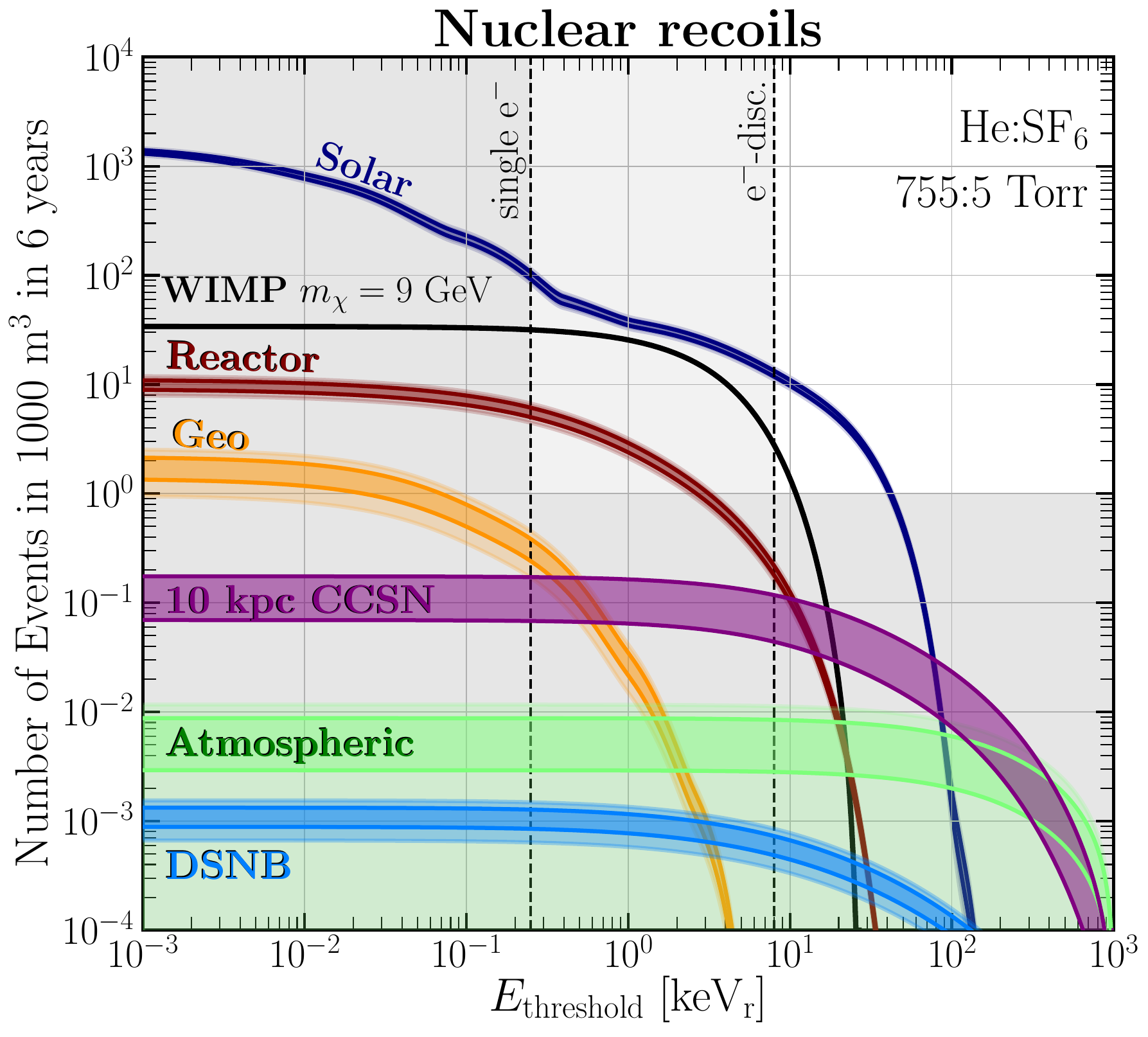}
\includegraphics[width=0.49\textwidth]{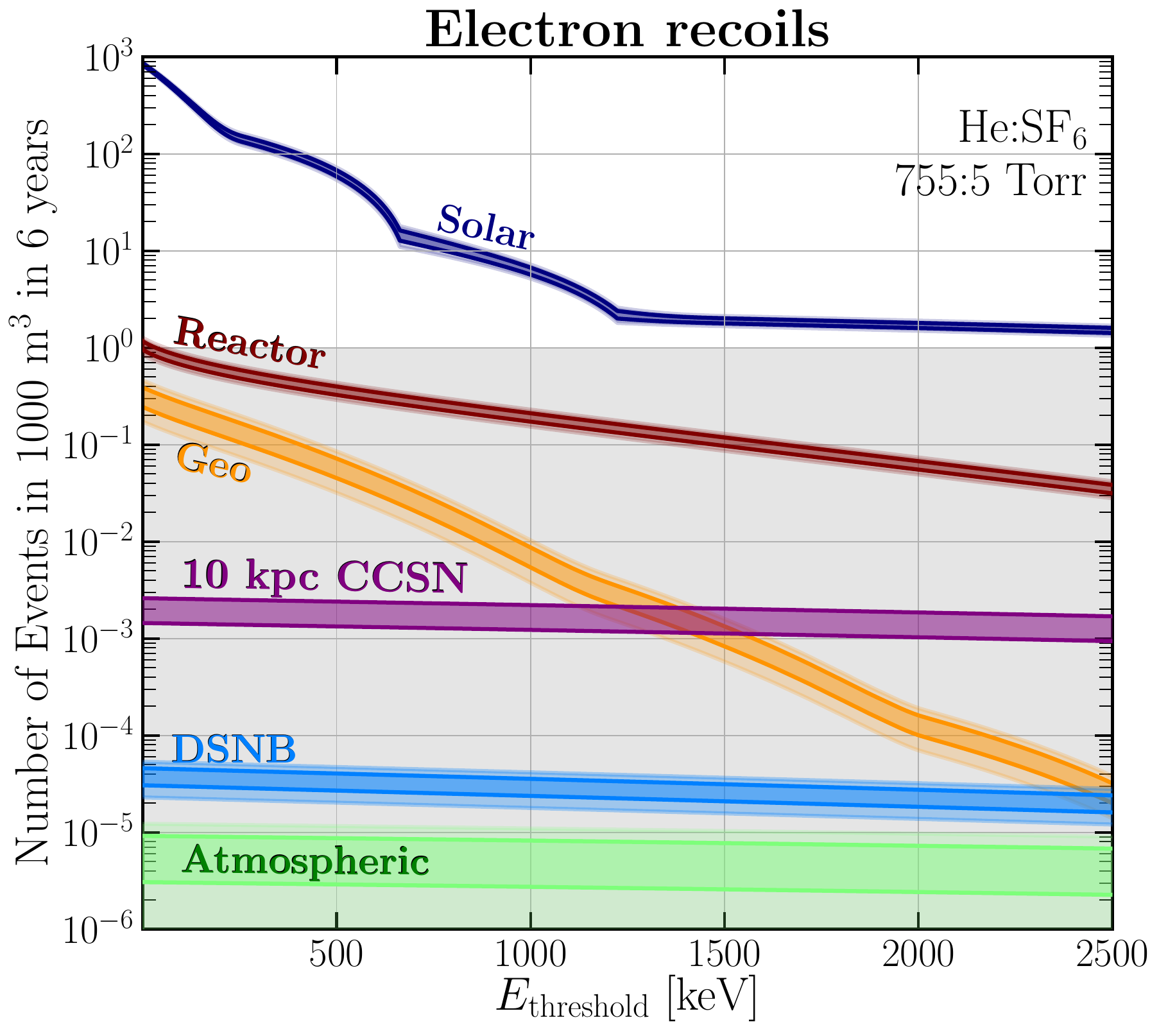}
\caption{Number of neutrino-nucleus (left) and neutrino-electron (right) recoil events observed in a \Cygnus-1000 m$^3$ detector filled with atmospheric pressure \hesfsix at a 755:5 Torr ratio (the event rates are summed over each target nuclei). We calculate the expected number of observed events by integrating the event rate for each background component above a lower energy threshold $E_{\rm threshold}$. The background components are shown as darker and lighter shaded regions indicating the 1 and 2$\sigma$ uncertainties from the predicted flux. For comparison we also show the nuclear recoil event rate expected from a $m_\chi = 9$\,\gevcc{} WIMP with a SI WIMP-proton cross section of $\sigma^{\rm SI}_p = 5\times10^{-45}$ cm$^2$ as a black line. For the reactor and geoneutrinos, we assume the entire 1000 m$^3$ is located at Boulby, UK. The purple region indicates the range of expected numbers of events from the neutrino bursts from 11--27 $M_\odot$ core-collapse supernovae located 10 kpc away from Earth. To add further clarity we shade in gray parts of the plot which give fewer than one event in this exposure. In the left panel we also show as dashed lines the 0.25 and 8 keV$_{\rm r}$ best-case and worst-case thresholds respectively.}\label{fig:NuRates}
\end{figure*}
Figure \ref{fig:limits} shows a selection of constraints from direct detection experiments along with our headline result: the WIMP reach of \Cygnus. Constraints exist for WIMPs with masses larger than $\sim 1$~\gevcc{} and SI cross sections larger than $\sim 10^{-46}$~cm$^2$. Underneath these limits lies the neutrino floor, below which WIMP models are rendered practically unidentifiable due to the saturation of their signal by the background from \cevns (to be discussed in Sec.~\ref{sec:floor}). 

We define the neutrino floor following a similar procedure to Ref.~\cite{Billard:2013qya}, but with a few additional steps. All neutrino floors combine two non-directional limits, at low and high masses. The low mass limit uses a  $\lesssim$~0.1 keV threshold and shows the impact of solar neutrinos, and the high mass limit uses a threshold just above where solar neutrinos become subdominant ($\sim 10$~keV depending on the target), and reflects mainly atmospheric neutrinos. We then vary the exposure, typically in the range 1 ton-year to 10$^4$ ton-year, and define the neutrino floor to be at the point when the rate of sensitivity improvement is the slowest. This ``slowing'' of the rate of sensitivity improvement with increasing detector mass is precisely the effect that \Cygnus aims to circumvent. The resulting floors typically correspond to limits which observe $\mathcal{O}(100 -1000)$ neutrino events. This is similar to Ref.~\cite{Billard:2013qya} who simply fixed an observed number of 500 for the low and high mass limits. However we have added this additional step to account for more recent improvements in the neutrino flux uncertainties, and to better reflect the cross sections at which the WIMP signal is most greatly impacted by the neutrino background. 

The limits we display are from the following experiments: CRESST-II~\cite{Angloher:2015ewa}, CDMSLite~\cite{Agnese:2017jvy}, COSINE-100~\cite{Adhikari:2018ljm}, EDELWEISS-III~\cite{Hehn:2016nll}, PICO-2L~\cite{Amole:2015lsj}, PICO-60~\cite{Amole:2015pla}, DarkSide-50~\cite{Agnes:2018oej}, PandaX~\cite{Tan:2016zwf}, XENON1T~\cite{Aprile:2017iyp}, PICASSO~\cite{Archambault:2012pm}, KIMS~\cite{Kim:2012rza} and COUPP~\cite{Behnke:2008zza}. For comparison, in the SD case we also display two competitive (but model dependent) limits from neutrino telescopes SK~\cite{Choi:2015ara} and IceCube~\cite{Aartsen:2016zhm}. These limits are based on searches for the annihilation of WIMPs captured by the Sun. In the SI case the closed detection regions correspond to the disputed DAMA/LIBRA~\cite{Bernabei:2013xsa} annual modulation signal. 

Also in Fig.~\ref{fig:limits} we display the final result of this paper: the potential sensitivities of \Cygnus. The gas mixture is assumed to be \hesfsix at 755 and 5~Torr respectively. This means that the relevant target nuclei are $^{19}$F and $^4$He for the SI search, and just $^{19}$F for the SD search\footnote{Fluorine is an attractive target for SD-proton searches since its most common natural isotope possesses a relatively high value of $\langle S_p \rangle$.}. We have assumed a detector with angular resolution, head/tail recognition efficiency, energy resolution, and charge detection efficiency derived for a strip-based readout (see Sec.~\ref{sec:technology_comparison}). The shaded regions indicate limits when varying the hard lower cut imposed on recoil energy. These range from 8 keV$_{\rm r}$ (above which the electron background can be definitively rejected by factors greater than $10^3$--$10^4$), down to 0.25 keV$_{\rm r}$ (the smallest physically possible threshold as it corresponds to the detection of a single electron). We emphasize that 8 keV is a worst-case scenario. As we will show, we expect electron rejection to be possible down to much smaller recoil energies. We show several possible volumes, including the benchmark 1000\,m$^3$ as used in our background study (see Sec.~\ref{sec:backgrounds}) and a further-future 100k m$^3$. The former would begin to break into the neutrino floor at low masses whereas the latter would allow us to further characterize the neutrino background and study in greater detail a possible WIMP signal if detected. Even an earlier-stage 10 m$^3$-scale experiment would already be able to set the most sensitive limits on SD-proton WIMP-nucleus interaction, thanks to the large number of fluorine nuclei. Beneath the largest 100k~m$^3$ cases we also show a hypothetical `search mode' limit. This corresponds to a mode in which the TPC contains entirely SF$_6$ at two atmospheres of pressure. This mode would have no directional sensitivity due to the high gas density, however the potential exclusion limits at high masses would greatly improve. The derivation of all the limits displayed here is the subject of the remaining sections of this paper.

\subsubsection{WIMP detection below the neutrino floor}\label{sec:floor}
\begin{figure*}[hbt]
\includegraphics[width=0.95\textwidth]{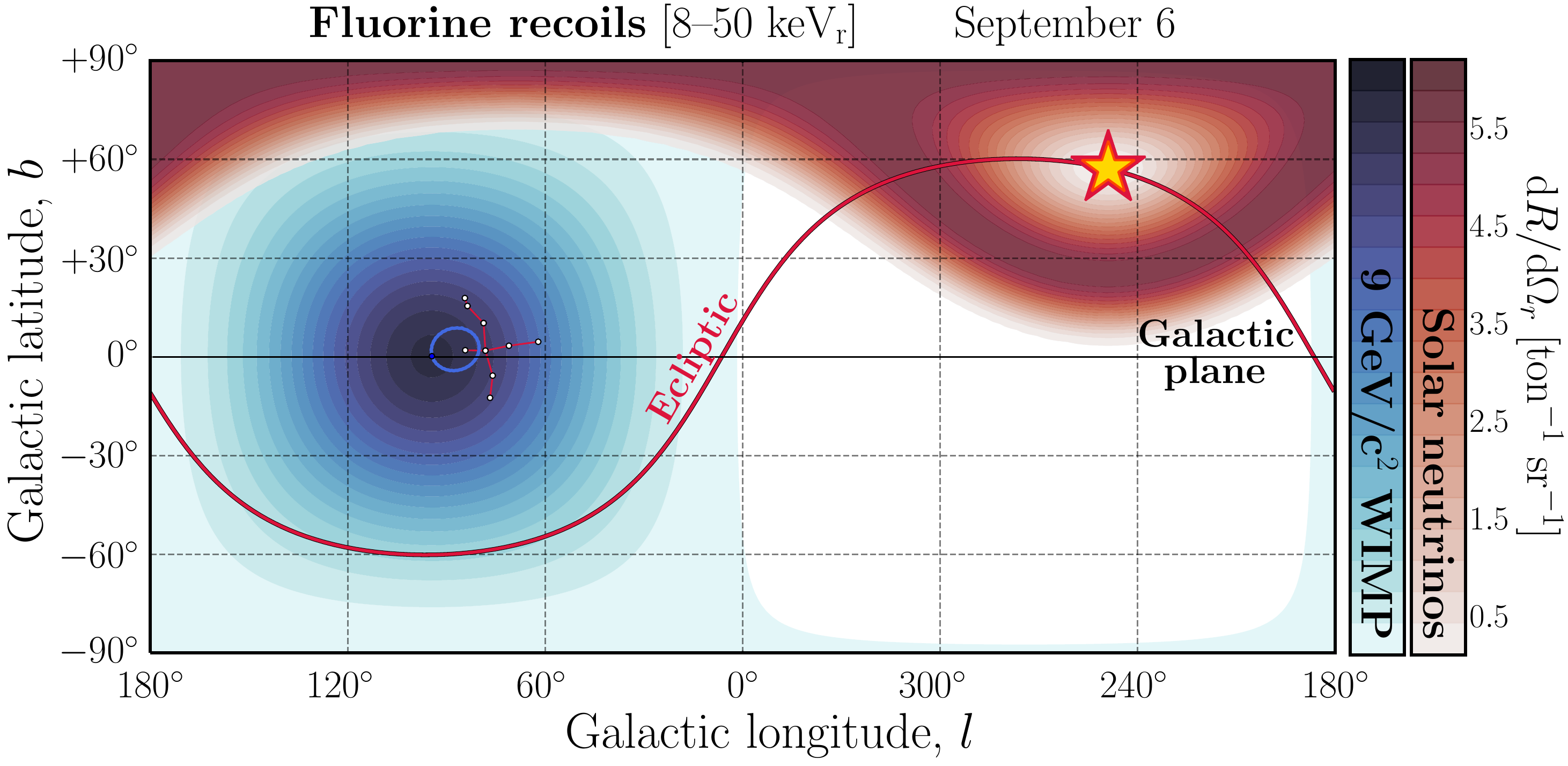}
\caption{Total angular distribution across the sky of WIMP-induced (blue contours) and neutrino-induced (red contours) fluorine recoils on September 6. We show the distributions in galactic coordinates $(l,b)$ where the line for $b=0$ corresponds to the galactic plane. Both distributions have been integrated over recoil energy between 8 and 50 keV$_{\rm r}$. We choose a WIMP mass of 9\,\gevcc{} and a SI cross section of $5\times 10^{-45}$~cm$^2$ so that its signal is of a similar size to that of the $^8$B neutrinos. For reference we also show the ecliptic in red: the path along which the Sun, shown by a star, moves over the year. Towards the center of the WIMP recoil distribution we also show the stars of the Cygnus constellation. The blue line which encircles the star Deneb shows the variation in the peak direction of the DM wind over the year.}\label{fig:mollweide}
\end{figure*}
It was anticipated in early work on direct DM detection that large detectors would eventually become sensitive to \cevns~\cite{Cabrera:1984rr}. For the keV nuclear recoil energies observed in direct detection experiments solar, diffuse supernovae and atmospheric neutrinos all constitute a significant background for detector exposures beyond the ton-year scale~\cite{Monroe:2007xp,Strigari:2009bq,Gutlein:2010tq}. Because neutrinos are impossible to shield against, they represent the ultimate background for the direct detection of WIMPs. Moreover, because the nuclear recoil energy spectra induced by \cevns mimic the spectra for WIMPs of certain masses, the discovery of these characteristic masses is limited due to the sizeable systematic uncertainty on the expected neutrino flux. We give more details on \cevns{} and the various neutrino backgrounds in Sec.~\ref{sec:neutrinos}. 

The limiting WIMP-nucleon cross section below which experimental sensitivities are impacted by the neutrino background is known as the ``neutrino floor''~\cite{Billard:2013qya}. The shape of the neutrino floor is dependent on the flux of each neutrino background component as well as, importantly, the uncertainty on this flux. The relevant neutrino backgrounds for WIMP searches using nuclear recoils are summarised in the left-hand panel of Fig.~\ref{fig:NuRates}. The most notable and threatening feature of the neutrino floor is the shoulder below $\sim$10\,\gevcc{} arising from the large flux and low energies of solar neutrinos\footnote{This shoulder extends down to very low WIMP masses, $< 1$\,\gevcc{}, where the lowest energy solar neutrino sources from $pp$ and CNO reactions become important. These neutrinos are important for DM searches which rely on electronic recoils~\cite{Essig:2018tss}, but are probably out of reach for nuclear recoil analyses for the foreseeable future.} (see Sec.~\ref{sec:SolarNu}). The most important of these are the neutrinos originating from $^8$B decay. In a fluorine experiment the nuclear recoil signal due to a 9\,\gevcc{} WIMP with an SI cross section around $5 \times 10^{-45}$~cm$^2$ is well matched by recoils from $^8$B neutrinos. Towards slightly larger masses (10--30\,\gevcc{}) the neutrino floor is set by the diffuse supernova neutrino background (DSNB): the cumulative emission of neutrinos from a cosmological history of supernovae. The expected flux of the DSNB is extremely low ($\sim 80~{\rm cm}^2~{\rm s}^{-1}$~\cite{Beacom:2010kk}) so the neutrino floor at these intermediate masses falls by several orders of magnitude in cross section. Towards masses beyond 100\,\gevcc{} the neutrino floor is induced by the low-energy tail of atmospheric neutrinos from cosmic ray interactions in the upper atmosphere. Atmospheric neutrinos (see Sec.~\ref{sec:atmonu}) are the only significant background contributing neutrino energies above 100 MeV. The low-energy tail of atmospheric neutrinos is difficult to both measure and theoretically predict~\cite{Battistoni:2005pd} so currently has uncertainties of around 20\%~\cite{Honda:2011nf}.

The shape of the neutrino floor is different depending on the target nucleus, because the \cevns energy spectrum also depends upon $m_A$. In this article, unless otherwise specified, we use the definition of the neutrino floor discussed in, for example, Refs.~\cite{OHare:2016pjy,OHare:2020lva}. As opposed to a limit which corresponds to a fixed expected number of neutrino events, we instead show a slightly lower floor which corresponds to the cross sections at which the rate of change in a discovery limit scales the slowest with increasing exposure. This allows us to account for the effect of the neutrino flux uncertainty and properly display which WIMP masses are most severely impacted by each background. Approximately though, the floor can be interpreted as a limit below which an experiment would observe more than $\sim$100 neutrino events.


Given that the next generation of ton-scale experiments are expected to become sensitive to \cevns, it is pertinent to search for alternative and more powerful methods of subtracting the background. The most basic approach to alleviate the background is to exploit the complementarity between target nuclei of differing masses and nuclear content. For the SI neutrino floor it has been shown that this approach only leads to a marginal improvement in discovery limits, but in the case of SD interactions the differences in nuclear spin contents make complementarity a more viable strategy~\cite{Ruppin:2014bra}. It was also shown that the use of event time information also allows the low-mass neutrino floor to be overcome with slightly lower statistics~\cite{Davis:2014ama,OHare:2016pjy}. This exploits both the annual modulation of the DM signal due to the relative motion of the Earth and the Sun, as well as the annual modulation in the solar neutrino flux due to the eccentricity of the Earth's orbit.

Directionality presents the most attractive prospect for circumventing the neutrino floor because the unique angular signatures of both DM and solar neutrinos allows optimum discrimination between signal and background~\cite{Grothaus:2014hja,O'Hare:2015mda}. This has also been explored in non-gas TPC directional experiments, such as with nuclear emulsions~\cite{Agafonova:2017ajg}, spin-polarized helium-3~\cite{Franarin:2016ppr} and electron-hole pair excitations in semiconductors~\cite{Kadribasic:2017obi}. The consensus is that: \emph{in a directional experiment there should be effectively no neutrino floor, provided that directionality is well-measured}.

We display the angular distribution of the DM and neutrino-nucleus recoils in Fig.~\ref{fig:mollweide}. The crucial factor that enables their discrimination is that over the course of the year the Sun does not pass through the constellation of Cygnus. The angular distance between Cygnus and the Sun undergoes a sinusoidal modulation which peaks in September at around $120^\circ$ and is a minimum during March at around $ 60^\circ$. Because solar neutrino recoils can only point with angles less than $90^\circ$ from the solar direction, this implies that over long periods during a year there are large patches of sky where the event rate of solar neutrinos is zero (before accounting for the finite angular resolution of the detector). The separation is large enough that it is possible to discriminate the two signals even if the recoil vectors cannot be oriented in galactic coordinates as shown here. This would be the case for experiments in which recoil vectors were only measured after being projected on to a plane~\cite{O'Hare:2015mda}, or if timing information was unavailable~\cite{OHare:2017rag}. 


\subsubsection{WIMP astrophysics}
\label{sec:astro}
Predicting WIMP signals requires astrophysical input in the form of the DM velocity distribution, $f(\mathbf{v})$, as well as the local density of DM $\rho_0$. While the former is not known concretely, the latter can be determined with astronomical data. Decades of attempts to constrain this value have begun to settle towards $\rho \approx 0.5$\,GeV cm$^{-3}$ (see, \eg{}~Refs.~\cite{Bovy:2012tw,Smith:2012,Bienayme:2014,Pi14,Sivertsson:2017rkp} and Ref.~\cite{Read:2014qva} for a review on methods). However the standard value used by experimental collaborations is $0.3$ GeV cm$^{-3}$. Upcoming analyses with the extremely high and precise statistics of the astrometric survey {\it Gaia}~\cite{GaiaDR2} are expected to lead to even more tightly constrained values in the next few years. Since the local dark matter density only amounts to a multiplicative factor which can be absorbed into the (also unknown) DM cross section\footnote{Although, see Ref.~\cite{Kavanagh:2020cvn} for a specific case where this is not true.}, its precise value is of low importance before a detection is made. 

On the other hand the great deal of uncertainty surrounding the velocity distribution of the DM is generally more important to understand. The benchmark model assumed since the very first direct detection experiments were conducted is the Standard Halo Model (SHM), in which the Milky Way's DM forms an isothermal sphere. The motivation is that it is the simplest model that gives rise to flat rotation curves. The SHM has a Gaussian distribution for $f({\bf v})$ (or Maxwellian distribution for $f(v)$),
\begin{equation}\label{eq:shm}
f(\mathbf{v}) = \frac{1}{(2\pi \sigma_v^2)^{3/2}N_\mathrm{esc}} \, \exp \left( - \frac{|\mathbf{v}|^2}{2\sigma_v^2}\right) \, \Theta (v_{\rm esc} - |\mathbf{v}|) \, .
\end{equation}
Under the isothermal SHM, the dispersion is related to the local rotation speed of circular orbits: $\sigma_v = v_0/\sqrt{2}$. The benchmark used for this speed is the now similarly out-of-date value of $v_0~\sim~220$\,\kms{}. A more recent analysis indicates a value of $v_0 = 235$\,\kms{}~\cite{Evans:2018bqy}. As is convention, the velocity distribution has been truncated at the escape speed $v_{\rm esc}$, with the constant $N_{\rm esc}$ used to renormalize the distribution after this truncation.  Experimental analyses typically have assumed $v_{\rm esc} = 544$ or $533$\,\kms{}, the latter being the more recent RAVE value~\cite{Piffl:2013mla}. Again with {\it Gaia} this is undergoing revision, however newly found high speed substructure is introducing additional complexity to its determination~\cite{Deason19}. The escape speed in effect controls the highest energy WIMP observable, but due to the exponential suppression this has a marginal impact on the event rate.

Astrophysical uncertainties in the form of $f(\mathbf{v})$ impact the reliability of signal modeling and hence feed into the measurements of DM particle properties~\cite{Peter:2011eu,Kavanagh:2013wba} and the calculation of exclusion limits~\cite{McCabe:2010zh,OHare:2016pjy}. Indeed much effort has been spent in devising methods to integrate out this uncertainty so that accurate limits can be made. See for example the extensive literature on ``halo-independent'' methods for the calculation of exclusion limits in Refs.~\cite{Fox:2014kua,DelNobile:2013cva,Feldstein:2014gza,Gelmini:2016pei,Ibarra:2017mzt,Fowlie:2017ufs,Gelmini:2017aqe} and many others. Much of this interest has been driven by previous claims that the anomalous DAMA/LIBRA result could be explained by particular choices for the speed distribution. 

Directional experiments, however, will offer a novel resolution to the uncertainty on $f(\mathbf{v})$ that is not present in conventional approaches. In addition to having a unique smoking gun signal that is difficult, or even impossible, to mimic with backgrounds, it has been shown that the prospects for directional detectors to \emph{measure} the DM velocity structure greatly exceed that of an equal-standing {\it non}-directional detector~\cite{Lee:2012pf,Kavanagh:2015aqa,Kavanagh:2016xfi}. This is primarily because recoil energy information alone cannot be used to access the full three-dimensional form of $f({\bf v})$ and is instead only sensitive to the one-dimensional speed distribution, $f(v)$. Since our study is largely comparative in nature we adopt the SHM and the out of date astrophysical parameter values for consistency with past studies and limits. However in the future as more ton-scale detectors begin taking data, it has been advised that the SHM be updated to include recent refinements. A summary of these recommendations can be found in Ref.~\cite{Evans:2018bqy}.

It should be emphasized additionally that the structure of the local halo of the Milky Way is itself also of great interest. The measurement of anisotropies in the velocity distribution may provide insights into the archaeology of the Milky Way's formation, as well as the fundamental properties of DM. In particular, directional detectors are well suited to detect kinematically localized substructures such as DM streams \cite{OHare:2014nxd}. Less prominent velocity substructures such as debris flow~\cite{Kuhlen:2012fz}, asymmetry in the velocity ellipsoid of the Milky Way~\cite{Evans:2018bqy}, and the possible influence of the Large Magellanic Cloud~\cite{Besla:2019xbx}, will all require many more events to detect, hence very large detectors will be essential. This is also true for very low-mass streams from smaller DM substructures or any other low-level structures that are likely to have undetectable levels of influence on luminous matter; an opportunity to better understand the level of phase space clumpiness and non-Gaussianity of the Milky Way DM halo is therefore a key potential advantage of directional detection.

While it is useful to remain speculative about the kinds of structures one could hope to see in a future experiment, in the advent of {\it Gaia}, our concrete knowledge about our local distribution is improving. For instance the expanding catalog of nearby populations of clumps and streams as well as the overall kinematic structure of the stellar halo is providing us greater insight into the formation of the Milky Way halo and the construction of its gravitational potential. In fact we already know that at a large chunk of the local DM halo is likely to be part of a highly radially anisotropic feature, variously called the Gaia Sausage or Gaia-Enceladus~\cite{Kr18,My18,Be18,Ma18,Evans:2018bqy,Bozorgnia:2019mjk,Mackereth:2019,Yu19,Naidu2020}. Some of the most interesting prospects for directional detectors are the tidal streams intersecting our galactic position~\cite{Myeong:Preprint,Myeong:2017skt}. These have been discovered recently thanks to {\it Gaia}, some prominent examples are the local streams S1~\cite{OHare:2018trr} and S2~\cite{OHare:2019qxc,Aguado:2020kki}, see also the examples studied in Ref.~\cite{Adhikari:2020gxw}. Streams will be subdominant contributions to the local density $\rho_0$, hence will be secondary signals. However their unique characteristics relative to the bulk of the halo DM are almost entirely washed out in their eventual nuclear recoil signals--unless one has directional information~\cite{OHare:2018trr,Evans:2018bqy,OHare:2019qxc}. The ability to resolve these substructures is another major advantage of such a detector~\cite{OHare:2014nxd}.

\subsubsection{Particle models and directionality}
\label{sec:models}

As well as simply detecting DM, we also require that a future large-scale experiment be able to uncover properties of the particle itself. This is particularly challenging as there are many competing models that may be degenerate with respect to the signals they produce in the usual direct detection schemes. Various classes of particle models give rise to unique directional signals that would go undetected in a conventional experiment. We outline a few of these here.

Inelastic DM (IDM) models are those in which DM can have a lower or higher energy excited state to which it can down- or up-scatter via a nuclear recoil event. IDM models were proposed to reconcile the DAMA annual modulation signal~\cite{TuckerSmith:2001hy}. The availability of excited states and the suppression of elastic scattering means that heavier nuclei are favored, the low energy recoil spectrum is modified, and the annual modulation is enhanced. It has also been shown that IDM models can give rise to enhanced signal discrimination power in directional detectors~\cite{Finkbeiner:2009ug}. This is because the recoils are more focused in the forward direction, since slower WIMPs cannot scatter with enough energy to induce an excited state. 

Directional detectors can also disentangle elastic and inelastic scattering events in DM models that allow for both~\cite{Lisanti:2009vy}, and if the detector is equipped with any photodetector technology then the luminous decay of the excited state may also be observed~\cite{Eby:2019mgs}. In this latter example, the sensitivity to this process is facilitated by the large volumes of gas-based detectors, as opposed to only their directional sensitivity.  This is also the case for proposals for super-heavy DM particles with masses extending up to the Planck scale and beyond~\cite{Bramante:2018qbc}.

Another possibility is if DM exists in the form of ``darkonium'' bound states composed of two or more particles (as is predicted in some configurations of asymmetric DM models). It has been shown that there may be angular signatures observable in directional experiments that can constrain the properties of these bound states~\cite{Laha:2015yoa}.

Several years ago a novel scheme to generalize the calculation of signals in direct detection experiments was developed. This uses a non-relativistic effective field theory description of the DM-nucleus interaction to posit a set of operators that describe general processes beyond the simple spin-independent and spin-dependent. The basic operators include all Hermitian, Galilean, and rotation-invariant interactions constructed out of the low energy degrees of freedom involved in a WIMP-nucleus interaction~\cite{Fan:2010gt}. Certain examples, those which are dependent on the DM particle's transverse velocity $\mathbf{v}^\perp = \mathbf{v} + \mathbf{q}/2\mu_{\rchi A}$), give rise to unique ring-like angular signatures~\cite{Kavanagh:2015jma,Catena:2015vpa}. This means that directional detectors would be more powerful than conventional experiments in distinguishing between these particular operators. An important consequence of these features is that it enables an experiment to distinguish spin-0 DM particles from spin-1/2 or 1~\cite{Catena:2017wzu}. The information which permits this turns out to be found in the angular dependence of certain effective field theory operators. However, many non-standard operators tend to be suppressed, meaning event rates are inherently low. Large scale detectors will therefore be required for studying the complete phenomenology of DM.

\subsubsection{Axions}
\label{sec:axions}

Direct detection experiments searching for WIMPs are not limited to a single class of DM candidate. Another very popular class of candidate is the axion and its generalization, the axion-like particle (ALP). The motivation for axions originates in the dynamical solution of Peccei and Quinn~\cite{Peccei:1977hh} (PQ) to the strong-CP problem of quantum chromodynamics (QCD), see \eg{}~Ref.~\cite{DiLuzio:2020wdo} for a recent review. The ALP is inspired phenomenologically by the QCD axion but could come from a variety of theoretical origins when spontaneously broken PQ-like symmetries are embedded in higher energy theories. This is the case most notably in string theory where both the QCD axion and a slew of phenomenologically similar particles are predicted~\cite{Arvanitaki:2009fg}. The masses of axions and ALPs can span many orders of magnitude but their extremely weak couplings to the Standard Model make them attractive DM candidates. Axions produced non-thermally in the early Universe via vacuum misalignment~\cite{Abbott:1982af,Dine:1982ah,Preskill:1982cy}, decaying topological defects~\cite{Chang:1998tb,Kawasaki:2014sqa} or in the form of axion stars~\cite{Liddle:1993ha} or miniclusters~\cite{Hogan:1988mp,Kolb:1994fi,Vaquero:2018tib} have been shown to be able to match the required properties and cosmological abundance of cold DM (see \eg{}~Ref.~\cite{Marsh:2015xka}). Axions and ALPs are by construction coupled to the Standard Model through quark loops which gives rise to a number of potentially observable interactions: for example, axion-photon conversion inside magnetic fields, absorption by atomic electrons (the axioelectric effect) and spin-precession of nuclei. In the case of the QCD axion the strength of the coupling and its linear relationship to the axion mass is prescribed by theory, but for the generalized ALP, the coupling may take any value. See Ref.~\cite{Irastorza:2018dyq} for a recent review of experimental searches for axions.

Existing WIMP direct detection experiments such as LUX~\cite{Akerib:2017uem}, XENON~\cite{Aprile:2014eoa,Aprile:2019xxb,Aprile:2020tmw}, EDELWEISS~\cite{Armengaud:2013rta}, CDMS~\cite{Ahmed:2009ht} and PandaX~\cite{Fu:2017lfc} have already constrained axions and ALPs in the search for their interactions with electrons via the axioelectric effect~\cite{Derevianko:2010kz,Derbin:2012yk,Ljubicic:2004gt}.  The coupling most accessible to a WIMP experiment is the axion-electron coupling $g_{ae}$, as opposed to the photon coupling measured in the most mature axion experiments using resonant cavities such as ADMX~\cite{Asztalos:2009yp} and CAST~\cite{Zioutas:2004hi} (although in QCD axion models these couplings are related). Note that most of these analyses can also be recast as constraints on vector bosonic dark matter particles, the dark photon being the most notable example~\cite{An:2014twa}.

ALPs may be searched for as both a DM candidate as well as solely a modification to the Standard Model. If ALPs with masses in the range $m_a \sim 1-40$ keV comprise a significant fraction of the local DM density, then they should stream into a detector and induce electron emission from target atoms with a sharp spectrum located at $m_a$. Since the spectral width of this signal would be well below the energy resolution of any detector, astrophysical uncertainties might only be resolved with the angular spectrum. \Cygnus may also be particularly advantageous in this regard because of the excellent electron/nuclear recoil discrimination that could be achievable at low energies. It should be mentioned that the keV-mass QCD axion couplings are already ruled out, so low-background recoil experiments would only be able to detect DM in the form of an ALP and not an axion. 

On the other hand, these experiments may in the future be able to see the QCD axion, since it is expected that they should be produced in the Sun with $\sim$keV energies, and their precise incoming flux and spectrum is well understood~\cite{Redondo:2013wwa}.\footnote{Recently an excess of electronic recoil events below 7~keV was reported by XENON1T~\cite{Aprile:2020tmw} with a similar spectrum to the solar axion flux. However the size of the excess would require couplings in violation of astrophysical bounds~\cite{Capozzi:2020cbu,Athron:2020maw,Croon:2020ehi}} As is the case with solar axion telescopes such as CAST, if the ALP mass is much less than a keV~c$^{-2}$ the axions are produced relativistically and the signal is dominated by the energy of the solar emission. This means an experiment is consistently sensitive over a large range of arbitrarily small masses. Even in the event of the detection of a DM or solar ALP, a large directionally-sensitive experiment could be novel because, as with WIMPs, there may be unique angular signatures. While directional sensitivity is desirable with regards to dedicated axion experiments~\cite{Garcon:2019inh,Knirck:2018knd}, in the case of the axioelectric effect this is yet to be studied in detail.

\subsection{Neutrinos}
\label{sec:neutrinos}

\subsubsection{Coherent neutrino-nucleus scattering}
\label{sec:neutrinoscattering}
\cevns was predicted over 40 years ago with the realization of the neutral weak currents~\cite{Freedman:1973yd}. This Standard-Model process went unobserved for many years due to daunting detection requirements: $\sim$keV nuclear recoil thresholds, kilogram to ton-scale target masses, and low backgrounds.  Recently COHERENT has made the first measurements of the \cevns{} cross section in agreement with the Standard Model~\cite{Akimov:2017ade,Akimov:2020czh,Akimov:2020pdx}. Due to the small weak charge of the proton, the coherence results in an enhanced neutrino-nucleon cross section that is approximately proportional to the square of the number of neutrons in the nucleus. A few years after the \cevns{} prediction, and ironically before the conception of the first DM direct detection experiments, the possibility of using this enhanced process to develop a ``neutrino observatory'' was put forward~\cite{Drukier:1983gj}. A cornucopia of physics searches was envisioned using neutrinos from stopped-pion beams, reactor neutrinos, supernovae, solar neutrinos, and even neutrinos of a geological origin. See \eg{}~Ref.~\cite{Vitagliano:2019yzm}~for a summary of natural sources of neutrino.

Shortly thereafter, the first generation of DM experiments began to search for the scattering of WIMPs, where the signature was a low-energy nuclear recoil. Today the irony lies in the fact that the unshieldable recoils that result from \cevns{} will soon be a source of background for the next generation of DM direct detection experiments. But an experiment that can successfully separate and identify these neutrino events can not only proceed past the neutrino floor, but can also realize the long-awaited vision of a ``neutrino observatory''. A detector with directional sensitivity has the potential to do just that.

In \cevns, coherence is only satisfied when the initial and final states of the nucleus are identical, limiting this enhancement to neutral current scattering. The coherence condition, where the neutrino scatters off all nucleons in a nucleus in phase, is also only maintained when the wavelength of the momentum transfer is larger than that size of the target nucleus. A high level of coherence across all recoil energies is only guaranteed for low energy neutrinos: less than tens of MeV, depending on the mass of the target nucleus. The Standard Model total cross section for the process can be approximated (neglecting subdominant axial-vector terms that arise from unpaired nucleons) as
\begin{equation}
\sigma = \frac{G_F^2}{4\pi}\left[A+Z(4\sin^2\theta_W-2)\right]^2E_{\nu}^2 F^2(E_r) \, ,
\end{equation}
where $G_F$ is the Fermi constant, $\theta_W$ is the Weinberg angle and $E_{\nu}$ is the energy of the incoming neutrino~\cite{Freedman:1973yd}. It is evident that the cross section increases with the square of the energy of the neutrinos; however, while the form-factor condition, which comes in as $|F(E_r)|^2$, is easily satisfied for solar neutrinos, the total cross section begins to suffer from decoherence for supernova and atmospheric neutrinos. As can be seen in Fig.~\ref{fig:NuRates}, a detector with an energy threshold of zero can expect to see several hundred to a few thousand recoils from solar neutrinos per ton-year of exposure, depending on the mass of the target nucleus~\cite{Drukier:1983gj}.

The differential cross section with recoil energy can be approximated as~\cite{Freedman:1973yd}:
\begin{equation}
\frac{\textrm{d}\sigma}{\textrm{d}E_r} = \frac{G_F^2}{8\pi}\left[Z(4\sin^2\theta_W -1)+N\right]^2m_A\left(2 - \frac{E_r m_A}{E_{\nu}^2}\right) \, .
\end{equation}
Assuming a $^{19}$F target, for example, and a 5 (10) keV threshold for observing nuclear recoils. This results in an expectation of $\sim$90 (15) background recoils per ton-year, from solar neutrinos alone~\cite{1742-6596-136-2-022037}.

\subsubsection{Solar neutrinos}
\label{sec:SolarNu}
On Earth, the most prominent source of neutrinos is our Sun with a total flux at Earth of $6.5\times10^{11}$~cm$^{-2}$~s$^{-1}$~\cite{Robertson:2012ib}. Due to the eccentricity of the Earth's orbit, the Earth-Sun distance has an annual variation leading to a modulation in the solar neutrino flux $\Phi$,
\begin{eqnarray}\label{eq:solarneutrinoflux_directional}
  \frac{\textrm{d}^2 \Phi(t)}{\textrm{d}E_\nu \textrm{d}\Omega_\nu}  &=&  \frac{\textrm{d} \Phi}{\textrm{d} E_\nu} \,\left[ 1 + 2 e \cos\left(\frac{2\pi(t- t_\nu)}{T_\nu}\right) \right]  \nonumber \\
 &\times& \delta\left(\hat{{\bf r}}_\nu+\hat{{\bf r}}_\odot(t)\right) \, ,
\end{eqnarray}
where $t$ is the time from January 1st, $e = 0.016722$ is the eccentricity of the Earth's orbit, $t_{\nu} = 3$ days is the time at which the Earth-Sun distance is minimum, $T_{\nu} = 1$ year, $\hat{{\bf r}}_{\nu}$ is a unit vector in $\textrm{d}\Omega_\nu$, and $\hat{{\bf r}}_\odot(t)$ is a unit vector pointing towards the Sun. The directional event rate is found by convolving this directional flux, with the directional cross section for \cevns{}. The cross section with respect to the lab-frame recoil angle $\theta$ can be written as~\cite{Freedman:1973yd,Drukier:1983gj}
\begin{equation}
\frac{\textrm{d}\sigma}{\textrm{d}(\cos \theta)} = \frac{G_F^2}{8\pi}\left[Z(4\sin^2\theta_W -1)+N\right]^2E_\nu^2(1+\cos\theta) \, .
\end{equation}
The resulting recoils are thus biased to the forward direction, away from the location of the Sun.

The spectra of solar neutrinos $\textrm{d}\Phi/\textrm{d}E_\nu$ come in various distinct forms depending on the nuclear fusion reaction involved in their production. Neutrinos from the initial proton-proton fusion reaction, $pp$, make up 86\% of the solar emission~\cite{Bellini:2014uqa}. Despite the huge flux of $pp$ neutrinos, they yield nuclear recoils well below the threshold of any direct detection experiment; however they would be the dominant source of electron recoils. Secondary fusion of $p+e^-+p$ and $^3 $He$~+~p$ produce neutrinos, labeled $pep$ and $hep$, extend to energies beyond $pp$ neutrinos but with lower flux. There are also two monoenergetic lines associated with $^7$Be electron capture with $E_\nu= 384.3$\,keV and 861.3\,keV. The latter of these is principally responsible for limiting WIMP discovery for $m_\chi<1$\,\gevcc{}~\cite{Strigari:2016ztv}. At higher energies there are neutrinos due to the decay of $^8$B which extend up to $E_\nu\sim10$\,MeV and within the reach of nuclear recoil WIMP searches, as already discussed. Finally, extending to relatively high energies but with much weaker fluxes, are neutrinos arising from the carbon-nitrogen-oxygen (CNO) cycle labeled by the decay from which they originate: $^{13}$N, $^{15}$O and $^{17}$F.

The theoretical uncertainties on the solar neutrino fluxes range from 1\% ($pp$ flux) to 14\% ($^8$B flux). For all except $^8$B, the theoretical uncertainty is smaller than the measurement uncertainty. The theoretical uncertainty originates largely from the uncertainty in the solar metallicity, and in order to establish a self-consistent set of solar neutrino fluxes one must assume a metallicity model. The Standard Solar models (SSMs) of Grevesse \& Sauval~\cite{Grevesse:1998bj} are generally split into two categories: ``high-Z'' and ``low-Z,'' based on the assumed solar metallicity. Both models have historically disagreed with some set of observables such as neutrino data, helioseismology, or surface helium abundance~\cite{Villante:2013mba}. More recent generations of SSMs~\cite{Vinyoles:2016djt} have a mild preference towards a high-Z configuration, though neither are free from some level of disagreement with the various solar observables. DM detection experiments may shed further light on the solar metallicity issue (see \eg{}~Refs.~\cite{Billard:2014yka,Strigari:2016ztv,Cerdeno:2016sfi}). The measurement of CNO neutrinos will be essential for this, and may be possible in future DM experiments~\cite{Cerdeno:2016sfi,Cerdeno:2017xxl}. Recently CNO neutrinos were measured for the very first time by Borexino~\cite{Agostini:2020mfq}, though currently with insufficient statistics to resolve the solar metallicity problem. The advantage of directional detection in performing these science goals is the vastly improved background rejection capabilities and the ability to reconstruct neutrino energies on an event-by-event basis.

In Fig.~\ref{fig:NuRates} we show the expected neutrino background for all components mentioned in this section, for both nuclear and electron recoils (although the latter do not enter into the main discussion of the paper). As in later examples we assume a 1000 m$^3$ gas TPC located at Boulby Underground Laboratory\footnote{The Boulby mine is the site of the DRIFT experiment, and is one of several candidate labs for future \Cygnus detectors, see Fig.~\ref{fig:map}.} with a \hesfsix gas mixture held at atmospheric pressure with a ratio of 755:5 Torr. Again, the justification for this choice of target, volume and site are the subject of the remaining sections of the paper. We show the expected number of events as a function of the lower limit of integration in recoil energy, \ie{}~a hard threshold but with detection efficiency and resolution ignored.

As we have discussed, solar neutrinos are expected---and indeed desired---to be the dominant background for \Cygnus. In the case of nuclear recoils, the majority of the event rate above reasonable $\mathcal{O}(1$--$10$ keV$_{\rm r}$) thresholds originates from $^8$B component of the flux.  One could expect up to 100 events in six years if the threshold can be lowered to 0.25 keV$_{\rm r}$.  Electron recoils are sensitive to much lower energy neutrinos, so all components of the solar flux contribute to the background leading to a very large event rate that persists even at high energies. If non-neutrino electron recoil backgrounds are able to be suppressed to low levels (see Sec.~\ref{sec:backgrounds}) then one would expect $pp$ and $^7$Be neutrinos to contribute significantly to the electron recoil signal in \Cygnus. In fact it is possible that these fluxes could be measured to even lower energies than in Borexino. We have shown the spectrum of neutrino-electron recoils for reference and future interest. 

The directionality of neutrino-electron recoils is not studied here, but will be crucial to understand if they are to be extracted from the large background of electron recoils from other sources. Nevertheless, the electron recoils at the energies shown in the right-hand panel of Fig.~\ref{fig:NuRates} will have very long tracks in a gas target, so their directions should be easily measurable~\cite{Seguinot:1992zu,Arpesella:1996uc}. A full investigation of neutrino-electron recoil signals in \Cygnus will be the subject of follow-up work.

\subsubsection{Supernovae}
\label{sec:supernovae}
Another potential signal of interest in directional DM detectors is the enormous burst of neutrinos from a core-collapse supernova (CCSN), which sheds the binding energy of the resulting compact remnant almost entirely in the form of neutrinos over a timescale of a few tens of seconds.  Such a collapse-induced burst is expected a few times per century in the Milky Way. The neutrinos in a supernova burst will have a few to a few tens of MeV of energy, and will include all flavors of neutrinos and antineutrinos with roughly equal luminosity~\cite{Janka:2006fh,Mirizzi:2015eza}.  

Dark matter detectors with low recoil energy thresholds are sensitive to supernova neutrino bursts via \cevns{}~\cite{Horowitz:2003cz}.  The order of magnitude is a handful of events per ton of detector material for a supernova at $\sim$10~kpc (just beyond the center of the galaxy, and close to the most likely distance to the supernova~\cite{Mirizzi:2006xx}). Observed numbers of events scale as the inverse square of the distance to the supernova. In Fig.~\ref{fig:NuRates} we showed the expected number of neutrino events in \Cygnus for such a typical CCSN. We source total neutrino luminosities and moments of the energy spectra from the 1d simulations of Ref.~\cite{SNarchive} and compute the neutrino spectra using the fitting formulae from Ref.~\cite{Tamborra:2012ac}. We show the range of expected event numbers for the explosions of stars with masses between 11 and 27 $M_\odot$ which covers most of the range of possible supernova-progenitor masses. For \Cygnus-1000~m$^3$ the typical galactic supernovae at 10 kpc would be just out of reach, however $>1$ nuclear or electron recoil events could be observed for supernovae closer than 3 or 0.5 kpc respectively. All of the neutrino burst events would be concentrated around a $\mathcal{O}(10\,{\rm s})$ period. For very close supernovae, at sub 100 pc distances, there may even be detectable pre-supernova neutrino events generated $\sim$10 hours prior to the explosion during the dying star's silicon burning phase~\cite{Raj:2019wpy}. 

A detection of a supernova explosion via \cevns{} would be valuable due to its sensitivity to the total, all-flavor burst flux. Most other detectors that are currently online are primarily sensitive to either the $\bar{\nu}_e$ (in water and scintillator detectors) or $\nu_e$ (in argon and lead detectors) components of the flux~\cite{Scholberg:2012id, Mirizzi:2015eza}. Furthermore, some neutrino spectral information and therefore properties of the supernova could be reconstructed from the measured recoil energy spectrum~\cite{Lang:2016zhv}.

The advantages of directionality for the detection of supernova burst neutrinos via \cevns{} are several: first, and most clearly, directional information about the source will be of value to observers in electromagnetic wavelengths and multimessenger channels who want to make prompt observations of the supernova event in real-time.  Currently, only detectors able to make directional measurements of elastic scattering on electrons have good pointing ability ~\cite{Beacom:1998fj,Tomas:2003xn} (and Super-Kamiokande is the only current instance)~\cite{Abe:2016waf}.  Even if there is no obviously bright supernova event (as may be the case for a failed supernova), directional information will be helpful to narrow down the possible progenitors.  Finally, if the direction to the supernova is known via astronomical measurements, then the neutrino direction information can be used to reconstruct neutrino energies on an event-by-event basis.

The diffuse supernova neutrino background (DSNB) is also of interest as an astrophysical target~\cite{Beacom:2010kk,Lunardini:2010ab, Barranco:2017lug,deGouvea:2020eqq}. A measurement of the DSNB flux would be an independent probe of cosmology, as well as representing an average of supernova bursts. Unfortunately, both the nuclear and electron recoil rates are expected to be immeasurably small in \Cygnus. This can be seen in light-blue curves in Fig.~\ref{fig:NuRates}; clearly an increase in detector size of several orders of magnitude from 1000~m$^3$ would be needed to realistically observe the signal. Even then, since the flux is expected to be very close to isotropic, isolating the DSNB would be difficult, and is therefore not a primary goal for \Cygnus. 

\subsubsection{Atmospheric neutrinos}
\label{sec:atmonu}
Cosmic ray interactions in the upper atmosphere have long been a reliable flux of 10 MeV--PeV neutrinos. Atmospheric neutrinos principally originate from the decays of $\pi$ and $K$ mesons in hadronic showers created by collisions between high-energy cosmic rays and air molecules. The emission of atmospheric electron and muon neutrinos and antineutrinos has been well-studied for energies above 100 MeV, and was important historically in the discovery of neutrino oscillations. The spectrum in the 10\,GeV -- \,TeV region is roughly a power law with $E_{\nu}^{-2.7}$. However at low energies, especially $\lesssim 100$~MeV, the geomagnetic field causes suppression in the flux~\cite{Battistoni:2005pd,Honda:2011nf}.  Current studies of atmospheric neutrinos are based on Monte Carlo simulations such as HKKM~\cite{Honda:2011nf}, Bartol~\cite{Barr:2004br} and FLUKA~\cite{Battistoni:2002ew}, which are informed by cosmic ray data and atmospheric nuclear interaction models. Recent observations by Super-Kamiokande~\cite{Richard:2015aua} and IceCube~\cite{Aartsen:2015xup} measure the spectrum between 100 MeV--10 TeV with uncertainties now comparable to the models, enabling tight constraints on sterile neutrinos~\cite{Abe:2014gda,TheIceCube:2016oqi} and non-standard interactions~\cite{Mitsuka:2011ty, Esmaili:2013fva, Salvado:2016uqu}. The limited availability of secondary muon data at low energies make the flux uncertainty for the 10\,MeV--1\,GeV neutrino tail much larger~\cite{Gutlein:2010tq}, but a nuclear recoil observatory would be most sensitive to atmospheric neutrinos in this very region. The nuclear recoil rate above these energies is very small due to the falling flux and nuclear form factor suppression. A future measurement of the low-energy flux by a nuclear recoil experiment would be beneficial for the improvement of atmospheric neutrino flux models. This is important as atmospheric neutrinos at these energies will be a background for future studies of the less well-understood diffuse supernova background. As with solar neutrinos, a key advantage of observation via nuclear recoils is the sensitivity to the three-flavor flux, improving the measurement of the total normalization.

In addition to spectral information, the angular distribution of atmospheric neutrinos is also of great interest. The flux is known to peak towards the horizon over the full energy range due to the longer flight path for primary cosmic rays through the atmosphere. The directionality of neutrinos is also controlled by the cutoff in the geomagnetic rigidity ($p c/Ze$ for a particle with momentum $p$ and atomic number $Z$), which sets the minimum energy required for a cosmic ray to have been deflected to a given direction. For the higher energies when cosmic rays are generally more energetic than the rigidity cutoff, the flux is symmetric in zenith angle about the horizon; however for low energies below the cutoff, the flux is expected to become enhanced in upward-going directions~\cite{Honda:2011nf}. The spatial dependence of the rigidity cutoff also induces an east-west dipole asymmetry in both $\nu_e$ and $\nu_\mu$ directions as well as a strong dependence in the flux normalization on latitude. The subsequent nuclear recoil distributions therefore also weakly depend on angle: a calculation relevant for directional detectors can be found in~\cite{OHare:2020lva}. A measurement of these angular features at new locations would serve as an additional test for flux simulations as well as again supplying an enhanced background rejection capability for a directional experiment. The science case for a nuclear recoil observatory and atmospheric neutrinos is therefore well motivated. However the relevant flux at low energies is still extremely small, $\sim 10$~cm$^{-2}$~s$^{-1}$~\cite{Honda:2011nf}, with roughly 10-20\% uncertainties (see the green curves in Fig.~\ref{fig:NuRates}). This would yield a nuclear recoil rate in a 5~keV$_{\rm r}$ threshold $^{19}$F target experiment around of 0.1 ton$^{-1}$ year$^{-1}$. Due to the high energies of the atmospheric neutrino flux, the electron recoil rate on the other hand is negligible. Atmospheric neutrinos may be measurable via nuclear recoils if further refinements to these calculations put the flux in the upper end of the current uncertainty, but this would still be a late-stage goal of a \Cygnus-like experiment, requiring DUNE-sized~\cite{Abi:2018dnh} $\sim$100,000 m$^3$ scale volumes.

\subsubsection{Geoneutrinos}
\label{sec:geonu}
The radioactive decay of uranium, thorium and potassium in the Earth's crust and mantle is believed to power a large fraction of the internal heat flow of the Earth~\cite{se-1-5-2010,Gando:1900zz}. Such decays will also be a significant source of antineutrinos for energies below 4.5 MeV. The impact of these geoneutrinos on DM experiments has been discussed in the past~\cite{Monroe:2007xp}, but as a background they are subdominant when compared with the large flux of solar neutrinos at similar energies. Recently however, Ref.~\cite{geoneutrinos} explored the science potential of \emph{directional} geoneutrino measurements in particular. Naturally, the flux of neutrinos originating from the Earth will be strongly anisotropic in an upward going direction, approximately azimuthally symmetric and constant in time. As with a DM search, the distinct angular signature and the lack of modulation for geoneutrinos (with respect to solar neutrinos for instance) would give a directional experiment strong background rejection capabilities in this context. 

As shown in Fig.~\ref{fig:NuRates} (left) the \cevns{} recoil rate from geoneutrinos in \Cygnus (assuming a location at Boulby) is expected to subdominant, and difficult to observe without lowering the threshold significantly. However there may be scope for using electron recoils as can be seen in Fig.~\ref{fig:NuRates} (right). The event rate for electron recoils from geoneutrinos is around 1--10 per 10,000 m$^3$ in six years. Boulby in fact has a relatively low flux of geoneutrinos compared with alternative sites~\cite{Huang:2013a}. The main issue in observing this flux would be the fact that the electron recoil rate is completely swamped by solar neutrinos. This is precisely why directional experiments would be desirable, since the geoneutrino and solar neutrino fluxes are almost completely separable in zenith angle during the daytime, assuming head/tail recognition.

The physics motivation for studying geoneutrinos is substantial. Past measurements of $^{238}$U and $^{232}$Th geoneutrinos by KamLAND~\cite{Araki:2005qa} and Borexino~\cite{Bellini:2010hy} have relied on antineutrino capture by protons. However the 1.8 MeV threshold for free proton inverse beta decay has rendered neutrinos from the decays of $^{40}$K and $^{235}$U undetectable in this manner. A measurement of the threshold-free neutrino-electron scattering from these lower energy sources could tighten the constraint on the radioactive contribution to the Earth's surface heat flow, which currently stands at $38\pm 23\%$~\cite{Usman:2015yda}. A better understanding of the heat flow from an improved knowledge of the abundance and spatial distribution of radioactive sources would be invaluable for tracing the thermal history of the planet. Since inverse beta decay measurements have a very weak angular dependence, a directional geoneutrino search would again be particularly advantageous in this regard. In Ref.~\cite{geoneutrinos} it was found that a 10 ton-scale detector operating for 10 years would be capable of a 95\%~CL measurement of the $^{40}$K flux. With even larger detectors the angular recoil spectrum of geoneutrinos would provide insight into the source of the Earth's magnetic field. Crucially this requires knowledge of the composition and distribution of radioactive elements in the core. Distinguishing neutrinos from the core and the mantle is not possible without directional information. Detecting geoneutrinos will therefore be an important secondary goal of a \Cygnus-100,000~m$^3$ scale experiment, particularly if this total volume is distributed over multiple sites.

\subsubsection{Science with source and detector}
\label{sec:source}
Artificial sources of neutrinos can be used in conjunction with a \cevns{}-sensitive  detector for multiple physics purposes.  Potential artificial sources include--- in approximately increasing order of neutrino energy---  radioactive sources (typically $\sim$MeV or less), nuclear reactors~\cite{Hayes:2016th} (several MeV), isotope decay at rest~\cite{Bungau:2012gj} ($\lsim 15$~MeV), stopped-pion neutrinos~\cite{Akimov:2015nza} (up to 52 MeV) and low-energy beta beams~\cite{Volpe:2009wk} (tunable, up to tens of MeV or more).  Many of these were first proposed in Ref.~\cite{Drukier:1983gj} in 1983.  For neutrino-nucleus interactions to be dominated by \cevns{}, the momentum transfer $Q$ must be less than the inverse nuclear size, which is largely the case for medium-size nuclei for neutrino energies up to around 50~MeV.

Physics possibilities with an artificial-neutrino-source \cevns{} experiment are extensive~\cite{Akimov:2015nza}.  They include Standard Model tests~\cite{Barranco:2005yy, Scholberg:2005qs, Barranco:2007tz, Coloma:2017ncl, Liao:2017uzy}, neutrino electromagnetic properties~\cite{Scholberg:2005qs, Papavassiliou:2005cs, Kosmas:2015sqa}, sterile neutrino oscillation searches~\cite{Formaggio:2011jt,Anderson:2012pn}, as well as nuclear form factor~\cite{Patton:2012jr, Patton:2013nwa} and neutron radius measurements~\cite{Cadeddu:2017etk}.  For some experimental setups, light DM produced at the source can be probed~\cite{deNiverville:2011it, deNiverville:2015mwa, Ge:2017mcq,Akimov:2019xdj}.

In Fig.~\ref{fig:NuRates} we showed the expected number of nuclear and electron recoil events in \Cygnus over six years from nearby nuclear reactors. The rates shown here are relatively large since the Boulby site benefits from its close proximity to the Hartlepool nuclear power station. As with geoneutrinos, the major advantage of a directionally sensitive experiment is in its ability to reject the otherwise irreducible background from other sources of neutrinos, namely the large rate of solar neutrinos at similar energies. As with natural sources, the measured recoil directions of artificial neutrinos, in combination with the known source location would enable event-by-event reconstruction of neutrino energies and improve background rejection. The improvements brought by directionality would in general enhance sensitivity to any of the array of physics measurements accessible to artificial-source \cevns{} experiments~\cite{Abdullah:2020iiv}.

Of the various artificial neutrino source possibilities, the currently-available ones are reactors, which offer huge fluxes of $\bar{\nu}_e$, and stopped-pion sources.  The latter has yielded the first measurements of \cevns{}~\cite{Akimov:2017ade,Akimov:2020czh,Akimov:2020pdx} by  the COHERENT experiment.  Stopped-pion sources produce neutrinos from the weak decays of charged pions at rest.  They emit neutrinos with a well-understood spectrum with a maximum energy of half the mass of the muon, $52.8$~MeV, overlapping well with typical supernova neutrino energies.  These neutrinos are produced copiously at accelerators when $\sim$GeV-scale protons collide with matter, producing pions; when the predominant $\pi^+$ decay after stopping in the matter, they yield monochromatic 30 MeV $\nu_\mu$ and muons; these at-rest muons decay on 2.2~$\upmu$s timescales to create $\bar{\nu}_\mu$ and $\nu_e$ with a few tens of MeV.  Stopped-pion sources have been used in experiments in the past~\cite{Armbruster:1998gk,Auerbach:2001hz}.  Currently the Spallation Neutron Source (SNS) at Oak Ridge National Laboratory,~\cite{Scholberg:2005qs,Bolozdynya:2012xv} is in use for \cevns{} measurements~\cite{Akimov:2015nza}, and other potential sources might be available in the future~\cite{Patton:2013nwa}.  The SNS produces about $5\times 10^{14}$ neutrinos per flavor per second with few-hundred-ns pulse width at 60~Hz, which results in a factor of $10^3-10^4$ background rejection.  While no current detector there has directional capability, this source would be suitable for such experiments. This setup could serve as a detector technology test, given that the expected recoil distributions are well known.

\subsubsection{Exotic models}
\label{sec:exoticnu}
DM experiments will also be able to explore novel neutrino sector physics. The recently measured \cevns{} cross sections~\cite{Akimov:2017ade,Akimov:2020czh,Akimov:2020pdx} appears to agree with the Standard Model prediction currently, but there may be additional non-standard interactions that would affect the recoil energy spectra observable in future DM experiments. For example Ref.~\cite{Cerdeno:2016sfi} explored the prospects for ton-scale experiments to perform novel solar neutrino measurements, such as measuring the $pp$ or $^8$B flux via the neutral current, as well as constraining the running of the electroweak mixing angle and the possible existence of additional mediators from some light dark sector. By modifying the recoil energy spectra, additional exotic interactions involved with both DM and neutrinos will affect the shape of the neutrino floor~\cite{Bertuzzo:2017tuf,Gonzalez-Garcia:2018dep}. Certain mediators may also increase the number of expected events from the neutrino background by several orders of magnitude~\cite{Boehm:2018sux}. This further emphasizes the need for directional experiments with low energy sensitivity. If the \cevns{} cross section is enhanced in this way by non-standard interactions then potentially a much greater range of WIMP cross sections at low masses are saturated by the neutrino background.  It was also shown in Ref.~\cite{Billard:2014yka} that direct detection experiments would be able to make complementary constraints on sterile neutrinos if both coherent nuclear and electron scattering of solar neutrinos is measurable. Again, the fact that the initial directions of solar neutrinos are known means that a directionally sensitive experiment can reconstruct the neutrino energy spectrum on an event-by-event basis.

\subsection{Summary of science case}
A large-scale directional nuclear recoil detector like \Cygnus will offer many opportunities in a great number of research avenues, both in particle physics and astrophysics. We have described the major goals for such an experiment functioning as an observatory for DM as well as neutrinos. As a DM detector, a recoil observatory offers one of the most sensitive tests of the putative WIMP in the sense that it can attain the most powerful background rejection capabilities and confirm the galactic origin on the detected signal. Additionally, we have discussed here that such an experiment will also be capable of searching for other candidate particles for DM, probe non-standard and exotic particle interactions and map the local DM phase space structure of the Milky Way galaxy around us. Furthermore, while we would desirably have the experiment detect the DM, we emphasize that the presence of the unavoidable neutrino background in nuclear recoil experiments brings with it novel opportunities for discovery in its own right.

\section{Existing directional detection technologies}
\label{sec:technology_choices}
The directions of nuclear recoils can be inferred in a detector by direct or indirect means. A direct reconstruction of the recoil track, such as in a tracking detector, can be achieved if the nuclear recoil geometry is measurable. An indirect measure on the other hand may be possible if there exists some proxy for the recoil direction, such as a detector response that depends upon some angle between the recoil and a detector axis. In this section we summarize the various available readout methods based on these two broad ideas. 	In doing so we outline and provide references to the experimental data that supports the simulation parameters used in the following Sec.~\ref{sec:technology_comparison}. 
For a more detailed and critical assessment of these technologies, including those not under consideration for \Cygnus, see Ref.~\cite{Battat:2016pap}.

\subsection{Detectors with recoil track reconstruction}
All currently active directional experiments aim to reconstruct the geometry of recoil tracks. Most of these make use of a low-pressure gas time projection chamber (TPC), in which the mm-scale track geometry is measured in 1, 2, or 3 dimensions. In addition to gas-based TPCs, track reconstruction at the 100\,nm scale has been demonstrated in solid-state nuclear emulsions~\cite{Aleksandrov:2016fyr}.  More exotic, and at this point unvalidated, technologies such as nitrogen-vacancy centers in diamond~\cite{Rajendran:2017ynw}, DNA strands~\cite{Drukier:2012hj}, and columnar recombination~\cite{Nygren:2013nda} have also been proposed.

\subsubsection{Gaseous TPCs}\label{subsec:gastpcs}
The low-pressure gas TPC is the most mature directional detection technology. In this scheme, the WIMP-induced recoil generates a track of ionization in the gas volume, and an electric field transports the resulting charge to an amplification and readout plane. The full three dimensions of a recoil track can be reconstructed by combining the 2d measurement of the ionization charge distribution on the readout plane, with the third dimension inferred by sampling the transported signal as a function of time. The projection of the track along this third dimension, parallel to the drift field, is found by multiplying the duration of the signal with the known drift velocity of the charge in the gas. In some designs, the ionization electrons can be transported directly. In others, a target gas with high electron affinity is used where the ionization electrons rapidly combine with surrounding gas molecules near the interaction site to form negative ions which instead drift to the readout plane. This latter method, so-called `negative ion drift' (NID), can help suppress the diffusion of the ionization track to the thermal limit and thus preserve more of the track geometry prior to readout~\cite{SnowdenIfft:2013iy}. Recent work has explored NID operation at atmospheric pressure as well~\cite{Baracchini:2017ysg}.

Given the $\mathcal{O}({\rm keV})$ energies expected for WIMP-induced nuclear recoils, and the typical energy required to create an electron-ion pair in gas targets, $W\approx20$\,eV, a recoil will produce $\mathcal{O}(10^2$--$10^3)$ primary ionization electrons. To enhance this signal, a gas amplification device is used, consisting of a carefully designed region of high electric field where avalanche multiplication occurs. In some cases, this amplification device is integral to the readout---Multi-Wire Proportional Chambers (MWPCs) and Micromegas for instance---while in others, the gas amplification and readout are distinct, as with Gas Electron Multiplier (GEM) amplification followed by charge readout via silicon pixel chips or by an optical readout.  A detailed overview of the avalanche gain and its resolution, and how the relevant experimental parameters contribute to each, for the specific case of GEM amplification, can be found in Ref.~\cite{thorpe_gain}. Multiple gas amplification techniques, and readouts, may be used together in the same detector. The recoil measurement may combine charge readout with other detection channels. For example, if the gas target scintillates during amplification, then optical readout can be used with, or in place of, charge readout.
\\
\paragraph{Multi-Wire Proportional Chambers}
Early work in directional detection used MWPCs, which provide both gas amplification and spatial information. The DRIFT collaboration pioneered this method~\cite{SnowdenIfft:1999hz,Martoff:2000wi}, and currently holds the leading DM limit set by a directional detector~\cite{Battat:2014van,Battat:2016xxe}. DRIFT-IId achieved zero-background operation over more than 100 live-days using a low-pressure gas mixture (30:10:1 Torr of CS$_2$:CF$_4$:O$_2$) in a back-to-back TPC configuration, each with drift length of 50\,cm. An energy resolution of 40\%, a threshold for head/tail recognition for fluorine recoils (in a statistical sense, not event-by-event) of 40\,\kevr, and a fluorine recoil detection energy threshold of 20\,\kevr{}, were all achieved experimentally. The energy threshold was ultimately limited by the achievable gas amplification of $\sim 10^3$~\cite{Burgos:2008jm,Battat:2016xaw}. The spatial resolution in the readout plane is given by the amplification wire spacing, or pitch, of 2\,mm. Mechanical instabilities arising from inter-wire electrostatic interactions prevent smaller wire spacings.
The spatial granularity along the drift direction, on the other hand, is substantially finer: $60\,\upmu$m, for a 1\,MHz sampling rate and negative ion drift speed of 60\,$\upmu$m/$\upmu$s.  In an MWPC, the capacitance per wire is very low ($\ll 1$\,pF for a wire length of 1\,m), and the noise will depend on the readout electronics. For example, using front-end ASICs developed for LAr TPCs \cite{DeGeronimo:2011zz} coupled to a custom digitizer, a noise of 250\,e$^-$ with a 10\,$\upmu$s averaging time has been demonstrated~\cite{gauvreauPrivate}.  Based on previous experience, a $2\times 2$\,m$^2$ MWPC would cost approximately US\$15k. A newer approach under investigation, closely related to MWPCs, is to combine GEM charge amplification with charge readout via wires~\cite{Ezeribe:2019tln}.
\\
\paragraph{Micro-Patterned Gas Detectors with strip readout}
The development of micro-patterned gas detectors (MPGDs)~\cite{Oed:1988jh,Sauli:1999nt} has enabled improved spatial resolution relative to MWPCs, which enhances the track reconstruction capability of gaseous TPCs. MPGDs are defined as high spatial granularity gaseous devices with sub-mm gaps between their anode and cathode electrodes, fabricated using microelectronics technology. Examples include the \mupic{}~\cite{ref:3uPIC},
and the Micromegas~\cite{Giomataris:1995fq}.

Large-area MPGDs have become commodity items.  
For example, a $2\times 1$\,m$^2$ resistive-strip micromegas with 2d strip readout (200\,$\upmu$m pitch) is available for purchase from CERN for US\$30k~\cite{CERNPersonal}. The measured strip capacitance is 500\,pF per meter~\cite{CERNPersonal}.  A vigorous R\&D effort is underway to produce low-background MPGDs for rare-event searches (see \eg{}~Ref.~\cite{Iguaz:2015myh}).  

MPGDs with segmented readout are in use for directional dark matter detection already. For example, the MIMAC collaboration~\cite{mimac,Riffard:2015rga} uses a low-pressure electron-drift gas mixture of \cff{}, CHF$_3$, and C$_4$H$_{10}$ (70\%/28\%/2\% at 50 mbar) in a back-to-back TPC configuration with 25-cm drift length.  MIMAC demonstrated an event detection threshold of 1\,\kevee{}, and an energy resolution of 10\% at 5.9\,\kevee{} for a gas amplification of 20,000~\cite{riffardThesis}.  In addition, the NEWAGE collaboration~\cite{Nakamura:2015iza}, uses a \mupic{} with 400\,$\upmu$m pitch and a readout sampling time of 10\,ns. To enhance the gas amplification, a Gas Electron Multiplier (GEM)~\cite{Sauli:1997qp} is used, for a typical total gas gain of 6000. NEWAGE has run a TPC with 40\,cm drift length and \cff{} gas at 100\,Torr. They have achieved an energy resolution of 16\% ($\sigma/\mu$), and an angular resolution of 40$^\circ$ at their recoil energy threshold of 50\,\kevee{} with an underground detector~\cite{Nakamura:2015iza}. NEWAGE has also demonstrated head/tail sensitivity down to 75\,\kevee{} with a test chamber \cite{Nakamura:2013hma}.
\\
\paragraph{MPGD with pad readout}
Driven by the needs of large-scale LAr neutrino detectors, the community has seen new developments in large-area MPGDs. For example, a segmented readout plane with integrated readout electronics has been developed~\cite{Dwyer}. The pad pitch of 3\,mm is large compared to the needs for directional DM detection, but the pad geometry could be optimized in the future. The model considered here has a pitch of 3\,mm, a pad capacitance of 0.25\,pF, and an rms readout noise of 375\,\elec{} in a 1\,$\upmu$s sampling window, with an estimated cost of US\$5k per square meter readout plane~\cite{DwyerPersonal}.
\\
\paragraph{Pixel chips}
Even finer spatial resolution than MPGDs can be obtained with pixelated silicon chips. Examples of pixel-based readouts include the FE-I4b ATLAS chips~\cite{GarciaSciveres:2011zz}, QPIX~\cite{ref:Khoa_Mth} and Medipix~\cite{Plackett:2010zz} pixel chip families, which provide spatial granularity as fine as 50\,$\upmu$m.  Rapid digitization of the readout plane provides spatial information along the drift direction.  The fine spatial resolution and low per-pixel capacitance (0.01--0.2\,pF) are clear strengths of this technology, though a drawback is the small size of each chip (the current-generation ATLAS pixel chip is $2.0 \times 1.68$\,cm).  Building a square-meter area readout plane would be highly costly and complex, though certainly possible. For example, the ATLAS Phase II pixel detector will employ enough chips to cover more than 10 square meters of readout area~\cite{Bates:2015fpj}. R\&D work has demonstrated the capability of pixel chips for directional DM detection. For example, in a TPC filled with one atmosphere of He:CO$_2$ (70:30) --- an electron-drift gas --- and using double GEMs for gas amplification, an energy resolution of 10\% and single electron threshold (for primary ionization) was demonstrated~\cite{Jaegle:2019jpx,Lewis:2014poa,Kim:2008zzi,Vahsen:2014mca,Vahsen:2015oya}. Presently, the cost of the ATLAS pixel chips (model RD53B) is US\$90k/m$^2$, not inclusive of the cost of a gas amplification device or readout system.
\\
\paragraph{Optical imaging}
The ionization from charge amplification in the gas may be accompanied by the emission of photons, which can be imaged optically (\eg{}~with a CCD or CMOS camera). This was the first approach used in 1994 for directional dark matter detection with TPCs~\cite{Buckland:1994gc}.  The technique was then revived in the 2000s~\cite{Fraga:2002uc}, and was employed by the DMTPC collaboration~\cite{Ahlen:2010ub,DMTPC_4shins,Deaconu:2017vam} who used four CCD cameras (Apogee Alta U6, $1024 \times 1024$ pixels, $24\times$\SI{24}{\micro\metre\squared} each) to image a TPC with 27\,cm drift length filled with \cff{} at 60\,Torr (typical). The proven energy resolution was 35\% at 80\,\kevr, with a recoil energy threshold of 20\,\kevee. The angular resolution was 15$^\circ$ at 20\,\kevee, with head/tail sensitivity above 40\,\kevee~\cite{DMTPC_4shins}. The DMTPC cameras had 10\,\elec{} readout noise.  More recent high-end commercially available cameras (\eg{}~Hamamatsu ORCA-Flash4.0 V3 CMOS, costing US\$9k) could offer substantial performance improvements.

Exposure and readout times of cameras are long compared to the signal generation in the detector, so a disadvantage of optical readouts is that only a 2d projection of a recoil track in the readout plane is measurable. One would need to combine this slow imaging technique with a faster readout (e.g. photomultipliers) to recover the third dimension of the track~\cite{Margato:2004iz,Fetal:2007zz,Antochi:2018otx}. 
Another drawback is that the ratio of photon production to charge production can be low, \eg{}~one photon per three electrons in \cff{}~\cite{Kaboth:2008mi}. More importantly, the geometric acceptance for photons is very small because very few photons produced in the amplification region make it onto the image sensor. Orders of magnitude more gas amplification are therefore required to compensate, relative to charge readout.  However, cameras do have advantages over charge-based readouts, in particular a lower burden on radiopurity since the cameras are located outside of the active volume of the detector. Additionally, high 2d spatial resolution is possible, and data acquisition is trivial, \eg{}~a USB cable to a PC.

Optical readouts are not a major focus of this paper. Instead, the CYGNO project~\cite{Pinci:2017goi,CYGNO:2019aqp} is working with and alongside \Cygnus to investigate separately the prospects of optically based readouts and electron drift. CYGNO employs a CMOS camera coupled to a TPC with triple thin GEMs for gas amplification in a 60:40 He:\cff{} mixture at 1~bar~\cite{Antochi:2018otx}. This configuration provides the necessary high gas gain of $\mathcal{O}(10^6)$, with about one photon produced for every ten electrons~\cite{Campagnola}. An energy threshold of 2\,\kevee{} and 20\% energy resolution at 5.9\,\kevee{} has been demonstrated~\cite{Costa:2019tnu}. In LEMOn, the largest CYGNO prototype, the drift length is 20\,cm, achievable thanks to the low electron diffusion in He:\cff. Fiducialization may also be possible because the high spatial granularity can allow the diffusion of ionization cloud to be measured, which in turn is dependent on the absolute track position along the drift direction. Preliminary measurements with LEMOn have demonstrated directional and head/tail sensitivity down to about 20\,\kevee~\cite{baracchiniPersonal}.

\subsubsection{Nuclear Emulsions}
The nuclear emulsion most well developed for low energy applications consists of silver halide (AgBr) crystals with a 2.7\,eV semiconducting band dispersed in a polymer. The crystal grains work as sensors of charged particles by producing several-nanometer diameter silver clusters in response to a nuclear recoil track. The emulsions need to be developed successively with a catalytic process, so that a two-dimensional projection of the track trajectory onto the surface can be eventually reconstructed with an optical or x-ray microscope. Recently, Nagoya University managed to produce a super-high-resolution nuclear emulsion called a Nano-Imaging Tracker (NIT), with a mean crystal size of 20-40 nm~\cite{FineGrained}. Currently no head-tail sensitivity has yet been demonstrated in nuclear emulsions.

The Nuclear Emulsions for WIMP Search project (NEWS) is a directional DM experiment with NIT and a fully automated optical scanning system. It demonstrated absolute tracking efficiencies for carbon recoils at 60, 80, and 100 keV of 30$\%$, 61$\%$ and 73$\%$, respectively, by implanting collimated monoenergetic ions (\eg{}~C, O, Kr, F, B etc.).
With a directional analysis based on the use of plasmon resonance~\cite{plasmon} and a prototype microscope, NEWS managed to establish a position accuracy of 10\,nm for a single grain~\cite{Aleksandrov:2016fyr} for 100\,keV C ions. Background discrimination methods are under investigation to suppress the high expected $^{14}$C contamination present in the gelatin producing $\beta$ decays: the current rejection power is 10$^{-6}$, but the required is 10$^{-8}$. An R\&D experiment of 10 grams of NIT is currently being prepared for installation at LNGS.  

Since the measurement of the nuclear emulsion must be performed after the exposure time of the experiment, any directional signal will be washed out by many rotations of the Earth during that time. NEWS have proposed mounting the experiment on an equatorial telescope but at great financial cost. It was shown in Ref.~\cite{OHare:2017rag}, however, that there is still a directional signal present after time-integration, but the strength of the anisotropy is reduced. A factor of 2--3 more events would be required to distinguish the WIMP signal from an isotropic background if the mounting strategy were not implemented.

\subsubsection{Crystal defect spectroscopy}
Nitrogen vacancy (NV) centers in diamond are quantum systems that are highly sensitive to nearby magnetic fields as well as to localized crystal strain. A recent proposal suggests that this emerging technology could be used for directional DM detection~\cite{Rajendran:2017ynw}. WIMP-induced nuclear recoils in the diamond target would create a damage trail in the crystal. The trail alters the strain pattern which could be measured using spectroscopic interrogation of nearby NV centers. This technology has the benefit of being solid-state, so a large potential target mass. With ultra-fine nanometer-scale spatial resolution, preliminary calculations suggest a sensitivity to a head/tail signature as well, though no experimental work has been reported yet. 


\subsubsection{Graphene targets}
Nuclear recoil directions in 3d (bulk) targets often get scrambled through multiple interactions with the surrounding medium. The recoil direction may be more directly measured in 2d targets. Such targets could be fabricated from semiconductor materials in which the excitation energy is on the order of $\sim 1$\,eV, allowing even MeV-scale WIMPs to initiate electronic excitations. A recent proposal suggests that 2d graphene could serve as a directional detector of sub-GeV WIMPs~\cite{Hochberg:2016ntt}. This is a particularly interesting idea, especially given that no other directional technology can probe this WIMP mass scale. There has not yet been an experimental demonstration of this technology, although it may be possible within the PTOLEMY relic neutrino search~\cite{Betts:2013uya}.

\subsection{Detectors that indirectly determine the recoil direction}
\subsubsection{Anisotropic scintillators}
Solid scintillators (\eg{}~NaI and CsI) are commonly used in particle detection, and specifically in DM detection. Because of their large target mass and high nucleon content, they are ideal for SI WIMP searches. Some scintillators, such as ZnWO$_4$ and stilbene, have been shown to exhibit a response that depends on the recoil ion direction relative to the crystal axes. In principle, this scintillation anisotropy can be used to infer the nuclear recoil track direction without direct reconstruction of the track geometry. Several groups have explored the possibility of using anisotropic scintillators for a directional dark matter search~\cite{1961ZPhy..162...84H,Belli:1992zb,Spooner:1996gr,Shimizu:2002ik,Cappella:2013rua,Sekiya:2003wf}, though the magnitude of the anisotropy is too small (less than 10\%) for anything particularly sensitive. Moreover, none of these studies have demonstrated anisotropic scintillation for nuclear recoils at low enough energies. Therefore, all quoted energy resolutions, thresholds, and general performances are for general detection of alpha, beta, and gamma radiation and not necessarily relevant for discussion in the context of DM.

\subsubsection{Columnar recombination}
When a recoil ionizes the detector medium, a track of electrons and ions is created. If no electric field is present, these ionization products recombine, producing scintillation light and suppressing the charge signal. With an external electric field, the amount of light produced will depend on the relative orientation of the recoil track and the field: a large angle results in a small recombination fraction, which enhances the charge signal relative to the light signal.  A precise measurement of the charge-to-light-ratio could be a proxy for recoil track direction~\cite{Nygren:2013nda,Cadeddu:2017ebu}. 

Evidence for columnar recombination with high-energy alpha tracks was observed long ago in dense xenon gas~\cite{jaffe}.  Recent simulations~\cite{Nakajima:2015dva} suggest that it could be present for tracks as short as 2\,$\upmu$m (corresponding to a threshold of 30\,\kevr{} at a pressure of 10 bar), if the ionized electrons are properly thermalized. This is not possible however in pure Xe due to the lack of inelastic scattering below 7 eV. Early experimental efforts to cool ionization electrons with trimethylamine strongly suppressed the primary scintillation light, so that a columnar recombination measurement was not possible~\cite{Nakajima:2015meb}. In liquid argon, however, there has been a marginal detection of the signature of columnar recombination for nuclear recoils with energies above 57\,keV~\cite{Cao:2014gns}. If the effect was shown to be strong enough in liquid noble experiments then it may marginally help in extending the discovery below the neutrino floor at WIMP masses around 100~\gevcc and above~\cite{OHare:2020lva}. R\&D work on columnar recombination is ongoing (see \eg{}~Ref.~\cite{Cadeddu:2017ebu}), but given the low technological readiness at this time, we do not evaluate it further here.

\subsubsection{Carbon nanotubes}
Single-wall aligned carbon nanotubes (CNTs) have been recently proposed as a DM target due to their expected anisotropic response to neutral particles~\cite{Capparelli:2014lua}. Under the right conditions, when a C ion scatters off of the CNT walls it sees the tube as empty and can travel with nearly no loss of energy: an effect known as channeling. 
Different orientations of the CNT axis with respect to the DM wind would give different channeling probabilities and therefore produce significantly different C ion currents at the end of the nanotube. 
The proposed detector concept in Ref.~\cite{Capparelli:2014lua} is a brush of CNTs array closed at one end and open at the other, inserted in a (low-pressure) TPC to detect the outgoing C ions down to $\sim$10 keV. An R$\&$D effort is currently underway in Italy to test the channeling hypothesis for neutral particle scattering and the TPC detector approach. 

\section{Determining the directional sensitivity of Cygnus}
\label{sec:technology_comparison}


To determine and optimize the directional sensitivity of a large scale experiment like \Cygnus, we will focus on gas TPCs, which constitute the most mature directional detection technology. In the previous section, we have seen that a large number of TPC readout technologies have already been successfully demonstrated on smaller scales. Our primary goal here therefore, is to determine the most appropriate TPC readout technology to realize the science goals set out in Sec.~\ref{sec:science_case}.

Previous theoretical work already compared the ability of directional detectors to discover galactic dark matter, and to set limits beyond the neutrino floor~\cite{Mayet:2016zxu, O'Hare:2015mda}. Here we go one step further, by simulating TPCs with specific readout technologies from the ground up, accounting for inevitable physics and detector effects such as nuclear straggling, diffusion of drift charge, readout quantization, and readout noise. This allows us to obtain an energy-dependent description of the detector performance including angular resolution, head/tail efficiency, energy resolution, and detection efficiency. The gradual variation of the detector performance with recoil energy is ultimately what will determine how well \Cygnus can utilize recoil directions to distinguish between dark matter and neutrinos, or between dark matter and an unexpected background of nuclear recoils. There is also the question of electron background rejection, which is also strongly energy-dependent, and which will determine the recoil energy threshold above which \Cygnus can be expected to remain background free. We will use the TPC simulations to estimate electron rejection factors, which will inform the later discussion (see Sec.~\ref{sec:backgrounds}) of backgrounds for a 1000 m$^3$ scale experiment. To further inform our readout choices for \Cygnus, we will also evaluate the cost of each simulated technology.
Ultimately the ideal detector is the one the maximizes science sensitivity per unit cost. This will be the manner in which we arrive at a final readout decision.

\subsection{Choice of gas}\label{subsec:gaschoice}

The parameter space for gas TPCs is large, as a number of readout technologies, gas mixtures, and operational parameters are possible. Furthermore, for each readout technology the spatial segmentation can be varied, for each gas mixture the pressure can be varied, and there are many operational parameters such as drift field strength, avalanche device voltage, and gas temperature.

Our goal here is to identify the optimal choice of readout, with the goal of uniting investigators in the field to pursue a single, optimal, detector design. This is complicated by the fact that the optimal choice of readout depends somewhat on gas mixture and operating parameters. For example, the gas mixture and pressure affect the maximum avalanche gain and recoil length, which affect the charge sensitivity and segmentation required of the readout, respectively.

To make progress, we partition the problem, and here compare multiple existing charge readout technologies with typical performance parameters while holding the gas mixture and detector avalanche gain constant. 

As mentioned in Sec.~\ref{subsec:gastpcs}, NID gases are well known to improve the performance of dark matter TPCs by limiting diffusion of the drift charge. Moreover, it has also been discovered that minority carriers in NID gases can enable fiducialization in the drift direction. Full 3d fiducialization in NID was first demonstrated with CS$_2$ gas~\cite{Snowden-Ifft:2014taa}, and more recently with SF$_6$ gas~\cite{Phan:2016veo}. This ability to locate events within the detector volume is crucial for background rejection. SF$_6$ gas also has a number of additional properties that when combined make it a particularly suitable TPC fill gas for a DM search: is non-toxic, non-corrosive, and contains $^{19}$F--one of the most powerful targets to set limits on the SD WIMP-proton cross section. For these reasons, the NID gas SF$_6$ has received substantial recent interest from the directional dark matter detection community \cite{Ikeda:2017jvy, Baracchini:2017ysg, phdthorpe, Ikeda:2020pex}. 

One potential drawback of SF$_6$, is that it tends to result in approximately two orders of magnitude lower gas avalanche gains than is typical for commonly used electron drift gases. As a result, fixing SF$_6$ as a design choice biases our comparison of TPC readouts in favor of more sensitive charge readout technologies. This is important to mention because alternative design strategies based on electron-drift gases may also be competitive. In detectors using electron drift, the substantially higher avalanche gain will generally lead to greatly improved charge sensitivity and performance of low mass WIMP and neutrino experiments. With a high-resolution charge readout, 3d fiducialization is also possible in electron drift detectors --- albeit with different techniques~\cite{Lewis:2014poa}. However charge tends to diffuse more in electron drift detectors which reduces directionality at the $\mathcal{O}(1$--$10$ keV) energies needed to detect light WIMPs and solar neutrinos. This aspect therefore favors NID gases. 

A new approach for this paper --- one which is already under experimental investigation by the authors~\cite{Jaegle:2019jpx, Costa:2019tnu} --- is the addition of helium to the fill gas. We expect this to greatly improve WIMP sensitivity and directionality at the lowest recoil energies. When used as a scattering target, helium has greater energy transfer than fluorine for particle masses near the proton mass. This means our DM mass sensitivity could be extended down to very light and unconstrained $\sim$1~\gevcc WIMPs, while increasing the solar neutrino event rate. Helium gas has a much lower mass density than fluorine at the same pressure, so its inclusion does not greatly affect the track lengths of fluorine recoils. Finally, helium ions have lower specific ionization than heavier nuclei and therefore longer tracks, further improving directionality at low recoil energies. For these reasons a \hesfsix or \hecffour mixture may be optimal as it could both improve the directional sensitivity in general as well as extend the dark matter search into unexplored parameter space.

In this work, we have focused on \hesfsix mixtures, and the final gas mixture was identified iteratively. Three different mixture were simulated, using the gas parameters in Table~\ref{tab:gas_specs}, and the performance was evaluated. The angular resolution for nuclear recoils depends strongly on these gas parameters (see Ref.~\cite{Vahsen:2015oya}, Eq.~5): the angular resolution is proportional to the readout point resolution and inversely proportional to the track length. The point resolution in turn grows with diffusion, and the track length is inversely proportional to the gas density. Hence angular resolution improves with smaller diffusion and with lower mass-density. In addition, recoil length at fixed energy also increases as the nuclear mass is lowered. Hence angular resolution can also improve when low-mass target nuclei are introduced, as long as other gas parameters do not degrade.

The first gas simulated, pure SF$_6$, has been used extensively experimentally, but we found here that the simulated fluorine recoils do not have satisfactory angular resolution at low recoil energies relevant to WIMP sensitivity near the solar neutrino floor. We then added helium to arrive at a 740:20 Torr \hesfsix mixture. This leads to two improvements: The helium recoils have substantially better angular resolution than the fluorine recoils, and an atmospheric pressure gas mixture would negate the need for low-pressure vessels, reducing cost. We found, however, that the addition of helium increased the mass density to a point where angular resolution for low energy recoils was still worse than desirable. For that reason, we performed a third set of simulations with a 755:5 Torr \hesfsix gas mixture. This mixture has the combined benefits of low mass density, high helium content, and atmospheric pressure, enabling good directional performance at low recoil energies while keeping cost low.

The final readout comparison and sensitivity analysis here is therefore performed with the 755:5 \hesfsix gas mixture. Due to substantial computing time required to perform detector simulations, not all the other studies below have been performed for all gas mixtures. Results with pure SF$_6$ and 740:20 \hesfsix should be considered lower limits on performance.

Improved performance and sensitivity per unit cost is likely to result from further optimization of the gas mixture. The final choice of gas mixture is also constrained by operational stability, which cannot be reliably simulated. An experimental investigation of gases suitable for directional recoil detection, for electron drift gases at atmospheric and sub-atmospheric pressures, can be found in Ref.~\cite{Vahsen:2014mca}.

\subsection{Simulation}
\begin{figure*}[hbt]
\begin{center}
\includegraphics[width=0.9\textwidth]{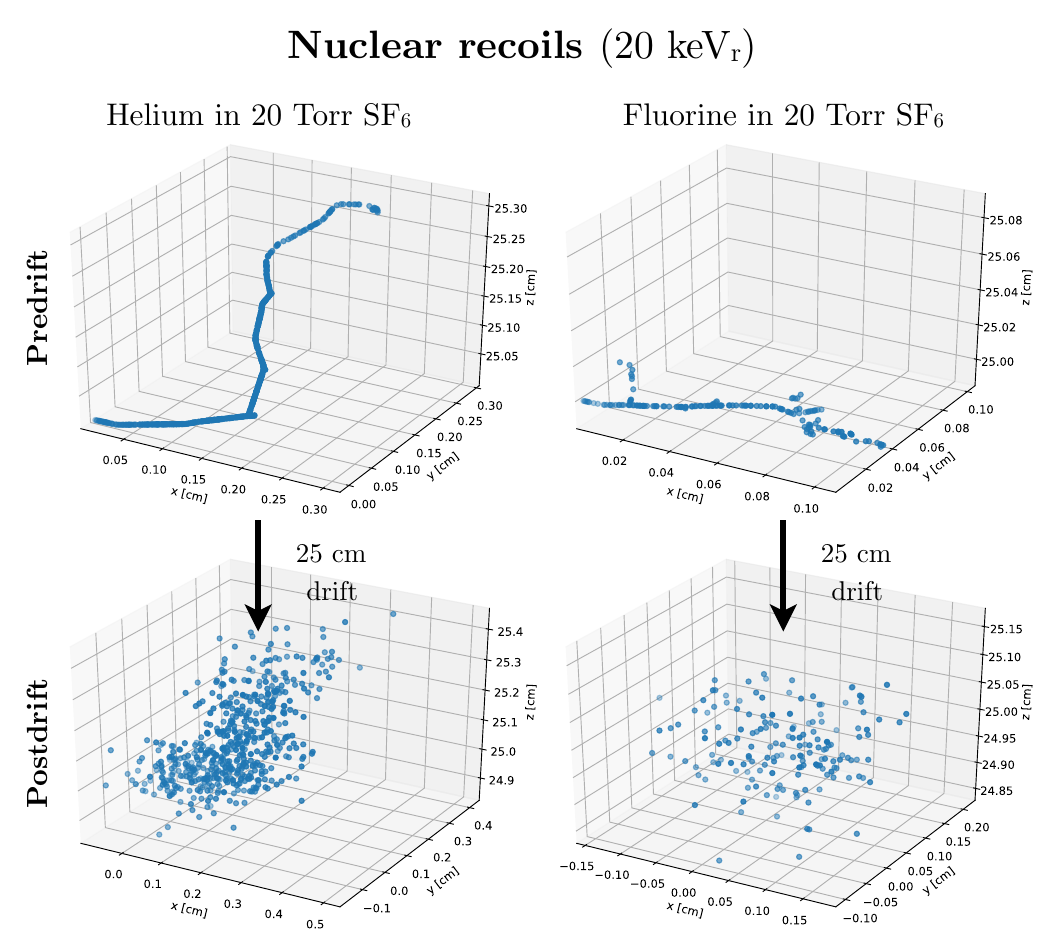}
\caption{Distribution of recoil ionization for 20~\kevr~helium (left column) and fluorine (right column) nuclear recoil events in 20 Torr of SF$_6$, generated with TRIM. Both recoils start out at $(x,y,z)=(0,0,\SI{25}{\cm})$ with initial velocity vectors in the direction of the positive $x$ axis. The top row displays the primary ionization prior to drift, and the bottom row displays the negative ion distribution after 25 cm of diffusion. The recoil direction of the helium nucleus is better preserved after diffusion of drift charge than the fluorine recoil. This is mostly because helium recoils are longer than fluorine recoils of the same energy. In addition, helium nuclei suffer smaller energy losses to nuclear collisions, which results both in more visible ionization and straighter recoils.}\label{fig:nuclear_recoils}
\end{center}
\end{figure*}

\begin{figure*}[hbt]
\begin{center}
\includegraphics[width=0.9\textwidth]{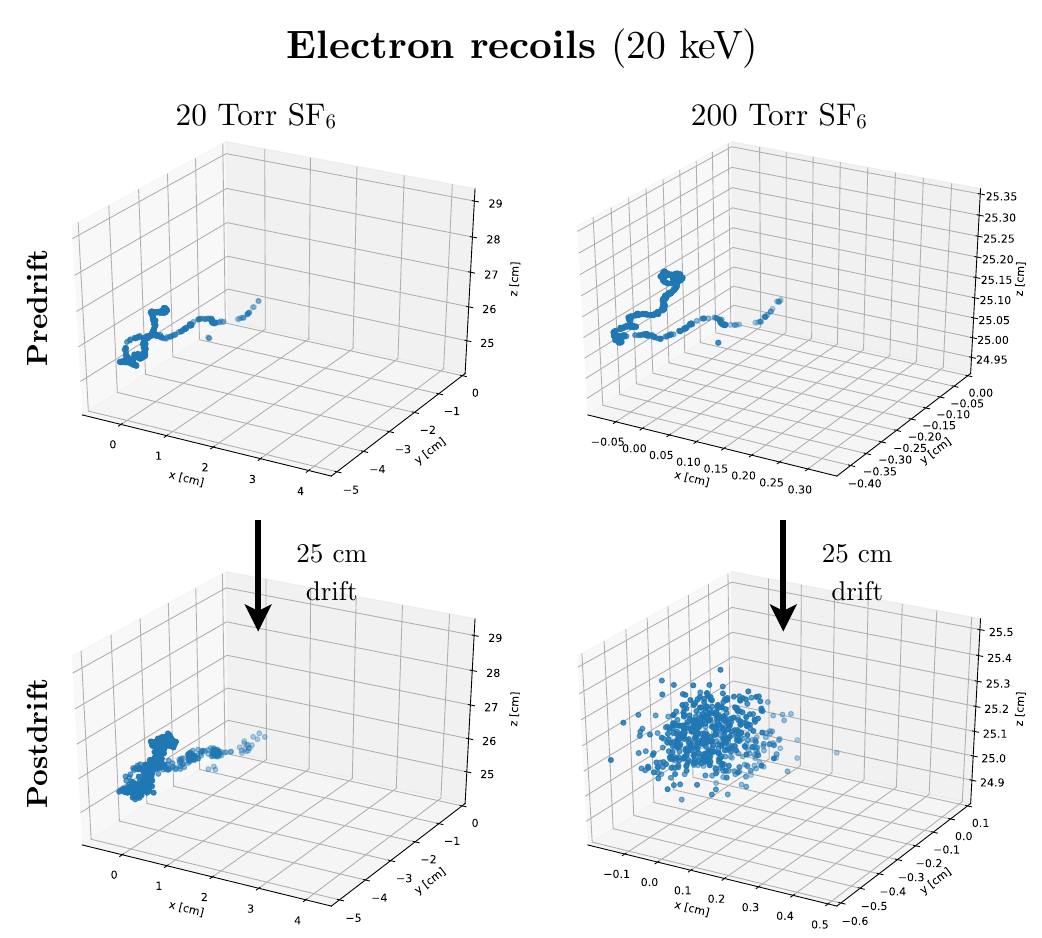}
\caption{As in Fig.~\ref{fig:nuclear_recoils} but now for 20~keV electron recoil events in SF$_6$, generated with DEGRAD. The left- and right-hand columns compare two different pressures: 20~Torr and 200~Torr respectively (note the different spatial scales between the two columns). Then, as in the previous figure, the top row displays the primary ionization prior to drift, and the bottom row displays the negative ion distribution after 25 cm of diffusion. In 20 Torr, the electron recoil is essentially preserved after the diffusion of drift charge, but when the pressure is increased to 200 Torr the topology is washed out. This illustrates that a gas with low mass density improves not only directionality but also particle identification.}\label{fig:electron_recoils}
\end{center}
\end{figure*}

Our simulation of a directional detector consists of the following stages: (1) nuclear and electron recoil momentum vectors are generated from the expected distribution of WIMP and neutrino event rates; (2) primary ionization distributions are generated for each generated recoil event; (3) the primary ionization is propagated through the detector; and finally (4) the simulation of the gain stage and charge readout. Each stage is described in more detail below. Once we have obtained a simulation of the signal measured by each readout technology under consideration, a fit to obtain a recoil direction for each event is performed, and the result is put through an analysis framework to derive physics results. These latter steps will be detailed in the subsequent sections.

\subsubsection{Momentum vector generation}
WIMP nuclear recoil vectors are generated assuming a Gaussian velocity distribution~Eq.(\ref{eq:shm}), with circular velocity $v_0 = 220$~km~s$^{-1}$, escape speed $v_{\rm esc} = 533$~km s$^{-1}$ and DM density $\rho_0 = 0.3$~GeV cm$^{-3}$ at the solar system. For SI recoils, we assume the standard Helm form factor (although in the case of helium this has a negligible impact on the distribution). For fluorine SD recoils, we take the two-body corrected values for $\langle S_p \rangle = 0.421$ and $\langle S_n \rangle = 0.045$ from Ref.~\cite{Cannoni:2012jq} combined with a shell model form factor calculation~\cite{Divari:2000dc}. We assume equal SI and SD couplings to protons and neutrons. Fluorine and helium recoils from coherent elastic neutrino-nucleus scattering, on the other hand, are generated using the analytic expression for the angular differential event rate calculated in Ref.~\cite{O'Hare:2015mda}. We again use the Helm form factor and assume a $^8$B flux normalization in accordance with a global analysis of solar and terrestrial neutrino data~\cite{Bergstrom:2016cbh}. 

WIMP and neutrino nuclear recoil vectors are generated in the lab (North-West-Zenith) coordinate system at the latitude and longitude of Boulby, UK (54.5534$^\circ$~N,~0.8245$^\circ$~W). We use the full computation of $\mathbf{v}_\mathrm{lab}(t)$ and $\hat{\mathbf{r}}_\odot$ in this system (see Ref.~\cite{Mayet:2016zxu}) which ensures the vector generation automatically includes annual and daily modulation signals for both WIMPs and neutrinos. The events are generated assuming uniform up-time throughout the calendar year. The simulated position distribution of recoils in the detector is uniform. Electron momentum vectors finally, are generated uniformly in the detector and isotropically with respect to this coordinate system.

\subsubsection{Generation of primary ionization distributions}
\begin{figure*}[hbt]
\begin{center}
\includegraphics[height=0.95\columnwidth]{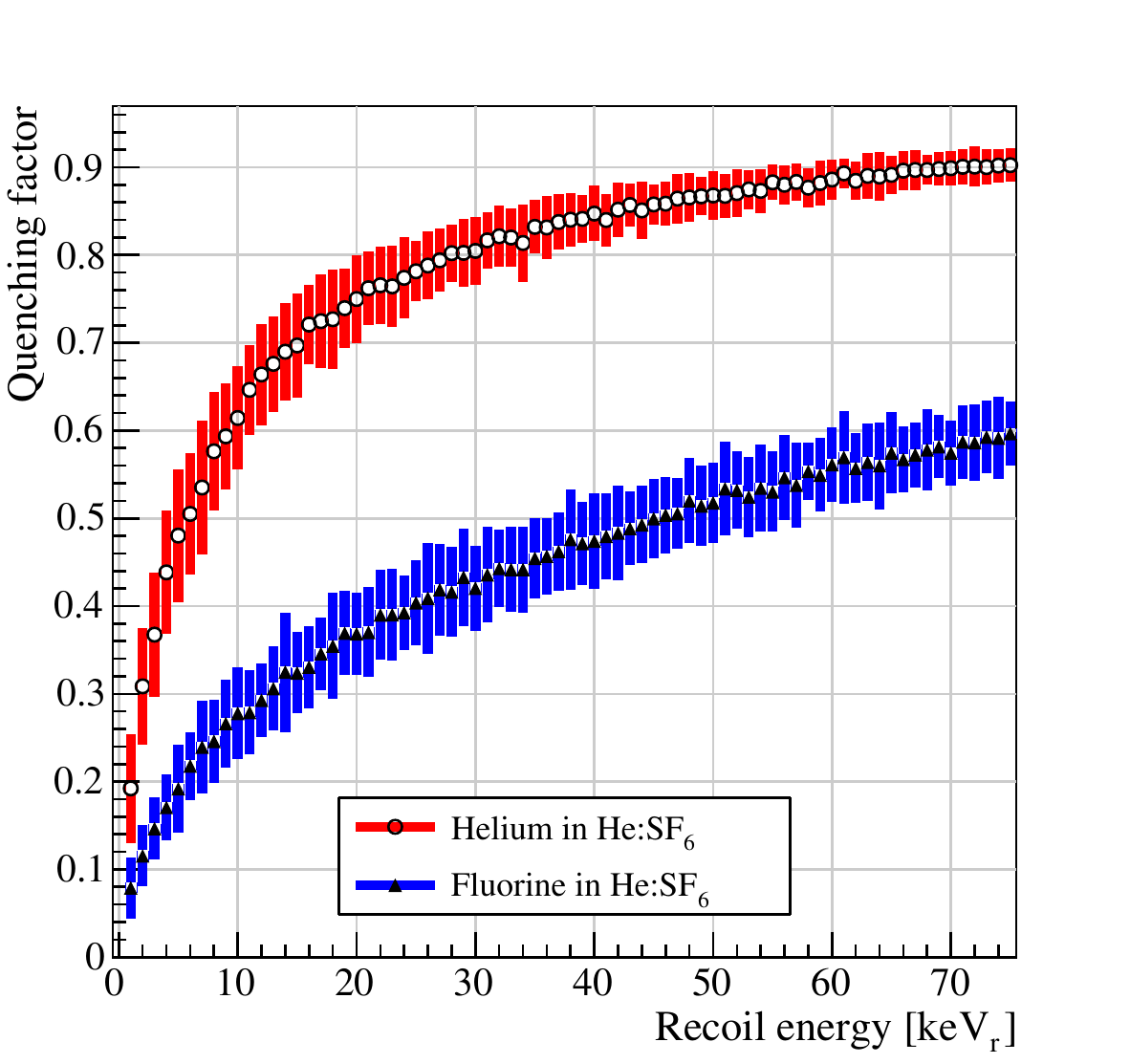}
\includegraphics[height=0.95\columnwidth]{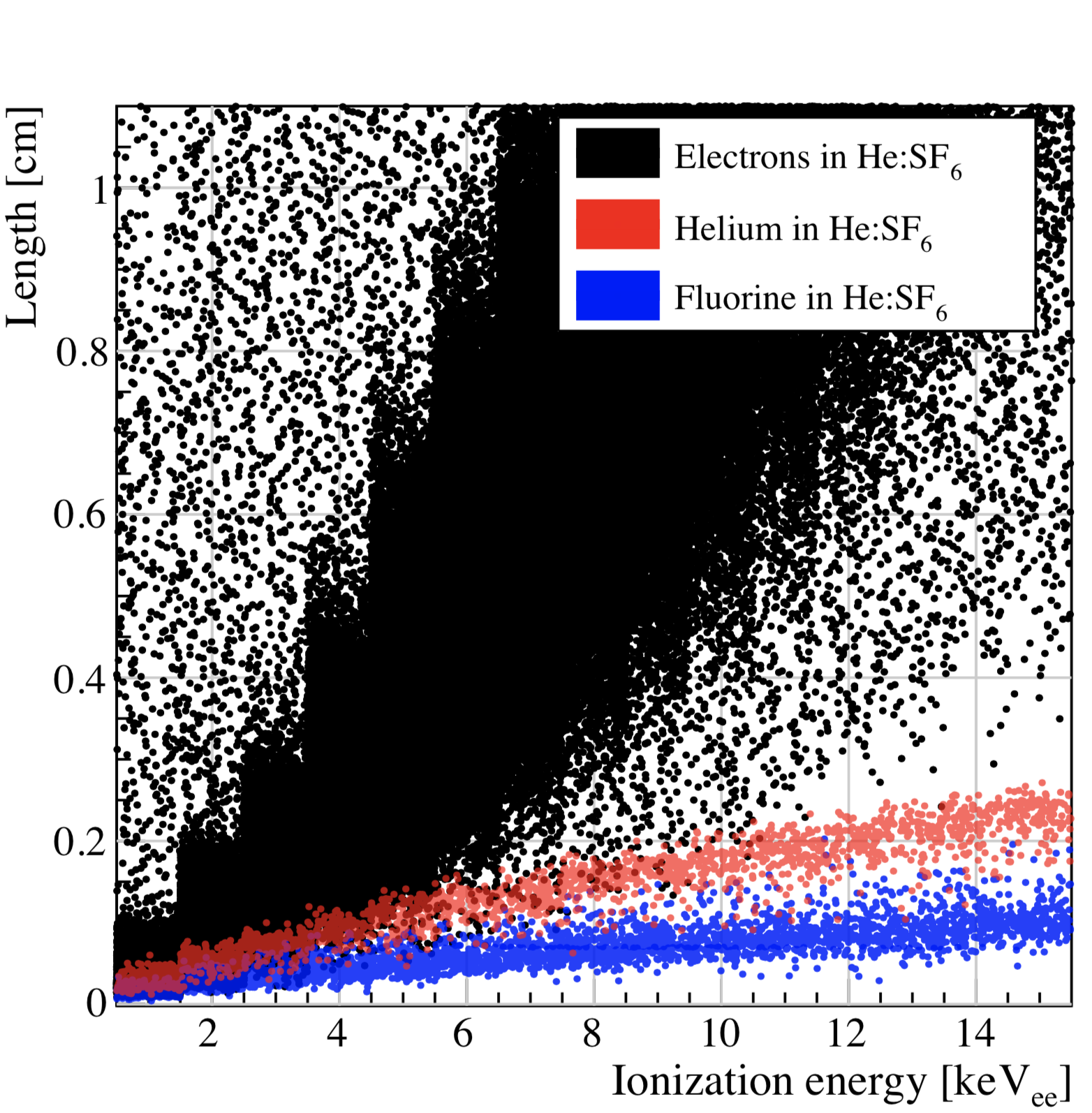}
\caption{Two different features of the primary ionization at generator level. {\bf Left}: quenching factors for TRIM fluorine (blue) and helium (red) recoils in 755:5 \hesfsix gas, versus recoil energy. Error bars indicate the standard deviation of the quenching factor, resulting from fluctuations of the primary ionization. {\bf Right}: track length versus ionization energy for DEGRAD electron recoils (black) and TRIM fluorine (blue) and helium recoils (red). Track length is defined as the projected length of the primary ionization distribution along a 3d track fit. Neither diffusion nor detector effects are included at this stage.}
\label{fig:srim_and_degrad2}
\end{center}
\end{figure*}

To generate the primary ionization distribution of nuclear recoils and electron recoils, we utilize the event generators TRIM~\cite{SRIM} and DEGRAD~\cite{DEGRAD}, respectively. Both generators take as input the momentum vector of the particle to be simulated, and configuration files that specify the gas mixture. DEGRAD directly outputs a 3d distribution of ionized electrons, while the TRIM output needs to be post-processed. At low energies, ionization from secondaries cannot be neglected. To estimate the full ionization distribution for individual recoils, we generate TRIM recoils in the most detailed mode and configure TRIM to output all primary and secondary collisions. From the detailed collision output, we reconstruct the trajectories and momenta of all particles in the recoil cascade, from which the 3d ionization is estimated. Further details on this procedure are available in Ref.~\cite{Deaconu:2015vbk}. The top rows of Figs.~\ref{fig:nuclear_recoils} and \ref{fig:electron_recoils} show examples of generated events as they appear at this stage. Figure~\ref{fig:srim_and_degrad2} then shows distributions of resulting event-level quantities that can be computed at this stage: the quenching factor for nuclear recoils (left), and the fitted length of the charge cloud versus the deposited ionization energy (right). The track lengths versus energy show clear separation between helium, fluorine and electron recoils. This will be important later in this Section when we discuss electron discrimination. 

\subsubsection{Simulation of diffusion and gain}
\begin{table}[ht]
	\centering
	\begin{tabular} {lccc}
	\toprule
	Gas mixture 						& SF$_6$ 	& \hesfsix	& \hesfsix  \\ 
	\hline
	Pressure [Torr]					& 20 			& 740:20 	&	755:5 	 \\
	Density [kg/m$^3$]				& 0.16		& 0.32	&	0.20		\\ 
	$W$ [eV/ion pair] 					& 35.5  		&  38.0   	& 	40.0			\\
	Trans. diffusion [\SI{}{\micro m/\sqrt{cm}}\,]	& 116.2	& 78.6  & 78.6  	\\
	Long. diffusion [\SI{}{\micro m/\sqrt{cm}}\,]	& 116.2	& 78.6   & 78.6 	 \\
	Drift velocity [\SI{}{mm/\micro s}]					& 0.140 	& 0.140	& 0.140  \\
	Mean avalanche gain					& $9 \times 10^3$ & $9 \times 10^3$  & $9 \times 10^3$   \\
	\bottomrule
	\end{tabular}
		\caption{Various gas-dependent parameters assumed in the TPC detector simulation. The values are sourced as follows: the $W$ factor for pure SF$_6$ is from a measurement with alpha particles \cite{0022-3727-20-7-023}, while the $W$ factors for the \hesfsix and \hecffour mixtures are calculated using Eq.(1) of Ref.~\cite{Vahsen:2014mca}. The diffusion values and drift velocity in 20~Torr of pure SF$_6$ were measured in Ref.~\cite{Phan:2016veo}. For the \hesfsix mixtures, no measurements or reliable simulations exist, so we use the 40~Torr pure SF$_6$ diffusion from Ref.~\cite{Phan:2016veo} and then assume the electric field can be adjusted to keep the drift velocity constant. The avalanche gain assumed for pure SF$_6$ has been achieved with THGEMs in Ref.~\cite{Scarff:2017ymw} and triple thin GEMs in Ref.~\cite{Ishiura2019}, and is also used for \hesfsix mixtures. 
}
\label{tab:gas_specs}
\end{table}

\begin{figure*}[htb]
\begin{center}
\includegraphics[width=0.8\columnwidth]{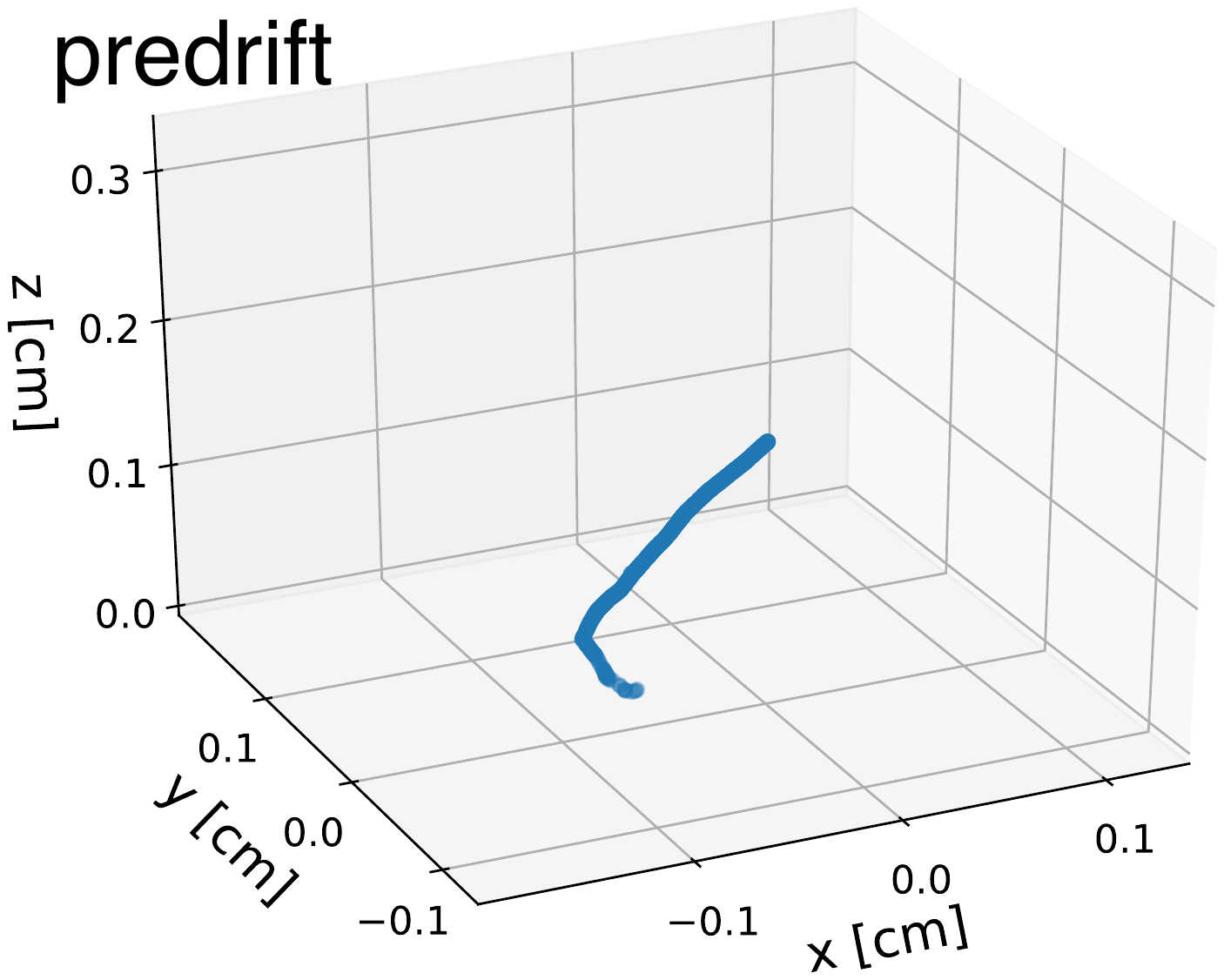}
\includegraphics[width=0.8\columnwidth]{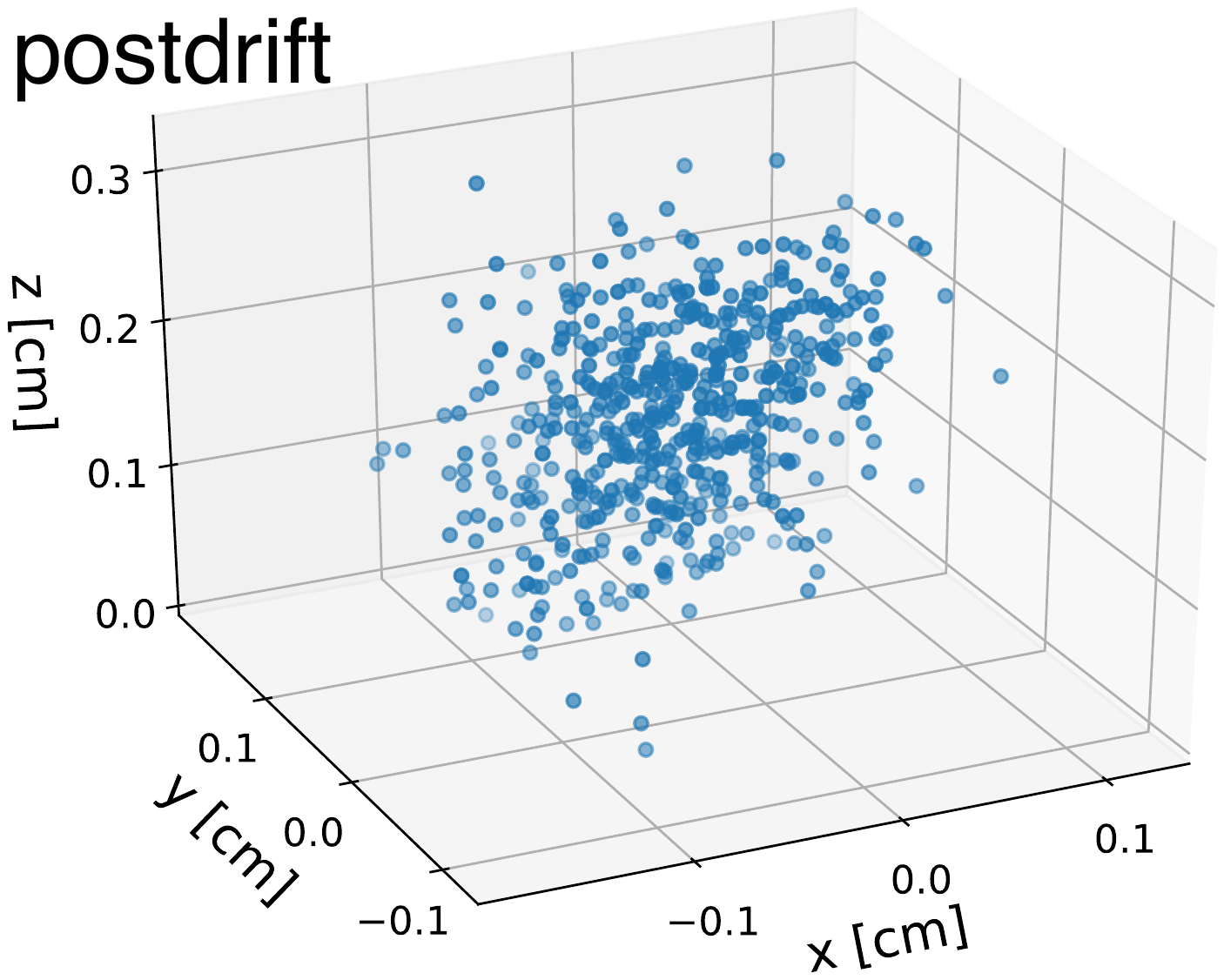}\\
\includegraphics[clip, trim=0.0cm 1cm 1.0cm 1cm, width=0.66\columnwidth]{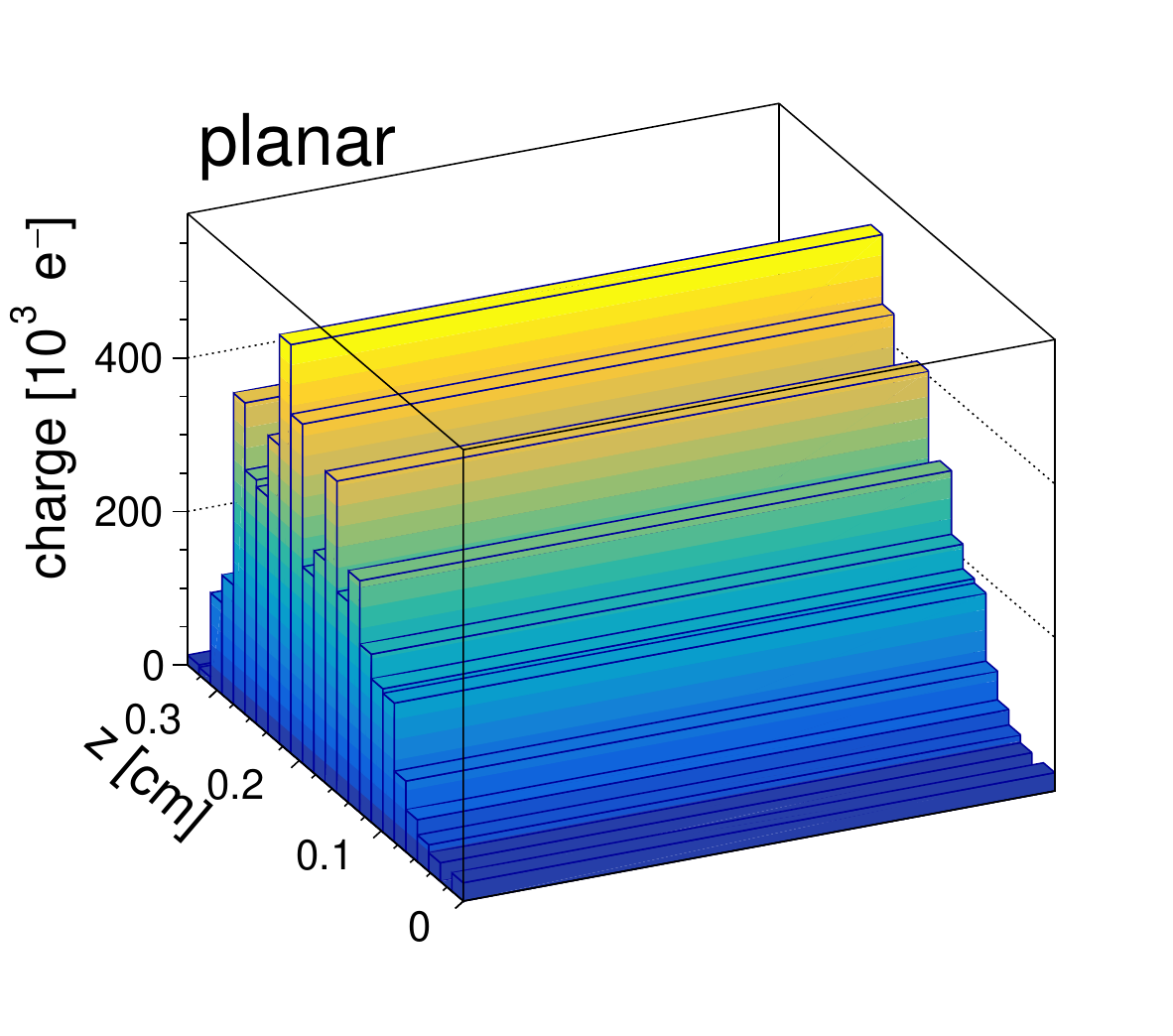}
\includegraphics[clip, trim=0.0cm 1cm 1.0cm 1cm, width=0.66\columnwidth]{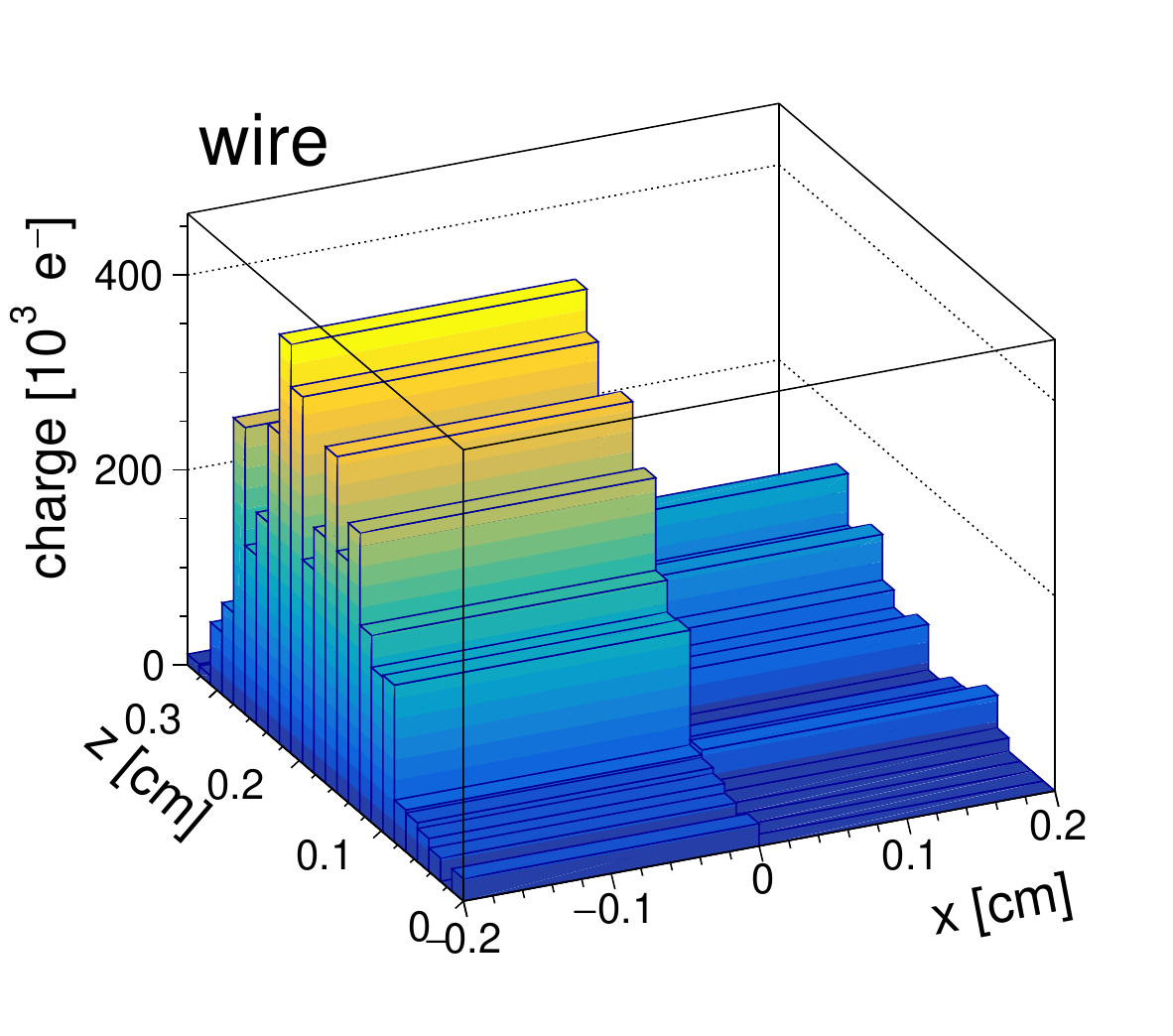}
\includegraphics[clip, trim=0.0cm 1cm 1.0cm 1cm, width=0.66\columnwidth]{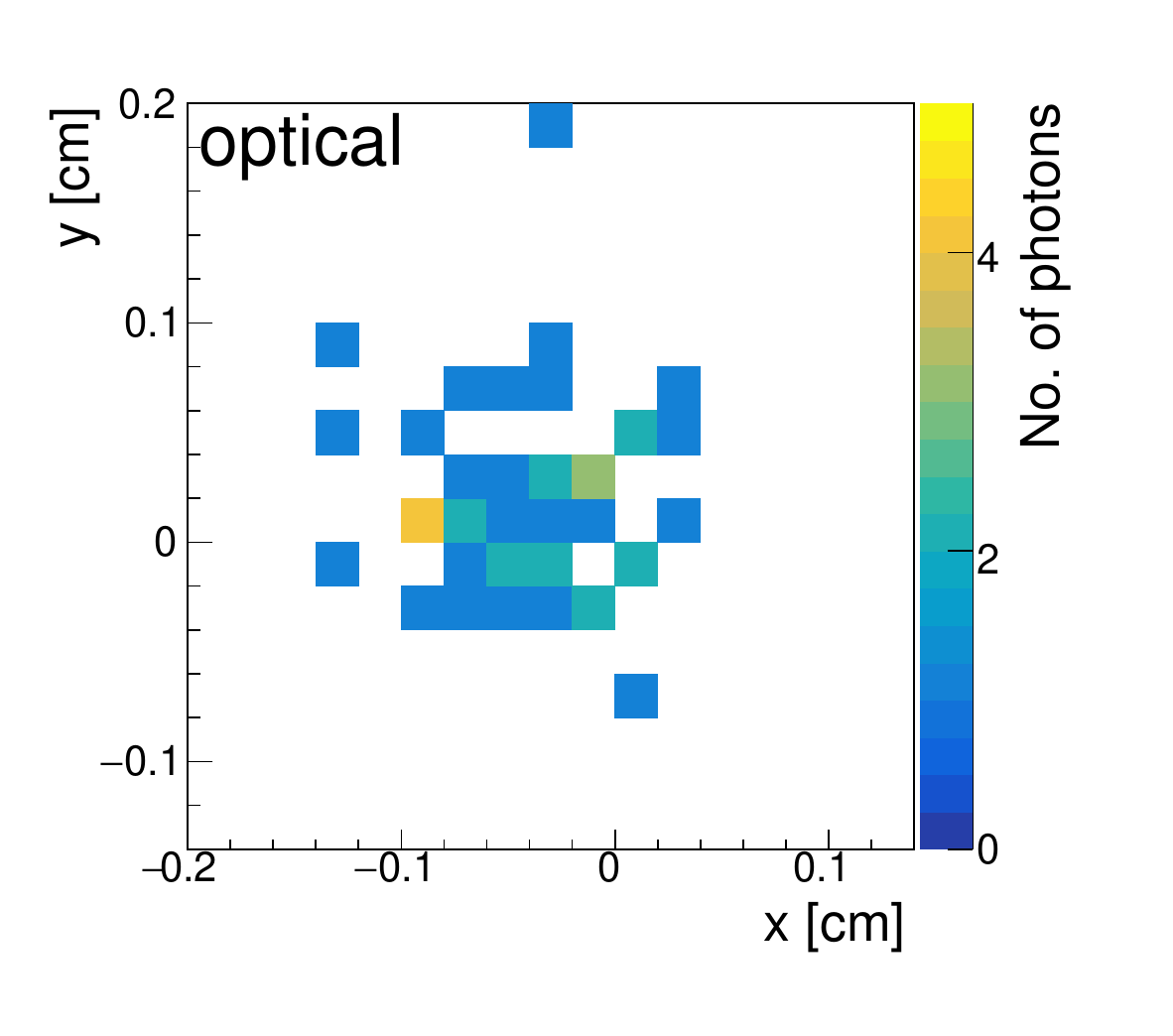}\\
\includegraphics[clip, trim=0.0cm 1cm 1.0cm 1cm, width=0.66\columnwidth]{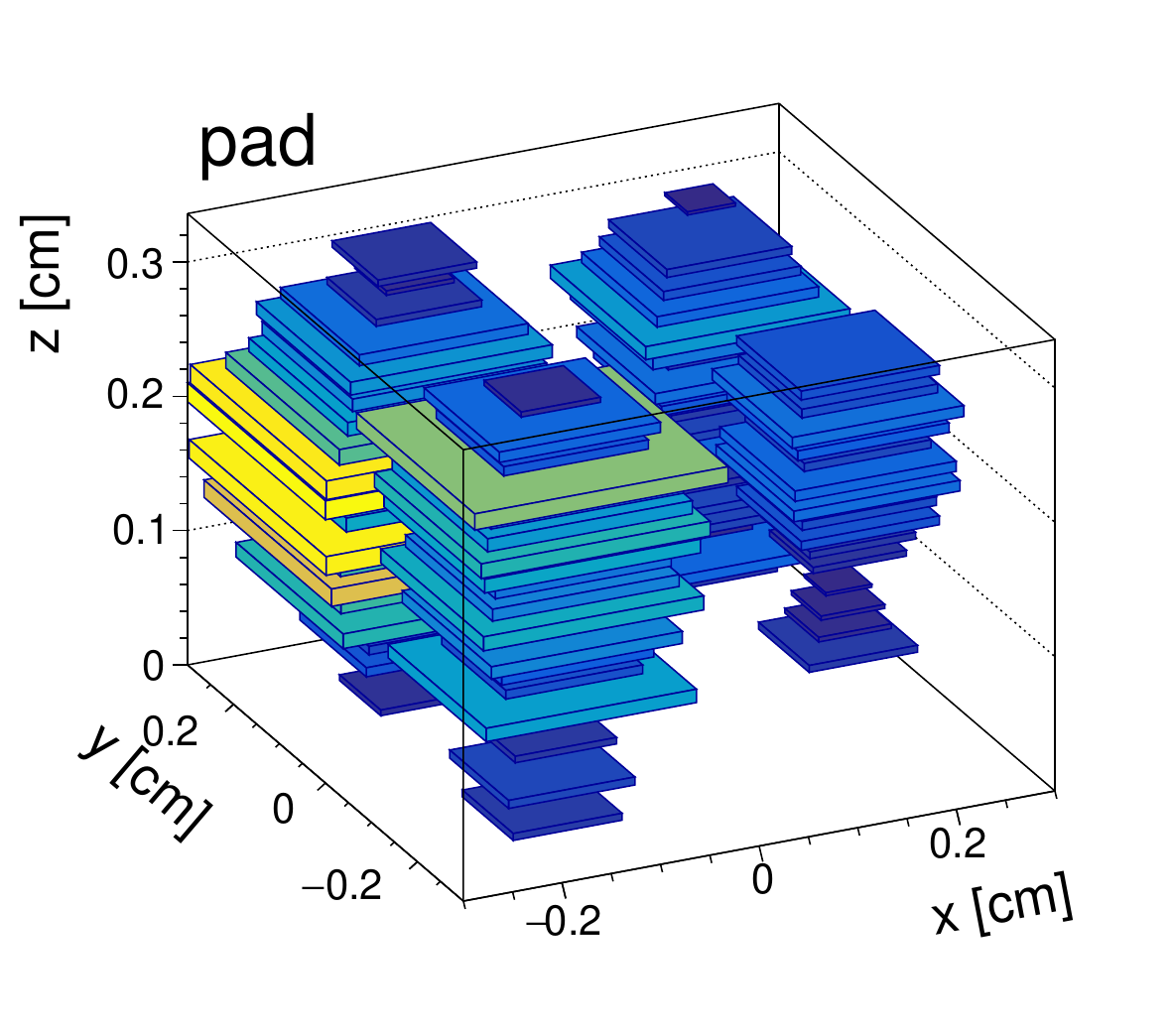}
\includegraphics[clip, trim=0.0cm 1cm 1.0cm 1cm, width=0.66\columnwidth]{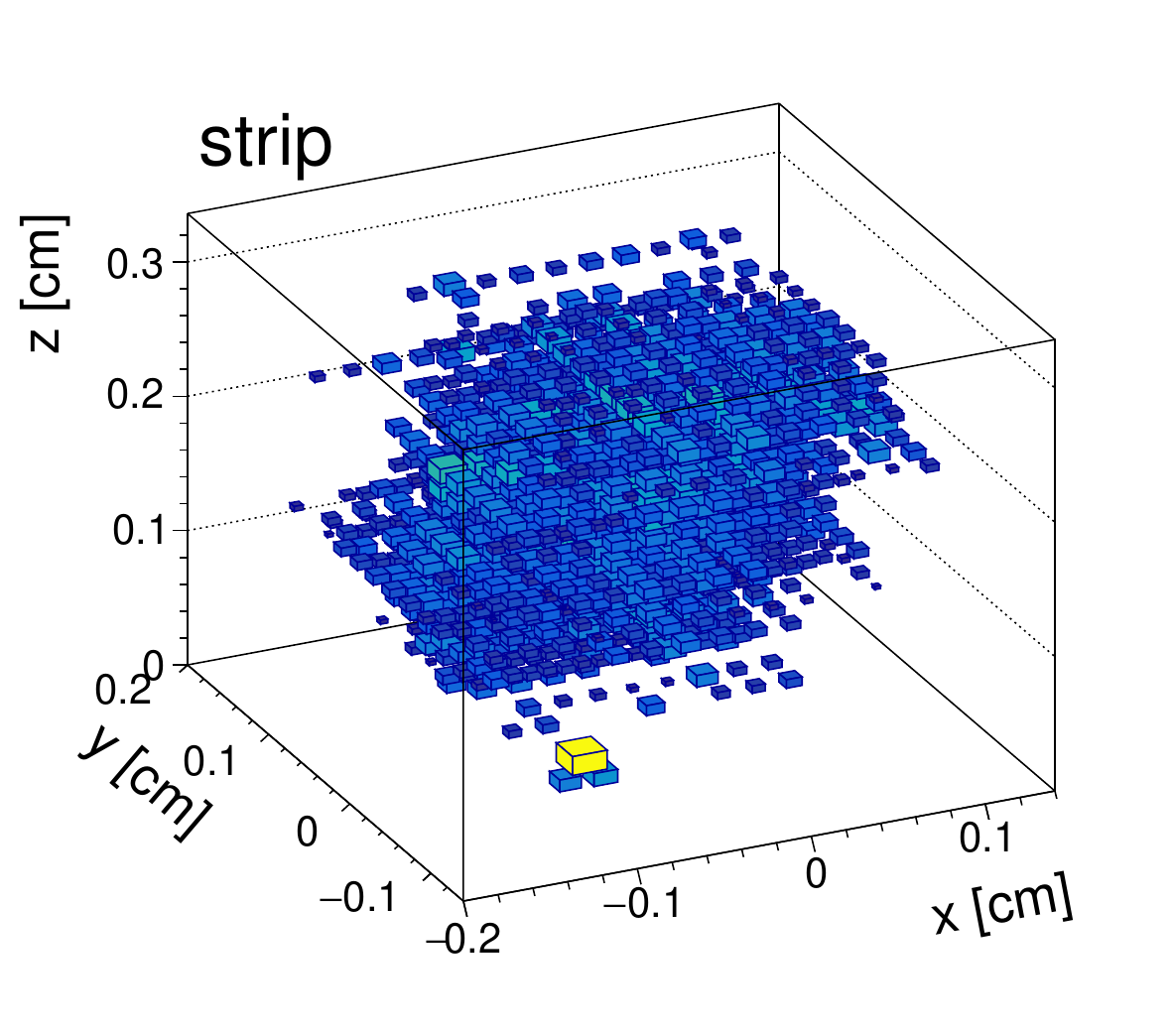}
\includegraphics[clip, trim=0.0cm 1cm 1.0cm 1cm, width=0.66\columnwidth]{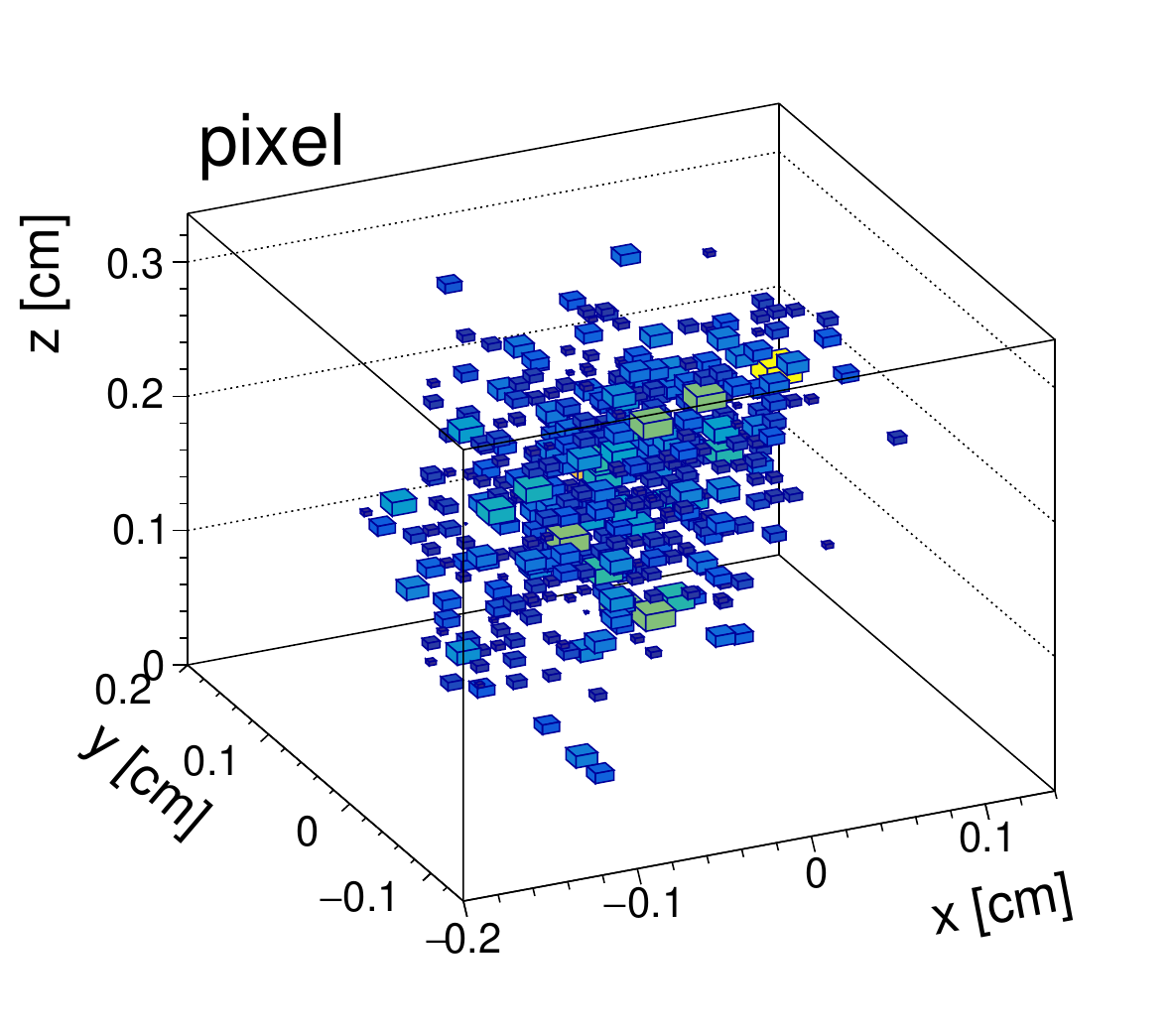}
\caption{Simulated 25~\kevr~helium recoil event in 740:20 Torr He:SF$_6$ gas before drift (top left), after 25 cm of drift (top right), and as measured by six readout technologies (remaining plots as labeled). Readout noise and threshold effects have been disabled.}\label{fig:recoil_event_displays}
\end{center}

\end{figure*}
\begin{figure*}[htb]
\begin{center}
\includegraphics[width=0.8\columnwidth]{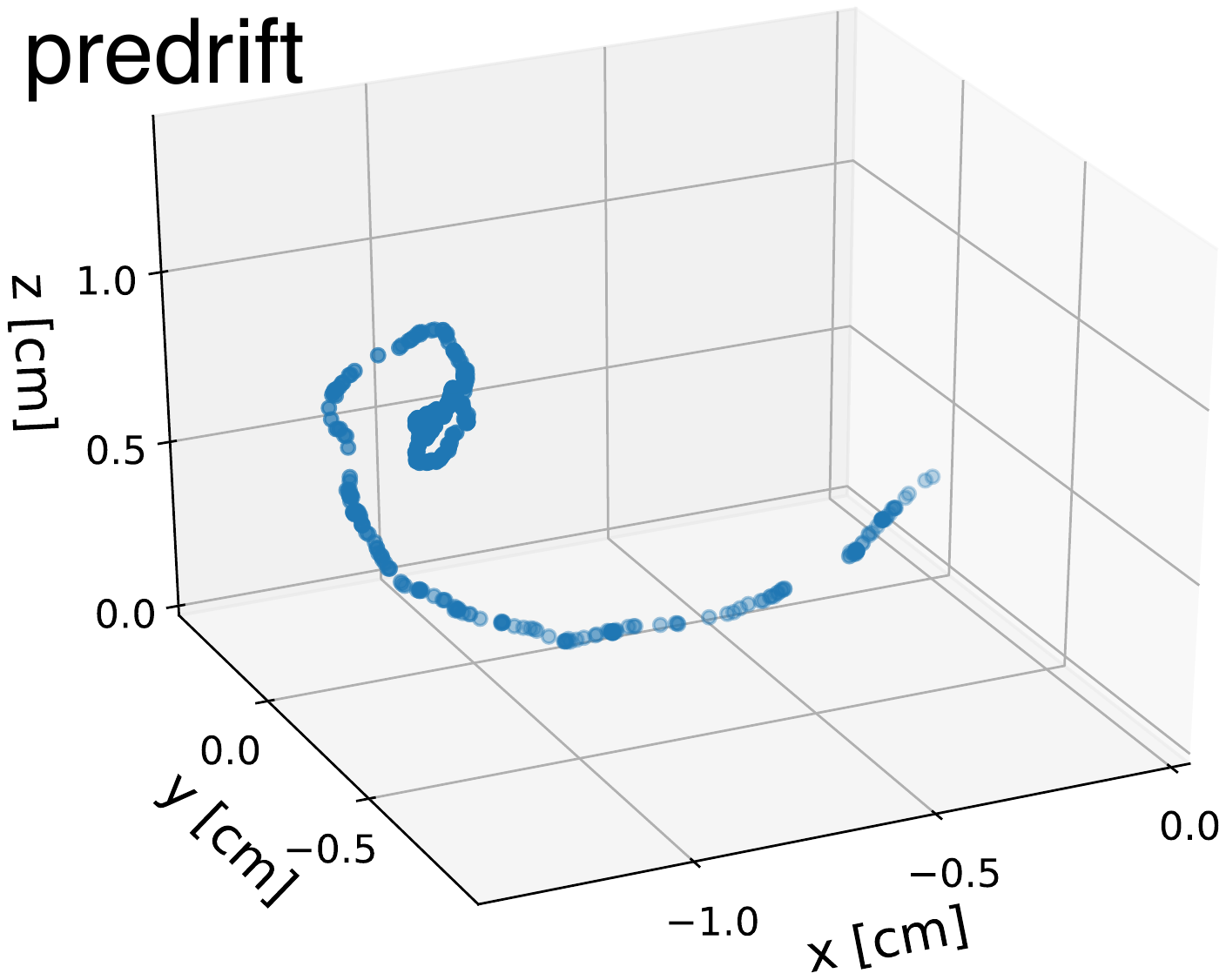}
\includegraphics[width=0.8\columnwidth]{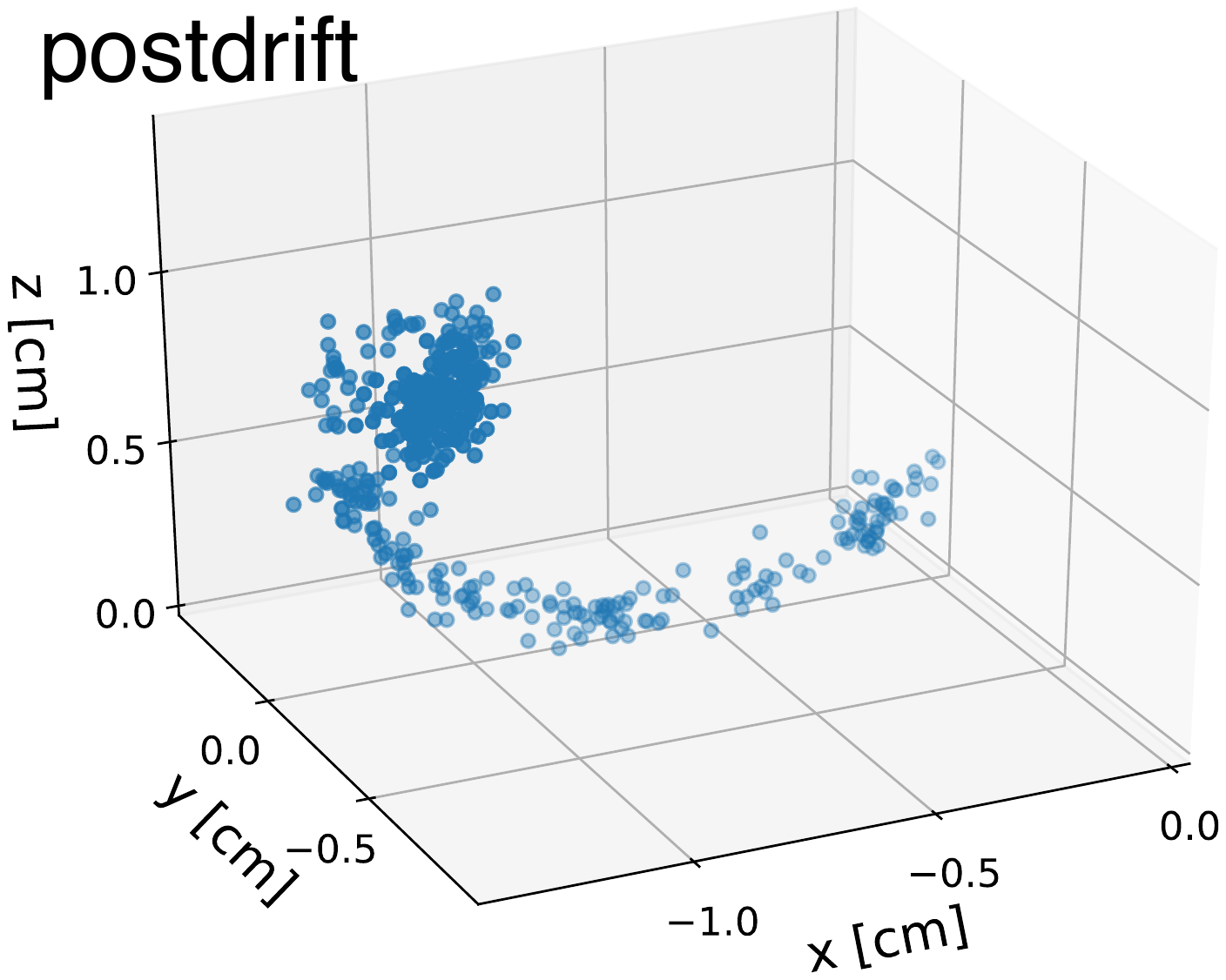}\\
\includegraphics[clip, trim=0.0cm 1cm 1.0cm 1cm, width=0.66\columnwidth]{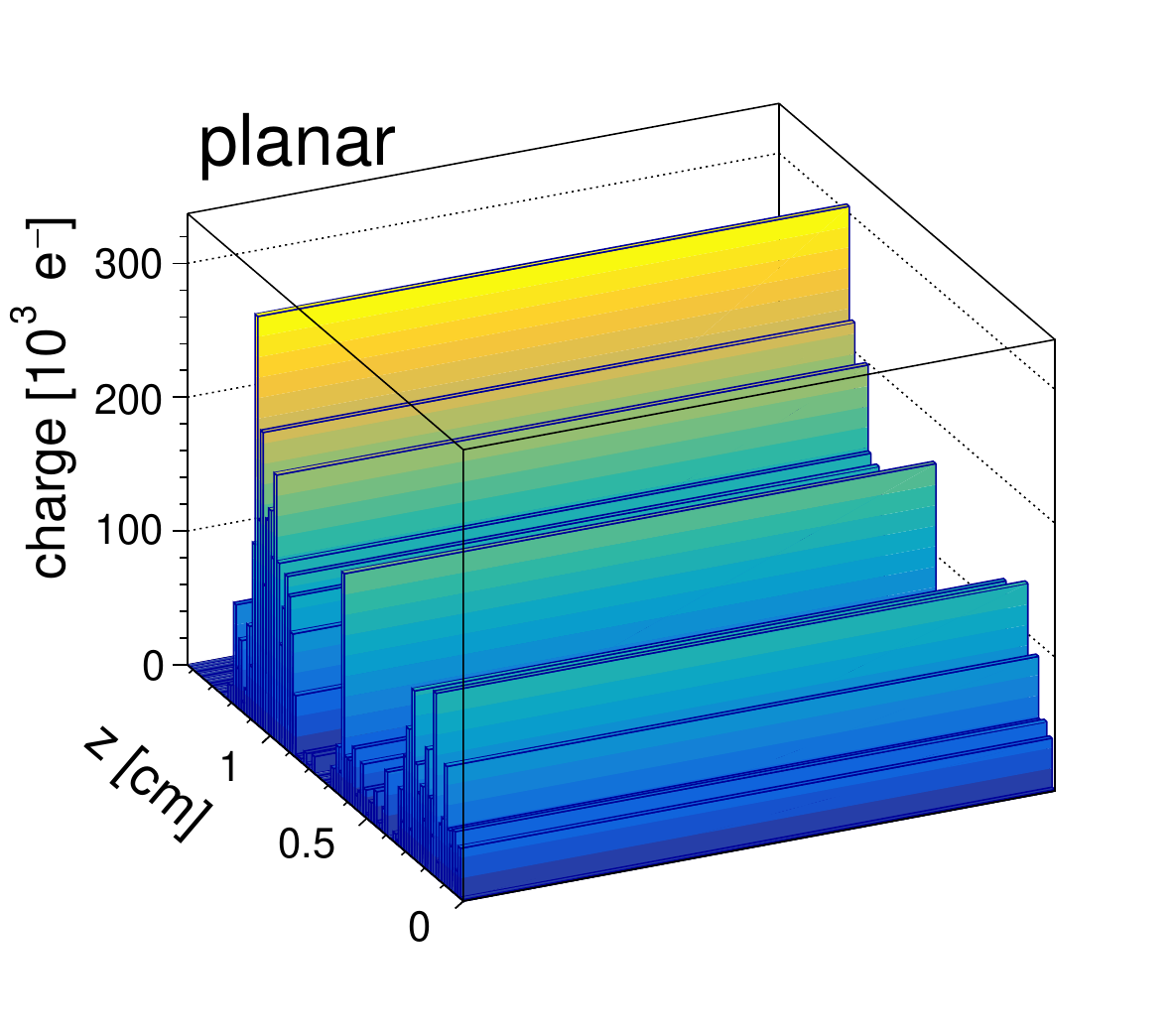}
\includegraphics[clip, trim=0.0cm 1cm 1.0cm 1cm, width=0.66\columnwidth]{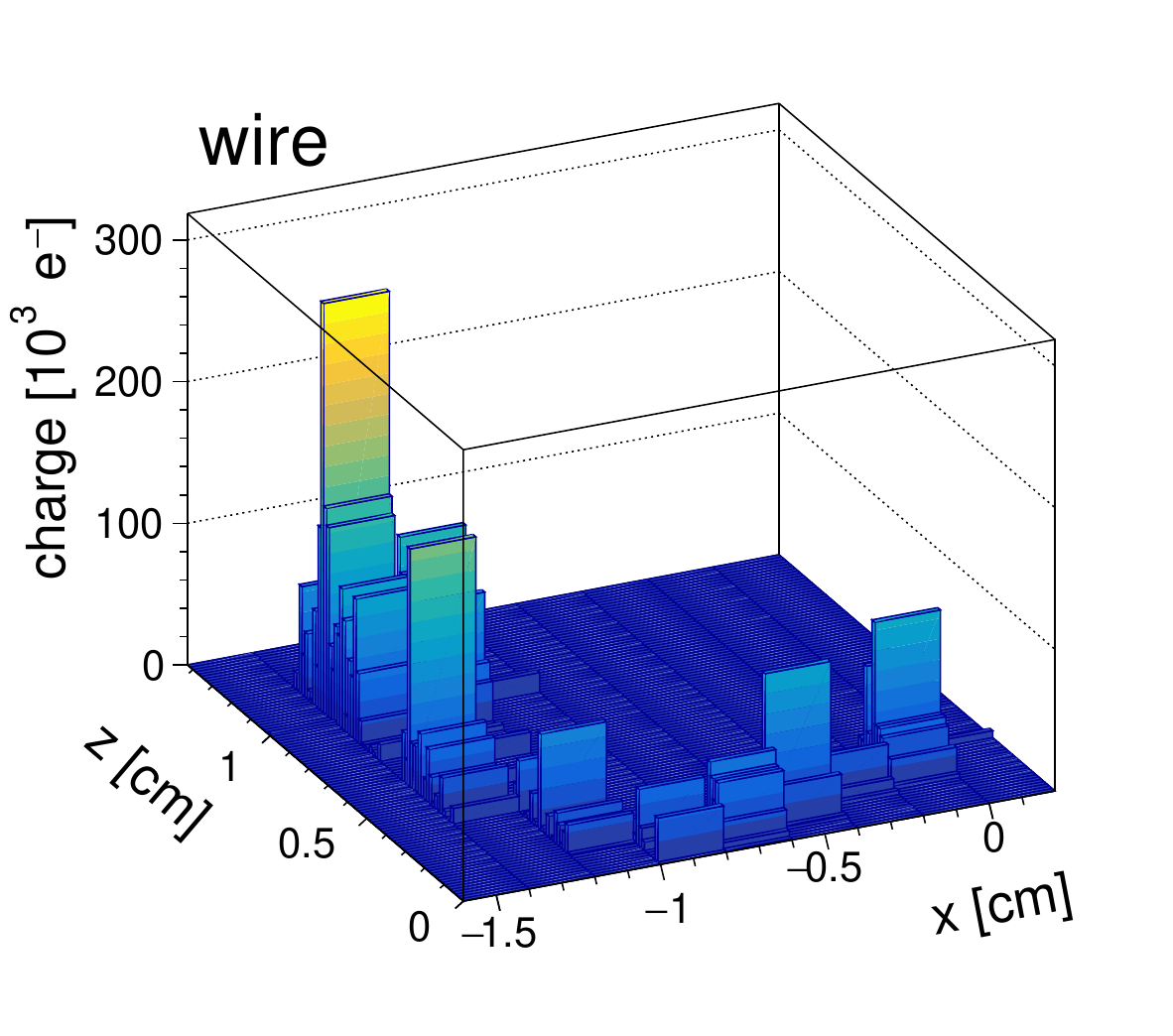}
\includegraphics[clip, trim=0.0cm 1cm 1.0cm 1cm, width=0.66\columnwidth]{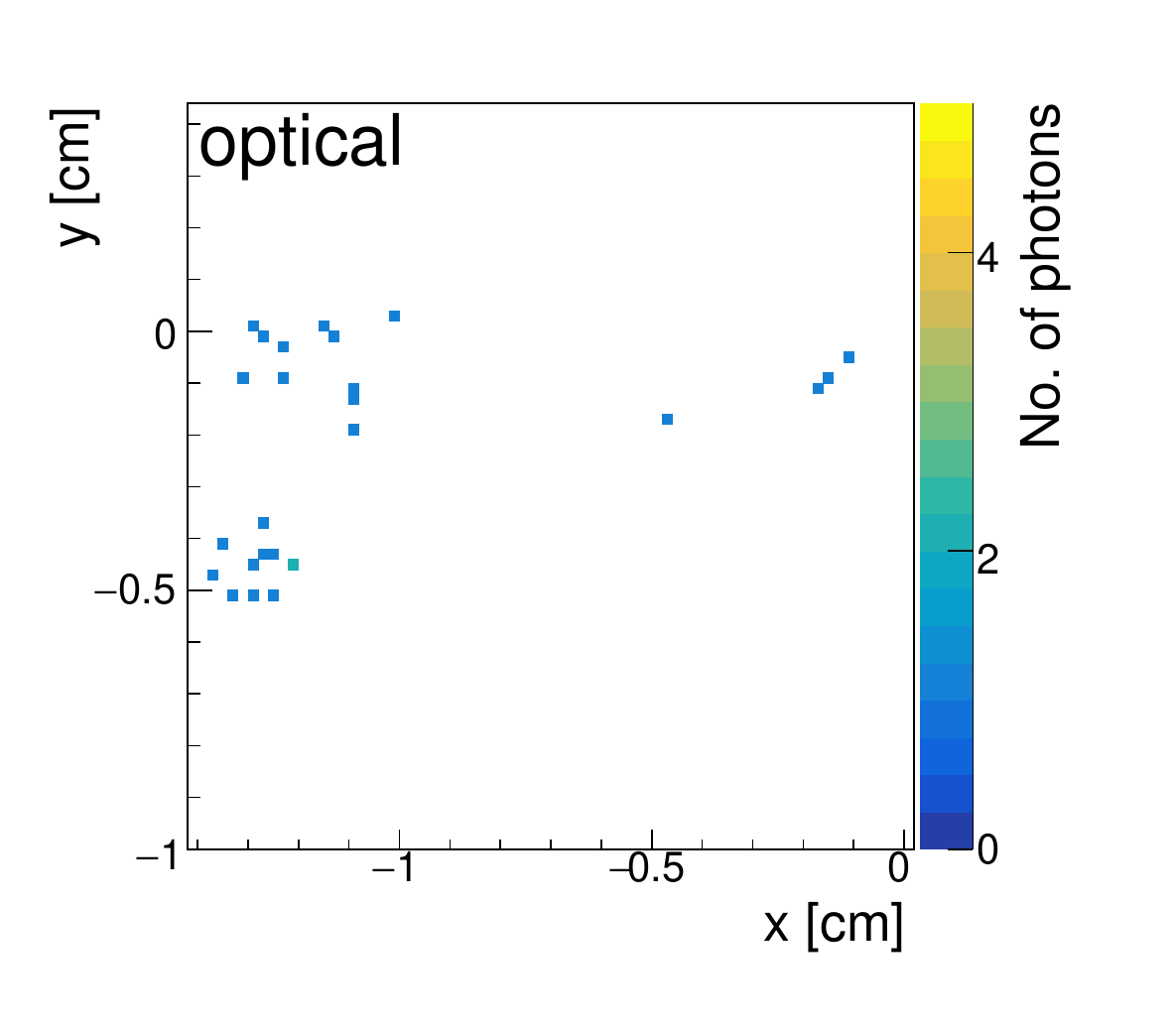}\\
\includegraphics[clip, trim=0.0cm 1cm 1.0cm 1cm, width=0.66\columnwidth]{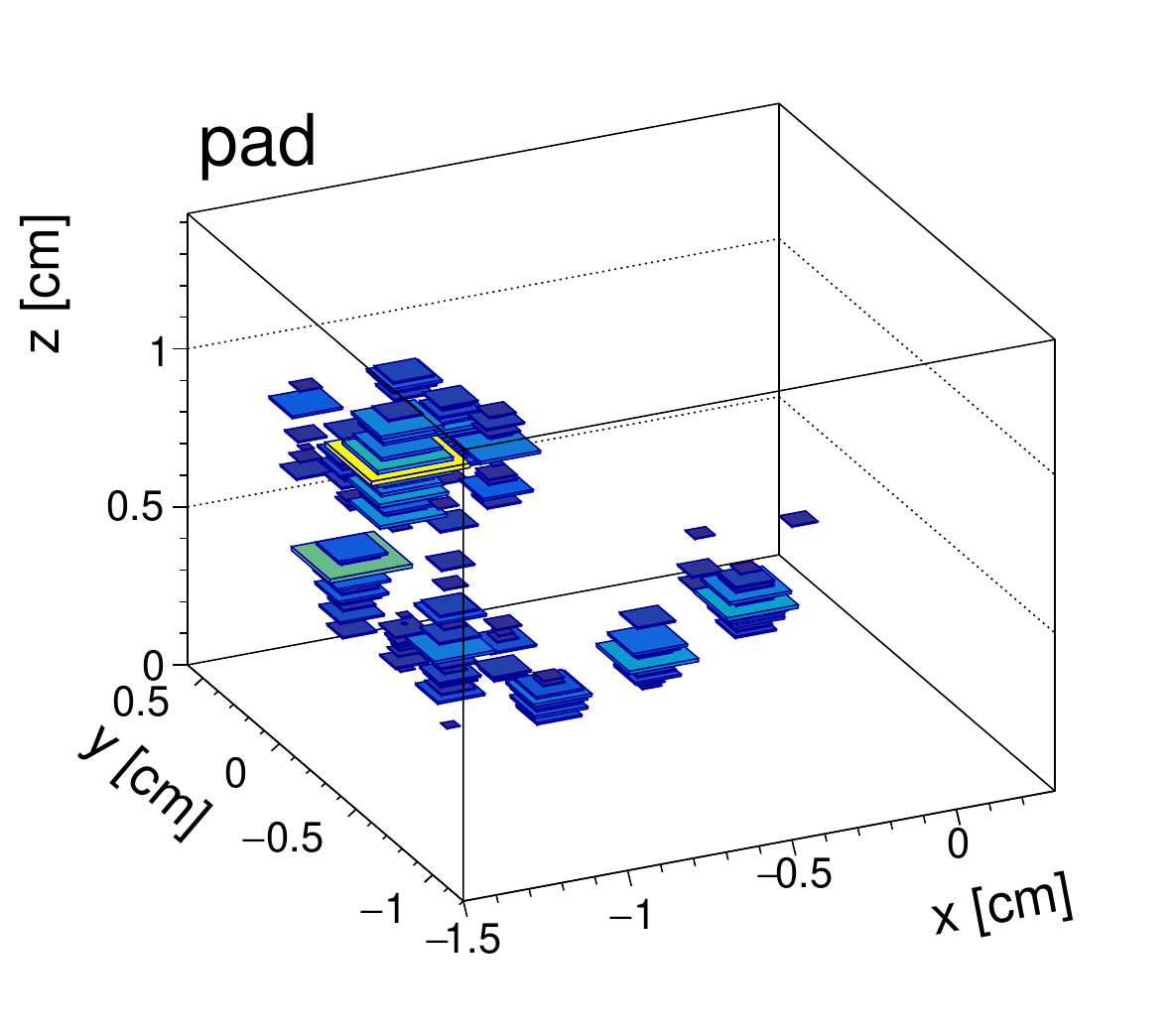}
\includegraphics[clip, trim=0.0cm 1cm 1.0cm 1cm, width=0.66\columnwidth]{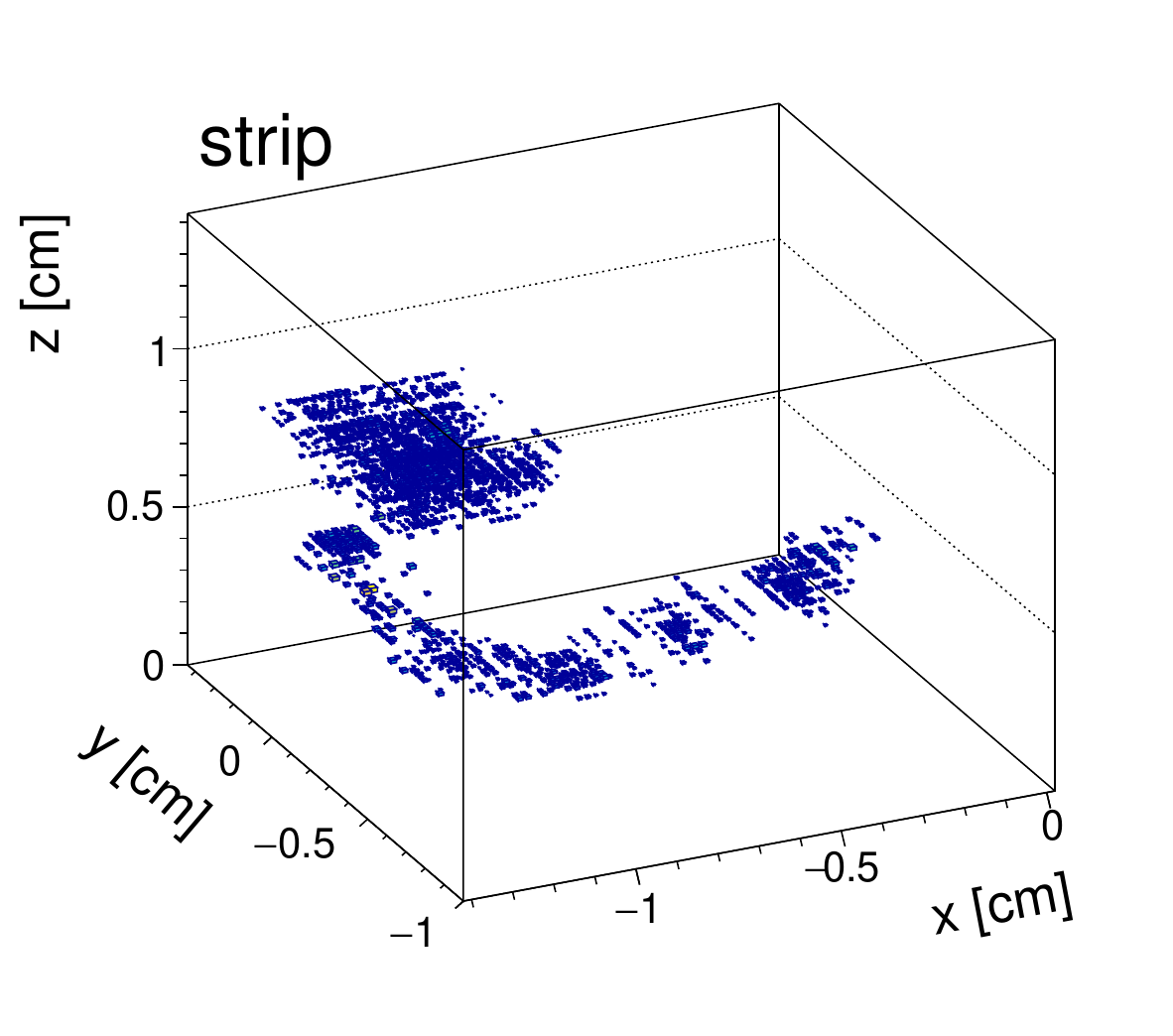}
\includegraphics[clip, trim=0.0cm 1cm 1.0cm 1cm, width=0.66\columnwidth]{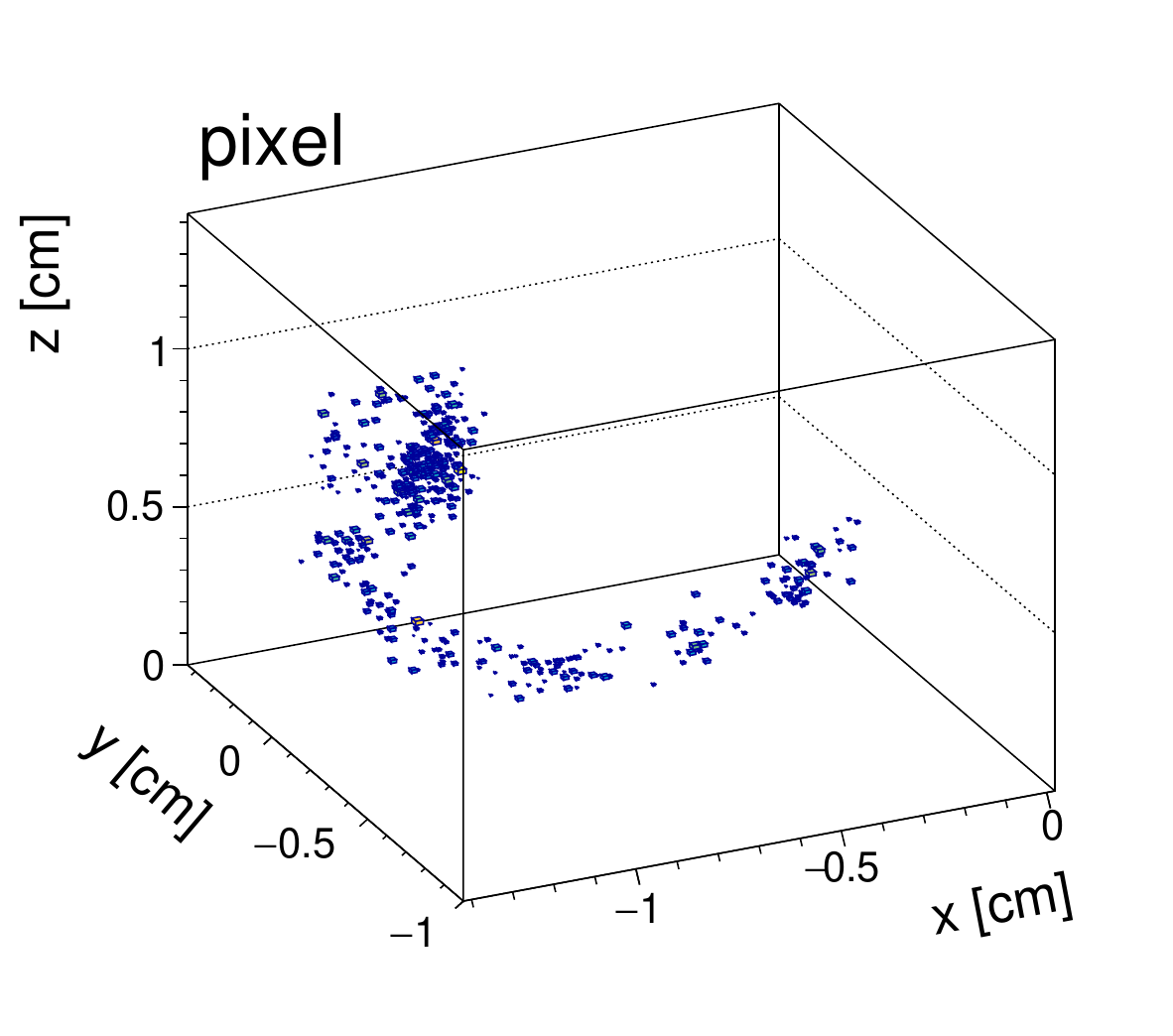}
\caption{Simulated 20~keV~electron event in 740:20 Torr He:SF$_6$ gas before drift (top left), after 25 cm of drift (top right), and as measured by six readout technologies (remaining plots as labeled). Readout noise and threshold effects have been disabled.}\label{fig:electron_event_displays}
\end{center}
\end{figure*}

After generating primary ionization distributions, we assume that all electrons attach to fluorine atoms, converting into negative ions, and then drift towards the readout plane. We simulate the combined effect of this attachment, the drift of individual ions, and the subsequent gas avalanche multiplication in the TPC using the diffusion and gain parameters listed in Table~\ref{tab:gas_specs}. For pure SF$_6$, these parameters have already been experimentally established. For the \hesfsix mixture, although the exact gas parameters have not been measured, SRIM predictions for recoil properties are similar to what is obtained by simulating helium and fluorine recoils in pure SF$_6$ of the same total density. In the diffusion and gas gain simulation, we assume that by adjusting the electric field strength and the gain stage, the same diffusion, drift velocity, and gas gains as measured in pure SF$_6$ can be achieved. Experimental work has shown the feasibility of obtaining gain in \hesfsix and He:CF$_4$:SF$_6$ mixtures, at both low ($\sim$ 100 Torr~\cite{gasmixgain}) and high ($\sim$ 600 Torr~\cite{Baracchini:2017ysg}) pressure.

The simulated drift length of each electron (negative ion in the case of NID) differs. The distance of the WIMP interaction vertex position to the readout plane is pulled from a uniform distribution ranging from 0 to \SI{50}{cm}. Each electron (ion for NID) in the generated recoil is then placed at the appropriate position in the detector and smeared with a Gaussian diffusion dependent on its distance to the readout plane. Edge effects in the drift direction are included by deleting charges that start outside the fiducial volume. Amplification is also simulated at the single electron (or ion) level, using an exponential distribution with mean equal to the avalanche gain. This produces an energy-dependent fractional gain resolution at the event level, which in turn results in an energy-dependent energy resolution, with $\sigma_E/E \sim \mathcal{O}(10\%)$ at \SI{5.9}{keV_{ee}}, typical for gas detectors. Readout noise broadens this resolution further, and is discussed later. For each gas, the diffusion and gain parameters from Table~\ref{tab:gas_specs} are fixed for all readouts to ensure a fair comparison. This means that we are essentially comparing detectors with identical gain stages, but different charge readout technologies. While many combinations of gain stages and readouts are possible, we keep the gain stage fixed so as to focus on how the readout affects the final directional performance.

Figure~\ref{fig:nuclear_recoils} shows 20~\kevr~helium and fluorine recoils in \SI{20}{Torr} of SF$_6$ gas. After \SI{25}{cm} of drift the fluorine recoil direction is rather washed out, but it can still be approximately determined by eye in the case of the helium recoil. Figure~\ref{fig:electron_recoils} shows a \SI{20}{keV_{ee}} electron event before and after the initial ionization distribution has drifted \SI{25}{cm} in the detector. For a gas pressure of \SI{20}{Torr}, the event topology is still preserved after diffusion of drift charge, and can be reconstructed with a detector capable of high-resolution charge readout. For a gas pressure of \SI{200}{Torr}, such an electron event would have a reduced range, and the topology is therefore mostly washed out by diffusion after \SI{25}{cm} of drift.

Comparing the bottom rows of Figs.~\ref{fig:nuclear_recoils} and \ref{fig:electron_recoils} it can be seen that nuclear and electron recoils with similar amounts of ionization (accounting for the quenching factor) differ both in the length and topology of the charge cloud. In particular  the charge cloud from the electron recoil is longer and less uniform. Together these results show that in \SI{20}{Torr} SF$_6$, the ionization distribution can be used to determine the recoil direction and discriminate between nuclear and electron recoils--even after charge diffusion from realistic drift lengths. Helium is a preferable target for extending our low WIMP mass reach. When adding helium, we must keep the overall mass density low, to preserve recoil lengths and hence directionality and particle identification capabilities. Here we accomplish this by maintaining atmospheric pressure and reducing the partial pressure of SF$_6$. An alternative strategy is to operate at sub-atmospheric pressure.

\begin{table*}[ht]
	\centering
	\begin{tabular} {l|cccccc}
	\toprule 
	Readout type	& Dimensionality	& Segmentation ($x \times y$)	&  \quad Capacitance  [$pF$]	\quad & \quad $\sigma_{\textrm{noise}}$ in 1 $\upmu$s   \quad	& \quad Threshold/$\sigma_{\textrm{noise}}$ \quad \\
	 \hline
	planar		& 1d ($z$)	& 10 cm $\times$ 10 cm		& 3000  				& 18000~$e^-$				& 3.09\\ 	
	wire			& 2d ($yz$)	& 1 m wires, 2 mm pitch 		& 0.25					& 800~$e^-$	 			& 4.11\\
	pad			& 3d ($xyz$)	& 3 mm $\times $ 3 mm 		& 0.25 		& 375~$e^-$				& 4.77\\
	optical 		& 2d ($xyz$)	& $200~\upmu \textrm{m} \times 200~\upmu \textrm{m}$ & n/a				& 2 photons			& 5.77\\
	strip			& 3d ($xyz$) & 1 m strips, 200 $\upmu$m pitch  		& 500		& 2800~$e^-$				& 4.61\\
	pixel			& 3d ($xyz$) 	& $200~\upmu \textrm{m} \times 200~\upmu \textrm{m}$	& 0.012 - 0.200 & 42~$e^-$					& 5.77\\
	 \bottomrule
	\end{tabular}
		\caption{List of readout-specific parameters that are used in the simulation of each technology we consider here. The capacitance, which determines the noise level, is listed as that for a single detector element. For the optical readout, a yield of $7.2\times 10^{-6}$ photons per avalanche electron is used to account for the combined effects of photon yield, geometric optical acceptance, optical transparency, and quantum efficiency.}\label{tab:readout_specs}
\end{table*}

\subsubsection{Simulation of charge readout}

A realistic detector requires a charge readout that is capable of preserving at least part of the topology of the charge cloud. We compare the performance of six charge readout technologies in doing this. We discuss the advantages and disadvantages of each technology below, in order of increasing cost, complexity and performance (the parameters used to simulate each one are also summarized in Table \ref{tab:readout_specs}).

\noindent {\bf Planar} denotes a simple but cost-effective readout where only the time-dependence of the avalanche charge is recorded, with 10 $\times$ 10 cm segmentation. This results in a signal that is effectively a 1d projection of the charge cloud. One implementation of such a readout would be to directly read the signal from the gain stage, for instance using a GEM with a digitizer. This approach has the advantage of having very low cost per unit area but also has several drawbacks resulting from the low segmentation transverse to the drift axis. This leads to reduced directionality, reduced electron rejection, and high capacitance, which in turn means high readout noise and high energy thresholds.

\noindent {\bf Wire} readouts are traditional MWPCs (see Sec.~\ref{sec:technology_choices}). They are of interest because they can be made highly radiopure, and have very small capacitance and hence low readout noise. One drawback however is that constraints on wire spacing limit the minimum segmentation of the readout plane. 

\noindent {\bf Pad} readouts with mm-scale feature size are under consideration for liquid-argon neutrino detectors~\cite{Dwyer}. The smaller segmentation leads to much lower noise, but also much higher channel counts and higher instrumentation cost than for planar readout. The transverse feature size is still larger than ideal for directional detection of WIMP and neutrino nuclear recoils.

\noindent {\bf Optical} readouts (CCDs or CMOS) can be used to image scintillation light from the amplification region. This approach may lead to lower backgrounds because the cameras can be placed outside optical ports in the gas vessel. Optical sensors provide exceptionally high segmentation and low noise but have comparatively poor temporal resolution. As a result, such readouts integrate the signal in the drift direction, resulting in a 2d projection of the charge cloud. TPCs with optical readout are cost-effective in that they need only a few cameras per unit readout area, but the effective charge collection efficiency consequently suffers due to geometric photon acceptance losses which can be as severe as $\mathcal{O}(10^{-4})$.

\noindent {\bf Strip} readouts, for example strip micromegas or $\upmu$PIC, use the coincidence of signals in orthogonal strips to create 3d hits. Such readouts are a good compromise between performance and cost. 

\noindent {\bf Pixel} readouts with application-specific integrated circuit (ASIC) chips generally have the lowest noise and highest performance, but are the most costly. Due to the relatively small $\mathcal{O}({\rm cm^2})$ area of typical ASICs, the full readout plane for \Cygnus would require a substantial number of chips and be quite labor-intensive to implement.

We note that readouts can be combined, for instance, Optical 3d readout can be achieved by combining the signal from 2d CCD or CMOS sensors with an independent 1d measurement of the time-dependence of the avalanche charge. See Section~\ref{sec:opt}.

Our aim here is to compare these quite different charge readout technologies in a fair and unified simulation framework, without invoking any technology-specific signal processing or reconstruction algorithms. In that spirit, we assume that the amount of charge arriving at each detection element can be recorded at 1~$\upmu$s intervals. For readout technologies that have already been constructed and characterized, we scale known readout noise levels to 1~$\upmu$s. For readouts that do not yet exist, such as the planar readout, we estimate the readout noise for this time interval from the typical capacitance of a detection element and the noise curves of a candidate charge sensitive preamplifier \cite{DeGeronimo:2011zz}. Capacitance values and the noise estimates used in the simulation are summarized in Table~\ref{tab:readout_specs}. We assume that before event reconstruction a hard threshold is applied to the recorded signal in each channel. This threshold is set so that for each readout the expected rate of noise hits is equal to $10^{-5} /($cm$^2\upmu$s). As a result, the ratio of the threshold to the noise level, threshold/$\sigma_{\textrm{noise}}$, varies from approximately three for the planar readout to about six for the pixel ASIC readout. The threshold/$\sigma_{\textrm{noise}}$ values resulting from these noise requirements are larger for the more segmented readouts, which agrees with common practice in the field. 

Since the optical readout cannot be sampled at the same speed as the other readouts, the same procedure cannot be used to determine an analysis threshold. In Table~\ref{tab:readout_specs} we instead set a requirement of $10^{-5} /$cm$^2$ noise hits per {\it event} on average, but the optical readout performance depends strongly on this threshold. For that reason, a fair comparison of optical and charge readout is not straightforward. In the comparisons that follows, we therefore focus only on charge readout, and have a short separate discussion of optical readout in Section~\ref{sec:opt}.

Figure~\ref{fig:recoil_event_displays} shows how a 25~\kevr~helium recoil in 740:20 Torr \hesfsix gas that has drifted 25~cm and is subsequently detected by each of the six readout types. The events are shown after simulation of gain, gain resolution, and spatial quantization into detection elements, using the parameters from Table~\ref{tab:readout_specs}. Noise has been disabled for this particular visualization, otherwise noise hits obscure the 3d event displays. 
Similarly, Fig.~\ref{fig:electron_event_displays} shows a 20 \kevee~electron in the same gas mixture and with the same detector simulation. In all cases, the detected event topology or waveform differs visibly from those of the typical nuclear recoil in Figure~\ref{fig:recoil_event_displays}. For a 755:5 Torr \hesfsix mixture, the main change is that the recoil lengths approximately double, improving directionality and discrimination between nuclear recoils and electron recoils. This suggests electron rejection is achievable in the $\mathcal{O}(1$--$10$ keV) energy range, perhaps even with very simple readouts like the planar. We extend this discussion quantitatively in Sec.~\ref{sec:electron_rejection}.

\subsection{Analysis methodology}\label{subsec:Analysis}

\begin{figure*}[hbt]
\begin{center}
\includegraphics[clip, trim=0.0cm 0cm 1.8cm 1.8cm, width=0.45\textwidth]{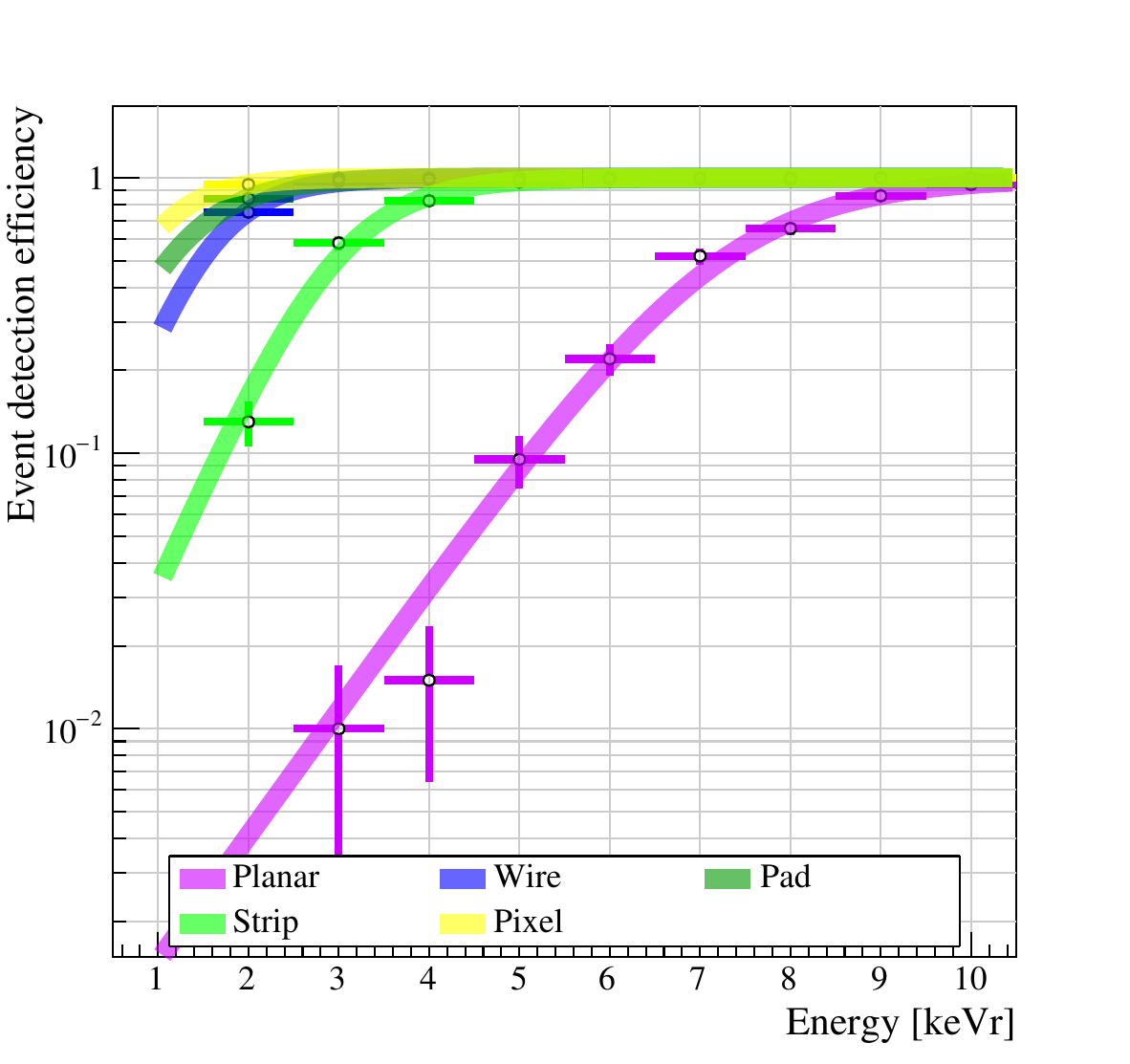}
\includegraphics[clip, trim=0.0cm 0cm 1.8cm 1.8cm, width=0.45\textwidth]{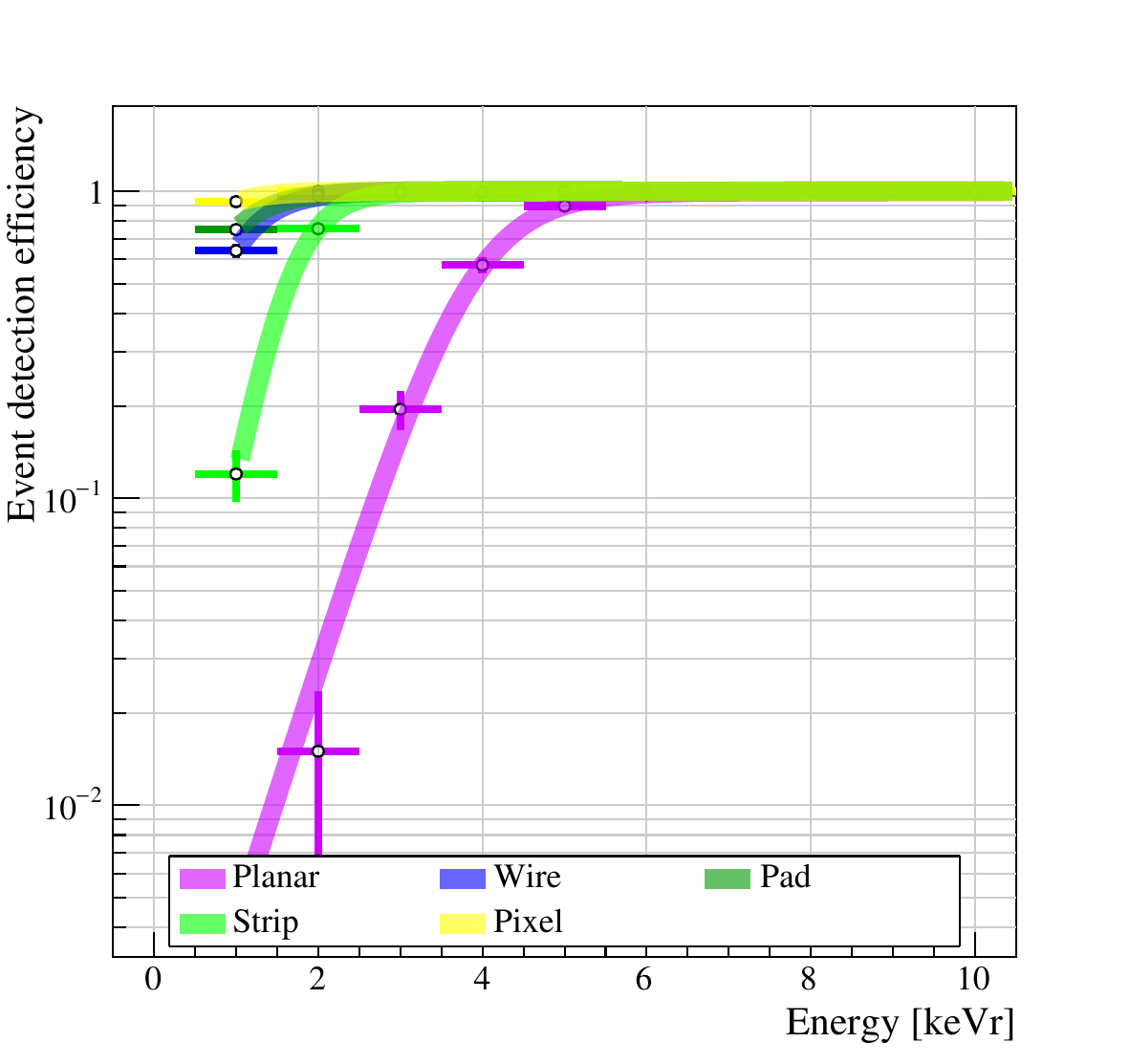}
\caption{Event-level detection efficiency of each readout for fluorine (left) and helium (right) recoils in a 755:5 Torr gas mixture of \hesfsix. Error bars are binominal standard deviations, assuming the efficiency measured is the true efficiency.}
\label{fig:event_efficiency}
\end{center}
\end{figure*}

\begin{figure*}[hbt]
\begin{center}
\includegraphics[clip, trim=0.0cm 0cm 1.8cm 1.8cm, width=0.45\textwidth]{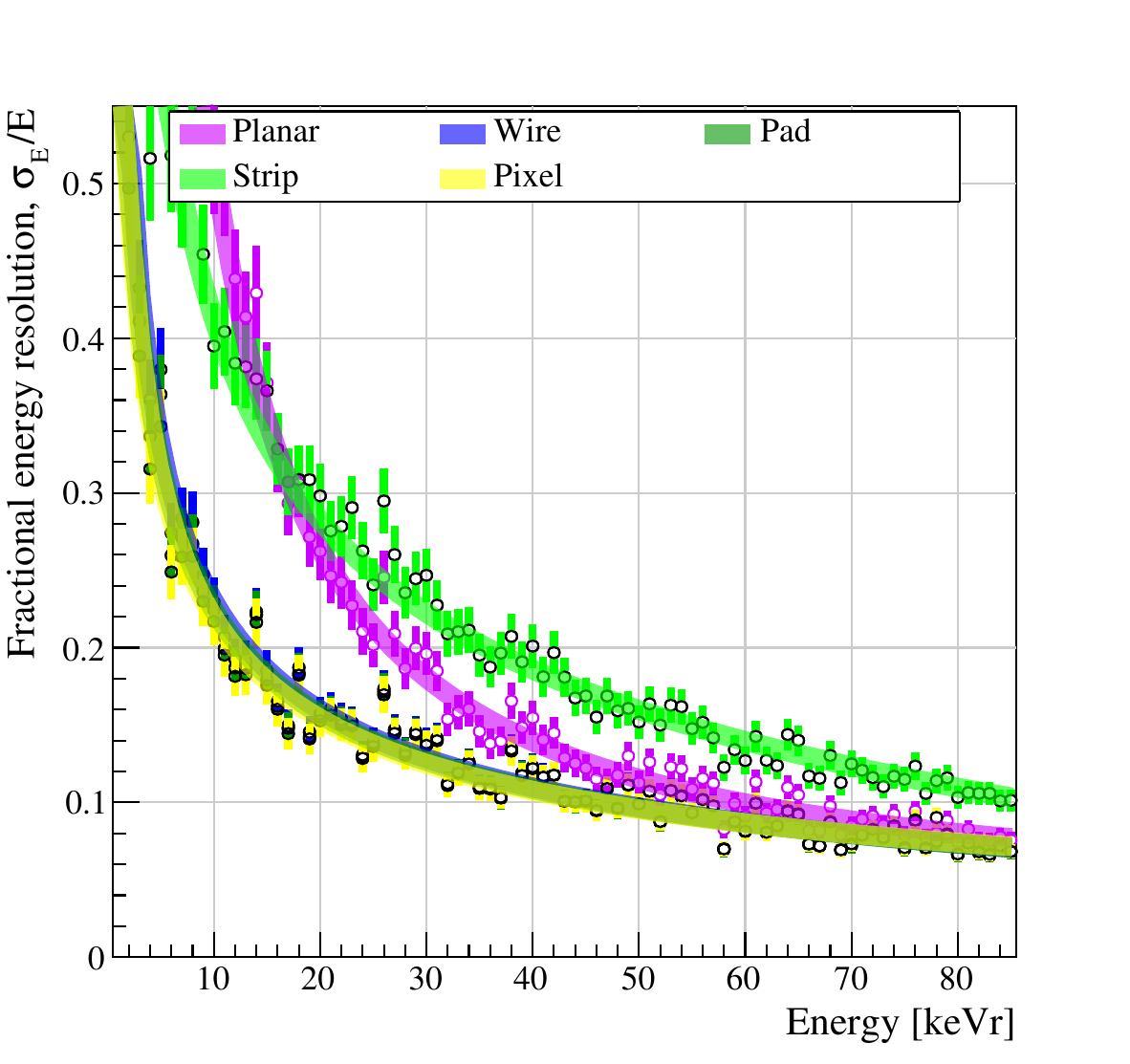}
\includegraphics[clip, trim=0.0cm 0cm 1.8cm 1.8cm, width=0.45\textwidth]{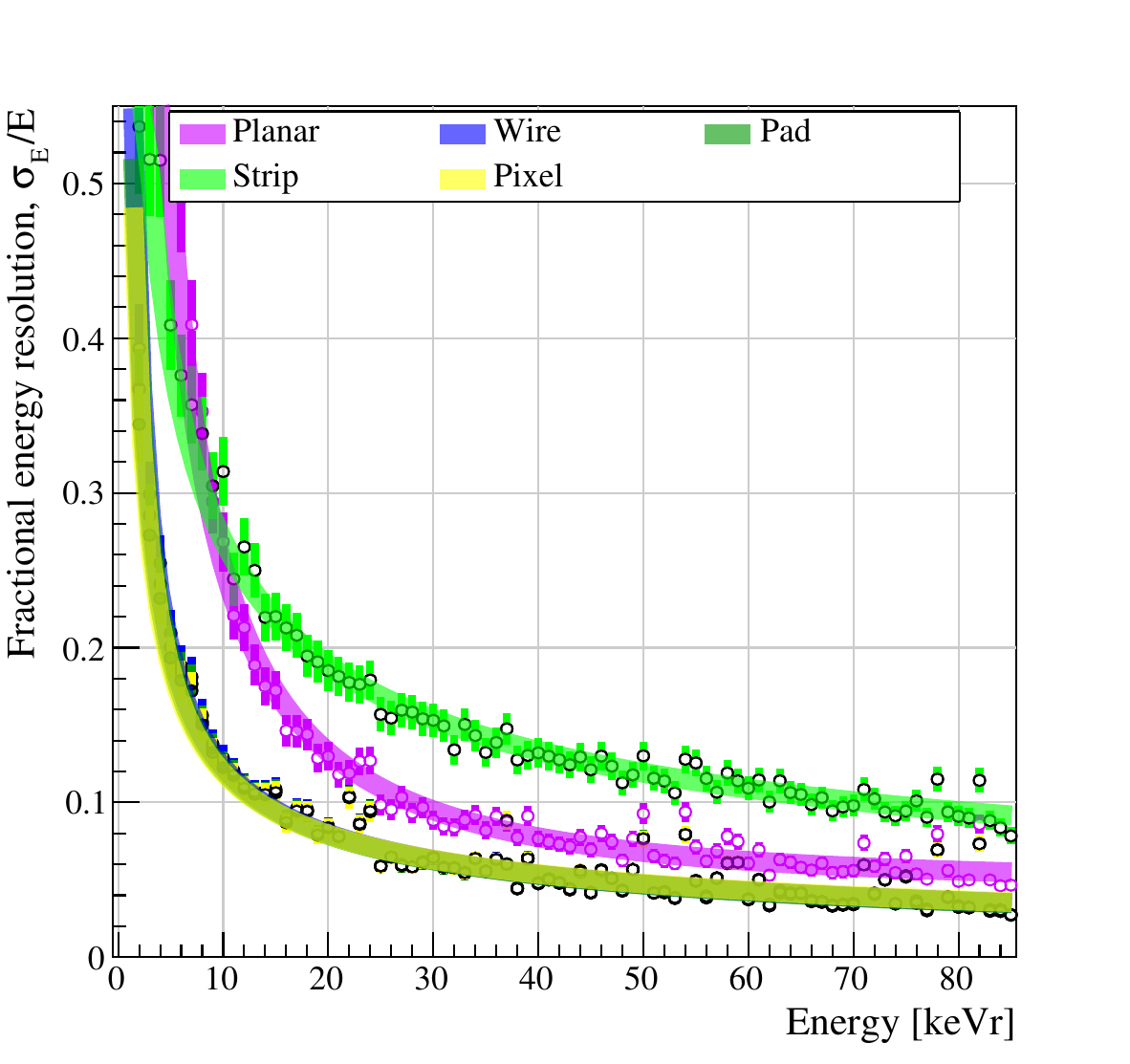}
\caption{Fractional energy resolution of each readout for fluorine (left) and helium (right) recoils in a 755:5 Torr gas mixture of \hesfsix. Error bars indicate the RMS variation in the recoil sample analyzed. Note that the curves for the pad readout are very close to the pixel and wire cases so are not easily visible here. }
\label{fig:energy_resolution}
\end{center}
\end{figure*}

\begin{figure*}[hbt]
\begin{center}
\includegraphics[clip, trim=0.0cm 0cm 1.8cm 1.8cm, width=0.45\textwidth]{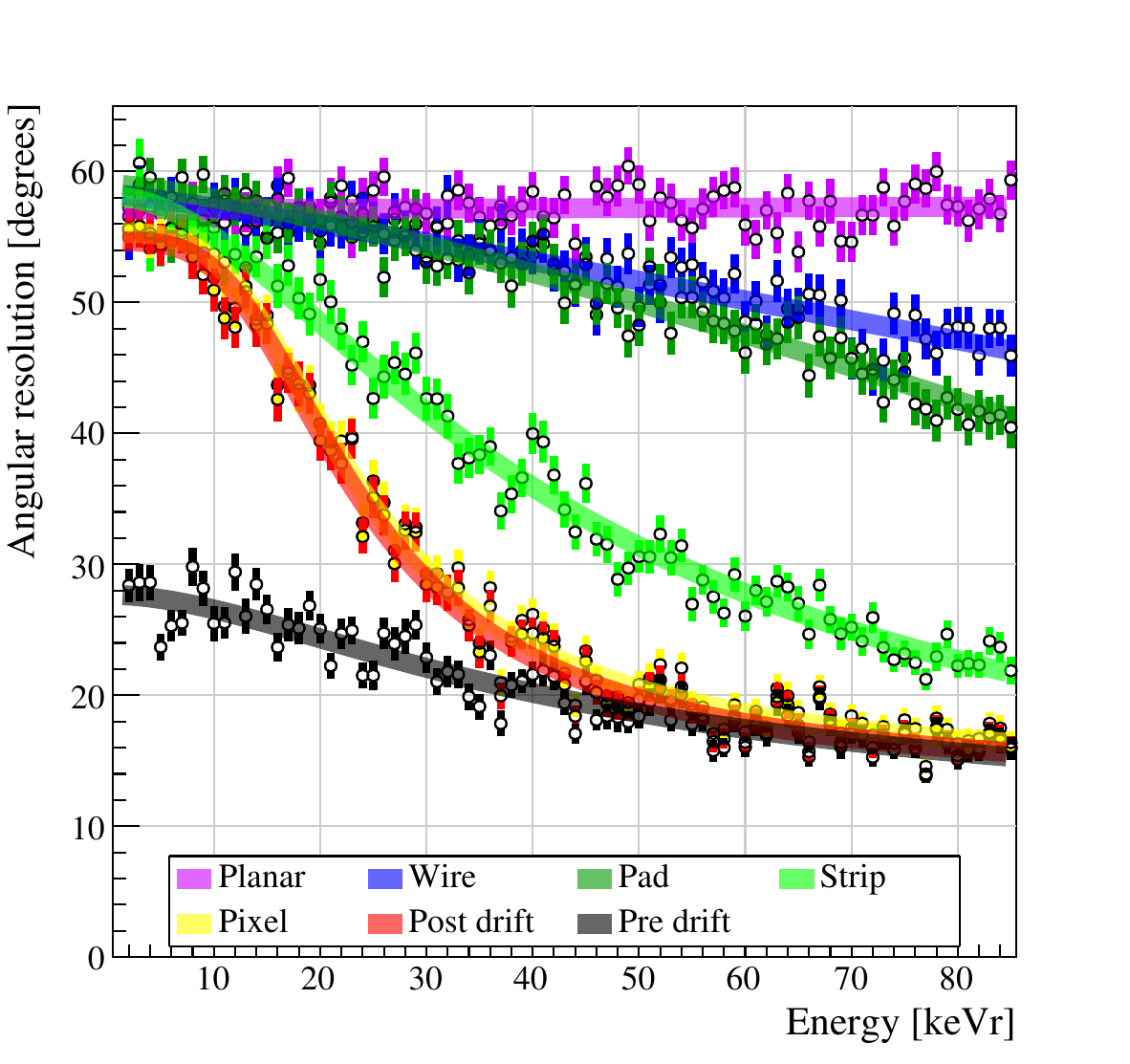}
\includegraphics[clip, trim=0.0cm 0cm 1.8cm 1.8cm, width=0.45\textwidth]{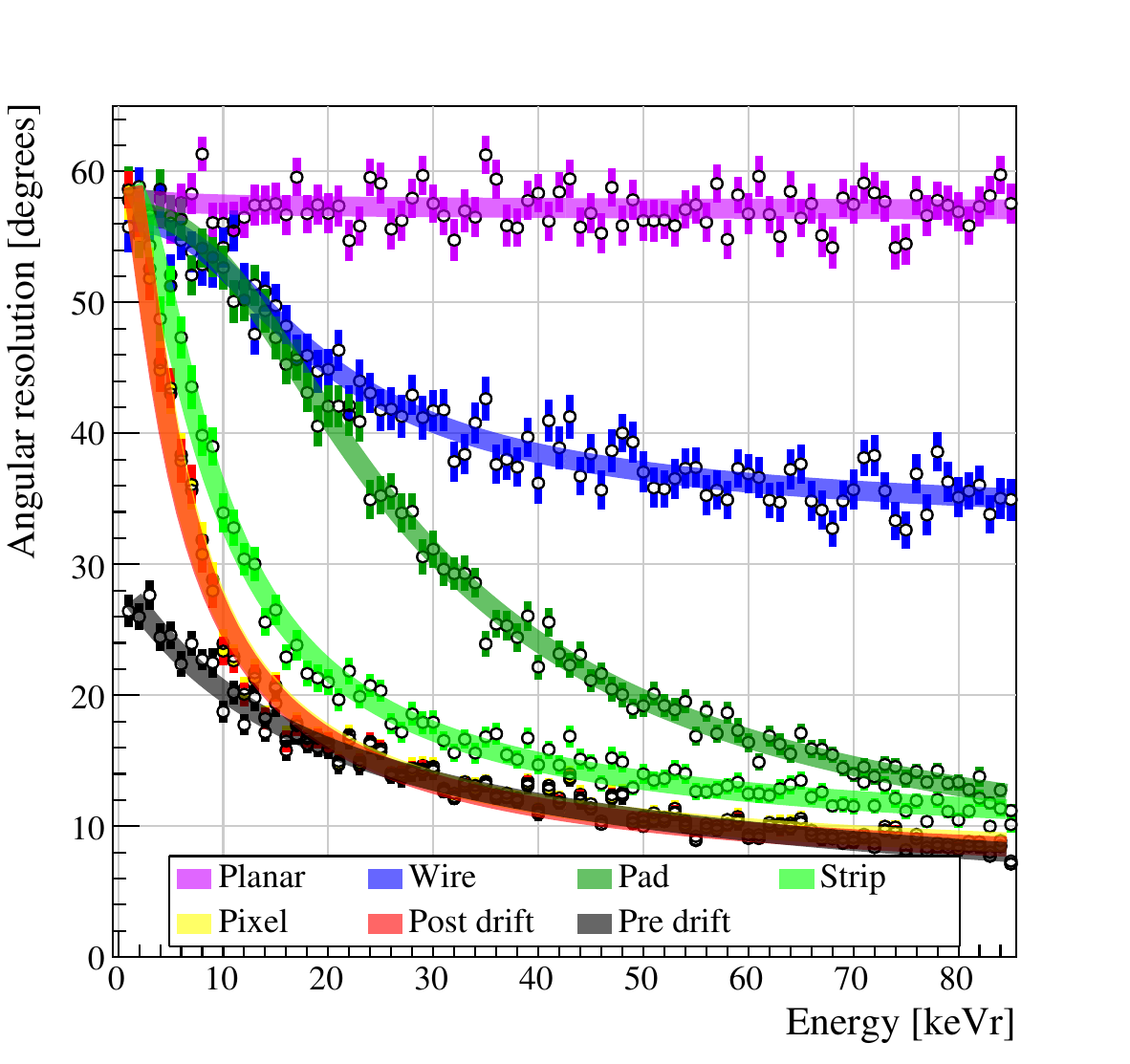}
\caption{Angular resolution of each readout for fluorine (left) and helium (right) recoils in a 755:5 Torr gas mixture of \hesfsix. Each color denotes a different readout as shown in the legend. Data points show the mean mismeasurement of the recoil axis direction determined from two hundred pseudo-experiments. Error bars show the statistical uncertainty of this mean. Lines are analytical functions fit to the data points, used to parameterize performance versus energy. The resolution that can be obtained with a perfect readout before (black) and after (red) charge diffusion during drift is also shown for comparison.}
\label{fig:angular_resolution}
\end{center}
\end{figure*}

In order to analyze the data from the different readout technologies in a uniform way, we turn the simulated output of each readout into 3d spatial points as follows. After the gain and noise simulation, charge above threshold in each detector element and temporal sampling interval is quantized by dividing the charge by the mean gain and turning the results into an integer. This quantized charge is then assigned an absolute $(x,y)$ (transverse to drift) coordinate based on the location of the center of the detector element, and a relative $z$ (parallel to drift) coordinate based in the detection time relative to the first charge above threshold in the event. 

Once the reconstructed spatial distribution of the charge cloud has been obtained, its primary axis is found via singular value decomposition (SVD). At this stage the reconstructed primary axis vector has an arbitrary sign, so we only have axial-vector information. The sign of the vector is reconstructed based on the projection of the positions of the quantized charges onto the primary axis. Each charge is assigned to one half of the track, based on whether the charge is closer to the charge with the minimum or maximum projection onto the reconstructed primary axis. We assume that the kinetic energy of the recoiling nucleus being reconstructed is sufficiently low so as to be below the Bragg peak; so we take the track end with the largest assigned charge as the tail. Our analysis methodology neglects all charge below threshold, but it could be partially recovered with a more sophisticated noise suppression algorithm. This would improve reconstruction performance further, therefore all results on the directional performance do stand to improve. However such an algorithm would likely need to be readout-specific, while here we aim to be readout-general to first establish the optimum technology for \Cygnus. The implementation and optimization of a bespoke track reconstruction algorithm would be the next step once this technology is chosen. 

\subsection{Low-level performance of electronic readouts}

We first look at the efficiency, energy resolution, and directional performance of each readout type, when detecting individual nuclear recoil events. All performance estimates are made at the end of the simulation chain, after the energy thresholds in Table~\ref{tab:readout_specs} have been applied, data have been digitized, and 3d space points have been reconstructed as discussed above. We evaluate the performance in 1-keV steps, and fit the results with analytical functions. These functions will allow us to perform high-statistics simulations of WIMP and neutrino detection scenarios, where the energy spectrum is continuous, without re-running the computationally expensive detector simulation. The parameterized detector performance versus energy may be useful for studies by the wider community, and the fit parameters are available from the authors by request. As mentioned earlier, a fair comparison of optical and electronic charge readout is not straightforward. We therefore compare only electronic readouts. Optical readout is discussed separately in Section~\ref{sec:opt}.

\subsubsection{Detection efficiency}

Figure~\ref{fig:event_efficiency} shows the event-level detection efficiency of each readout. Here, an event is counted as detected if the reconstructed avalanche charge above threshold is equivalent to at least one electron of ionization before gain, and if at least three space points have been reconstructed. Such criteria should ensure that some minimal directional information can be inferred. A very loose requirement, but one that is simply intended to give an indication of the lowest possible recoil energy sensitivity that could be achieved with each technology at the simulated gain. The energy dependence is obtained by fitting the function $a/(1+ e^{-b(E-c)})$ to the data points via $\chi^2$ minimization. Here, $a$, $b$, and $c$ are floating fit parameters and $E$ is the recoil energy.

We find that all readouts except planar have an efficiency higher than 50\% above 3\,\kevr, even for the low gain of 9000 simulated (with electron drift, gains of order $10^{5}-10^{6}$ and drastically lower energy thresholds are expected.) The efficiency is primarily determined by the noise level of each readout, so that the pixel readout ends up being the most sensitive at low recoil energies, followed by the pad, wire, strip, and planar readouts. This begins to inform us of the hierarchy in our collection of readouts, but electron background rejection and obtaining good event-level directional information will typically require significantly more detected charge per event, as we discuss below.

\subsubsection{Energy resolution}

Figure~\ref{fig:energy_resolution} shows the energy resolution for each readout. Since each readout has been simulated with the same gain stage, the energy resolution differences seen are due to the charge readout. The energy dependence is obtained by fitting the functional form $\sigma_{E}/E= \sqrt{(a^2/E^2+b^2/E+c)}$ to the data points via $\chi^2$ minimization. Here, $a$, $b$, and $c$ are floating fit parameters and $E$ is the recoil energy. For pixel, wire, and pad readout, the low readout noise does not noticeably affect the energy resolution. As a result, these three readouts have identical energy resolution, limited by the resolution of the gain stage. For strip and planar readout, the higher noise floor increases energy resolution above the gain resolution. It is possible that this could be improved with a more clever energy determination at the algorithm level. If not, somewhat higher gain than simulated would be beneficial for these two readouts. 

\subsubsection{Angular resolution and head/tail recognition}
\begin{figure*}[hbt]
\begin{center}
\includegraphics[clip, trim=0.0cm 0cm 1.8cm 1.8cm, width=0.45\textwidth]{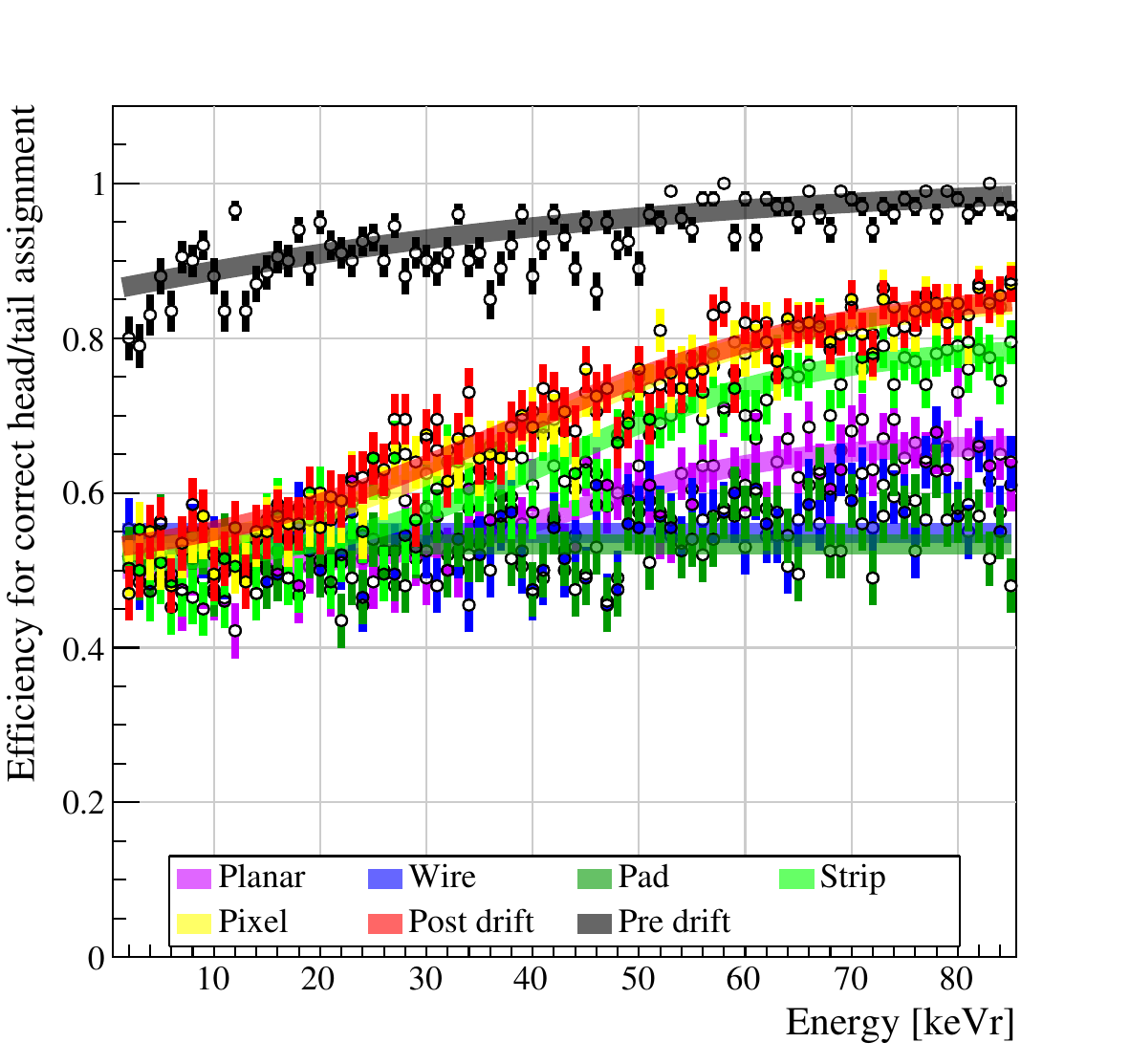}
\includegraphics[clip, trim=0.0cm 0cm 1.8cm 1.8cm, width=0.45\textwidth]{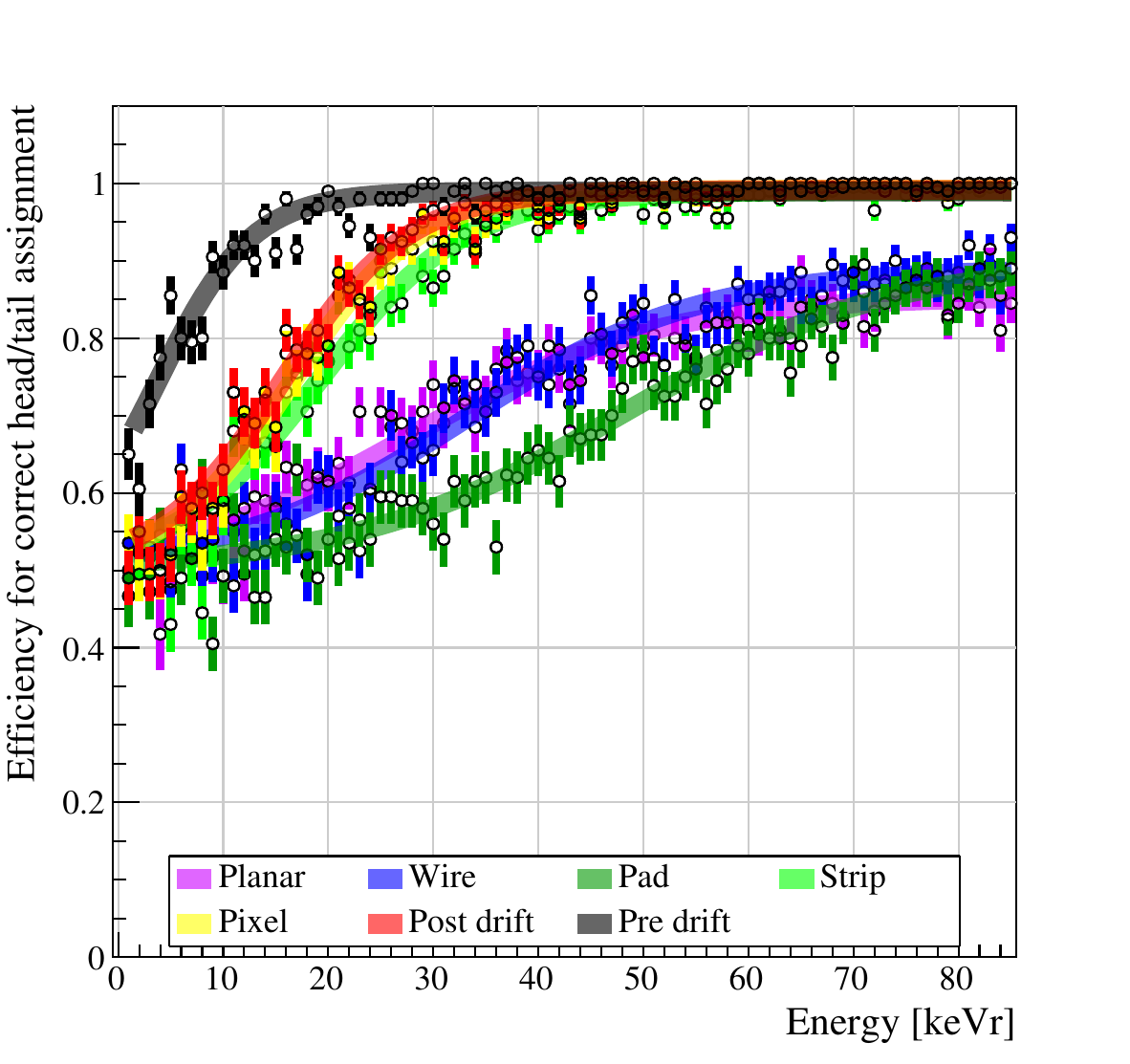}
\caption{Head/tail recognition efficiency versus recoil energy and readout type, for fluorine (left) and helium (right) recoils in a 755:5 Torr gas mixture of \hesfsix. Each color denotes a different readout as shown in the legend. Data points show the efficiency determined from two hundred pseudo-experiments. Error bars are binomial standard deviations, assuming the efficiency determined is the true efficiency.  Lines are analytical functions fit to the data points, used to parameterize performance versus energy. The performance that can be obtained with a perfect readout before (black) and after (red) charge diffusion during drift is also shown, for comparison.}
\label{fig:readout_headtail}
\end{center}
\end{figure*}
We quantify and compare the directional performance of each readout by computing their angular resolutions and head/tail recognition efficiencies. 

We define angular resolution to be the mean difference between the initial recoil axis and the reconstructed recoil axis. Two randomly oriented 3d axial vectors will on average be separated by an angle of 1~radian or 57.3$^\circ$. Hence this value corresponds to no angular sensitivity. The energy dependence is obtained by fitting the functional form $a/\sqrt{(b+E^c)}+d$ to the data points via $\chi^2$ minimization. Here, $a$, $b$, $c$, and $d$ are floating fit parameters and $E$ is the recoil energy. The angular resolution (Fig.~\ref{fig:angular_resolution}) of the pixel readout is superior. For other readouts, the angular resolution gradually gets worse with increasing feature size and noise floor, as expected from the performance ordering in Table \ref{tab:readout_specs}. The performance of the pixel readout is essentially as good as the performance obtained by analyzing the primary charge distribution immediately after diffusion (indicated as postdrift in Fig.~\ref{fig:angular_resolution}). This implies that there is no need for any finer segmentation of this readout. 

At low energies, the effective resolution obtained by fitting the charge cloud {\it before} diffusion (pre drift) is substantially better than the post drift performance.  This tells us that the performance at low energies becomes limited by the track length being short compared to the diffusion scale. This occurs below $\sim$20~\kevr~for helium recoils and below $\sim50$~\kevr~for fluorine. The difference arises because helium recoils are longer than fluorine recoils of the same energy. In this comparison, helium recoils also benefit from two additional effects: they deposit more ionization due to a higher quenching factor, and they exhibit less straggling. The latter limits the angular resolution before diffusion (pre drift). Further improvements may be achievable with a curved track fit to account for straggling, but this is not explored here.

Figure~\ref{fig:readout_headtail} shows the head/tail recognition efficiency of each simulated readout type versus recoil energy. A random head/tail assignment will be correct in 50\% of the cases, so a value of 0.5 corresponds to no sensitivity. The energy dependence is obtained by fitting the functional form $a/(1+ e^{-b(E-c)})+d$ to the data points via $\chi^2$ minimization. Here, $a$, $b$, $c$, and $d$ are floating fit parameters and $E$ is the recoil energy. In the case of fluorine, the efficiency is quite poor for all readouts, even at the higher recoil energies. We see that before diffusion, some head/tail information is still present in the charge cloud, but most of this information appears to be lost after diffusion. In contrast for helium recoils, the head/tail information at a given energy is much higher both before and after diffusion, and the readouts are able to measure head/tail with some significance down to energies of order 5~\kevr.

Because head/tail recognition and angular resolution at low energies are critical for detecting a WIMP signal, the findings here support utilizing helium as a target nucleus. Even with helium, the directionality gradually becomes diffusion-limited at the lowest recoil energies. This effect starts below 20~\kevr~for the angular resolution, and below approximately 35~\kevr~for the head/tail recognition. This suggests that the design simulated here is not yet fully optimized. We anticipate further improvements in performance at low recoil energies by lowering the diffusion and/or increasing the recoil length further. Lowering the gas pressure, lowering the drift length, implementing readout-specific reconstruction algorithms, and using hydrogen target nuclei are examples of possible strategies that should be pursued. Improving this low energy performance will be crucial for maximizing the WIMP/solar neutrino discrimination.

\subsection{Optical readout}
\label{sec:opt}
For optical readout, we find that the simulated performance depends strongly on the exact value of the optical loss factor, the noise rejection signal threshold, and the reconstruction algorithm. Our overall finding is that optical readout looks challenging with the specific assumptions given in Table~\ref{tab:readout_specs}. Optical readout can be competitive, however, if the yield of detected photons per primary electron is increased. This increase can be achieved with improved amplification structures for negative ion drift gases, by utilizing an electron drift gas, or by bringing the optical sensor closer to the amplification plane, thereby reducing the optical loss factor. It is also possible to increase the photon yield via electroluminescence \cite{Baracchini:2020dib}.

\begin{figure*}[hbt]
\begin{center}
\includegraphics[clip, trim=0.0cm 0cm 1.8cm 1.8cm, width=0.45\textwidth]{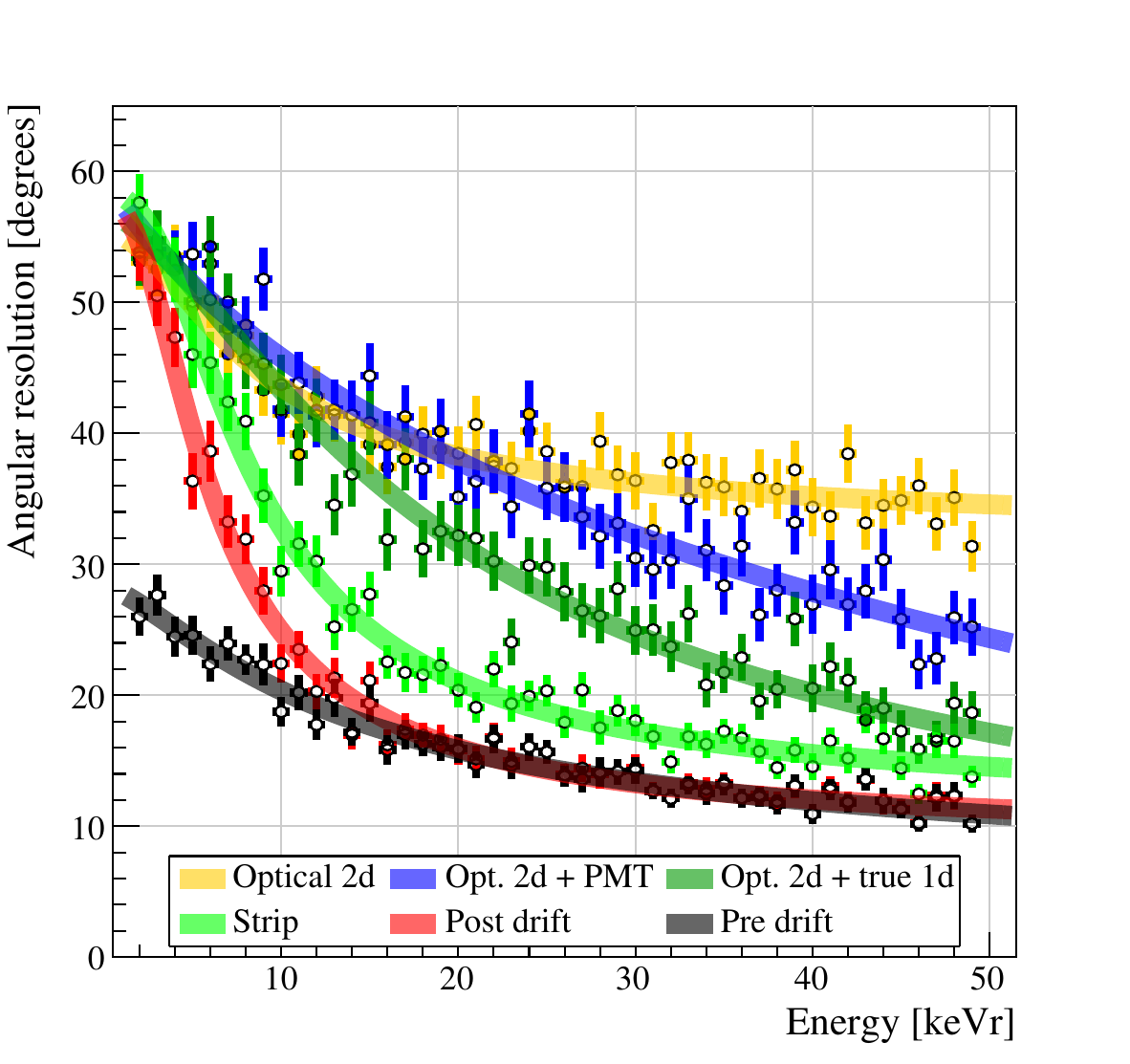}
\includegraphics[clip, trim=0.0cm 0cm 1.8cm 1.8cm, width=0.45\textwidth]{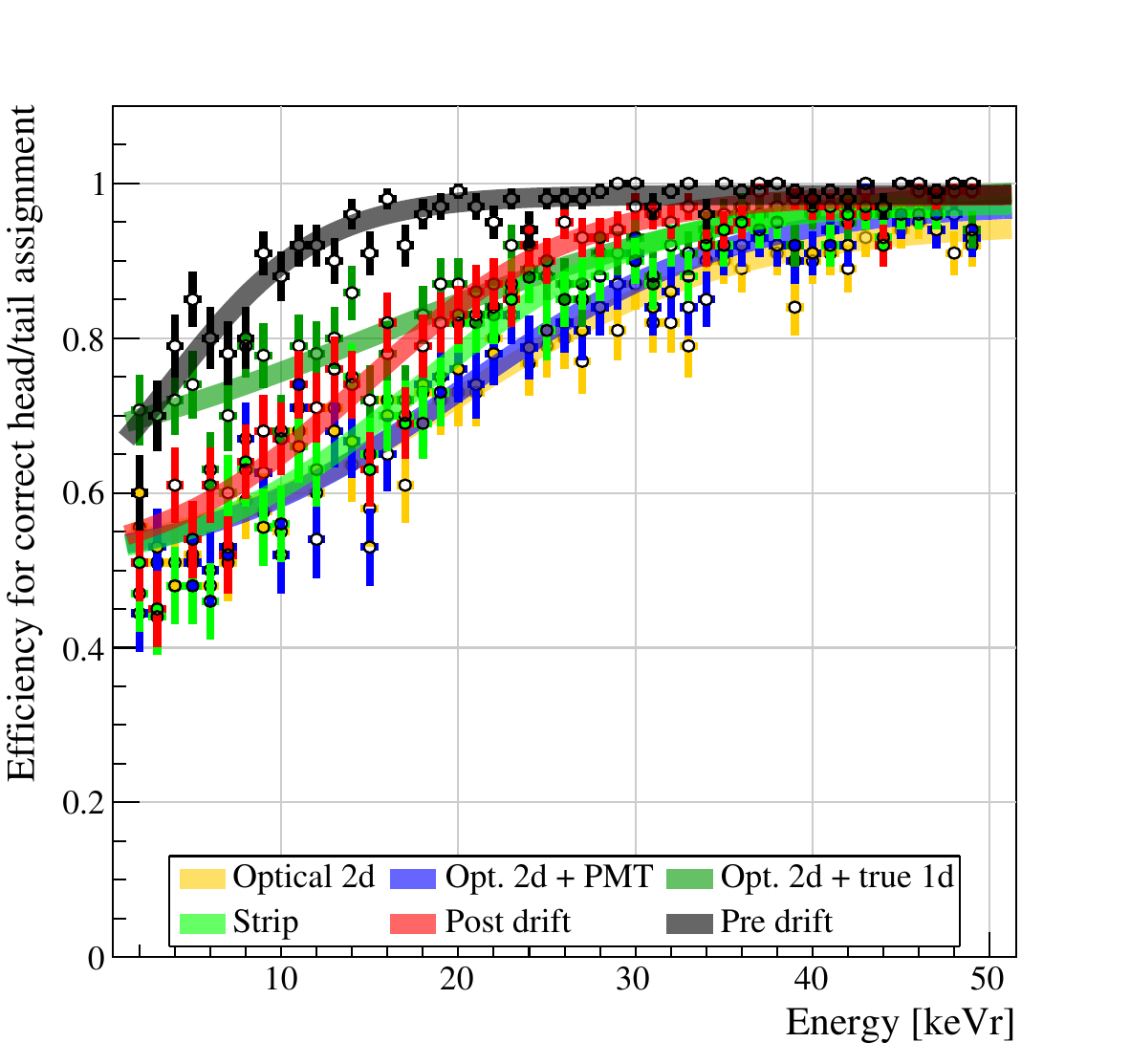}
\caption{Angular resolution (left) and head/tail recognition efficiency (right) for helium recoils in 755:5 Torr \hesfsix gas, detected with various optical readout technologies, with optical loss factors and detector noise disabled in the simulation. Strip readout and maximum achievable performance before and after diffusion are also shows, for comparison. See text for discussion.}
\label{fig:optical}
\end{center}
\end{figure*}

To get a feel for the capabilities of optical readout with more favorable assumptions, Fig.~\ref{fig:optical} shows the simulated performance when the loss factor and noise are both disabled in the simulation. Then, 2d optical readout by itself has similar head/tail recognition as charge readout, but reduced angular resolution when compared to three-dimensional charge readout. This is as expected due to the 2d nature of the sensor. The angular resolution can be improved by combining the 2d optical signal with a 1d signal from a PMT. If we assume the PMT perfectly measures the true $z$-width and the true sign of the $z$ component of the recoil vector (green data points and lines in Fig~\ref{fig:optical}), this 2d+1d optical reconstruction performs quite well in terms of angular resolution, though not quite as well as true 3d reconstruction. Note that the head/tail performance in unrealistically good when assuming this perfect z-measurement. If we simulate a realistic PMT, with performance loosely based on Ref.~\cite{Antochi:2018otx}, the performance somewhat degrades, but is still substantially better than optical 2d by itself. It may be possible to improve the optical 2d+1d readout performance substantially with more sophisticated algorithms, which would utilize not only the PMT signal width, but its detailed time structure. This has already been demonstrated by the CYGNO collaboration for electron tracks \cite{Antochi:2018otx}.

If we turn on either the optical loss factor or the signal noise, performance does not degrade substantially. However, if we turn on both effects, the optical 2d signal suffers strongly. For such a scenario, a readout specific noise-rejection algorithm will be required. This is currently being investigated by the CYGNO collaboration \cite{Baracchini:2020iwg}.

In summary, we find that the optical readout performance is strongly dependent on the algorithm that combines the PMT and optical 2d signal, and on the algorithm that rejects the noise. Optical readout also favors substantially higher gain than what is simulated here. For these reasons we conclude that a fair comparison against charge readout is not easily possible, and leave the evaluation of optical readout for future work.

\subsection{Directionality threshold}

\begin{figure*}[hbt]
\begin{center}
\includegraphics[clip, trim=0.0cm 0cm 1.8cm 1.8cm, width=0.45\textwidth]{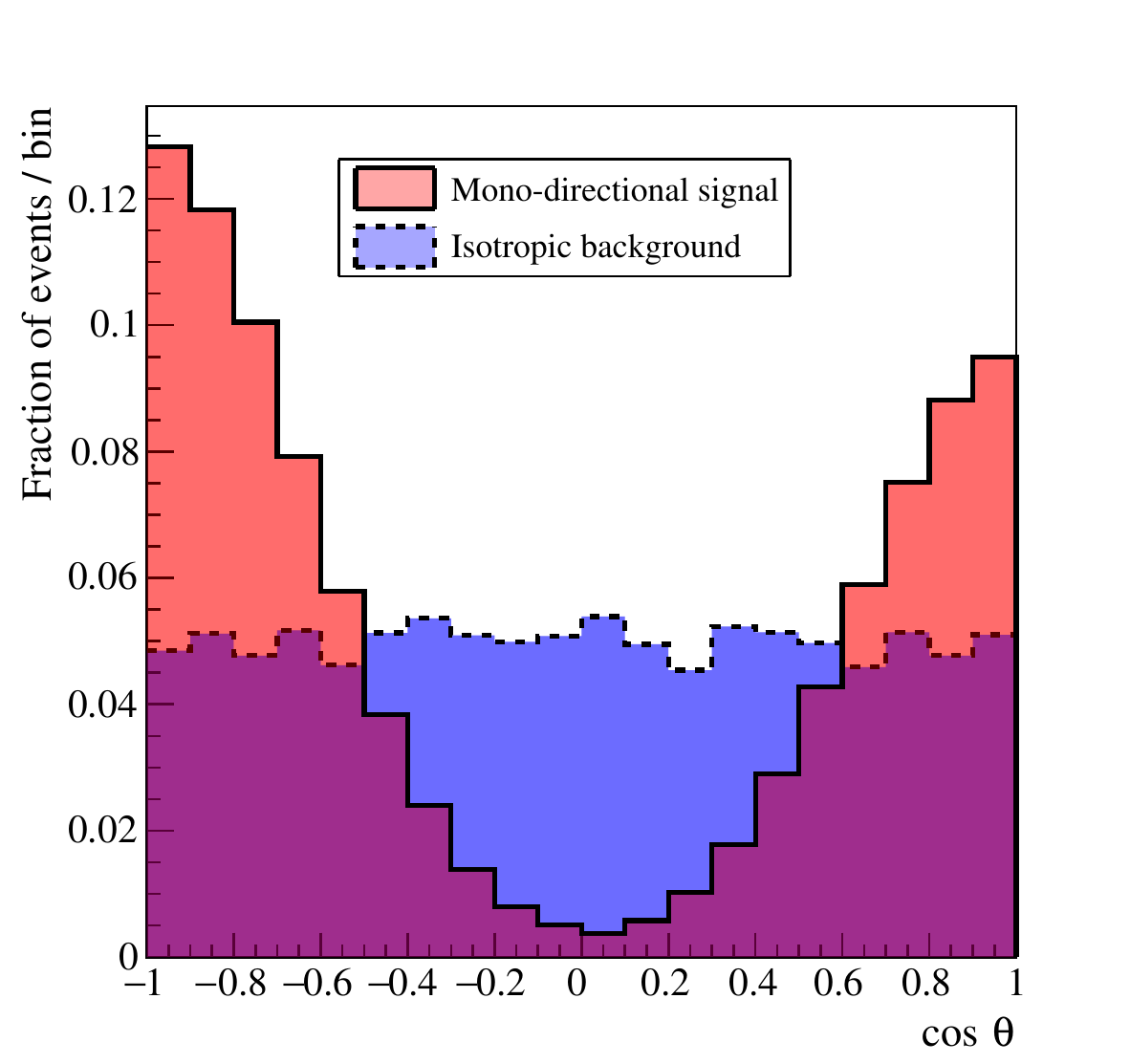}
\includegraphics[clip, trim=0.0cm 0cm 1.8cm 1.8cm, width=0.45\textwidth]{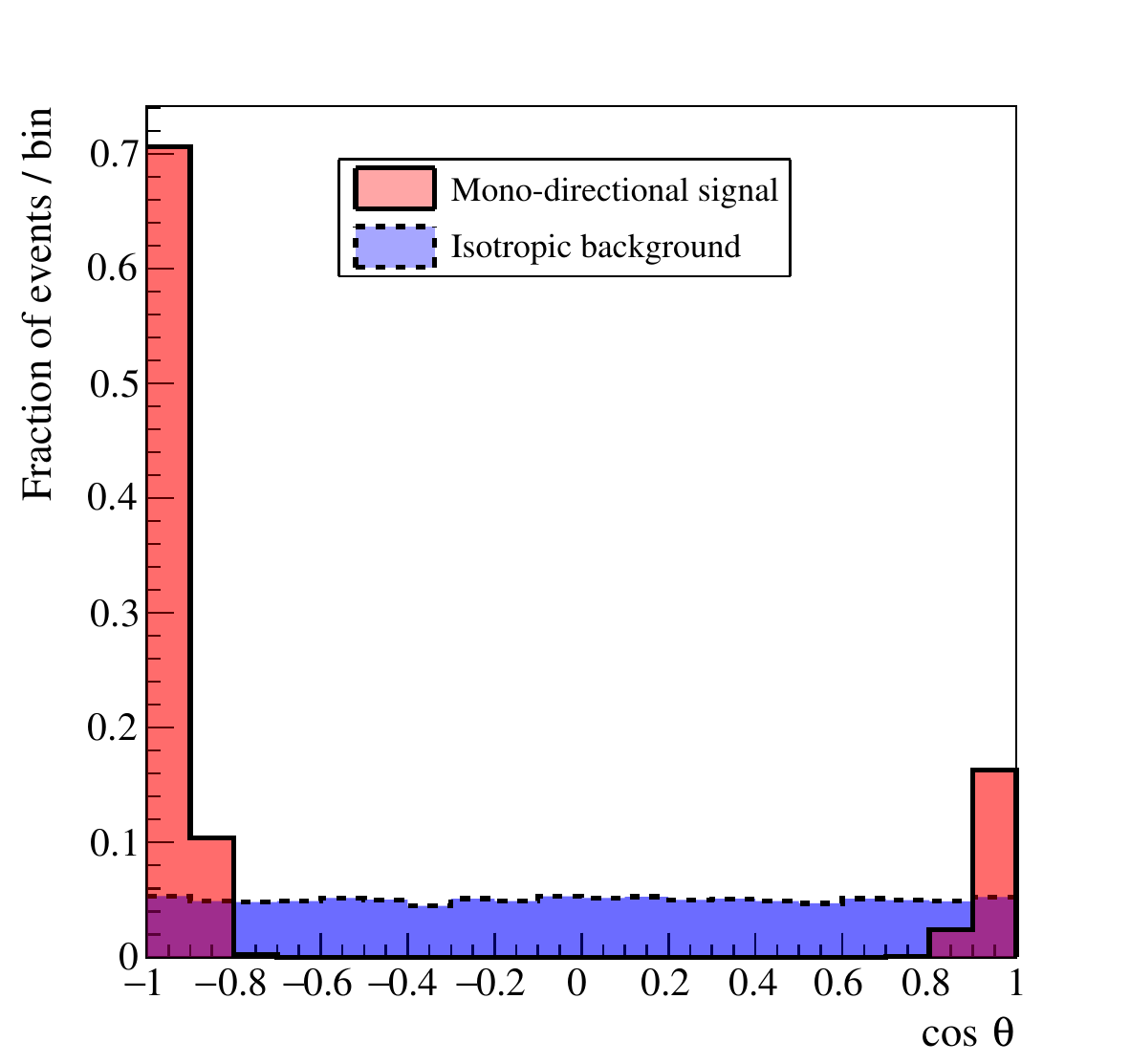}
\caption{Reconstructed polar angle distribution for 20~\kevr~fluorine (left) and helium (right) recoils detected with a pixel readout for a hypothetical mono-directional recoil signal (red) and isotropic recoil distribution background (blue).}
\label{fig:delta_pix}
\end{center}
\end{figure*}

\begin{figure*}[hbt]
\begin{center}
\includegraphics[clip, trim=0.0cm 0cm 1.6cm 1.7cm, width=0.45\textwidth]{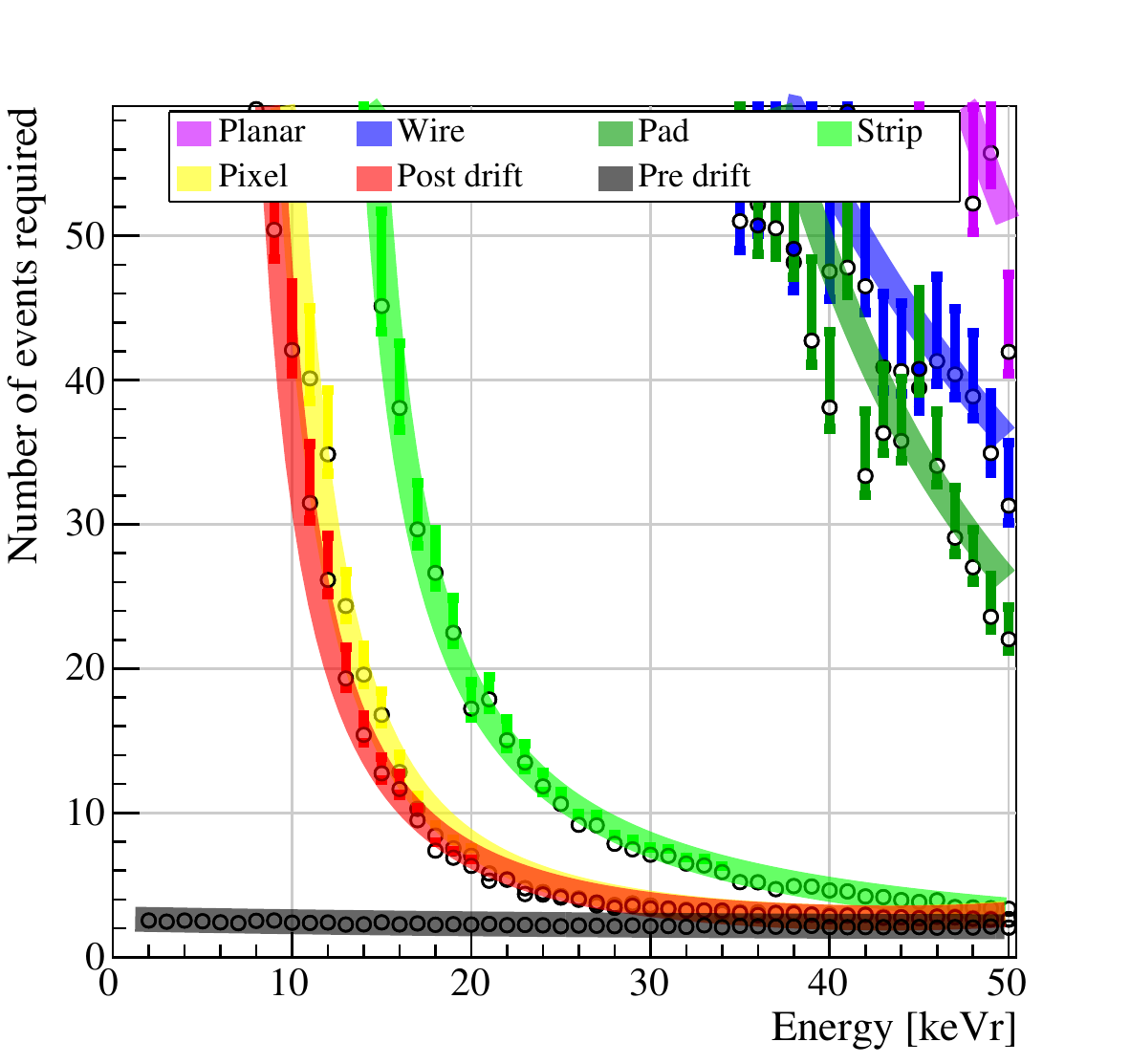}
\includegraphics[clip, trim=0.0cm 0cm 1.6cm 1.7cm, width=0.45\textwidth]{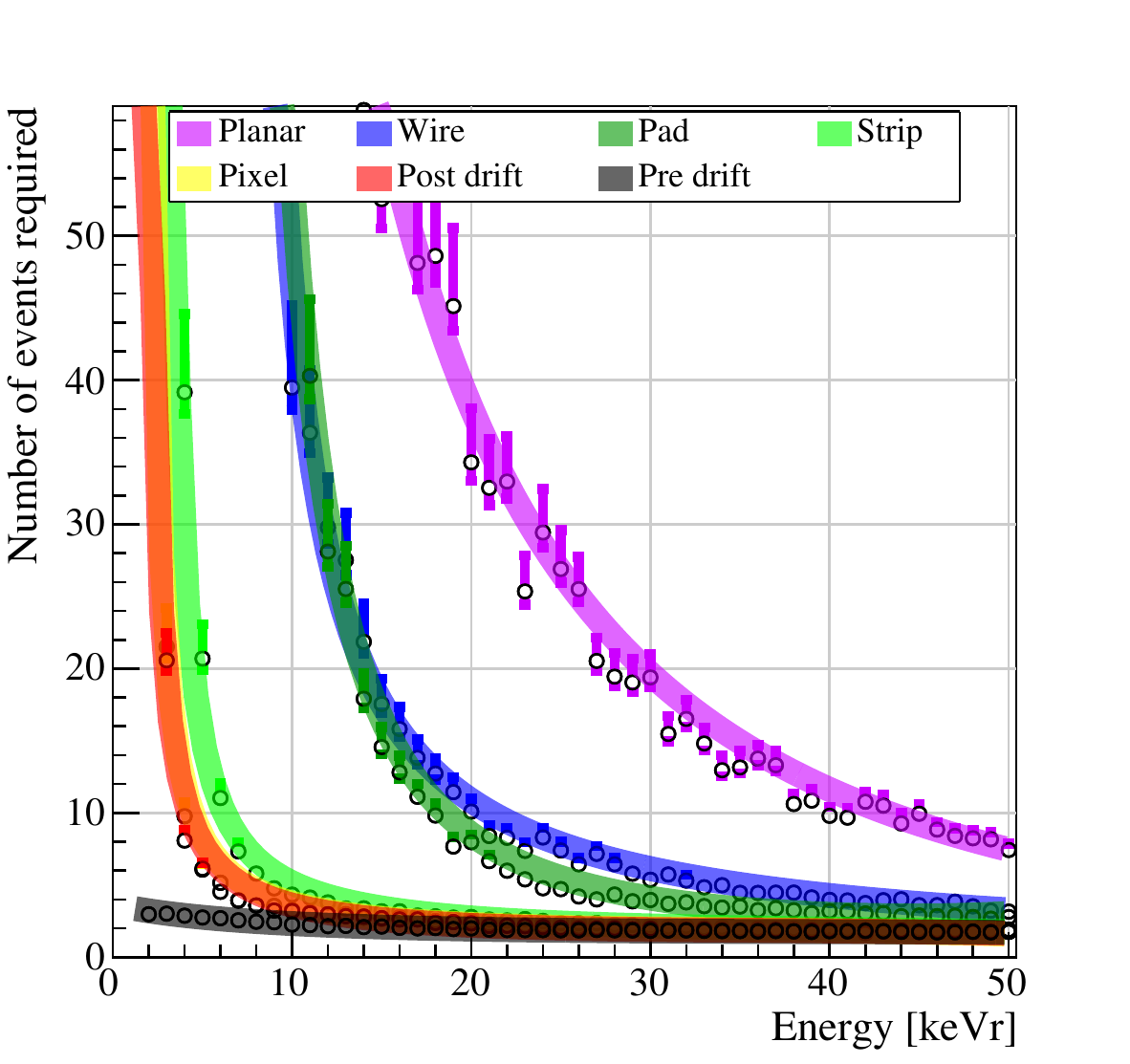}
\caption{Mean number of mono-directional fluorine (left) or helium (right) recoil events required to reject an isotropic recoil hypothesis at 90\% CL, in the case of zero background events, for a 755:5 Torr gas mixture of \hesfsix. Each color denotes a different readout as shown in the legend. Data points show the mean number of recoils required, based on pseudo-experiments, with errors indicating the uncertainty due to finite simulation statistics. The performance that can be obtained with a perfect readout before (black) and after (red) charge diffusion during drift is also shown, for comparison.}
\label{fig:directional_threshold}
\end{center}
\end{figure*}

\begin{figure*}[hbt]
\begin{center}
\includegraphics[clip, trim=0.0cm 0cm 1.6cm 1.7cm, width=0.45\textwidth]{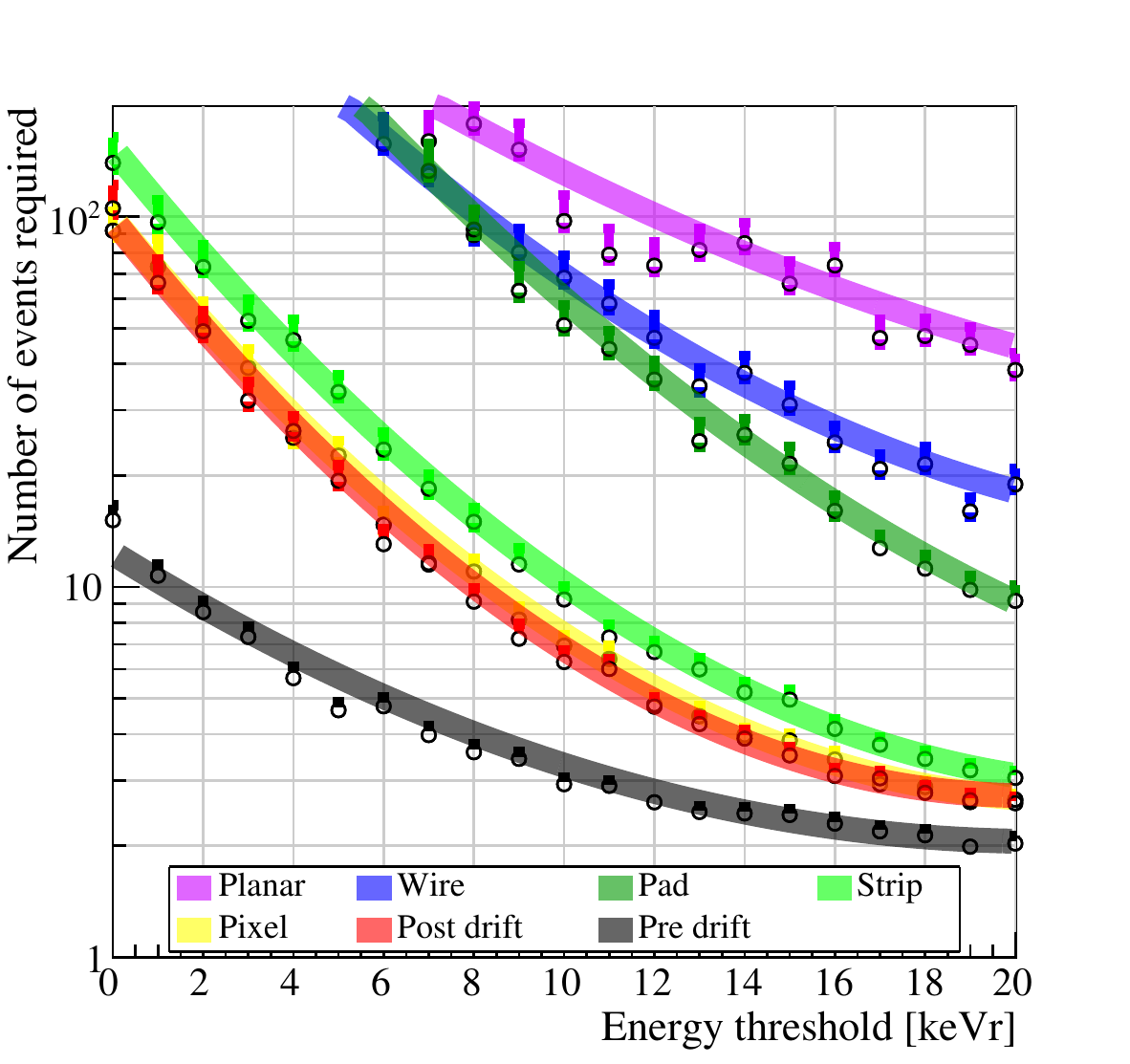}
\includegraphics[clip, trim=0.0cm 0cm 1.6cm 1.7cm, width=0.45\textwidth]{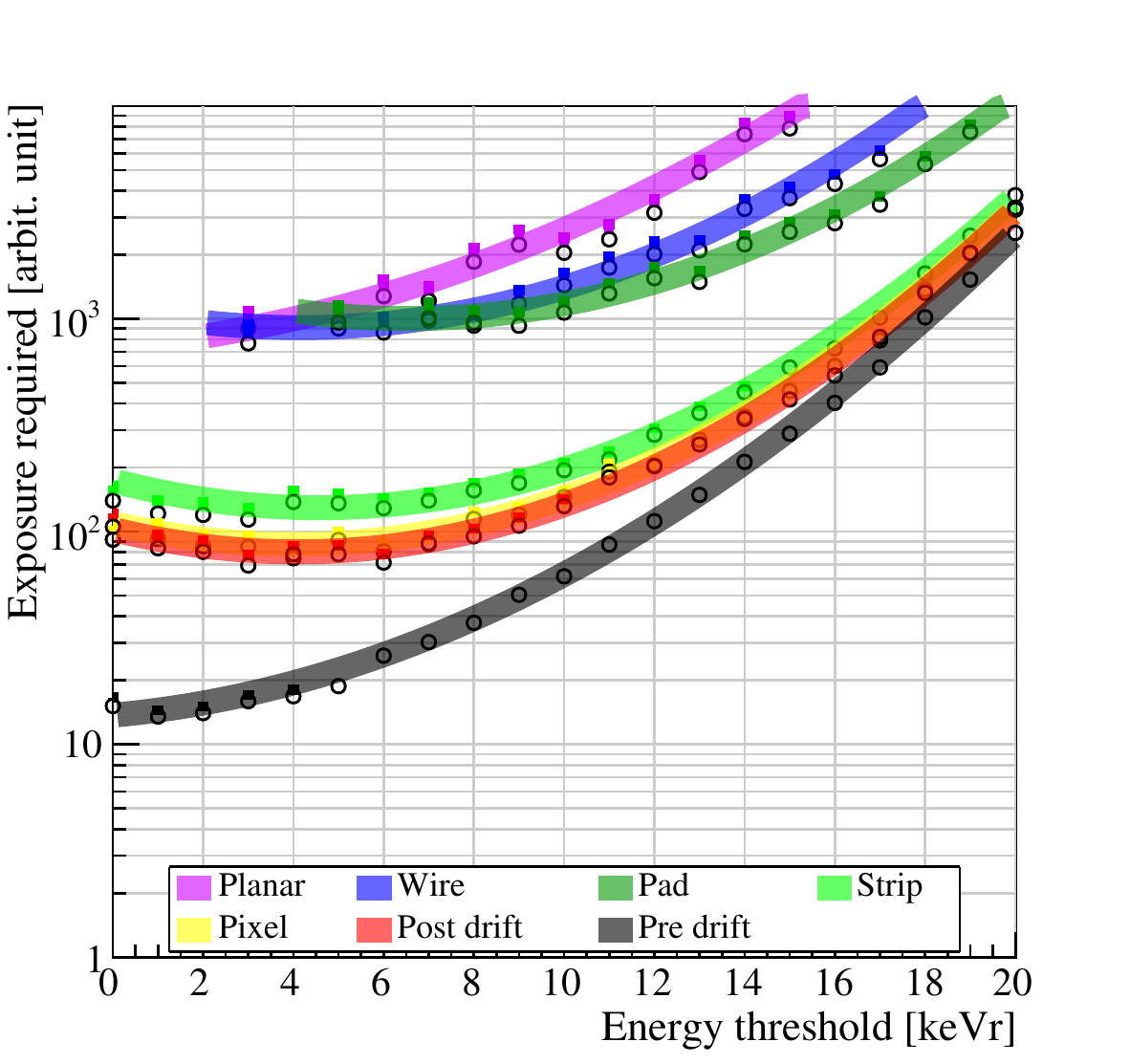}
\caption{Mean number of 10~\gevcc-WIMP-helium recoils (left) and exposure (right) required, in order to exclude isotropy in galactic coordinates at 90\% CL, versus energy threshold, for a 755:5 Torr gas mixture of \hesfsix. Each color denotes a different readout as shown in the legend. Data points show the mean number of recoils required, based on pseudo-experiments, with errors indicating the uncertainty due to finite simulation statistics. See text for further details on this and following plots, including why some data points are not drawn.}
\label{fig:n_90cl_isotropy_10gev}
\end{center}
\end{figure*}

\begin{figure*}[hbt]
\begin{center}
\includegraphics[clip, trim=0.0cm 0cm 1.6cm 1.7cm, width=0.45\textwidth]{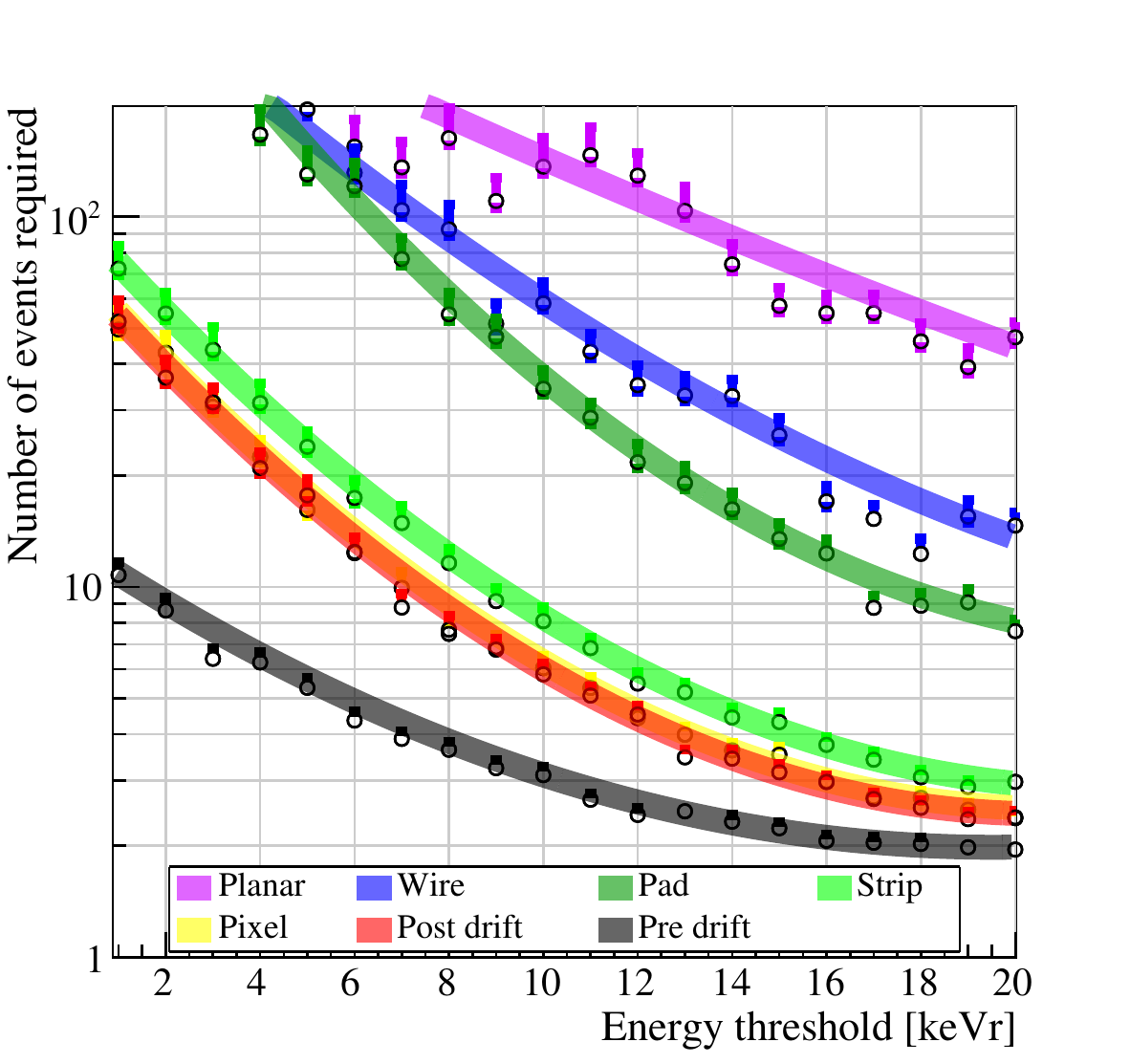}
\includegraphics[clip, trim=0.0cm 0cm 1.6cm 1.7cm, width=0.45\textwidth]{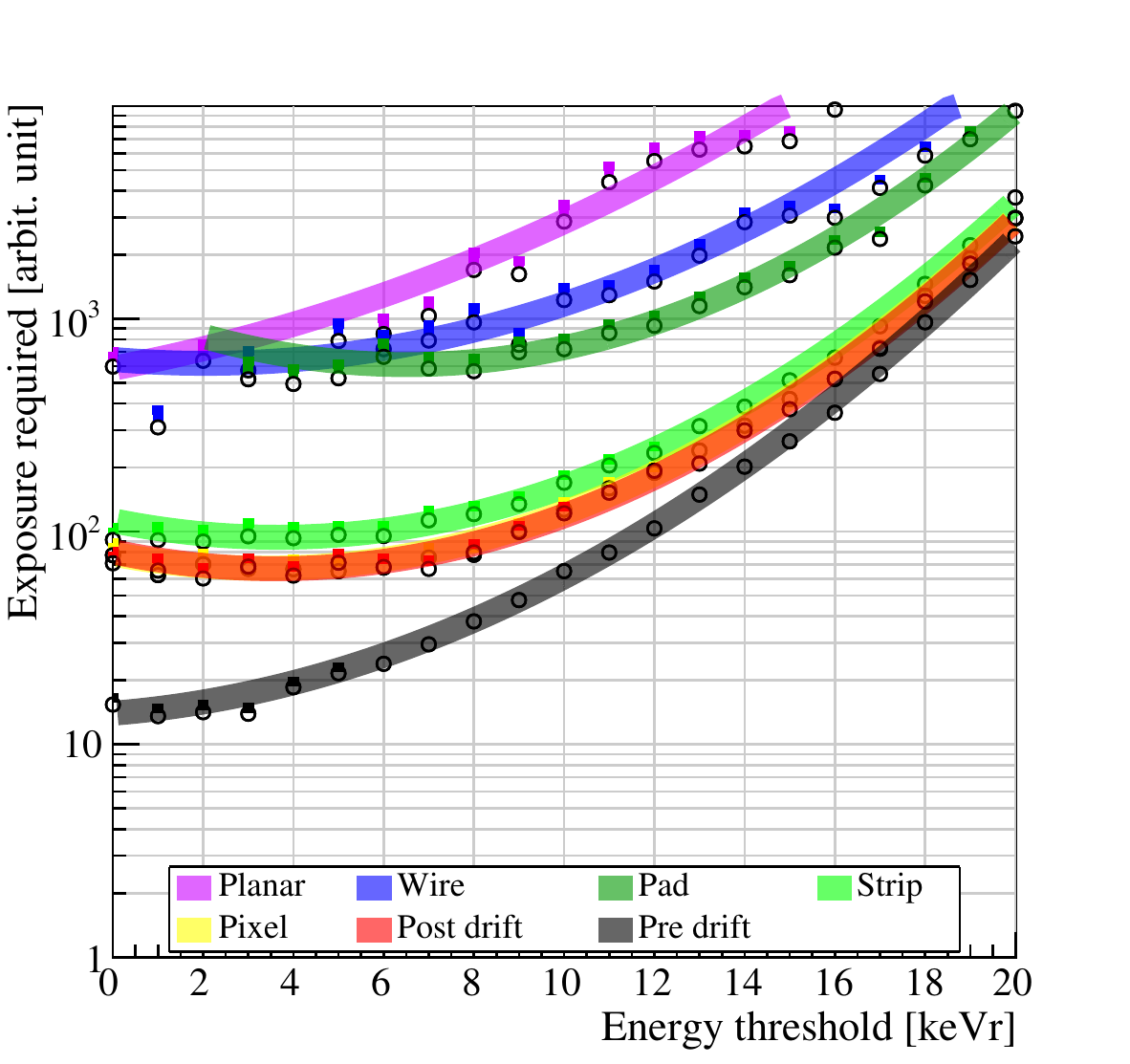}
\caption{Mean number of 10~\gevcc-WIMP-helium recoils (left) and exposure (right) required to exclude a neutrino background hypothesis at 90\% CL, versus energy treshold, in the case of no detected neutrino background events, for a 755:5 Torr gas mixture of \hesfsix. Each color denotes a different readout as shown in the legend. Data points show the mean number of recoils required, based on pseudo-experiments, with errors indicating the uncertainty due to finite simulation statistics.}
\label{fig:n_90cl_neutrinos_vs_threshold}
\end{center}
\end{figure*}

\begin{figure*}[hbt]
\begin{center}
\includegraphics[clip, trim=0.0cm 0cm 1.6cm 1.7cm, width=0.45\textwidth]{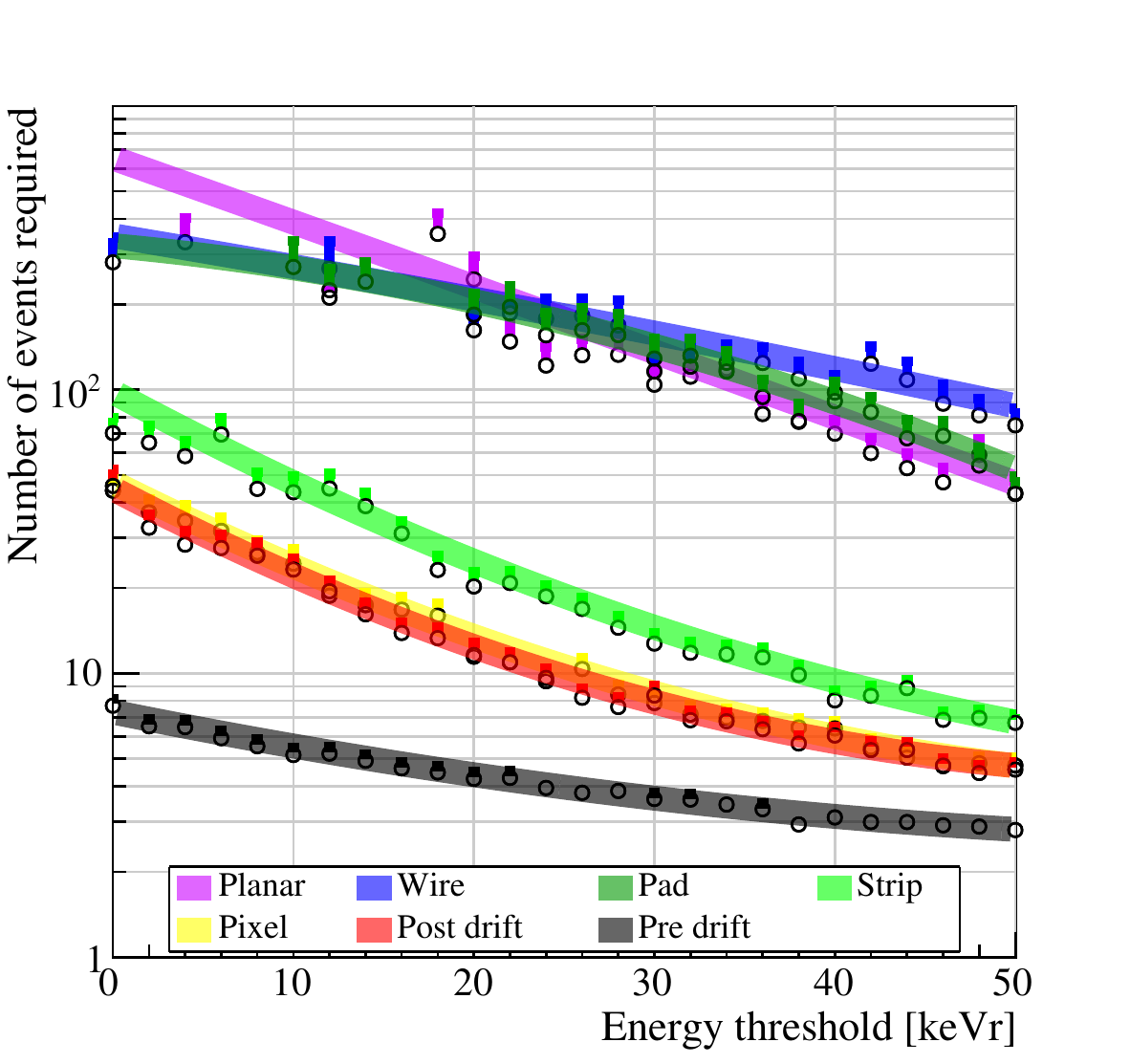}
\includegraphics[clip, trim=0.0cm 0cm 1.6cm 1.7cm, width=0.45\textwidth]{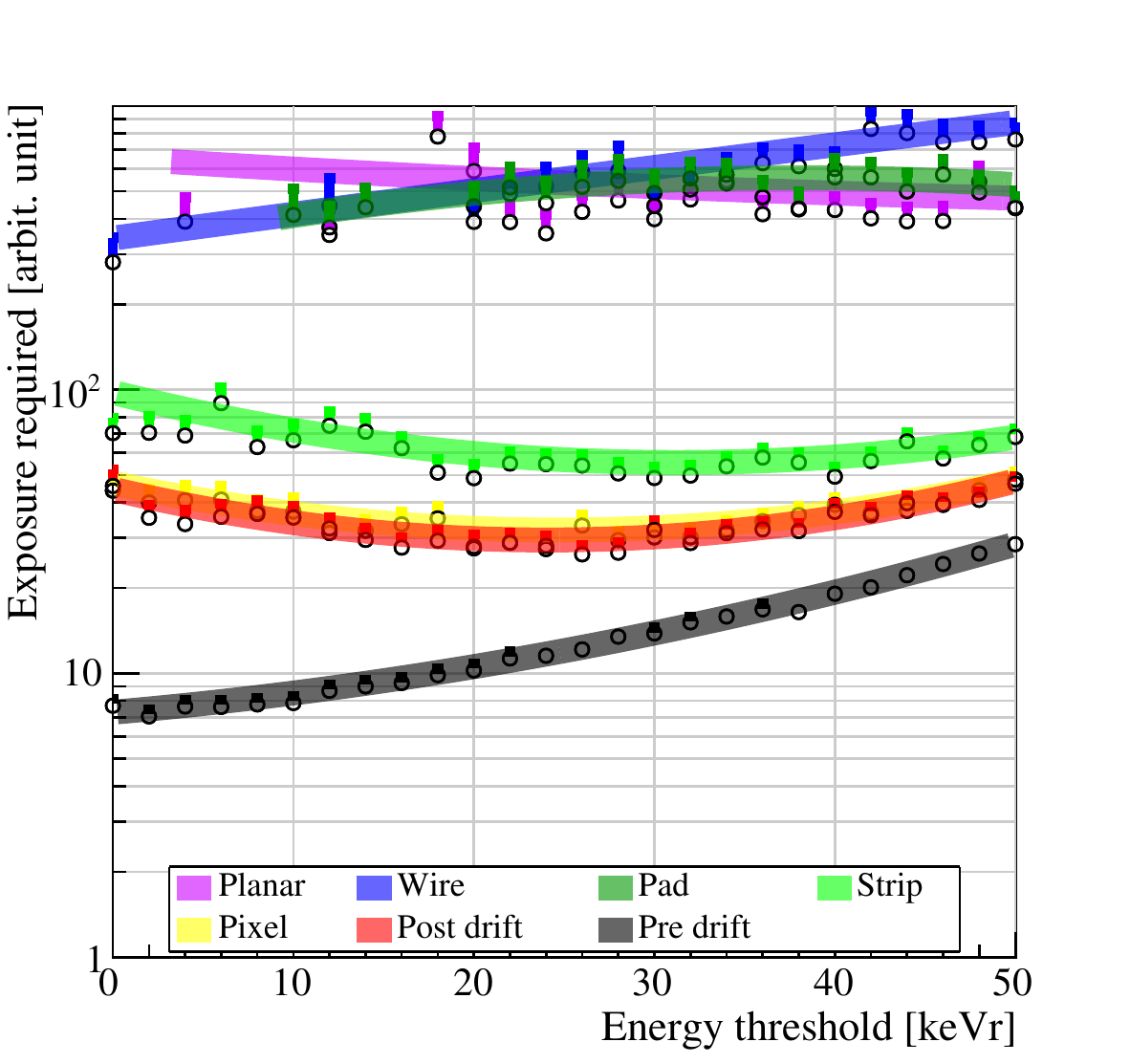}
\caption{Mean number of 100~\gevcc-WIMP-fluorine recoils (left) and exposure (right) required, in order to exclude isotropy in galactic coordinates at 90\% CL, versus energy threshold, for a 755:5 Torr gas mixture of \hesfsix. Each color denotes a different readout as shown in the legend. Data points show the mean number of recoils required, based on pseudo-experiments, with errors indicating the uncertainty due to finite simulation statistics.}
\label{fig:n_90cl_isotropy_100gev_helium}
\end{center}
\end{figure*}

\begin{figure*}[hbt]
\begin{center}
\includegraphics[clip, trim=0.0cm 0cm 1.6cm 1.7cm, width=0.45\textwidth]{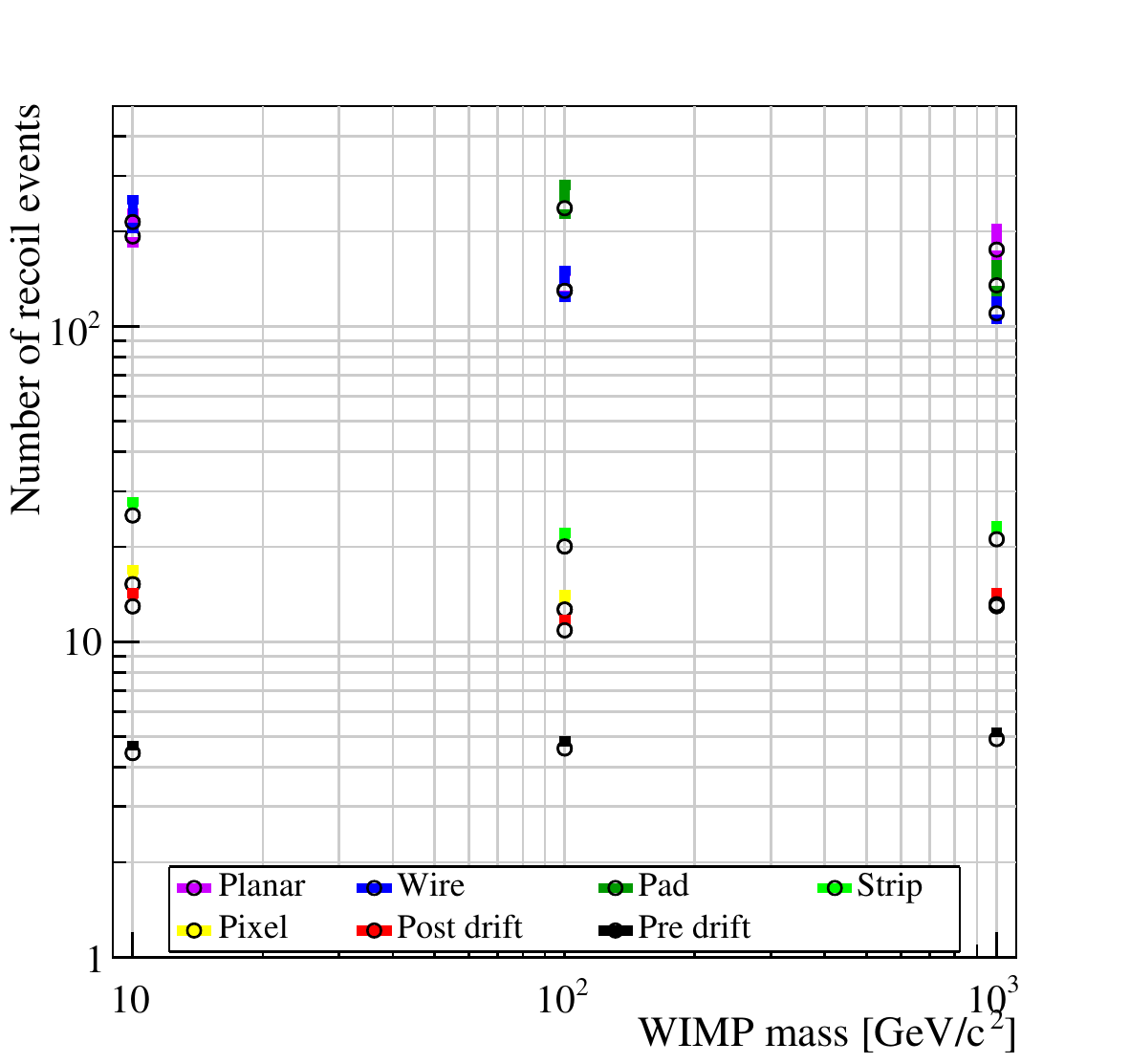}
\includegraphics[clip, trim=0.0cm 0cm 1.6cm 1.7cm, width=0.45\textwidth]{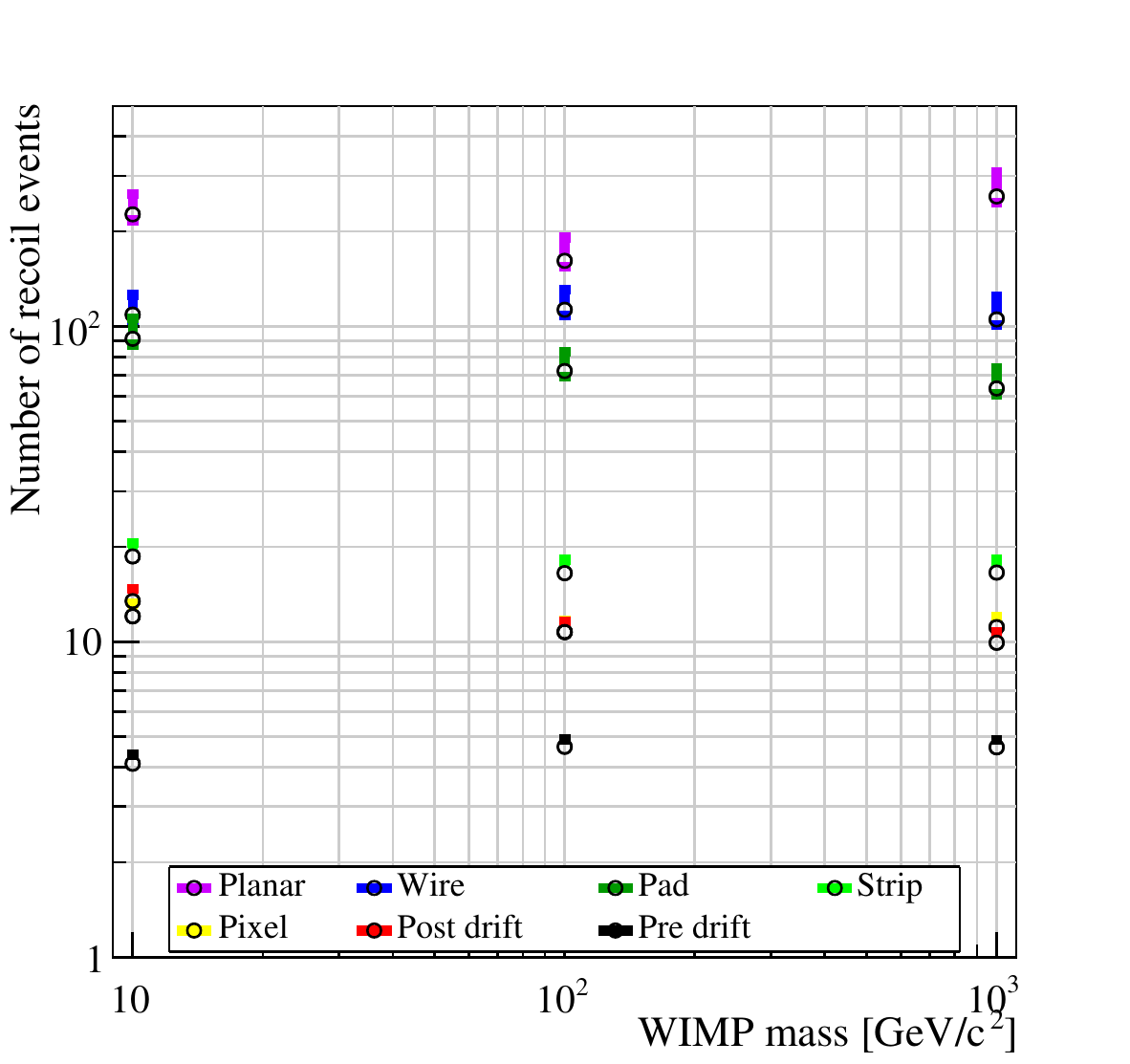}
\caption{Mean number of WIMP-helium recoil events with energy greater than 6~\kevr~required to exclude an isotropic (left) and neutrino (right) background hypothesis at 90\% CL, versus WIMP mass, in the case of no detected background events. The target gas simulated is a 755:5 Torr mixture of \hesfsix. Each color denotes a different readout as shown in the legend. Data points show the mean number of recoils required, based on pseudo-experiments, with errors indicating the uncertainty due to finite simulation statistics.}
\label{fig:n_90cl_isotropy}
\end{center}
\end{figure*}

\begin{figure}[hbt]
\begin{center}
\includegraphics[clip, trim=0.0cm 0cm 1.6cm 1.7cm, width=0.45\textwidth]{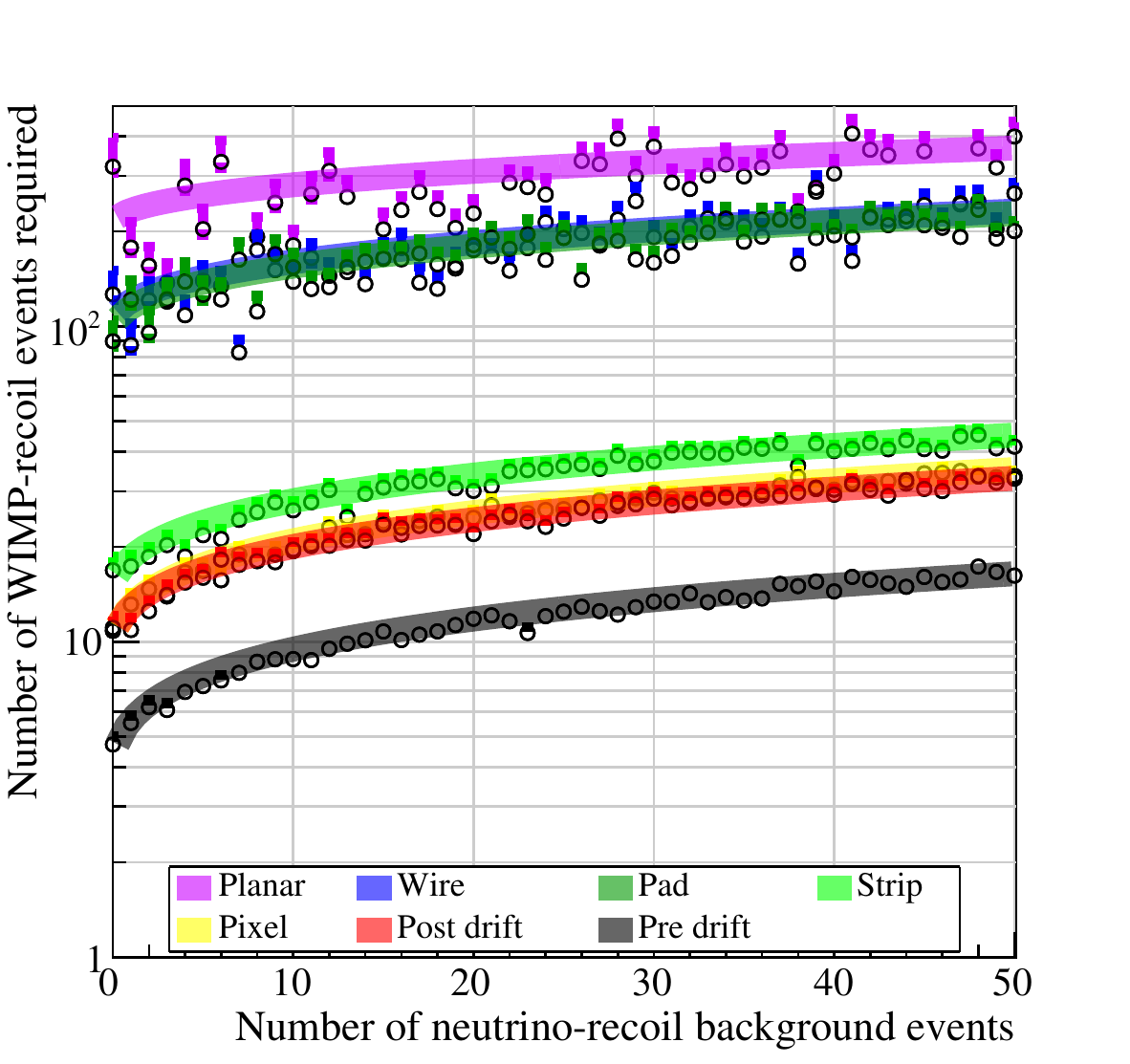}
\caption{Mean number of 10~\gevcc-WIMP-helium recoils with energy greater than 6~\kevr~required to exclude a neutrino background hypothesis at 90\% CL, versus number of neutrino background events in the detected event sample. The target gas simulated is a 755:5 Torr mixture of \hesfsix. Each color denotes a different readout as shown in the legend. Data points show the mean number of recoils required, based on pseudo-experiments, with errors indicating the uncertainty due to finite simulation statistics. The lines show $\chi^2$-minimizations of the functional form $a+b\sqrt{n}$ to the data points, with $a$ and $b$ being floating fit parameters and $n$ being the number of neutrino background events.}
\label{fig:n_90cl_neutrinos_vs_nneutrino}
\end{center}
\end{figure}

\begin{figure*}[hbt]
\begin{center}
\includegraphics[clip, trim=0.0cm 0cm 1.6cm 1.5cm, width=0.45\textwidth]{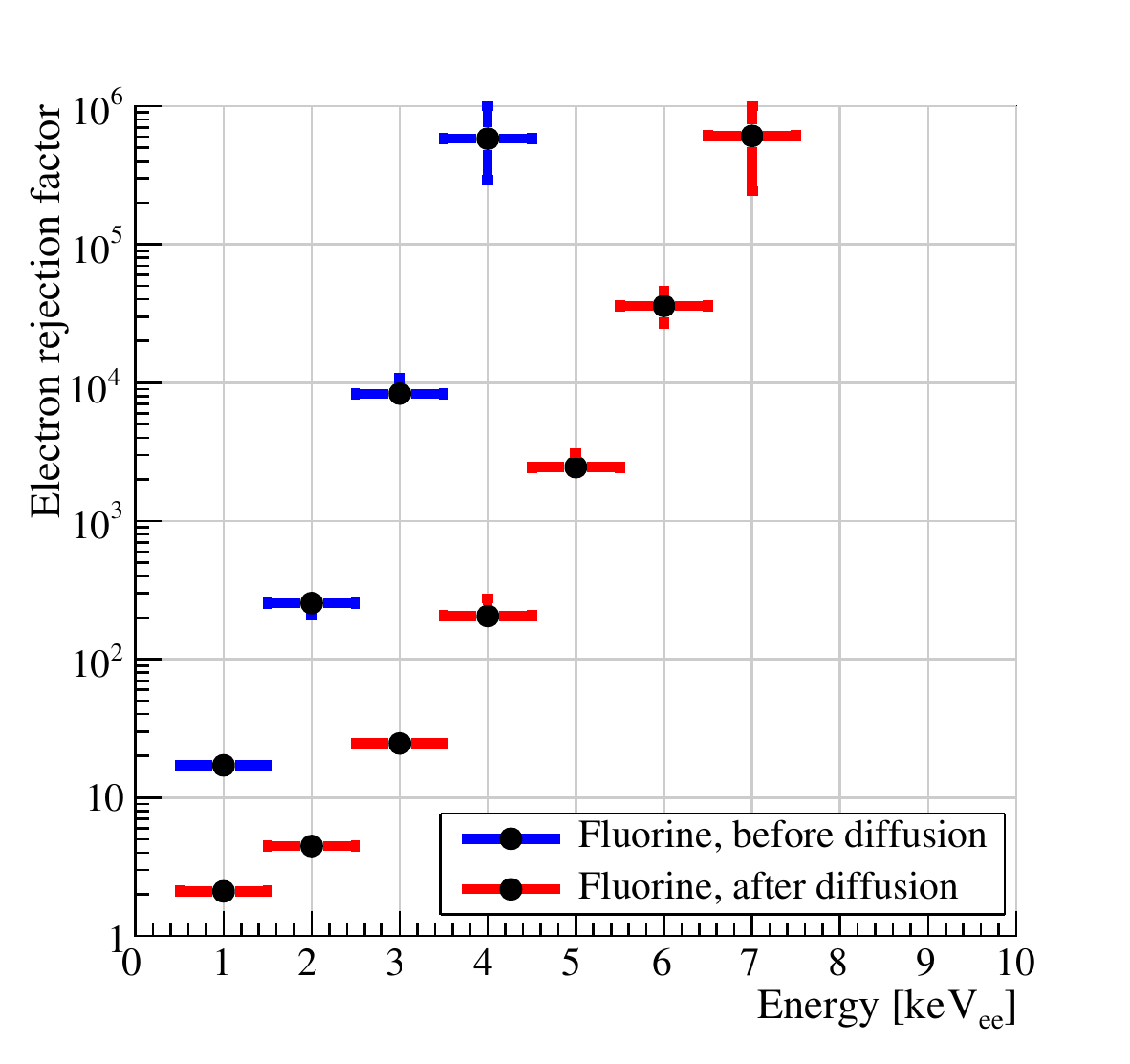}
\includegraphics[clip, trim=0.0cm 0cm 1.6cm 1.5cm,width=0.45\textwidth]{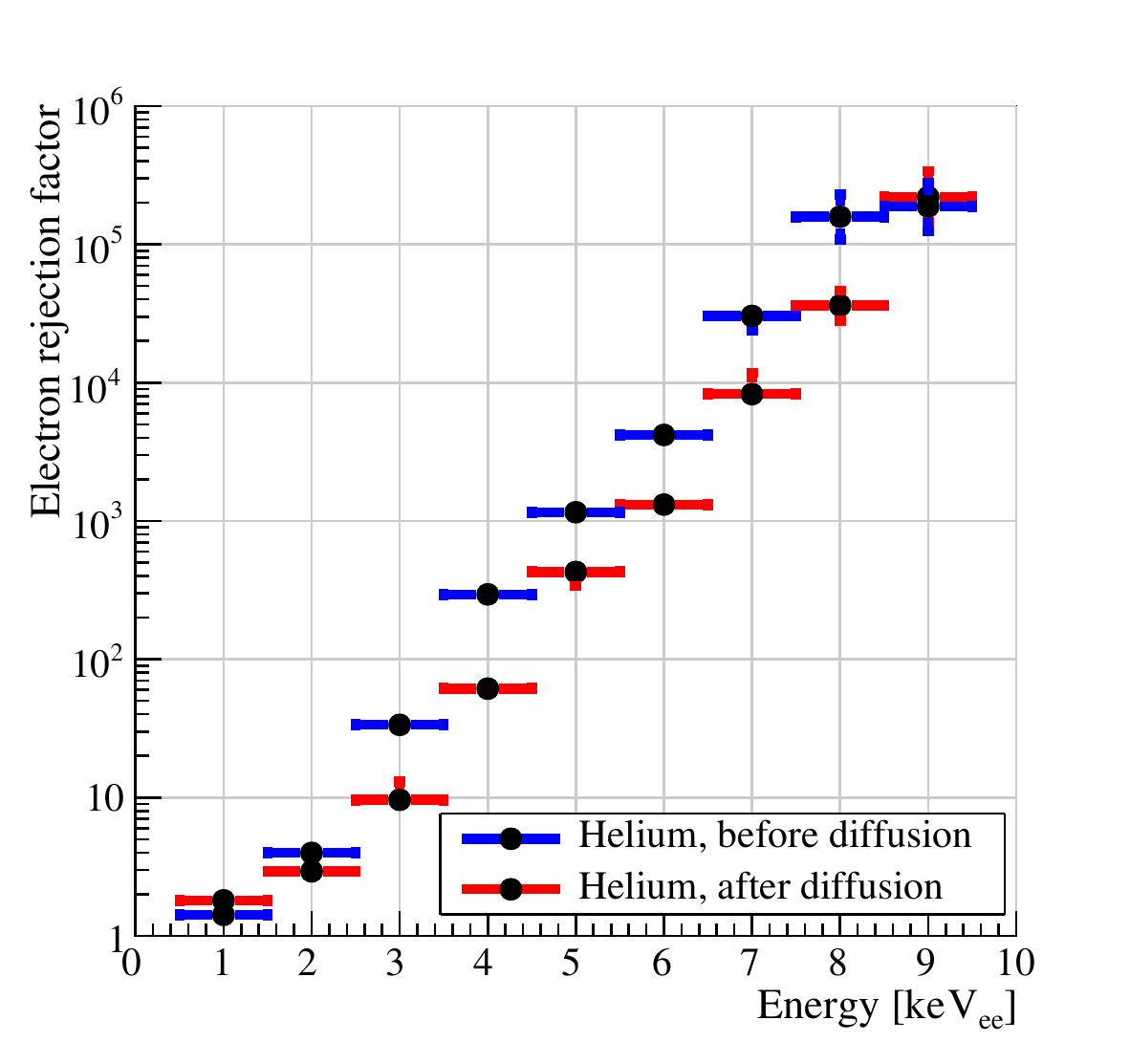}
\caption{Electron rejection factors for fluorine recoils (left) and helium recoils (right) in 755:5 Torr of \hesfsix. Only the simplest observable, the fitted-track length at a given ionization energy, is used to discriminate between electrons and nuclear recoils. Within each energy bin we apply a variable minimum track length that retains 50\% of the nuclear recoils. Error bars show combined uncertainties due to finite simulated electron and nuclear recoil statistics. The electron rejection rises exponentially with energy. For the higher energy bins, where no data point are shown, {\it all} simulated electron events were rejected, and the electron rejection factor exceeds $10^6$.}
\label{fig:electronrejection}
\end{center}
\end{figure*}

We have seen that there is substantial variation in the energy-dependent angular resolutions and head/tail recognition efficiencies of the different readouts considered. The optimal readout, however, is not the one with the best performance, but rather the one with the best compromise between performance and cost, so that the physics reach per unit cost is maximized. Different signals will have different ranges of relevant recoil energies, so the best compromise may depend on the specific physics goal.

As a first step towards a final cost-performance analysis, we quantify the directional performance versus recoil energy in a way that combines both the angular resolution and head/tail recognition efficiency. This is done by estimating how many recoils each readout needs to detect in order to discriminate between a monodirectional (``signal'') delta function (all recoils have initial momentum vectors in the same 3d direction) and an isotropic (``background'') recoil distribution. The goal here is to provide an intuitive result that clearly shows the recoil energy range where each technology has good directionality. Note that a 3d delta function is a hypothetical scenario with maximum directionality. In any real physical scenario, the recoil distribution is always broadened by the non-zero scattering angles. 

Several discovery variables and test statistics have been proposed in the literature for detecting anisotropic recoil distributions for WIMP searches~\cite{Mayet:2016zxu}. The most powerful variables are those which depend on the angle between the recoil direction and the direction of Cygnus (or potentially the direction of the Sun when the principal background is neutrinos). Not all of our simulated readouts, however, have a dimensionality which allows these angles to be measured at all times. Our 3d delta function points along the negative $z$ direction, which is the direction that drift charge is traveling in the detector. This is the most sensitive direction for the less segmented readout technologies. Our discovery variable in this case is the polar angle, $\theta$, with respect to the vertical $z$-axis. Our generated signal peaks at $\cos{\theta}=-1$, while our background is flat in $\cos{\theta}$. The {\it detected} $\cos{\theta}$ distribution for the pixel readout is shown in Fig.~\ref{fig:delta_pix} for fluorine and helium recoils. As discussed earlier, the helium recoils are more strongly directional than fluorine recoils of the same energy. This is a consequence of the performance difference we saw for the two recoils species in Fig.~\ref{fig:angular_resolution}, and primarily caused by helium recoils producing longer tracks than fluorine.

Also note that the helium mono-directional signal recoil distribution in Fig.~\ref{fig:delta_pix} is significantly more asymmetric than the equivalent fluorine distribution, because the head/tail is assigned correctly more often for helium, as previously shown in Fig.~\ref{fig:readout_headtail}: at 20~\kevr, the simulation predicts a correct head/tail reconstruction of only $\sim 55\%$ in fluorine, while it is $\sim80\%$ for helium.

In general the reconstructed signal will differ even more from the background at higher recoil energies when the directional performance is better.  The $\cos{\theta}$ signal distribution is also more asymmetric at higher energies. For lower energies than 20~\kevr~the contrast between the signal and background distributions becomes gradually less defined, and at the lowest energies simulated here they are essentially indistinguishable. We saw this in Figs.~\ref{fig:angular_resolution} and~\ref{fig:readout_headtail} where all readout curves converged to 1~rad angular resolution and 0.5 head/tail efficiency respectively, corresponding to no directional sensitivity.

To quantify the physics reach corresponding to the directional performance of each readout, we perform a Kolomogorov-Smirnov test~\cite{smirnov1948} comparing each signal and background $\cos{\theta}$ distribution. Figure~\ref{fig:directional_threshold} shows how many monodirectional signal events are required to reject the isotropic background hypothesis at the 90\% confidence level (CL) when there are zero background events present. To fairly compare readouts, we count the number of {\it interacting} events, meaning that we also count those that go undetected due to finite reconstruction efficiency. For higher recoil energies, as little as two-three events are sufficient to exclude isotropy for the highest-performing readouts. 

For lower recoil energies however, the number of required events diverges as the directional sensitivity and event-level detection efficiency deteriorates in all readouts. To quantify these observations with a single number, we define a {\it directionality threshold}: the recoil energy at which ten or more interacting events are required to exclude isotropy. Note that the choice of ten events is an arbitrary choice made to allow an easy comparison of readouts --- therefore the directional threshold is not to be interpreted as a hard threshold, and there is still directional sensitivity below this threshold.

With those disclaimers made, the directionality thresholds for pixel, strip, pad, wire, and planar readouts are approximately 4~\kevr, 6~\kevr, 17~\kevr, 22~\kevr, and 42~\kevr, respectively, for helium recoils. We note that to get such good performance at low recoil energies, the low mass density of the 755:5 Torr \hesfsix gas mixture was required. For a 740:20 mixture (not shown here), the directional thresholds were approximately a factor of two higher. 

Comparing Fig.~\ref{fig:directional_threshold} left and right, we see that for fluorine recoils, the directional thresholds are more than a factor of three higher than for helium recoils. Since our ``signal'' is maximally anisotropic here, these thresholds estimate the limit for good directional sensitivity. It should be noted however we have used a very simple measure of directionality. Directional sensitivity can be improved by performing statistical tests that use all recoil angles and their correlation with recoil energy, for example in a likelihood-based analysis. Furthermore, as noted earlier, the pixel and to some extent the strip readout, are diffusion-limited below 20~\kevr~for helium, and 50~\kevr~for fluorine recoil. Hence the directionality threshold and WIMP sensitivity should improve even further with a detector operating point (gas pressure and drift field) fully optimized for these high-resolution readouts. That optimization is beyond our scope, but will be an important part of the next stage, a technical design for a large directional detector.

\subsection{Directional WIMP/neutrino discrimination and energy threshold requirements}\label{sec:direction_nu_sensitivity}
Given that WIMP and neutrino recoils produce keV-scale energies and steeply falling energy spectra, it is clear that the directionality thresholds of different readouts will have a major effect on the WIMP and neutrino sensitivity. Figure \ref{fig:n_90cl_isotropy_10gev} (left) shows the mean number of WIMP-helium recoil events required to exclude a background recoil distribution that is isotropic in galactic coordinates, versus recoil energy threshold. The number of events required for exclusion falls with energy, because the angular resolution and head/tail efficiency both improve with higher energy, so that the WIMP recoil distributions become more asymmetric. Because the WIMP recoil rate falls with energy, however, the best strategy for an experiment is not simply to raise the energy threshold. To get a feel for the energy threshold that minimizes the exposure need to exclude isotropy, we also show the exposure required for exclusion versus energy threshold in Fig.~\ref{fig:n_90cl_isotropy_10gev} (right). Note that the unit we use as a proxy for exposure is actually the number of recoil events before the energy threshold - so while these exposure units can be used to compare readouts, they are arbitrary in the sense that they cannot be used to compare scenarios with different target nuclei or different WIMP masses, \ie~different figures in this article.

An important result in Fig.~\ref{fig:n_90cl_isotropy_10gev} (left) is that only three to four recoils above 20~\kevr~are sufficient to exclude isotropy for the two highest-performing readouts, pixels and strips, respectively. This appears consistent with previous idealized studies such as Ref.~\cite{Morgan:2004ys}, which found that 5 to 9 events are required for exclusion of isotropy at 90\% CL in 90\% of the cases, which is a stricter statistical requirement than ours. Fig.~\ref{fig:n_90cl_isotropy_10gev} (right) shows that the exposure required to exclude of isotropy plateaus below 6~\kevr for all readouts, but the required exposure is one order of magnitude lower for pixels and strips than the other readout technologies. This leads to two major conclusions: First, directional detectors with strip or pixel readout have about one order of magnitude higher directional WIMP sensitivity per unit of exposure than other readouts studies here. Second, to maximize directional sensitivity to 10~\gevcc WIMPs, we should aim to achieve an energy threshold of 6~\kevr~or lower. 

Before moving on, we give more detail on exactly how Fig.~\ref{fig:n_90cl_isotropy_10gev} is produced, as the same procedure is also used to produce Figs.~\ref{fig:n_90cl_neutrinos_vs_threshold} through \ref{fig:n_90cl_neutrinos_vs_nneutrino}. We perform 400 pseudoexperiments for each readout and energy threshold combination. For each experiment, we record the number of events required to achieve exclusion at 90\% CL, using only the observed recoil direction $\cos \theta$ as a discriminant. The number of events required to achieve exclusion at 90\% CL is a highly asymmetric distribution, with a significantly larger high-side than low-side tail. In Fig.~\ref{fig:n_90cl_isotropy_10gev} we conservatively plot the mean, which tends to be significantly larger than the median. The error bars shown are the asymmetric statistical uncertainties on that mean. For cases with limited directionality and low energy thresholds, sometimes a very large number of events is required for exclusion, requiring impractically long computations. To reduce computational time, we skip these scenarios: if any of the 400 pseudoexperiments require more than 5000 detected events to achieve exclusion, we abort that particular simulation and do not draw a data point. This is the reason, for example, why there are no data points for the less directional readouts in Fig.~\ref{fig:n_90cl_isotropy_10gev} (right) for the lowest energies. In this simulation, we orient the detector so that the WIMP wind is aligned with the TPC drift axis ($z$). This is equivalent to putting the detector on an equatorial mount, and continuously rotating the drift axis towards the expected WIMP wind direction. For the 3d pixel and strip readouts, this subtlety is unimportant as these readouts have close to isotropic performance. However this is a choice that will artificially improve the sensitivity of less segmented readouts with anisotropic performance or lower dimensionality, though only mildly.

Figure \ref{fig:n_90cl_neutrinos_vs_threshold} shows the analogous analysis, but now with $^8$B neutrino recoils as the background hypothesis. We maintain the recoil angle distribution in galactic coordinates as the WIMP/neutrino discriminant. Compared to the isotropic background case, slightly fewer events are required to exclude the neutrino background hypothesis. Improved discrimination is possible by including the energy spectrum, event time, and signal normalization as discriminants, to be included in Sec.~\ref{sec:wimp_reach}.

Figure \ref{fig:n_90cl_isotropy_100gev_helium} shows the case of 100~\gevcc WIMPs. For the readouts with the best performance, five or fewer fluorine recoils at high energy are sufficient to exclude an isotropic background hypothesis at 90\% CL. Note that in this case the plateau in the exposure required for exclusion occurs around 30~\kevr, while for the other scenarios studies here this plateau occurred around 6~\kevr. This means that for a directional detector optimized specifically for high-mass WIMPs, a higher energy threshold is acceptable. 

In summary, we have found that a 3d readout with good vector tracking capability, \eg~pixel or strip readout, can rule out an isotropic background or neutrinos with as few as three to ten recoils above an energy threshold of 20 to 50 \kevr, depending on the exact scenario. The required exposure for exclusion is typically minimized, however, with an energy threshold of about 6~\kevr.  Figure \ref{fig:n_90cl_isotropy} provides a compact summary of the number of helium recoils required to exclude the isotropic or neutrino background hypotheses for 10, 100, and 1000~\gevcc WIMPs, using this 6~\kevr~energy threshold. In that case typically between ten and twenty recoil events are required for exclusion via directionality only. The findings here are consistent with past theoretical works, summarized in Ref. \cite{Mayet:2016zxu}. 
 
Finally, Fig.~\ref{fig:n_90cl_neutrinos_vs_nneutrino} shows how the number of WIMP events required to exclude the neutrino-only hypothesis increases with neutrino background events present in the detected event sample. For a 10~\gevcc WIMP, the number WIMP-recoils required (in this case for an energy threshold of 6~\kevr) grows only slowly with the number of neutrino events, illustrating the power of directionality in discriminating against even large a neutrino background. 

\subsection{Electron background rejection\label{sec:electron_rejection}}

Electron backgrounds (i.e. gamma-recoils) are a key issue for \Cygnus in that they will effectively determine the energy threshold. In our background simulations of a strawman \Cygnus-1000 design (Section~\ref{sec:backgrounds}) we find that the electron energy spectrum is dominated by Compton scattering, with a flat energy spectrum in the $\mathcal{O}({\rm keV})$ region where we expect the WIMP recoil signal. Our design goal, which we show can be met for the lowest-background readouts, is an electron background rate of $10^4$ ${\rm keV_{ee}}^{-1}$year$^{-1}$. We will quantify the ability to reject electrons via the electron rejection factor, $R$, defined as

\begin{equation} R = N_{all}/N_{surv}, \end{equation} where $N_{all}$ is the number of electron background detected in a given energy range, and $N_{surv}$ is the subset of those events that survive an electron-veto algorithm that retains 50\% of nuclear recoils. $R$ is thus the factor by which we expect to suppress detected electron background via offline selections.
We will see that the electron rejection factor in \Cygnus is high and rises exponentially with energy. Assuming a six-year exposure and a flat electron background energy spectrum at a rate of $10^4$ ${\rm keV_{ee}}^{-1}$year$^{-1}$, if we assume a hard energy threshold at the center of the first 1-\kevee~energy bin where $R >  6\times 10^4$, then (due to the steep rise of $R$ with energy) \Cygnus will essentially be free of electron backgrounds. This is somewhat conservative; the electron rejection factor also rises strongly if we accept lower nuclear recoil efficiency, so that a real experiment can probably remain background free at even lower energies by accepting some efficiency loss. To quantify this lower-energy performance, we also define the {\it electron rejection turn-on} as the lowest energy where >90\% of electrons can be rejected while retaining half of the nuclear recoils. This corresponds to the energy where $R=10$.

The charge density distributions of electronic and nuclear tracks in gas are very different in both shape and scale, as demonstrated by Figs.~\ref{fig:nuclear_recoils} and \ref{fig:electron_recoils}. This suggests that good electron discrimination should be possible. As a first step, the linear charge density is a simple, yet powerful discriminant than can be obtained from the charge distributions. Lines of constant charge density correspond to lines of constant slopes through the origin in Fig.~\ref{fig:srim_and_degrad2} (right). Furthermore, we can expect the discrimination between electrons and fluorine recoils to be better than between electrons and helium recoils. Fluorine recoils have higher specific ionization and therefore differ more from electron recoils than helium recoils do. In Fig.~\ref{fig:electronrejection}, we show the electron rejection factor $R$ that can be achieved using only this simplest linear charge density. Before diffusion (data points with blue error bars), we obtain significant discrimination down to the lowest energy studied (1~\kevee) for fluorine. The electron rejection factor rises quickly with energy, and exceeds $6\times 10^4$ at approximately 3.5 and 7.5~\kevee~for fluorine and helium recoils respectively. We also see that the electron rejection turn-on occurs at order 1~\kevee~and 2.5~\kevee~for fluorine and helium, respectively.

The data points with red errors bars in Fig.~\ref{fig:electronrejection} show the result after diffusion corresponding to \SI{25}{cm} of drift. We see that the discrimination power remains high even after this amount of diffusion, and that the electron-rejection turn-on threshold moves up to about 2.5~\kevee and 3~\kevee for fluorine and helium, respectively. It should be noted that the diffusion simulated here is {\it more} than the average for a \SI{50}{cm} long maximum drift length (due to the $\sqrt{{\rm distance}}$ scaling), and that for recoils with short drift length the electron rejection will be closer to the pre-drift case. This result is remarkable because conventional direct detection experiments based on ionization signals typically run out of discrimination power at these low energies. 

An important limitation of this study worth remarking upon can be seen in the absence of pre-diffusion data points above 4~\kevee~(fluorine) and 9~\kevee~(helium). This is an artificial effect due to the finite Monte Carlo samples; in fact {\it all} simulated electron events were rejected for these bins. In reality we expect the exponential increase of the electron discrimination factors to continue. It is clear from Fig.~\ref{fig:srim_and_degrad2} (right) that the discrimination should continue to rapidly improve with energy, but studying this further will require substantial CPU time. For similar reasons, we have not studied the readout dependence of electron discrimination, unlike previous results in this section. Nevertheless, from results such as Fig.~\ref{fig:angular_resolution}, we can expect that the most segmented readouts (pixel and strip) will have rejection factors quite close to what we report here.

While already promising, we expect these electron rejection factors to only improve from here. Preliminary experimental and simulation work by some of the authors has already achieved improved discrimination with more sophisticated shape analysis of the detected recoils~\cite{dinesh_electrons, vahsenPrivate}. For example, deep learning neural net studies at the University of Hawaii achieved an impressive electron rejection factor of $R=10^6$ at 3~\kevee~for 740:20~\hesfsix mixtures and helium recoils with 100~$\upmu$m of diffusion included. We expect this to improve by one additional order of magnitude in the lower-density 755:5~\hesfsix mixture. In these preliminary results, the neural networks are not fully optimized, but already we see background rejection better than chance at 1~\kevee. With additional tuning, we expect to see good discrimination at energies less than 1~\kevee~\cite{PeterSadowski}. Putting all this together, we expect to achieve $R$ of order $10^7$ at 3~\kevee for helium recoils, which corresponds to 7.5~\kevr, for a 50\% helium recoil efficiency. Based on this estimate, we should remain background free at substantially lower energies, but more work is required to establish the exact threshold electron-background-free operation. In our reach curves, we draw 8~\kevr~as the highest possible threshold line. Because we expect high electron rejection at substantially lower recoil energies, and some electron rejection even at 1~\kevr, we also show a number of lower energy thresholds. 

Future work will therefore be devoted to experimental verification of the electron rejection factors reported here, followed by the implementation of discrimination methods using the detailed charge density distribution.

\subsection{WIMP sensitivity below the neutrino floor\label{sec:wimp_reach}}
\begin{figure*}[ht]
\includegraphics[width=0.49\textwidth]{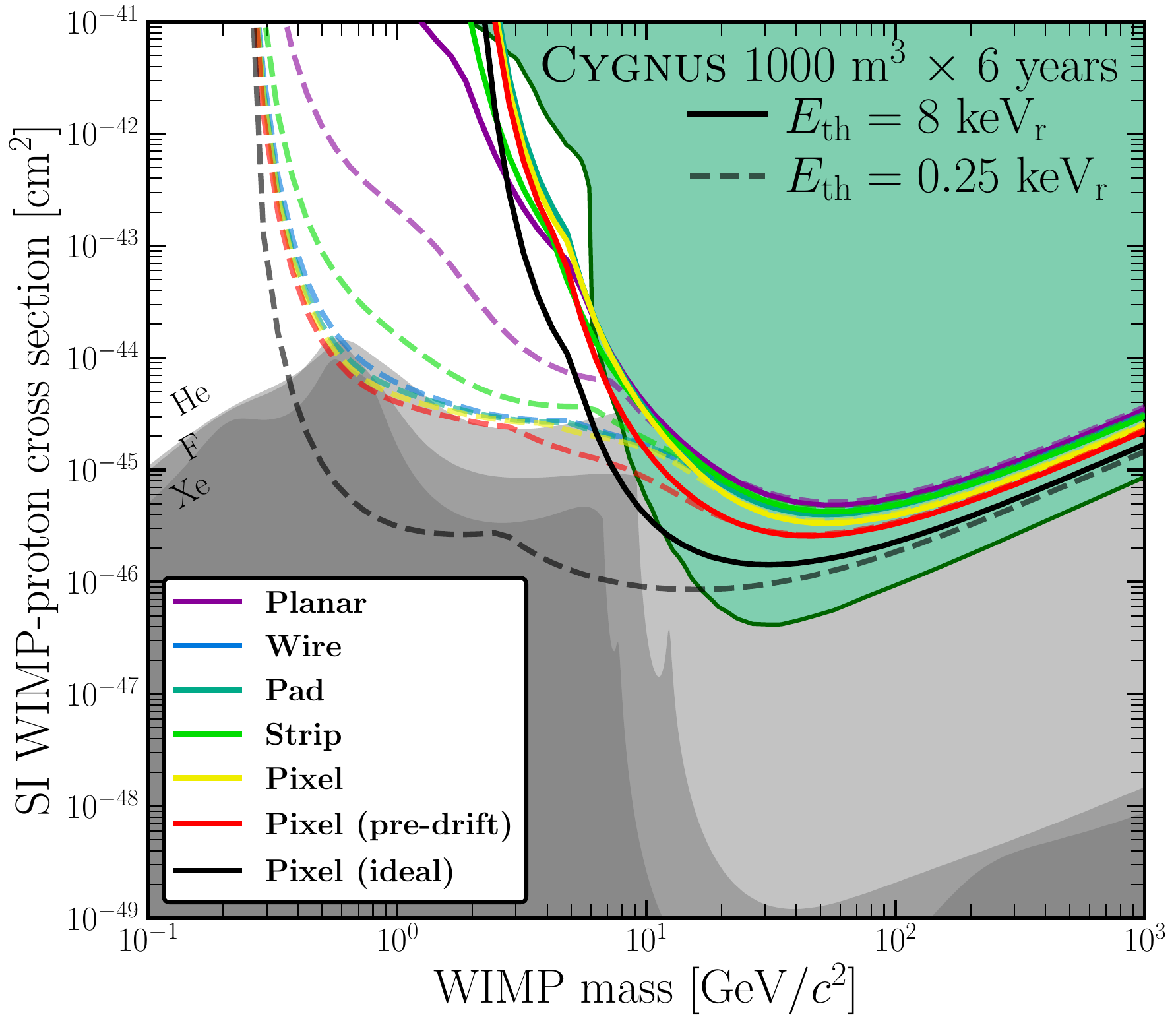}
\includegraphics[width=0.49\textwidth]{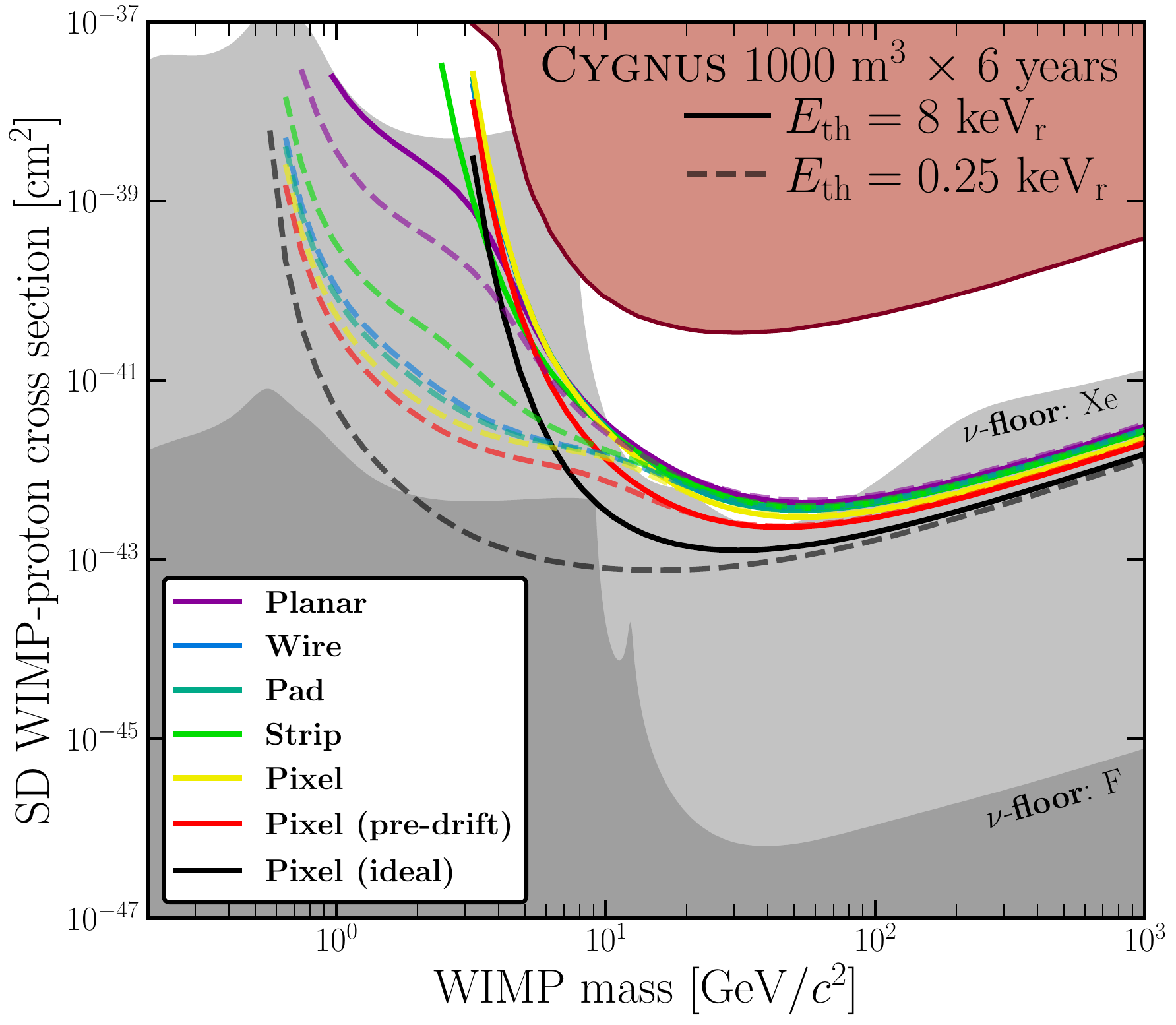}
\caption{Expected SI-nucleon (left) and SD-proton (right) 90\% CL exclusion limits for \Cygnus-1000 using \hesfsix at 755:5 Torr over an exposure of six years ($\sim 1$ ton-year exposure). The solid lines all correspond to limits assuming an 8 keV$_{\rm r}$ threshold whereas the dashed are for 0.25 keV$_{\rm r}$. We compare all readouts that have been simulated in this section, as well as two theoretical limits labeled ``predrift'' and ``ideal'' which are included to highlight the role of the directional performance in limiting the sensitivity. The two additional examples both assume the charge detection efficiency and resolution of the pixel readout, but have their directional performance modified. The ``predrift'' limits correspond to cases already shown in previous results, \ie~the angular resolution and head/tail efficiency of the track as if it were measured before diffusion. The ``ideal'' limit corresponds to an angular resolution of 0$^\circ$ and a head/tail efficiency of 100\%. The combination of all existing SI and SD exclusion limits are shown as a green and red filled region respectively. We also show the neutrino floors for helium, fluorine and xenon targets in grey.}
\label{fig:wimp_reach_allreadouts}
\end{figure*}

\begin{figure*}[ht]
\includegraphics[width=0.49\textwidth]{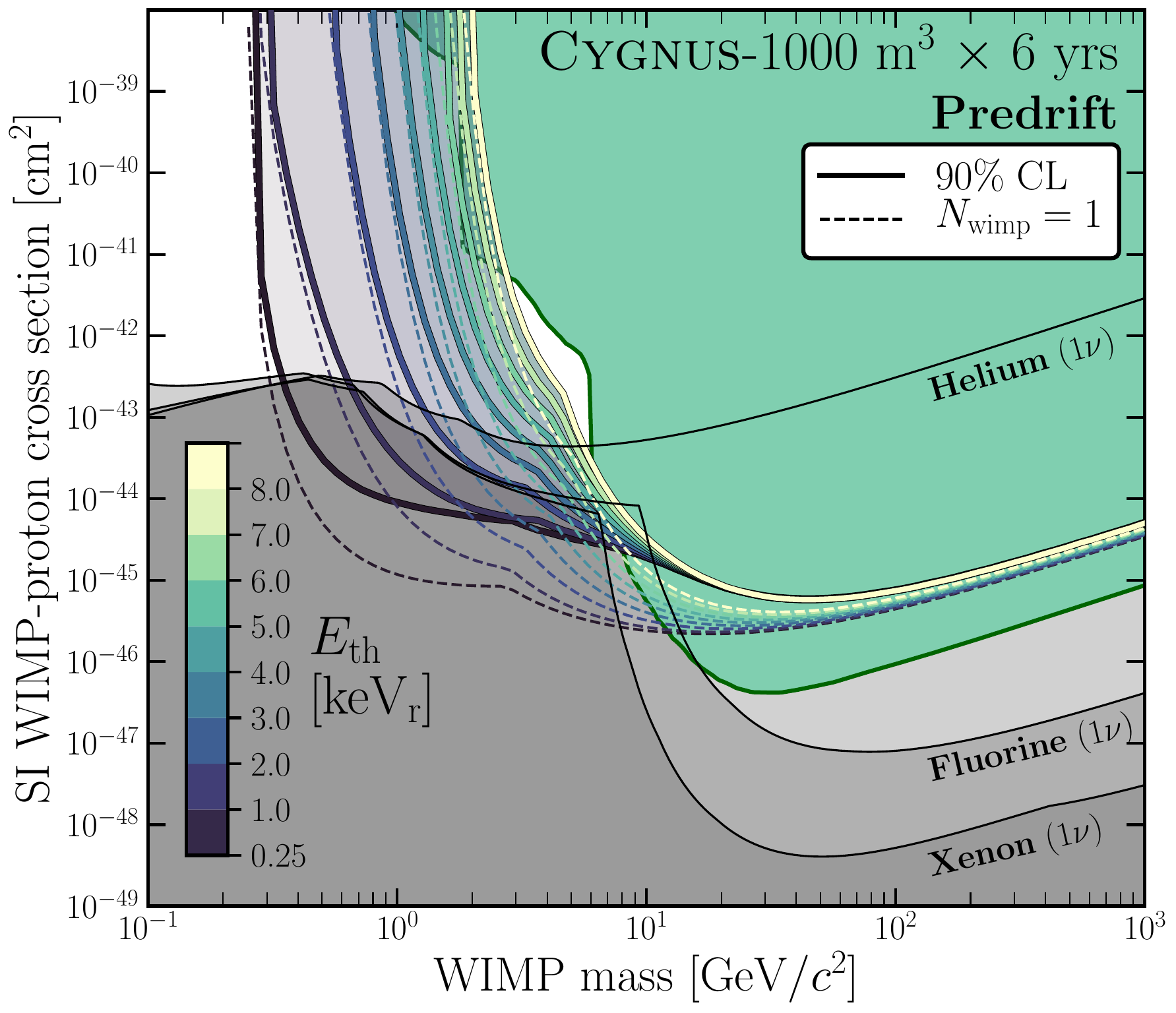}
\includegraphics[width=0.49\textwidth]{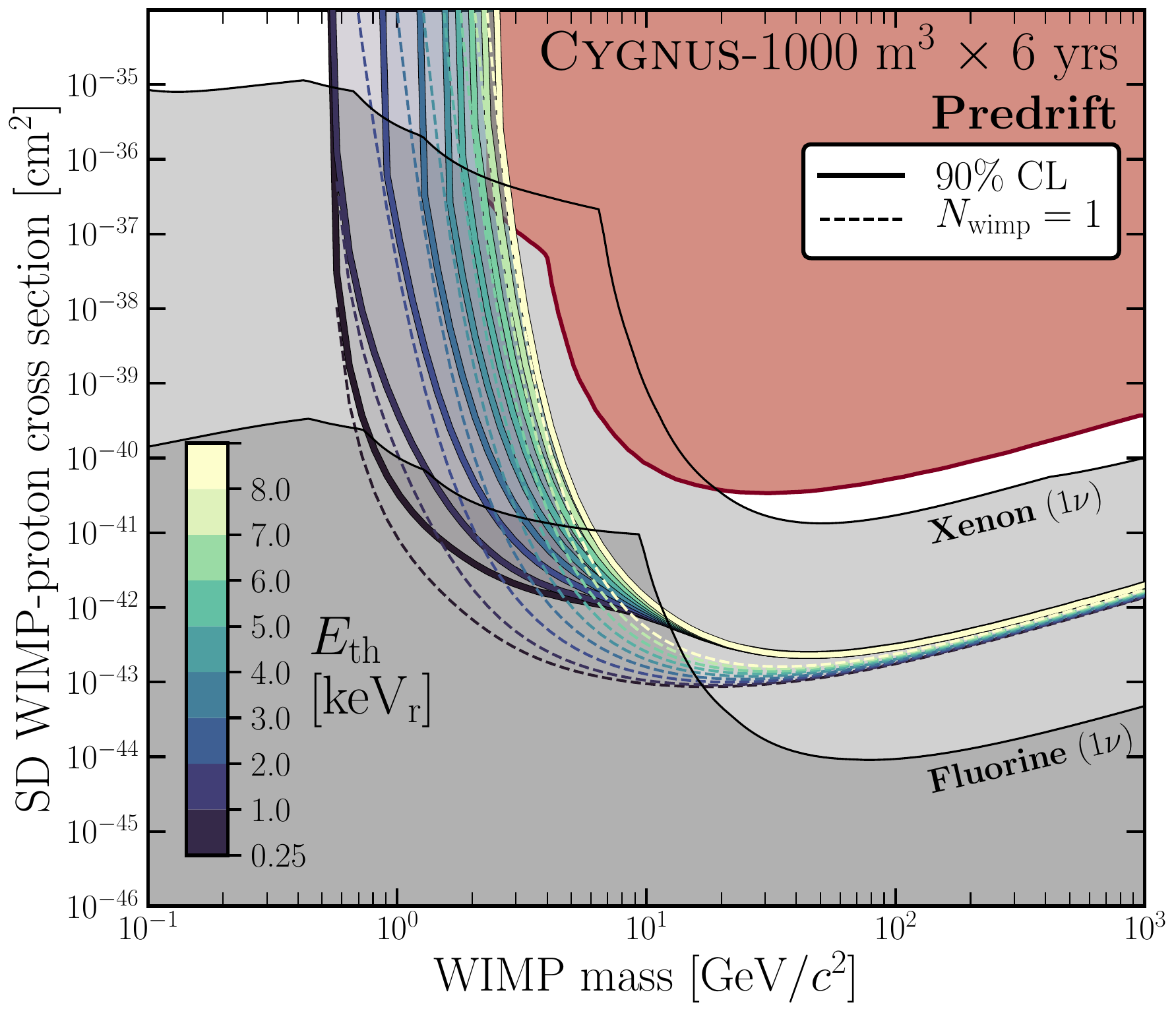}
\includegraphics[width=0.49\textwidth]{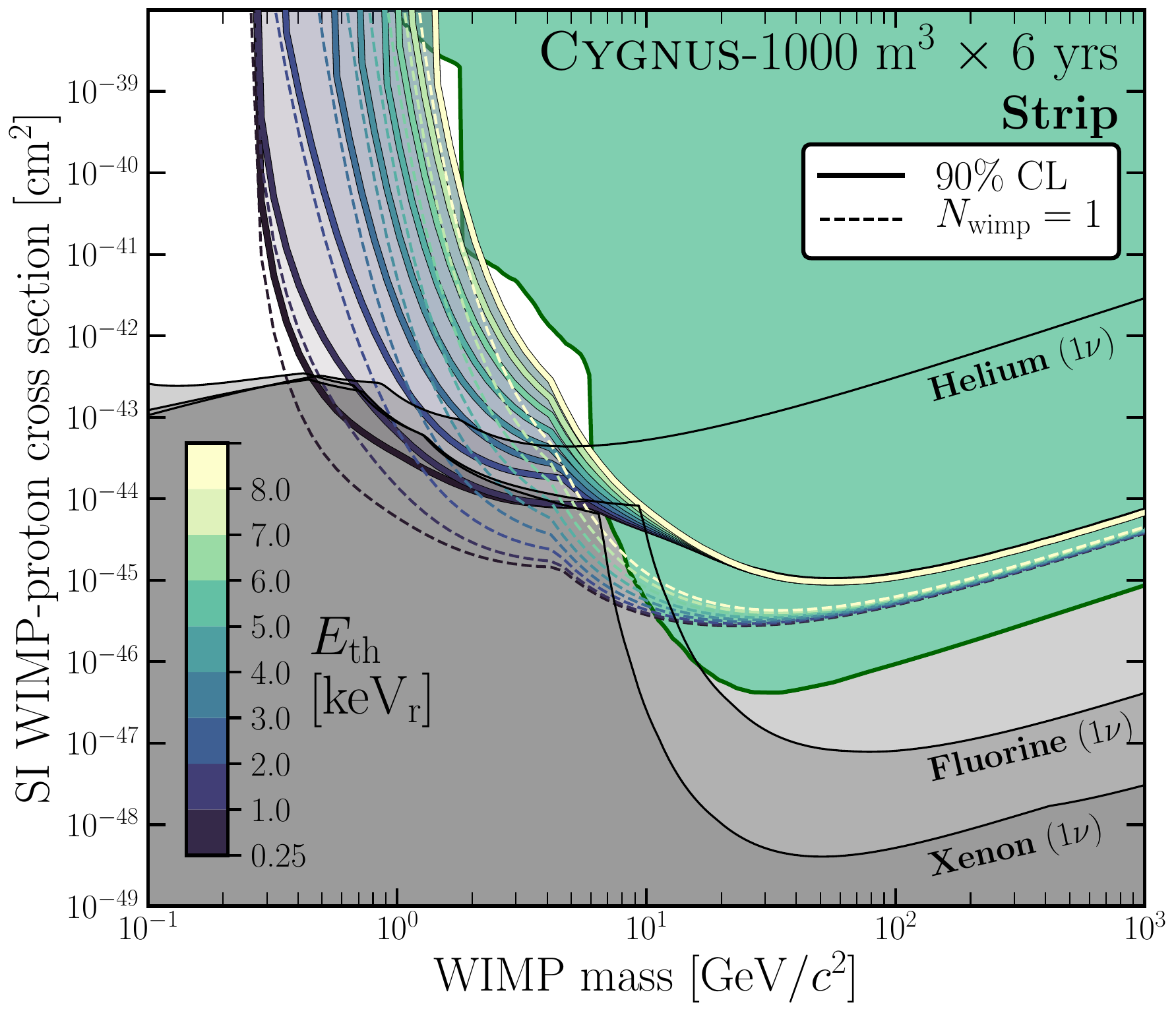}
\includegraphics[width=0.49\textwidth]{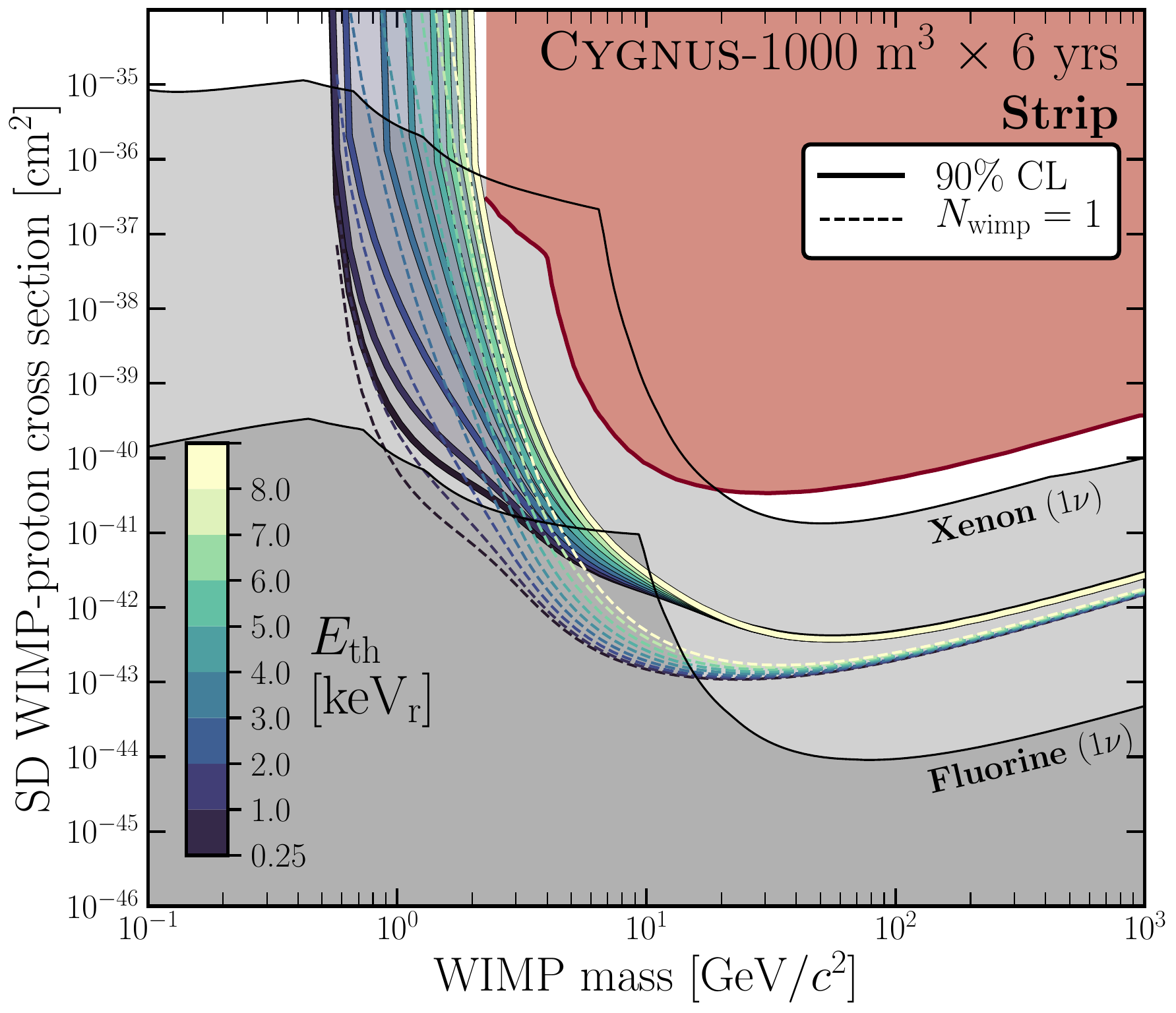}
\caption{Expected SI-nucleon (left column) and SD-proton (right column) 90\% CL exclusion limits for \Cygnus-1000 as a function of the hard recoil energy threshold. We focus on the analyses corresponding to the predrift directional sensitivity (top row) and the strip readout (bottom row). We vary the threshold again from 0.25 keV$_{\rm r}$ (single electron) to 8 keV$_{\rm r}$ (worst-case threshold for electron-background-free operation). In contrast to the previous figure we now show the 1-neutrino event lines for helium, fluorine and xenon targets in grey, as opposed to the neutrino floor. This allows us to see that the 8 \kevr~threshold only just observes a non-zero neutrino background--highlighting the importance of improving electron discrimination.}
\label{fig:wimp_reach_vs_threshold}
\end{figure*}

\begin{figure}[ht]
\includegraphics[width=0.49\textwidth]{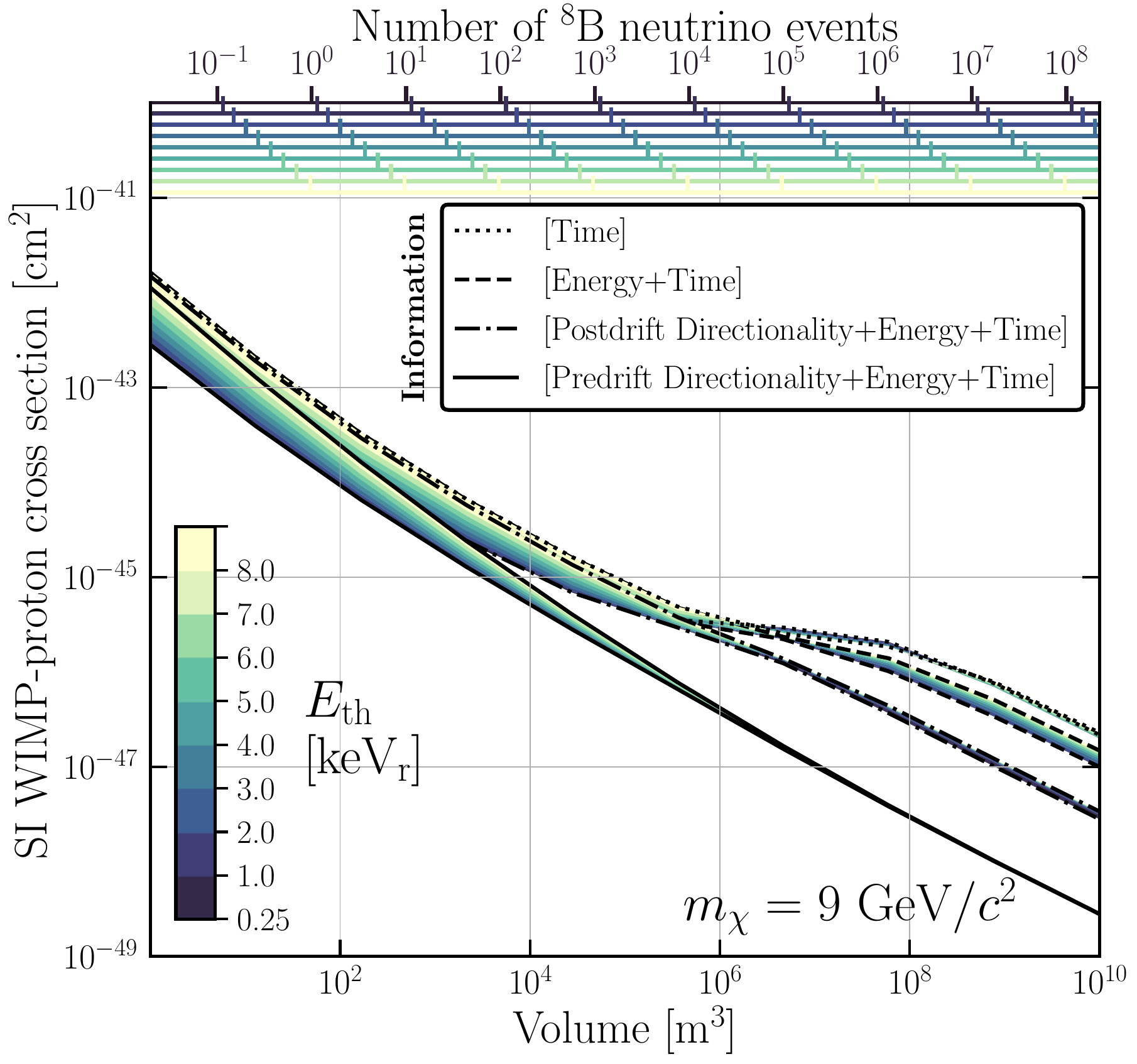}
\caption{Minimum excludable SI WIMP-nucleon cross section at 90\% CL, for a 9~\gevcc WIMP, as a function of detector volume. The colored bands correspond to energy thresholds in the same range used in the previous two figures. The number of $^8$B neutrino events observed for each threshold as a function of volume is displayed in the upper colored horizontal axes. The various regions correspond to the performance of the pixel readout when different levels of information are used in setting the limits. The four sets of limits display various levels of information used from the least powerful to the most powerful. The solid lines correspond to the theoretical optimum for a gas detector whereas the dotted lines use only the minimum amount of available information (event times).}
\label{fig:wimp_reach_vs_number}
\end{figure}
Following our discussion of the directional performance of each readout technology, we now examine how this performance translates into sensitivity to WIMP cross sections. To do this we build a signal and background model for the WIMP and neutrino recoil distributions based on the theoretical input described in Sec.~\ref{sec:science_case}. Then, all energy-dependent angular resolutions and head/tail efficiencies from Figs.~\ref{fig:angular_resolution} and~\ref{fig:readout_headtail} are applied, as well as the event-level energy resolutions and efficiencies from Figs.~\ref{fig:energy_resolution} and~\ref{fig:event_efficiency}. For the threshold, a hard cut on recoil energy is imposed at a range of values up to 8~keV$_{\rm r}$, but the Gaussian energy resolution is applied to the underlying $\textrm{d}R/\textrm{d}E_r$ before this cut is made. A harder cut of 0.25~\kevr~is applied first however, to address the uncertainty in determining the readout performance for energies below the single electron level. 

Then we use our signal and background models to calculate median 90\% CL exclusion limits using the standard profile likelihood ratio test (see \eg~Ref.~\cite{Cowan:2010js} for the statistical formalism, but the WIMP+neutrino analysis here closely follows Ref.~\cite{O'Hare:2015mda}). Each neutrino background flux normalization is accounted for as a nuisance parameter using a Gaussian parameterization. No other uncertainties or backgrounds are accounted for, so our hard threshold effectively enforces the assumption that perfect electron discrimination can be achieved above it. 

The resulting limits on the SI and SD parameter spaces are shown in Fig.~\ref{fig:wimp_reach_allreadouts} for a 1000 m$^3$ experiment and for each readout. We also include several additional theoretical limits with improved directional sensitivity to highlight the room for improvement. The limits labeled `predrift' are similar to those introduced earlier. We assume the energy resolution and efficiency of the pixel readout, but the angular resolution and head/tail efficiency is as though we were able to measure the track topology before diffusion (this emphasizes the impact of diffusion on the sensitivity). 

We also introduce in this plot a limiting case labeled ``ideal'', which takes the predrift sensitivity and simply sets the angular resolution to zero and the head/tail recognition to 100\%. This is a theoretical ideal, not possible in practice, but our fully simulated readouts are promisingly close already. This limit corresponds to almost perfect discrimination between WIMPs and neutrinos, a feat that should be possible in principle based on their angular distributions (see Fig.~\ref{fig:mollweide}). We ignore the `postdrift' curve here (which was shown in previous results), since its corresponding limits are nearly identical to the pixel readout. As demonstrated by the fact that most of the curves converge at high masses, the directional information is playing a subdominant role in setting limits when compared with statistics alone. Interestingly the poor energy resolution of several readouts does enable them to set limits at lower masses than those with superior resolution, but at the cost of poorer sensitivity over larger masses due to the lack of ability to perform signal characterization. This demonstrates that for a \Cygnus-1000 experiment the sensitivity is much more sensitive to the number of events than the overlap between the signal and background distributions. Therefore the hard threshold is something that should be considered carefully.

Fig.~\ref{fig:wimp_reach_vs_threshold} shows in more detail how the limits of two separate cases, predrift (top) and strip (bottom), depend on the imposed electron discrimination threshold. Lowering the threshold down to 3--4 keV, can claim a generous portion of unconstrained low-mass WIMP parameter space. Attempting to lower this threshold even a small amount is therefore highly motivated, even though the directionality at low energies is poorer. Unfortunately, for 1000~m$^3$ of \hesfsix at atmospheric pressure, the low target mass only allows the experiment to just reach the neutrino floor, although there is an expected neutrino event rate greater than one. To see this explicitly we show 1-neutrino event lines rather than the neutrino floor (which typically corresponds to $\mathcal{O}(100)$ expected neutrino events). Limits intersecting these boundaries (for the relevant target) observe more than one neutrino event. The current electron discrimination threshold of 8 keV$_{\rm r}$ in fluorine is extremely close to the tail of the solar CE$\nu$NS recoil spectrum; anything slightly larger would not observe the background and not cut into the floor. In fact the improvement brought about by directionality, comparing the top and bottom row, is relatively minor in comparison to the substantial benefit at low masses when lowering the threshold. In other words, the 1000 m$^3$ exposure is still statistics limited. For larger experiments this conclusion will change, as we show next.

In Fig.~\ref{fig:wimp_reach_vs_number} we show two discovery limits for individual WIMP masses as a function of the number of expected neutrino background events. The readout assumed in each case is the same (pixel) but we compare several analyses which use all, or a subset of the available information. The region outlined by dotted lines assumes only the recoil event time information is used, so that only event counts and the annual modulation enable sensitivity. The dashed lines assume that recoil energy information is also used. The dot-dashed lines assume that the directional information is used with directional performance curves simulated for the pixel readout. The lowest, solid lines correspond to the pixel readout if it were able to measure the track before diffusion, so as to highlight how much of the sensitivity is limited by diffusion.

Until now we have considered a \Cygnus-1000 benchmark volume. Comparing the limits at this point in Fig.~\ref{fig:wimp_reach_vs_number} we can see that, as hinted by the previous figures, we are still in a regime where the sensitivity gains more from the improved threshold via statistics, than from directionality. For larger experiments however the directional sensitivity adds a considerable benefit. In an experiment without directional information that also observes a sizable neutrino background, the sensitivity scales with detector volume at a rate much slower than the standard Poissonian background subtraction of $V^{-1/2}$. However, directional limits scale much faster. This information can improve limits by almost an order of magnitude when the directionality is good (see the predrift solid lines). This is also the case for the more realistic directional performance of the full readout accounting for diffusion (dot-dashed lines), however the conclusion again seems to be that diffusion is the dominating factor in restricting better sensitivity. We reiterate that the only difference between the predrift and postdrift lines is an improved angular resolution and head/tail recognition. This figure makes clear how much our sensitivity could stand to improve.

\subsection{Summary of readout performance and conclusion on cost-optimal readout choice}

\begin{table*}[]
\begin{tabular}{@{}lrrrrrrr@{}}
\toprule
Charge readout                            &  predrift 	& postdrift & \quad pixels 	& \quad strips	& \quad pads 	&\quad wires 	& \quad planar  \\ \hline
Event detection threshold (F) [\kevr]	& n/a     	& n/a 			& $<1$	&   1.5 		& $<1$	& $<1$ 		& 4	       		       \\
Event detection threshold (He) [\kevr] & n/a                      	& n/a 	& $<1$	&   3 			& $<1$	& $<1$ 		& 7	       	    \\          
Directionality threshold (F) [\kevr]	& $<1$                     	& $<1$	& 14		& 25		&  $>50$	& $> 50$ 	& $>50$		\\
Directionality threshold (He) [\kevr] 		& $<1$		& $<1$	& 4		& 6		& 17				& 22			&42\\
Electron rejection turn-on (F)[\kevee]    &  $1$                      &   $2.5$               &   -            &  -                          &   -                     &  -                        &        -                  \\
Electron rejection turn-on (He)[\kevee]   &  $2.5$                   &   $3$             &  -            &        -                  &       -               &            -              &             -                \\
Exp. penalty, exclude isotropy, 10-1000~\gevcc WIMPs (F)  	  & 4.7	& 12		& 14		& 22		& 186	& 151 	& 	166        \\                                  
Exp. penalty, exclude neutrinos, 10-1000~\gevcc WIMPs (He) & 4.5  	& 11 		& 11		& 17		& 76		& 109	& 	215        \\
Average relative exposure penalty factor      			&    n/a     &    n/a	& 1      	&  1.6 	&  10   &    10       	&  15                 \\
Approx. cost per unit readout area [US \$/m$^2$]	&    n/a 	&    n/a     		& 180k     	& 22.5k    	&    5k       &         5k        	&         0.050k                   \\
Total readout cost (US \$)						&    n/a    	&    n/a		& 360M   	&  71M  	&  105M     &   104M        	& 2M                                      \\ 
Total volume cost  (US \$) 						&    n/a    	&    n/a	& 25M      	&  39M  	&  262M     &     261M        	& 382M   \\ 
total detector cost, constant WIMP sensitivity (US \$) 	&  n/a	&  n/a 	& 385M	&  111M	&  367M	&     365M		& 384M	\\ 
total detector cost, \SI{1000}{m^3} volume (US \$) 		&  n/a 	&  n/a 	& 385M	&  70M	&  35M	&     35M		& 25M					\\ 
 \bottomrule
\end{tabular}
\caption{Summary of main performance parameters and estimated detector cost at equal directional sensitivity, for the simulated TPC charge readout technologies. Results assume a 755:5 \hesfsix gas mixture at atmospheric pressure, room temperature operation, a gain of 9000, charge diffusion of \SI{78.6}{\micro\meter\per\sqrt\centi\metre}, and maximum drift length of \SI{50}{cm}.}
	\label{tab:readout_summary}
\end{table*}

Table \ref{tab:readout_summary} summarizes the main detector performance parameters resulting from the detector and readout simulation. We define each of these parameters in turn. 

We define the {\it event detection threshold} as the recoil energy at which the event-level detection efficiency exceeds 50\%. This threshold is mainly determined by the avalanche gain and the readout noise floor, and the noise floor is in turn determined by the capacitance of the readout elements. Even with the modest gain of 9000 assumed in this study, the readouts with the lowest capacitance per readout element (pixels, pads, wires) have high efficiency at 1~\kevr, the lowest recoil energy simulated. The strip and planar readouts have larger detector elements with larger capacitance and noise, and require higher thresholds to compensate. The threshold are slightly lower for fluorine than helium recoils, because the former have higher charge density.

The {\it directionality threshold}, defined as the recoil energy where ten events are sufficient to identify a maximally directional source, is determined by the recoil track length, energy threshold, readout segmentation as well as the nuclear straggling and diffusion of drift charge. We note that without diffusion or readout effects simulated, this directionality threshold is below 1~\kevr. After diffusion but before detection, the directionality threshold is about 15~\kevr~for fluorine and 4~\kevr~for helium, mainly because the helium recoils are longer, which helps overcome the diffusion. Comparing the different readouts, we see that the pixel readout directional threshold is nearly identical to what is seen after diffusion. This means that the pixel readout essentially extracts all relevant information from the diffused ionization distribution. The segmentation of this readout is thus sufficient, and finer segmentation would not improve the directional performance further, unless we also achieve lower diffusion. This may be possible with NID at larger electric field strengths, which should be explored. Note that the performance ordering of the detectors is different for directionality threshold than for energy threshold, especially for strip readout. The reason is that strip readout has high capacitance per readout element, which raises the energy threshold, but excellent segmentation once x/y strip coincidence is utilized, which lowers the directionality threshold.

We define the {\it electron rejection turn-on} threshold as the detected energy at which at least 90\% of electron background events can be rejected while maintaining a 50\% nuclear recoil detection efficiency. This threshold is less than 1~\kevee (fluorine) and 2.5~\kevee (helium) before diffusion, and approximately 4~\kevee~after diffusion for the 755:5 Torr \hesfsix gas mixture where we have simulated electrons.  At higher energies, the electron rejection power improves exponentially. We take these electron rejection turn-on thresholds as upper limits, as we expect improved electron rejection, and consequently even lower energy thresholds, are possible with more sophisticated selections. Preliminary work on this is highly encouraging, see Section~\ref{sec:electron_rejection}. A full analysis of electron rejection capabilities of different readouts, with more sophisticated discriminants, is important for future experiments, and should be followed up in future work. 

We have seen that ruling out an isotropic background (in galactic coordinates) or neutrino background hypothesis using only the directional distribution of detected events, is possible at 90\%~CL with as few as three recoils with energy $E$ greater than 20~\kevr~or of order ten recoils with $E >  6$~\kevr, for the highest performance readouts (pixels and strips), for WIMP masses of 10, 100, and 1000~\gevcc. To fairly compare readouts with different energy thresholds and directional performance, we report in Table~\ref{tab:readout_summary} the {\it exposure penalty}, defined as the number of WIMP interactions with $E>6$~\kevr~that must take place, in order to detect sufficient events to rule out the background hypothesis. This would be the first step towards confirming the cosmological origin of a tentative WIMP dark matter signal. In order to reach identical WIMP sensitivity in this context, detectors with different readouts need to have an exposure proportional to their respective exposure penalty. We find that this penalty does not depend very strongly on whether the background is isotropic or due to neutrinos. The penalty is also roughly the same for 10, 100, and 1000~\gevcc WIMPs. 

We see that it instead increases with the directionality threshold, which makes sense, given the falling energy spectrum of WIMP recoils. Quantitatively, we find that a pixel readout needs the lowest exposure to rule out a background via directional measurements. Hence we average the penalty factors for the six scenarios (three WIMP masses and two background hypotheses for each), and normalize them to the pixel readout, to arrive at the {\it average relative exposure penalty} factor. We find that strip, pad, wire, and planar readouts require on average 1.6, 10, 10, and 15 times higher exposures, respectively, to reach the same directional WIMP sensitivity as the pixel readout.

\begin{figure}[hbt]
\begin{center}
\includegraphics[width=0.49\textwidth]{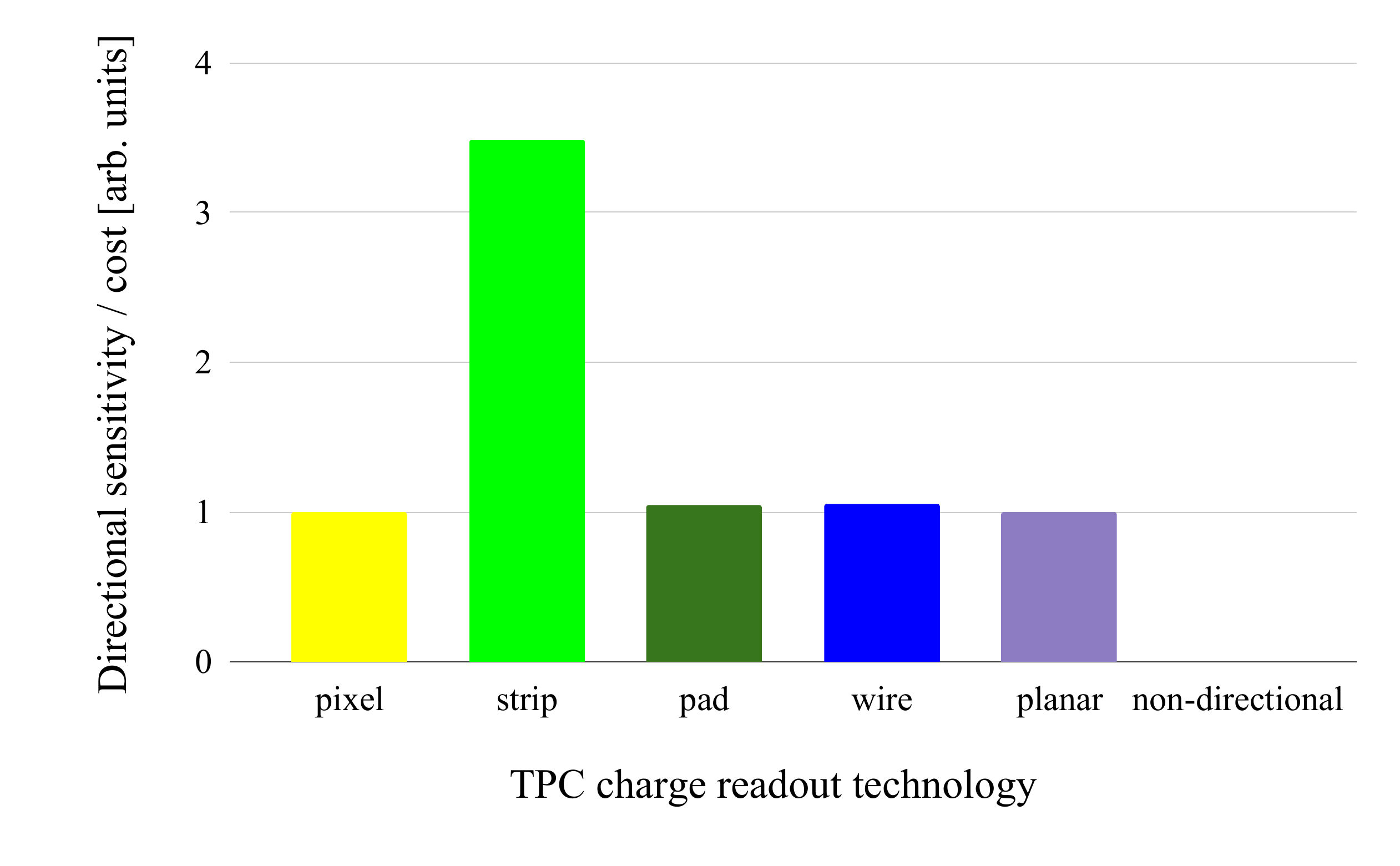}
\caption{Estimated directional sensitivity per unit cost for different TPC readout technologies. Pixel readout, which has the highest performance, is taken as the reference, with directional sensitivity per unit cost equal to unity by definition. We find that strip readout provides the optimal tradeoff between cost and performance. Note that non-directional nuclear recoil detectors score zero on this performance metric.}
\label{fig:cost_shootout}
\end{center}
\end{figure}

Next, we estimate cost. Given that pixel readout has the lowest exposure penalty, we take a detector with a volume of \SI{1000}{m^3} and with pixel charge readout as a baseline scenario. For each of the other readout technologies, we set the volume equal to the product of \SI{1000}{m^3} and the average relative exposure penalty factor for that particular technology. This way, we are comparing designs with identical expected directional sensitivity. For all designs, we assume a drift length of \SI{50}{cm}, so that there are two readout planes per \SI{1}{m^3} unit cell of the detector.

We consider two cost components: The first component is {\it total readout cost}, which we take to be the product of three factors: \SI{2000}{m^2}, the {\it readout cost per square meter}, and the average relative exposure penalty factor. The two latter factors both depend on readout choice, and vary substantially. Interestingly, the two effects cancel to some degree, because the readouts with the best performance have the lowest exposure penalty but are also the most costly.  As a results, the variation in total readout cost is not as large as one might have expected. For pixel readout, we assume a readout cost of \$180,000/${\rm m^2}$, which is a conservative factor of two higher than the raw pixel ASIC cost, to account for readout integration cost and non-recurring engineering costs, both of which are high for this technology. For the strip readout, the cost is based on \$12,500/${\rm m^2}$ for strip micromegas or similar charge-readout and \$10,000/${\rm m^2}$ for readout ASICs (10,000 readout channels per m$^2$ at a cost of \$1/channel). We find that strip readout and planar readout have the lowest total readout cost.

The second cost component considered is cost per unit volume, which is independent of readout type and assumed to be US \$ 5000 (\$5k). This cost term accounts for all volume-dependent cost other than the charge readout itself: the vacuum vessel, field cages, shielding, gas, the charge avalanche device, and downstream data acquisition. For the pixel readout, we base the cost on a volume of  \SI{1000}{m^3}, resulting in a total volume cost of US \$ 5 million (\$5M). For the other readouts, we increase the volume (and thus the volume cost) by the relative exposure penalty factor. The resulting estimated total volume cost is quite high for the less directional readout choices, such as pad, wire and especially planar readout. While cost-effective per unit readout area, these readout choices would require expensive and large detector volumes to reach the desired directional sensitivity.

Strip readout emerges as the technology with the lowest cost at equal directional sensitivity, or equivalently, as the technology with the highest directional sensitivity per unit cost. The latter is shown graphically in Fig.~\ref{fig:cost_shootout}, normalized to pixel readout. A strip readout TPC with a fiducial volume of about \SI{1000}{m^3}, would have total cost of about \$70M, with about two thirds of that cost being associated with the specific charge readout. This total cost is similar to the cost of a typical HEP detector, but much less than the cost scale of a particle collider. This seems like a reasonable scientific investment compared to others in the field, given that it would allow us to make unique experimental progress on one of the most significant unsolved problems in physics (dark matter).  Such a detector could rule out isotropy of a detected recoil distribution from 10 to 1000~\gevcc WIMPs at 90\%~CL with 22 WIMP-helium-recoils above 6~\kevr, or with only 3-4 WIMP-helium-recoils above 20~\kevr.

A detector based on pixels would be approximately 3.5 times more expensive, but have better low-level performance. With the lowest exposure penalty, it would require the smallest detector for a given physics reach, but the cost is currently dominated by the relatively expensive pixel ASICs. The number of pixel chips, and hence cost of pixel chip readout planes, could potentially be greatly reduced via so-called charge focusing -- electrostatic focusing of the drifting ionization before detection~\cite{Ross:2013bza}. Alternatively, the cost of pixel chips may be considerably lower with future semiconductor manufacturing processes or chip designs. In either case, pixel chip readouts, or similar ultra-high-resolution charge readouts could then become the most cost-effective readout options. With further reductions in diffusion, which would need to be demonstrated, a pixel-based design is likely to be the most competitive options at WIMP masses of order 1~\gevcc and lower. 

As it stands, pad, wire, and planar score similarly in terms of directional sensitivity per unit cost. These readouts are performance-limited due to relatively coarse segmentation. For planar readout, large segmentation is inherent to the design and cannot be reduced. Planar readout may still prove somewhat more competitive by utilizing a dedicated discriminant, as opposed to the general event reconstruction utilized in the comparison here. A detector based on wires or pad technologies could become substantially more cost competitive if a reduction in feature size is possible, which would lower the directional threshold and exposure penalty, while maintaining the relatively low readout cost. Here, pads have the benefit of offering full 3d directionality, while wires have the benefit of lower backgrounds. 

This brings us to out next topic: we note that the cost/performance analysis here does not directly consider the effect of electron discrimination. As discussed above, it appears likely that pixel and strip readout will have the best electron background rejection capabilities due to their high segmentation. However the wire readout has less intrinsic radioactive backgrounds, so that a smaller electron rejection factor will be required. 

Taking a step back, we note in closing that a higher-gain TPC with electron drift gas could also be competitive. That option should be studied further, but is beyond the scope of this article.

\section{Zero background feasibility}
\label{sec:backgrounds}

Direct DM search experiments typically strive to control backgrounds to less than one expected event within the fiducial volume over the anticipated exposure time and energy range (so-called zero background). As we have discussed in previous sections, directional detectors---more-so than non-directional detectors---are in principle able to tolerate a non-zero level of nuclear recoil background, for example from neutron or neutrino interactions, while still being able to positively identify a DM signal. However, any backgrounds will negatively impact the sensitivity, so this level of tolerance will depend on the capabilities of the detector technology. 

For \Cygnus we aim for zero electron background from radioactivity of the detector and surrounding materials, and neutron recoil background that is a factor of four smaller than the expected nuclear recoil count from neutrinos. This section describes the feasibility of this for a 1000 m$^{3}$ TPC. The simulations and preliminary result with machine learning previously discussed suggest that an electron background of order 10$^{4}$ yr$^{-1}$ keV$^{-1}$ could be tolerated even for energies below 10 keV. Given the quick rise of electron rejection with energy, this means that for a six year exposure, \Cygnus will be background free above the energy at which the electron rejection crosses 6$\times{10^{4}}$. We also saw in earlier sections that we expect approximately 40 nuclear recoils above a 1\kevr~threshold over a six year exposure, or about 4 per year (see Fig.~\ref{fig:NuRates}).

We therefore set the following, preliminary, background design goals for \Cygnus: (1) an intrinsic electron background rate of 10$^{4}$ yr$^{-1}$ keV$^{-1}$, (2) one nuclear recoil above 1~\kevr~from neutrons per year, and (3) any other backgrounds from other sources, notably radon, should be controlled to at least this level. This then leaves neutrino induced recoils as the only non-negligible nuclear recoil background. In what follows we must verify if these design goals are achievable. To streamline a detailed discussion of all anticipated backgrounds for \Cygnus we set a benchmark nuclear recoil threshold of 1~\kevr and calculate background rates above this value. This will inform the amount of shielding required and the level of material radiopurity needed to reach the design goals.

As discussed in Section~\ref{subsec:gaschoice}, we have been iterating on the \Cygnus target gas during the conceptual design process. While the final results presented here are for a 755:5 Torr He:\sfsix mixture, the discussion on backgrounds below will also iterate through a number of gases, and scale backgrounds from one to the next.

To start out, we make a few simplifying assumptions. For the vessel and gas we assume that the TPC is a $10\times10\times10$~m cube filled with 20 Torr of SF$_6$. This choice allows detailed exploration of a basic, large, low-density negative ion gas configuration, but one for which the data can reasonably be extrapolated to estimate backgrounds in other gas mixtures and pressures, including those used elsewhere in this paper.   We assume a single average thickness for the vessel wall, rather than individually simulating the support structures that would be needed in a real vessel. Then, for the readout we assume it is configured as in the TPC design of Fig.~\ref{fig:TPCdiagram}, with a 50~cm drift distance (following the previous section). The total readout area is therefore 2000~m$^2$. The laboratory environment is based on the salt rock found at Boulby, UK. This site has certain advantages such as a low rock gamma flux, but we leave the discussion of the various site considerations to Sec.~\ref{sec:sites}.


Sections~\ref{subsec:Neutron-Backgrounds}--\ref{cosmogenic activation} detail all neutron, gamma, radon-related and cosmogenic backgrounds respectively. We then describe in Sec.~\ref{atm_backgrounds} how the results can be extrapolated to the atmospheric pressure \hesfsix gas mixture currently envisaged for \Cygnus. We give our final background summary and recommendations in Sec.~\ref{background conclusion}.



 
\subsection{Neutron backgrounds \label{subsec:Neutron-Backgrounds}}
Neutrons produce nuclear recoil events in the same range of energies as those expected from WIMPs. The neutron background is therefore one of the most problematic for all direct detection experiments. However, mitigating against it in a low-density gas TPC is distinct from the methods used in solid or liquid-based detectors. The low density of gases means that neutrons are less likely to undergo double or multiple scattering, reducing the potential to veto on this basis. Estimating neutron backgrounds by extrapolation from existing background simulations for massive xenon or bolometric detectors~\cite{Aprile:2013} for example is therefore not appropriate. Instead, we must perform a dedicated Monte Carlo simulation for \Cygnus, for which we use Geant4~\cite{Agostinelli:2002hh}. We simulate the neutron background from the rock, shielding, vessel, and internal readout components using neutron energy spectra obtained from SOURCES~\cite{osti_976142}. The spectra include neutrons originating from the $^{238}$U and $^{232}$Th decay chains due to spontaneous fission and $\left(\alpha,\text{n}\right)$ reactions. For neutrons due to cosmic ray muons interacting with the detector or surrounding materials, we use the simulation code MUSUN~\cite{2009CoPhC.180..339K}. We now detail the rate calculation for each of these components in turn.


\subsubsection{Rock neutrons and passive shielding\label{subsec:rock_neutrons}}
\begin{figure}
\noindent \begin{centering}
\includegraphics[width=\columnwidth]{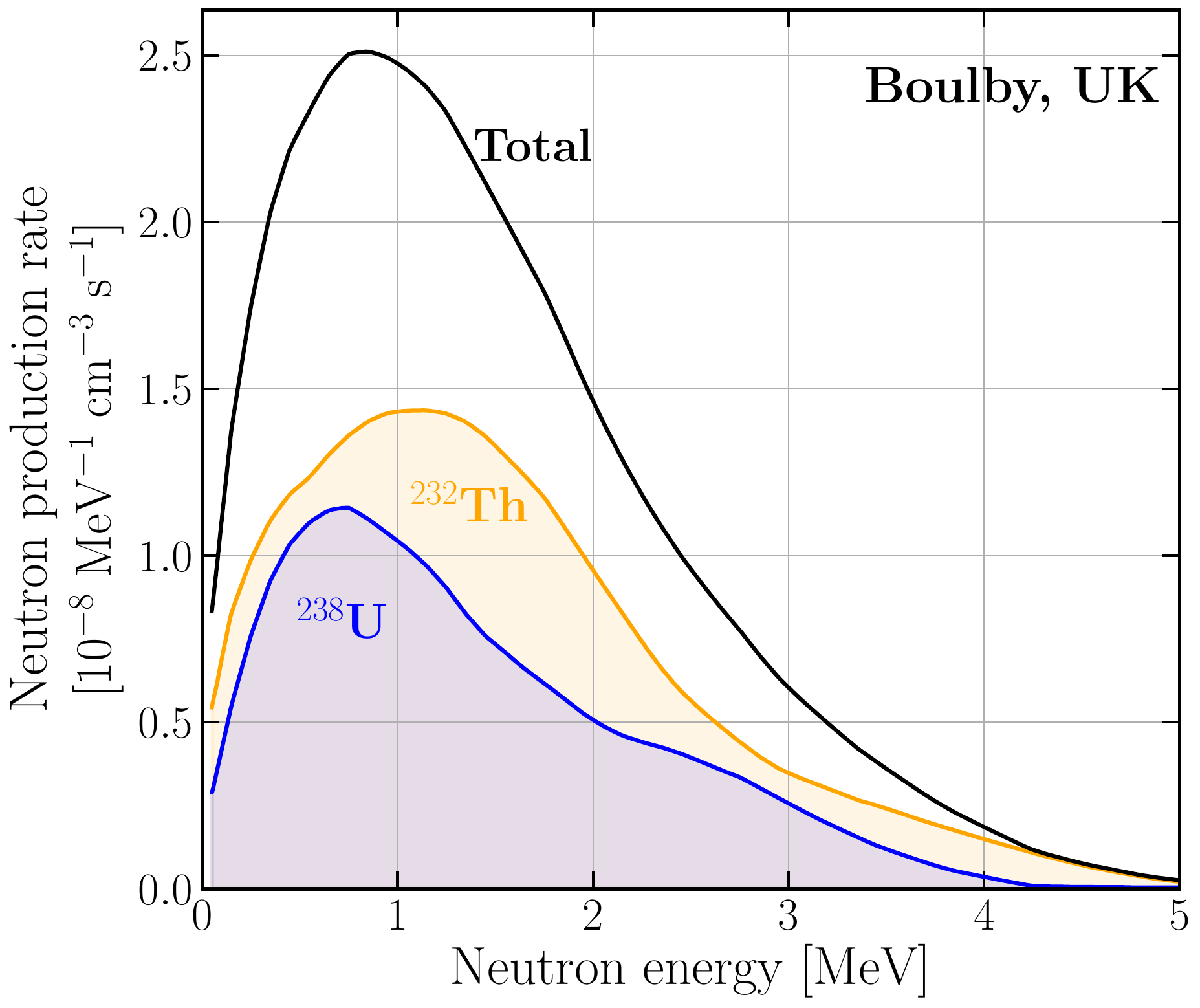}
\par\end{centering}
\caption{Neutron energy spectrum from the salt rock cavern surface at Boulby. The blue and orange curves show the spectra due to $^{238}$U and $^{232}$Th respectively, and the black line shows the sum of the two. \label{fig:rock_neutron_spectrum}}
\end{figure}

Concerning the salt rock at the Boulby site we use the measured values of $70\pm10$ ppb of $^{238}\text{U}$ and $125\pm10$ ppb of $^{232}\text{Th}$~\cite{UKDMC}. Figure~\ref{fig:rock_neutron_spectrum} shows the energy spectrum of the neutron flux from isotopic decays of these nuclei.  Since the flux saturates beyond 3~m, it is sufficient to only simulate the rock to this depth to accurately replicate the laboratory neutron flux. We then simulate the resulting nuclear recoils in the gas volume. To reduce the computational burden the simulations assumed 600~Torr of gas and then used the linear relationship between the nuclear recoil rate and the gas pressure to scale each result to 20 Torr. This should hold well assuming that the double scattering of neutrons in the gas remains negligible. Nevertheless, we have crosschecked that the rates at the nominal and increased pressure are consistent with this scaling. Water was selected for the neutron shielding and is simulated at different thicknesses surrounding the \Cygnus gas volume. The rock, water and gas material is simulated using a cubic geometry enclosing each respective volume. No space is simulated between the volumes as this was seen to have a negligible effect on the results. We find that a 75~cm thickness reduces the rate to an acceptable $\lesssim 0.93$~yr$^{-1}$ (90\% CL).\footnote{Possible contamination of the water by radionuclei is not accounted for here as the background rate is likely to be subdominant. See for example Ref.~\cite{UKDMC}, which lists water radioactivity measured as low as~10$^{-4}$ ppb for both $^{238}\text{U}$ and $^{232}\text{Th}$.}

\subsubsection{Vessel neutrons\label{subsec:Vessel Neutrons}}
The vacuum vessel---the TPC component with by far the largest mass---has the potential to dominate the neutron background. Here we consider the possibilities of vessels constructed of steel, titanium, copper or acrylic. For each material, we consider a thickness, $T$, ranging from 5-30 cm, and compute the neutron recoil rate, $R_n$, in the gas volume using the formula,
\begin{equation}
R_n = \frac{N_r}{N_{\rm tot}} \Phi_n \, V\, a\, .
\label{eq:neutron_rate}
\end{equation}

Where $N_r/N_{\rm tot}$ is the ratio between the number of neutron recoils that occurred within the gas volume and the total number simulated. The rate is proportional to the material volume, $V(T)$, and the neutron production rate within the vessel material, $\Phi_n$, which is expressed as $V^{-1}$ $s^{-1}$ $a^{-1}$, where $a$ is the contamination of the vessel material by radioisotopes. The values of $a$ used here for $^{238}$U or $^{232}$Th isotopes within each material under consideration are listed in Table~\ref{tab:mat_back}. 

 \begin{table}[ht]
	\begin{centering}
	\begin{tabular}{lcccl}
	\toprule
	Material & $^{238}$U  & $^{232}$Th  & $^{40}$K & Reference \\
	\midrule
	Steel & 0.27 & 0.49 & 0.40 & LZ \cite{Mount:2017qzi} \\
	Titanium & <0.09 & 0.23 & < 0.54 & LZ \cite{Mount:2017qzi} \\
	Copper & < 0.012 & < 0.0041 & 0.061 & NEXT-100 \cite{Alvarez:2012as} \\
	Acrylic & 0.029 & 0.039 & 2.1 & SNO+ \cite{Hallin:2017} \\
	Silicon & < 12.35 & < 4.07 & < 6.81 & UKDMC  \cite{UKDMC} \\
	Aluminum & < 0.52 & 1.94 & < 6 &  \cite{1742-6596-718-6-062050} \\
	Polyimide & < 36.79 & < 27.52 & < 410 & NEWAGE \cite{Hashimoto:2017hlz,KentaroPrivate} \\
	Kapton &  < 98.77 & < 36.59 & 58.82 & UKDMC \cite{UKDMC} \\
	\bottomrule
	\end{tabular}
	\par\end{centering}
		\caption{Published material contamination levels and upper limits (in mBq kg$^{-1}$), used for our background simulation.}
	\label{tab:mat_back}
\end{table}

In some cases the published upper limits on $a$ for certain materials will likely give background rates exceeding our requirements. In these cases we rearrange Eq.(\ref{eq:neutron_rate}) to find the value that would be required to give $R_n<1$~yr$^{-1}$. The values of $R_n$ and required purities are summarized in Table \ref{tab:vessel_backgrounds}. Note that in this and subsequent tables the upper limit is inherited from the isotope measurement and the error is the statistical error from the Monte Carlo simulation.
\begin{table}[ht]
	\begin{centering}
	\begin{tabular} {lllcc}
	\toprule
	Material & {$\text{Thickness}$} & Rate & $^{238}$U limit & $^{232}$Th limit \tabularnewline
	 & (cm) & (yr$^{-1}$) & (mBq kg$^{-1}$) & (mBq kg$^{-1}$) \tabularnewline
	\midrule
	Steel & 5 & 21$\pm$4 & 0.016 & 0.019\tabularnewline
	& 10 & 50$\pm$9 & 5.5$\times{10}^{-3}$ & 9.4$\times{10}^{-3}$\tabularnewline
	& 20 & 177$\pm$25 & 1.8$\times{10}^{-3}$ & 2.3$\times{10}^{-3}$\tabularnewline
	& 30 &  242$\pm$36 & 1.5$\times{10}^{-3}$ & 1.5$\times{10}^{-3}$\tabularnewline \hline
	Titanium & 5 & < 11$\pm$3 & 0.013 & 0.015\tabularnewline
	& 10 & < 45$\pm$8 & 4.2$\times{10}^{-3}$ & 3.2$\times{10}^{-3}$\tabularnewline
	& 20 & < 88$\pm$15 & 2.9$\times{10}^{-3}$ & 1.5$\times{10}^{-3}$\tabularnewline
	& 30 & < 200$\pm$28 & 1.2$\times{10}^{-3}$ & 6.9$\times{10}^{-4}$\tabularnewline \hline
	Copper & 5 & < 0.39$\pm$0.07 & \tick & \tick\tabularnewline
	& 10 & < 1.0$\pm$0.16 & \tick & \tick\tabularnewline
	& 20 & < 2.0$\pm$0.3 & 5.6$\times{10}^{-3}$ & \tick \tabularnewline
	& 30 & < 2.6$\pm$0.4 & 4.1$\times{10}^{-3}$ & \tick \tabularnewline \hline
	Acrylic & 5 & 0.11$\pm$0.01 & \tick & \tick \tabularnewline
	& 10 & 0.17$\pm$0.02 & \tick & \tick \tabularnewline
	& 20 & 0.21$\pm$0.04 & \tick & \tick \tabularnewline
	& 30 & 0.17$\pm$0.04 & \tick & \tick \tabularnewline
	\bottomrule
	\end{tabular}
	\par\end{centering}
		\caption{Neutron recoil rate for different vessel materials at different thicknesses. We display a checkmark where the existing $^{238}\text{U}$ and $^{232}\text{Th}$ limits give rates of less than one recoil per year. In cases where they do not, we instead give the value of radiopurity that would.}\label{tab:vessel_backgrounds}
\end{table}

Some initial conclusions can be extracted from the results in Table~\ref{tab:vessel_backgrounds}. Based on the most recent low background limits reported, both acrylic and copper produce significantly fewer neutron recoils than either steel or titanium of the same width. Acrylic in particular has a neutron self-shielding property due to its relatively high hydrogen content. However, while the construction of a 5--10 cm thick copper vessel or a 5--30 cm thick acrylic vessel would keep the neutron recoil rate within 1 per year, this may not be practical. In reality, a steel or titanium frame could be needed to support the vessel. The exact structure of such a frame is not explored here but from a background perspective, if this frame were to average 5~cm in width, then the  $^{238}$U and $^{232}$Th content would need to be reduced by a factor of 17 and 26, respectively for steel, or 7 and 15 for titanium. It is worth remarking that the $^{238}$U activities for titanium and copper are upper limits, so further screening tests may reveal them to be within the limit.

An alternative solution for improving the radiopurity with steel or titanium would be to use an internal acrylic shield of around 10~cm. This width would not contribute significantly towards the total neutron background but it would help to shield against neutrons produced by a steel or titanium support frame. Running at atmospheric pressure may also reduce the vessel neutron background since it would require thinner walls and less support structure, as we explain in Sec.~\ref{atm_backgrounds}.  Overall we can summarize that there is a good prospect for reaching a design of TPC vessel that can achieve the required level of neutron-induced background from that source.

\subsubsection{Internal TPC neutron background \label{subsec:readout neutrons}}
We now consider the unshieldable neutron backgrounds from readout components within the TPC. We simulate a 1~m$^2$ sheet readout positioned centrally inside the TPC gas volume, and the simulated rates are then scaled up to the required 2000~m$^2$.  We focus on THGEM and GEM amplification stages, with readout stages comprising a generic future strip readout, wire-based readout or pixel chips.  

There are some technologies which are neglected in this analysis due to the relatively high background of current devices, such as the micro-RWELL---which combines a GEM-like structure and printed circuit board readout--- and the micromegas, see Refs.~\cite{UKDMC} and~\cite{Cebrian:2010ta} respectively. These are of high interest for \Cygnus, but would both need improved material radiopurities to be incorporated in a zero-background \Cygnus-1000. Similarly, current \mupic readout backgrounds exceed what is acceptable for a 1000~m$^3$ detector, but for illustration we do include this readout in our simulation. Extensive work by authors here is underway to reduce \mupic backgrounds. We point out that even the current version of these higher-background readouts are still of interest for near-term smaller-exposure detectors. We also do not simulate readouts based on optical technology such as CCD or CMOS cameras. These would need to be placed outside of the vessel with low-activity transparent windows between the cameras and the TPC volume, thus requiring a major rethinking of the vessel design. For these reasons we focus only on the technologies listed in Table~\ref{tab:readout_list}. We briefly describe each below.

\begin{table}[ht]
\begin{centering}
\begin{tabular}{llll}
\toprule
Technology & Material & Thickness & Total mass\\
 &  & (mm) &  (tons) \\

\midrule
Gain stages: & &\\
THGEM & Acrylic & 1.0 & 2.36  \\
THGEM & Copper & 0.1 & 3.6  \\
GEM & Kapton & (0.05) & 0.0142  \\
\hline
Readout stages:& &\\
Strip & Acrylic-Cu & 1.05 & 4.16  \\
MWPC (wires) & Steel & 0.05 & 1.94$\times{10^{-3}}$ \\
MWPC (frame) & Acrylic & 10 & 0.236 \\
Pixel chip & Silicon & 0.400 & 1.86 \\
Pixel chip & Copper & 3.9$\times{10^{-3}}$ & 0.07 \\ 
Pixel chip & Aluminum & 4.5$\times{10^{3}}$ & 0.024 \\
\hline
Other:& &\\
\mupic & Polyimide & 1.0 & 2.84 \\

\bottomrule
\end{tabular}
\par\end{centering}
\caption{The simulated readouts, material components, thickness and total mass.}
\label{tab:readout_list}
\end{table}

Starting with the amplification stage, a typical GEM has a $50\upmu$m-thick Kapton layer coated with a thin, $\sim5\upmu$m copper layer on either side. We only simulate the Kapton layer here as it is the main mass contributor and has a relatively high background compared to copper. In order to achieve high gas gains of $\sim10^4$--$10^5$, the stacking of GEMs may be required~\cite{PHAN201682}, doubling or even tripling the Kapton background. A Thick GEM (THGEM) on the other hand has a thicker Kapton or acrylic layer, usually between 0.4 and 1~mm thick. We use the upper end of 1~mm for these simulations as the THGEM can then be made of low background acrylic at this thickness~\cite{Oliveira:2017}. We also simulate 0.1~mm of copper on either side of the acrylic because the copper layers of the THGEM are significantly thicker than that of an ordinary GEM.

Turning to the readout technologies, for the strip readout we assume a single layer of $x$ and $y$ copper strips on a low background acrylic substrate and use the same radiopurity values as those adopted for the THGEM components detailed above. Regarding the wire-based readout (MWPC), this has two main sources of background: the acrylic frame supporting the wires, and the wires themselves. The DRIFT-IId detector consists of two MWPCs each with three planes of 552 steel wires, where the middle plane is made from $20~\upmu$m thick wire and the two outer planes $100\upmu$m thick wire~\cite{Battat:2016xxe}. DRIFT-IId uses the field between the middle plane, which is grounded, and the two outer planes, which are set to around $-3000$~V, to provide the signal amplification. If used in \Cygnus however, the MWPC readout would be grounded and a GEM or THGEM would provide the amplification stage. Therefore, with no voltage being applied, the wires do not need to be $100~\upmu$m thick. However, to ensure greater stability at larger scales, they do need to be slightly thicker than $20~\upmu$m. We simulate $50~\upmu$m thick wires with a total steel mass of 1.94~kg (based on the total steel wire mass of 1.94~g that makes up the two MWPCs in the 1~m$^{3}$ DRIFT-IId detector). We assume two 1~m$^2$ wire planes separated by 1 cm with a 2~cm thick border as the acrylic frame. The pixel chip readout is based on the ATLAS FE-I4 pixel chip~\cite{GarciaSciveres:2011zz}, made of silicon with metal and dielectric layers. Copper and aluminum make up the bulk of the metal layers, in addition to small amounts of tantalum, chromium and titanium. Here we consider a simplified model made from a 400$~\upmu$m thick block of 98\% silicon, 1\% copper and 1\% aluminum by mass. Finally, the \mupic readout as used by NEWAGE~\cite{NAKAMURA2015737} is composed of a double-sided circuit board separated by a 1~mm-thick polyimide substrate, the latter being the main background contributor. 

A large-scale TPC readout would require structural support, most likely made of acrylic. We do not explore the potential design and the impact on the background contribution because the focus here is on the base materials. We remark however that the inclusion of a large scale support structure would further increase the background inside the vessel for all readouts and that the exact amount of support required would depend on the readout requirements. For example, with an MWPC the forces and torques exerted due to the tension in the wires may mean that a more rigid support structure would be required than for other readouts. The basic acrylic MWPC frame investigated in this section is not for support but to simply hold the wires in position. For the same reasons, the background contribution from readout electronics downstream of the primary charge sensitive device is not studied here.  However, in general this could be reduced by placing the electronics outside of the vessel or by using internal shielding. For the pixel chip readout, backgrounds from front end electronics are already accounted for, as they reside on the chip itself.

For the field cage and cathode, the acrylic frame from both of these structures provides the dominant mass, conservatively estimated to be 86.4 tons (1000$\times$DRIFT-IId). However, the highest source of background is from alumina found inside resistors that make up the resistive divider used to define the drift field between the cathode and readout. Around 700 resistors would be required in a 10~m long \Cygnus vessel, based on extrapolation from the DRIFT-IId field cage. We assume radiopurity levels of $0.4\pm0.2$ and $<0.023$~mBq/pc for $^{238}$U and $^{232}$Th respectively, following the levels reported by TREX-DM~\cite{Iguaz:2015myh} for their field cage SM5D resistors. These resistors have dimensions $6.4\times3.2\times0.55$~mm~\cite{FinechemSM5D}. A block of alumina with these dimensions is simulated at the center of the TPC gas volume and the resulting neutron background rate is scaled up to the required number of resistors.

\begin{table}[ht]
\begin{centering}
\begin{tabular}{llll}
\toprule
Technology & Material & Neutron recoils & U/Th limit\\
&&(yr$^{-1}$)& (mBq kg$^{-1}$) \\
\midrule
Gain stages: & &\\
THGEM & Acrylic  & 0.122$\pm$0.002 & \tick \\
THGEM & Copper & 2.4$\pm$1.6$\times$10$^{-3}$ & \tick \\
GEM & Kapton & < 9$\pm$1 & 13.6/18.7 \\
\hline
Readout stages:& &\\
Strip & Acrylic-Cu  & 0.123$\pm$0.002 & \tick \\
MWPC (wires) & Steel & 4$\pm$0.5$\times{10^{-4}}$ & \tick \\
MWPC (frame) & Acrylic & 0.048$\pm$0.004 & \tick \\
Pixel chip & Silicon & 25$\pm$3 & 0.31/0.40 \\
Pixel chip & Copper & 2.9$\pm$0.3$\times{10^{-4}}$ & \tick \\ 
Pixel chip & Aluminum & 0.29$\pm$0.03 & \tick \\
\hline
Other:& &\\
\mupic & Polyimide & <160$\pm$16 & 0.185/0.20 \\
Resistors & Ceramic & < 0.35$\pm$0.23 & \tick \\
\bottomrule
\end{tabular}
\par\end{centering}
\caption{Table of neutron rates and required radiopurities for the readout background. In the final right-hand column, a tick indicates that the U and Th levels are indeed seen to be sufficiently low for that material to reach the $< 1$~yr$^{-1}$ goal.  Where this is not true we give an estimate of the maximum allowed $^{238}$U and $^{232}$Th radiopurity levels (termed U/Th limit) that must be reached to obtain the $< 1$~yr$^{-1}$ goal.}
\label{tab:readout_neutrons}
\end{table}

The predicted rates and required radiopurities for all the generic internal TPC components studied in this section are listed in Table~\ref{tab:readout_neutrons}. A full readout plane requires both a gain stage and a readout stage, with background rates from both parts contributing to the total background. An example would be a THGEM gain stage combined with a strip readout.  In Table~\ref{tab:readout_neutrons} (and similarly Table~\ref{tab:readout_gammas}) the first three rows refer to components of possible charge gain elements.  For a single THGEM gain stage, values in the first two rows need to be summed.  The various components for strip, wire and pixel readouts, are given in the following rows of the table. Values for the strip readout assume a 2d copper strip format on an acrylic support. The final rows give typical values for the necessary field cage resistors and an example  \mupic readout ~\cite{NAKAMURA2015737}. Current values for Micromegas strip readout \cite{TREX_micromegas2019} are similarly high. 

Assuming the radiopurity measurements of acrylic and steel, made by SNO+ and LZ respectively (see Table \ref{tab:mat_back}), it appears that only the THGEM amplification stage combined with either wire readout or future acrylic-based strip readout can likely satisfy the \Cygnus background criteria without significant improvements to intrinsic radiopurity. Although recent \mupic design developments have allowed for a significant decrease in their alpha emission~\cite{Hashimoto:2017hlz}, a better understanding of the polyimide radiopurity will be required to estimate the true neutron-induced recoil rate more accurately. This is also the case for the GEM amplification stage and the pixel chip readout, where the activity levels of Kapton and silicon respectively are still only upper limits. For all these cases, further study could possibly reveal a lower neutron background, but if the true levels are close to the upper limits alternative materials would need to be found. Finally, readouts aside, the field cage fortunately has a neutron background within limit, based on using the TREX-DM resistor values. New resistive sheet technology \cite{Miuchi:2019qdp}, may also provide an alternative to conventional field cage structures in the future and possibly contribute a total lower background. Note also that in this study no allowance has been made for additional internal detector support structures.

These results are useful benchmark values. Ultimately however an important developmental step for \Cygnus will be to perform extensive material screening to resolve these uncertainties further and to develop new ways to install the readout planes that minimize material use. The background from all readouts could be lowered by using more radiopure materials or by attempting to improve the radiopurity of the same materials. This is particularly relevant to the \mupic readout and GEM amplification. For instance, better-refined glass polyimides are being developed for the \mupic, and a Kapton replacement for the GEM made from high purity G10 insulator is already being investigated by some of the authors. The background for pixel chips could also be reduced if charge focusing~\cite{Ross:2013bza} becomes a possibility, as this would reduce the number of units required to fill the readout plane.

\subsubsection{Muon-induced neutrons and active vetoing \label{subsec:Muon-induced-neutrons}}
Neutrons produced by muon interactions in and around the detector are another source of background. To compute the energy spectrum, flux and angular distribution of incoming muons we use the MUSUN code~\cite{2009CoPhC.180..339K}. The simulation takes into account the angular profile and the composition of the rock overburden for the transportation of cosmic-ray muons. The muon energies, positions, and momenta are then put through Geant4, which transports them through $20\ \text{m}$ of rock before reaching the detector. More than 200 million muons are generated at the surface of the rock volume. Those reaching the rock-cavern boundary are then normalized to the measured muon flux at Boulby: $\left(4.09\pm0.15\right)\times10^{-8}\ \text{cm}^{-2}\text{s}^{-1}$~\cite{Robinson:2003zj} (with a vertical rock overburden of $2805\pm45\ \text{m.w.e.}$). Over a period of $\left(2.8\pm0.1\right)\times10^{7}\  \text{s}$, no events are recorded within the fiducial volume, providing us an upper limit on the rate of muon-induced neutron nuclear recoils of 2.75 yr$^{-1}$. More extensive simulations will be required to see if the actual rate might be lower. If not, these remaining few events could be identified and tagged using either an active muon veto, such as a plastic scintillator, or by looking for coincident electromagnetic events in the gas. This could be made possible by installing PMTs in the water shielding described earlier.

\subsection{Gamma backgrounds \label{gamma backgrounds}}

The challenges in performing electron/nuclear recoil discrimination below 10 keV$_{\rm ee}$ were discussed in Sec.~\ref{sec:electron_rejection}. The electron rejection rises exponentially with energy, and is expected to exceed 10$^{4}$ between 1 and 10 keV$_{ee}$, even after diffusion. The exact threshold where this occurs will be readout-dependent, and must be determined in the future. We focus therefore on the gamma-electron recoils in this range, aiming to find vessel and readout materials with intrinsic radiopurity levels that can allow the detected background rate to be limited to $\lesssim$10$^{4}$ keV$^{-1}$ yr$^{-1}$.

\begin{figure}
\begin{centering}
\includegraphics[width=\columnwidth]{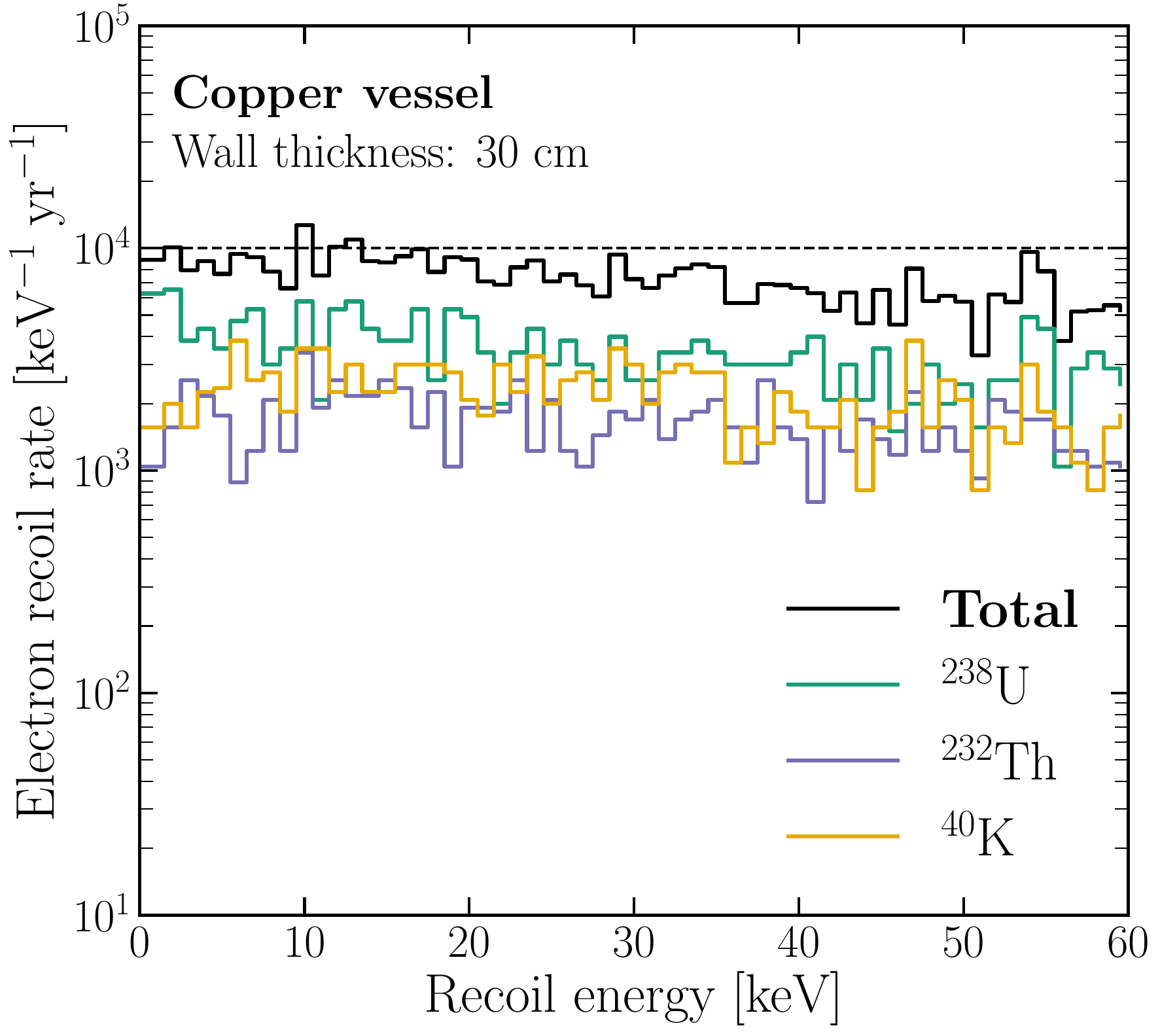}
\par\end{centering}
\caption{Gamma-induced electron recoil background contribution observed in the 20 Torr SF$_{6}$ gas volume, originating from a 30~cm thick copper vessel. We see clearly that each nuclide has an essentially flat recoil spectrum that lies below $\lesssim$10$^{4}$ keV$^{-1}$ yr$^{-1}$.}\label{fig:copper_gam_spec}
\end{figure}
Using Geant4, we homogeneously populate the rock, vessel, and sheets of readout materials with $^{238}\text{U}$, $^{232}\text{Th}$, and $^{40}\text{K}$, assuming secular equilibrium. The 75~cm water shield is also included for these simulations (see Sec.~\ref{subsec:Neutron-Backgrounds}). The corresponding gamma fluxes are then computed following the simulation of the decay chains of each isotope. For materials with relatively small thicknesses, such as the readouts, the homogeneous distribution of the isotopes allows for the simultaneous simulation of the bulk and surface background. For each isotope, the electron recoil rate due to gammas was computed as,
\begin{equation}
R_\gamma = \frac{N_r}{N_{\rm tot}} M a \, .
\label{eq:gamma_rate}
\end{equation}
Again $N_r$ is the observed number of recoils and $N_{\rm tot}$ is the total number of decays simulated for the isotope in question. As before, $a$ is the material radioactivity per unit mass for that isotope and $M$ is the total material mass. The recoil spectrum is flat to a good approximation over the 1--10 keV range because the majority of gamma recoils are due to Compton scattering. Figure~\ref{fig:copper_gam_spec}, shows an example of this: the recoil energy spectrum for gammas originating from a 30 cm copper vessel and scattering within the fiducial volume. This example is also below our electron recoil target rate, as we discuss next. 

\subsubsection{Rock, vessel and shielding gammas}\label{subsec:vessel gammas}
As in our discussion of the rock neutron background, we assume the measured contamination levels of the salt rock at Boulby, now with the inclusion of $1130\pm200$~ppm~$^{40}$K~\cite{UKDMC}. For rock gammas, an even smaller $25$~cm depth is all that contributes to the background, beyond which the rock is self-shielding. Since we found that 75~cm of water shielding was needed to shield the rock neutrons, we now include this also in our rate calculation.

The rock and vessel gamma-induced electron recoil rates for each material and thickness are listed in Table~\ref{tab:vessel_gammas}. As before for the neutron results (see Table~\ref{tab:readout_neutrons}) in the right-hand columns a tick indicates where we find the levels sufficiently low for that material to reach the $< 10^{4}$~keV$^{-1}$~yr$^{-1}$ goal for the listed isotope.  Where this is not true we give an estimate of the maximum allowed radiopurity levels (termed U/Th/K limit) that must be reached to obtain this goal. 

\begin{table*}
\begin{centering}
\begin{tabular}{llllllll}
\toprule
Material & Width & Rock $\gamma$ recoils & Vessel $\gamma$ recoils & Total $\gamma$ recoils & $^{238}$U limit & $^{232}$Th Limit & $^{40}$K Limit \\
 & (cm) & (keV$^{-1}$ yr$^{-1}$) & (keV$^{-1}$ yr$^{-1}$) & (keV$^{-1}$ yr$^{-1}$) & (mBq kg$^{-1}$) & (mBq kg$^{-1}$) & (mBq kg$^{-1}$)\\
 \toprule
Steel & 5 & 3.8$\pm$0.3$\times$10$^{6}$ & 6.6$\pm$0.6$\times$10$^{5}$ & 4.4$\pm$0.4$\times$10$^{6}$ & 0.003 & 0.0045 & 0.08 \\
& 10 & 6.0$\pm$1.0$\times$10$^{5}$ & 7.2$\pm$0.9$\times$10$^{5}$ & 1.32$\pm$0.19$\times$10$^{6}$ & 0.003 & 0.004 & 0.06 \\
& 20 & 2.1$\pm$0.6$\times$10$^{4}$ & 7.3$\pm$1.4$\times$10$^{5}$ & 7.5$\pm$1.5$\times$10$^{5}$ & 0.0027 & 0.0042 & 0.075 \\
& 30 & 4.6$\pm$3.0$\times$10$^{3}$ & 6.3$\pm$1.5$\times$10$^{5}$ & 6.3$\pm$1.5$\times$10$^{5}$ & 0.003 & 0.0053 & 0.053 \\
\midrule
Titanium & 5 & 1.0$\pm$0.2$\times$10$^{7}$ & < 2.9$\pm$0.2$\times$10$^{5}$ & < 1.0$\pm$0.2$\times$10$^{7}$ & 0.003 & 0.0046 & 0.06 \\
& 10 & 3.8$\pm$0.9$\times$10$^{6}$ & <4.13$\pm$0.36$\times$10$^{5}$ & < 4.2$\pm$0.9$\times$10$^{6}$ & 0.0022 & 0.0031 & 0.05 \\
& 20 & 6.6$\pm$1.1$\times$10$^{5}$ & < 4.17$\pm$0.53$\times$10$^{5}$  & 1.08$\pm$0.16$\times$10$^{6}$ & 0.002 & 0.0035 & 0.041 \\
& 30 & < 4.8$\pm$3.1$\times$10$^{4}$ & < 5.11$\pm$0.71$\times$10$^{5}$ & < 5.6$\pm$1.0$\times$10$^{5}$ & 0.0017 & 0.0027 & 0.041 \\
\midrule
Copper & 5 & 2.3$\pm$0.2$\times$10$^{6}$ & < 1.57$\pm$0.17$\times$10$^{4}$ & 2.3$\pm$0.2$\times$10$^{6}$ & 0.0057 & \tick & \tick \\
& 10 & 4.0$\pm$0.9$\times$10$^{5}$ & < 1.60$\pm$0.24$\times$10$^{4}$ & 4.1$\pm$0.9$\times$10$^{5}$ & 0.0058 & \tick & \tick \\
& 20 & 9.5$\pm$4.0$\times$10$^{3}$ & < 1.58$\pm$0.33$\times$10$^{4}$ & < 2.53$\pm$0.73$\times$10$^{4}$ & 0.0056 & \tick & \tick \\
& 30 & 5.1$\pm$3.3$\times$10$^{2}$ & < 1.58$\pm$0.43$\times$10$^{4}$ & < 1.6$\pm$0.5$\times$10$^{4}$ & 0.0053 & \tick & \tick \\
\midrule
Acrylic & 5 & 2.5$\pm$0.3$\times$10$^{8}$ & 3.44$\pm$0.32$\times$10$^{5}$ & 2.5$\pm$0.3$\times$10$^{8}$ & 0.0002 & 0.0017 & 0.037 \\
& 10 & 1.90$\pm$0.19$\times$10$^{8}$ & 5.97$\pm$0.57$\times$10$^{5}$ & 1.90$\pm$0.19$\times$10$^{8}$ & 5.7$\times$10$^{-4}$ & 9.3$\times$10$^{-4}$ & 0.024 \\
& 20 & 9.7$\pm$1.4$\times$10$^{7}$ & 1.14$\pm$0.12$\times$10$^{6}$ & 9.8$\pm$1.4$\times$10$^{7}$ & 3.4$\times$10$^{-4}$ & 5.4$\times$10$^{-4}$ & 0.011 \\
& 30 & 4.1$\pm$0.9$\times$10$^{7}$ & 1.12$\pm$0.14$\times$10$^{6}$ & 4.2$\pm$0.9$\times$10$^{7}$ & 3.2$\times$10$^{-4}$ & 4.9$\times$10$^{-4}$ & 0.013 \\
\bottomrule
\end{tabular}
\par\end{centering}
\caption{Gamma recoil rates in a 20 Torr SF$_{6}$ target volume, for vessel wall materials of varying thicknesses, and the required radiopurities to obtain a rate of 10$^{4}$ keV$^{-1}$ yr$^{-1}$.}
\label{tab:vessel_gammas}
\end{table*}

For metal vessel materials, rock gamma recoils are reduced by $\sim$3 orders of magnitude as the material thickness is increased from 5 to 30~cm. For acrylic, the corresponding reduction is less than 1 order of magnitude, so that acrylic by itself is insufficient to shield the rock gammas to acceptable levels. With the exception of acrylic, the radioactivity from the vessel materials does not vary significantly with material thickness. This is due to the gamma self-shielding properties of metals which cause the gamma flux to saturate at relatively low thicknesses. Therefore, there is an unavoidable gamma flux in the metals that cannot be mitigated by increasing the thickness. This occurs at a flux of $\sim$7$\times$10$^{5}$~keV$^{-1}$~yr$^{-1}$ at 20 cm, $\sim$5$\times$10$^{5}$ at 30 cm, and $\sim$10$^{4}$ at 30 cm for steel, titanium, and copper, respectively. This effect does not occur in acrylic. Indeed, Table~\ref{tab:vessel_gammas} shows that acrylic is not comparatively effective as gamma shielding and as such would not be an appropriate vessel material, at least not by itself. The only way the saturated gamma flux from the metals can be addressed further is if the radioactivity levels are lowered from what we are assuming here. Since some of the values from Table~\ref{tab:mat_back} are in fact only upper limits, further screening tests may reveal certain materials to be within limit in addition to those already shown. For example, with 30 cm thickness, copper gives a rate only slightly above $10^{4}$~keV$^{-1}$~yr$^{-1}$ and a $^{238}$U activity of around one half of the upper limit assumed here would be acceptable. A purely copper vessel could be a viable option but as we discussed in Sec.~\ref{subsec:Vessel Neutrons} if a steel or titanium frame were needed in addition to the internal acrylic shield then every material would need a reduction in radioactivity, if actual rates are close to the limits seen.

\subsubsection{Internal TPC gamma background}
Next, we consider the gamma background due to the same readout systems and field cage resistors listed in Sec.~\ref{subsec:readout neutrons}. For the latter, we assume the same radiopurity levels of $^{238}$U and $^{232}$Th from before, but we now also include the 0.17$\pm$0.7 mBq pc$^{-1}$ of $^{40}$K~\cite{Iguaz:2015myh}. 

\begin{table*}
\begin{centering}
\begin{tabular}{lllllllll}
\toprule
Readout & Material (width) & $\gamma$ recoils & U limit & Th limit & K limit \\
&& (keV$^{-1}$ yr$^{-1}$) &(mBq kg$^{-1}$)&(mBq kg$^{-1}$)&(mBq kg$^{-1}$) \\
\midrule
Gain stages: & &\\
\midrule
THGEM & Acrylic (1 mm) & 3.3$\pm$0.7$\times$10$^{4}$ & \tick & \tick & 0.54 \\
 & Copper (0.1 mm $\times$2) & < 1.5$\pm$0.3$\times$10$^{3}$ & \tick & \tick & \tick \\
GEM & Kapton (50 microns) & 1.57$\pm$0.02$\times$10$^{5}$ & \tick & \tick & 3.65 \\
\midrule
Readout stages: & &\\
\midrule
Strip & Acrylic-Cu (1 mm) & 3.4$\pm$0.7$\times$10$^{4}$ & \tick & \tick & 0.54 \\
Wires & Steel (50 $\mu$m) & 1.8$\pm$0.3 & \tick & \tick & \tick \\
 & Acrylic (2 cm $\times$ 1 cm) & 2.4$\pm$0.1$\times$10$^{4}$ & \tick & \tick & 0.88 \\
Pixel chip & Silicon (400 $\upmu$m) & < 2.55$\pm$0.19$\times{10^{5}}$ & 0.26 & 0.29 & 0.46 \\
 & Copper (3.9 $\upmu$m) & < 24$\pm$2 & \tick & \tick & \tick \\
 & Aluminum (4.5 $\upmu$m) & < 937$\pm$77 & \tick & \tick & \tick \\
 \midrule
Other: & &\\
\midrule
\mupic & Polyimide (1 mm) & < 1.3$\pm$0.2$\times$10$^{7}$ & 0.12 & 0.09 & 0.12 \\ 
Resistors & Ceramic & 2.5$\pm$1.3$\times$10$^{4}$ & 0.13 & \tick & \tick \\
\bottomrule
\end{tabular}
\par\end{centering}
\caption{Readout gamma recoil rates in 20 Torr SF$_{6}$ for different readout materials and the $^{238}$U, $^{232}$Th, and $^{40}$K radiopurity required to achieve 10$^{4}$ recoils keV$^{-1}$ yr$^{-1}$ (as well as the materials which already satisfy this requirement).}
\label{tab:readout_gammas} 
\end{table*}

Table~\ref{tab:readout_gammas} shows the gamma-induced electron recoil rates, given here in the same manner as Table \ref{tab:readout_neutrons}, split between gain stage and readout components. We can see clearly that the $^{40}$K radioactivity levels of acrylic causes the dominant gamma background for the THGEM and wire readout, both of which produce a background rate only slightly above limit. Cutting the current $^{40}$K activity of acrylic in half would bring the wire frame gamma background to within $10^4$ keV$^{-1}$ yr$^{-1}$. Similarly, a factor of four would be needed for a THGEM gain stage and for the generic strip readout. For a single GEM gain stage however one would need an order of magnitude reduction; as with acrylic, the $^{40}$K activity of Kapton is the dominant factor here. Additionally, based on the most current measurements of micromegas readouts \cite{TREX_micromegas2019} and of \mupic readouts \cite{Hashimoto:2017hlz}, we confirm that current typical radioactivity levels are, respectively, a factor of ~10$^{6}$ and ~~10$^{3}$ away from matching the activity produced by the 1 mm thick acrylic, as used here for the conceptual strip readout example.  For the polyimide in the \mupic readout, further screening is underway to better understand its radiopurity levels. In the pixel chip, silicon is the main source of gammas, since they only contain a small quantity of copper and aluminum. Again, since the silicon radiopurity is an upper limit, true levels could be lower, and more sensitive tests are required.  Also, charge focusing~\cite{Ross:2013bza}---as mentioned earlier---could help to reduce the number of pixels needed and therefore the background. Finally, the gamma background caused by the field cage resistor chain is twice the tolerable limit and a reduction in the $^{238}$U activity of ceramic by a factor of four is needed. Again, resistive sheets \cite{Miuchi:2019qdp} should be evaluated as a potential resistor replacement. 

In summary, the general conclusion for the gamma-induced backgrounds internal to the TPC is that some existing technologies are already close to meeting the design goal of $10^4$ keV$^{-1}$ yr$^{-1}$.  This appears particularly true for the THGEM-wire combination and generic strip readout.  Most other technologies and materials will require further screening and/or further reductions in background levels before they can be shown to be viable for \Cygnus.

\subsection{Radon and radon progeny backgrounds \label{radon backgrounds}}
\begin{figure}
\noindent \begin{centering}
\includegraphics[width=\columnwidth]{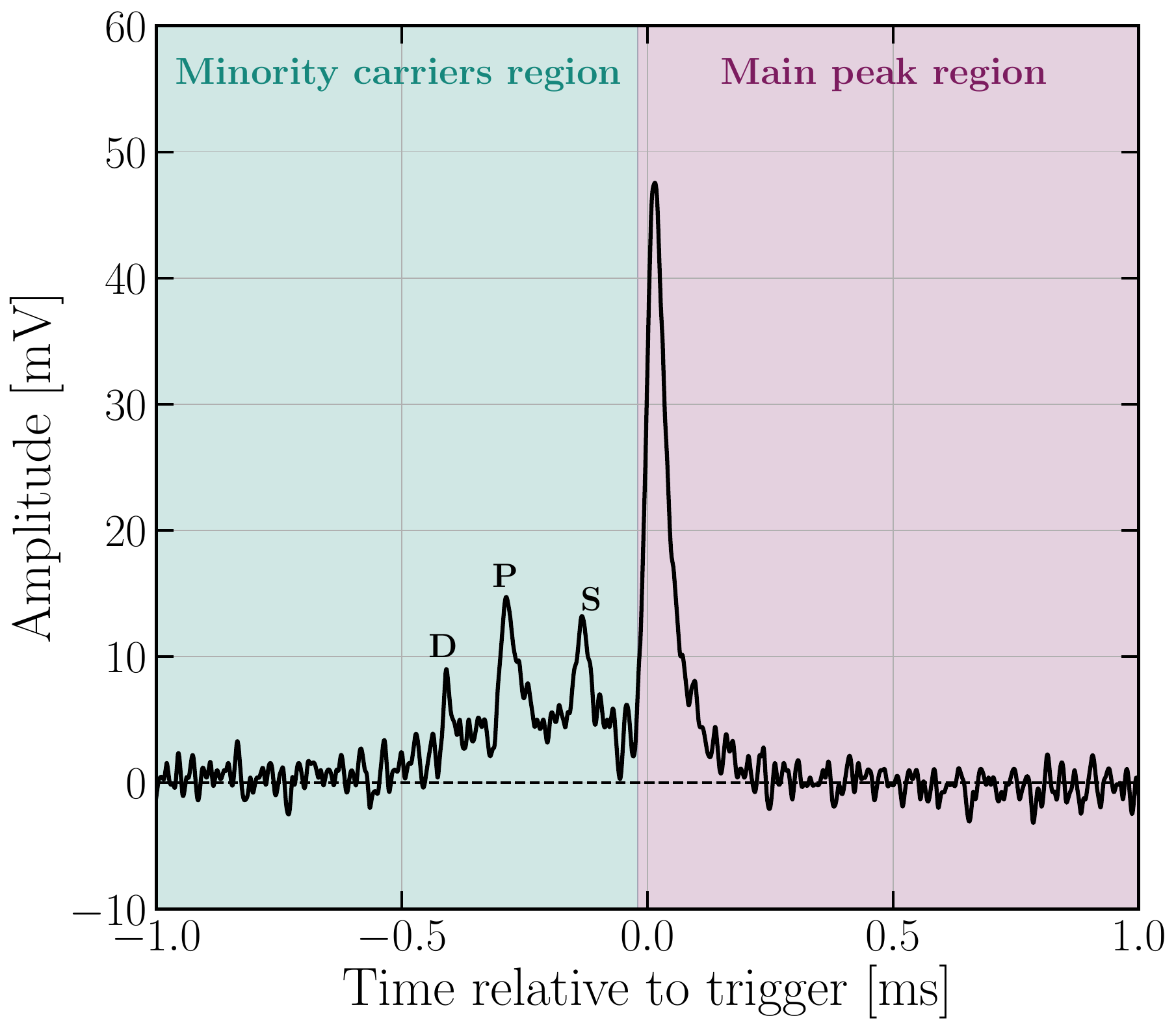}
\par\end{centering}
\caption{Amplitude as a function of time of the main signal peak as well as several minority peaks from the DRIFT-IId gas mixture, adapted with permission from Battat et al.~\cite{Battat:2014van}. The minority peaks are labelled D, P and S by convention.\label{fig:minority_peaks}}
\end{figure}

Radon gas emanating from materials is a further source of background for rare-event experiments. In particular, the low energy ($\sim 100$ keV) of radon progeny recoils (RPRs) can mimic WIMP interactions~\cite{Battat:2014oqa}. Being a noble gas, $^{222}$Rn has a low chemical reactivity making it particularly difficult to filter. Moreover, its 3.8 day half-life allows enough time for the radon gas to spread from the materials in which it is produced, resulting in a rather widespread source of background. 

Alpha particles from radon decays are relatively easy to identify. For a back-to-back TPC (such as that shown in Fig.~\ref{fig:TPCdiagram} of Sec.~\ref{sec:introduction}), events that trigger on both the left and right readout planes of the detector can be spotted via their coincidence in time. The only events with tracks long enough to achieve this are alpha particles originating from an RPR event fully contained within the fiducialized gas volume. These types of events can be used to estimate the level of radon present in the detector~\cite{Battat:2014oqa,Riffard:2015rga}. However, not all RPR events originate within the fiducial volume as the positively charged radon daughter nuclei drift towards the central cathode and become stuck within the material. A recoil event caused by the decay of RPRs can prove difficult to reject as the associated alpha particle may not be identified if it remains trapped. One method of reducing the number of trapped particles is to have smaller cathode widths, thus increasing the transparency. This approach was successfully implemented in DRIFT with the use of a submicron thick cathode. This helped to reduce the RPR rate from approximately 130 events per day to less than one~\cite{Brack:2014pwa}. In DRIFT-IId, RPRs originating from the cathode surface without a detected alpha track are vetoed by defining the fiducial region such that any recorded signal within 2~cm of the cathode is rejected~\cite{Battat:2016xxe}. A similar technique is also used by MIMAC~\cite{Riffard:2015rga} where an analysis cut based on the $z$ positions of events is used to reject RPRs originating from the cathode. Ideally, a similar $z$-cut to veto RPRs would be used in \Cygnus.

Many techniques for dealing with RPRs rely on fiducialization: the ability to fully locate in three dimensions the source of ionization within the TPC gas volume. This is of vital importance for rare event search TPCs as it allows for background events from material surfaces to be distinguished from potential signal events occurring within the target volume. Fiducialization in the TPC readout plane ($x,y$) can be done at the reconstruction stage for the more segmented readouts. For less segmented readouts one can use a surrounding veto area that triggers when events enter from outside of the fiducial volume. Fiducializing parallel to the drift field ($z$) presents more of a challenge as no active veto can be placed in front of either the cathode or readout. However, $z$-fiducialization has been achieved in previous work~\cite{Snowden-Ifft:2014taa,phdthorpe,Lewis:2014poa}. DRIFT-IId uses a gas mixture of CS$_{2}$, CF$_{4}$ and O$_{2}$ at a ratio of 30:10:1 Torr to produce distinct groups of ``minority carriers'' within the gas that drift at different speeds. The signal then has a main peak adjacent to three minority peaks as shown in Fig.~\ref{fig:minority_peaks}. The separation in time between these peaks is proportional to the drift distance traveled and therefore the $z$ position at which the primary event occurred. Reference~\cite{Phan:2016veo} showed that the same fiducialization method is possible with SF$_{6}$ due to the presence of an SF$^{-}_{5}$ minority peak; so we believe fiducialization should be achievable in \Cygnus. There are other options available, such as using timing differences between the signals at the cathode and induced signals at the anode as in MIMAC~\cite{1748-0221-12-11-P11020}, or by measuring diffusion of drift charge transverse to the recoil axis as demonstrated by D$^3$~\cite{Lewis:2014poa,phdthorpe}. Several of these techniques may also be combined to further improve performance. Note that fiducialization via minority carriers is only possible with negative ion drift gases, as proposed here, so that we expect the best fiducialization in such gases.


\subsection{Cosmogenic activation \label{cosmogenic activation}}
When materials are present at the surface they are subject to a much higher flux of cosmic rays than when underground. If materials spend a significant time above ground cosmic ray interactions can activate different long-lived isotopes, which when part of the detector become an additional source of background. As rare event searches increase in both size and sensitivity, this cosmogenic activation is going to have an increasing influence on the ultimate background rate. See for example Ref.~\cite{Cebrian:2017} for further details.

The activity of a cosmogenically activated isotope is expressed as,
\begin{equation}
a = R[1 - e{^{-{\lambda}t{_{\rm exp}}}}]e{^{-{\lambda}t_{\rm cool}}} \, ,
\label{eq:activation}
\end{equation}
where $R$ is the production rate, $\lambda$ is the decay constant, $t_{\rm exp}$ is the time the material spent exposed to the cosmic ray flux and $t_{\rm cool}$ is the `cooling off' time the material spent underground.

Activation can occur in all materials, but we use copper here as an example since it is our best candidate vessel material so far. Reference~\cite{Baudis:2015kqa} showed that cosmic ray interactions with copper at sea level can produce multiple unstable isotopes. The longest-lived is $^{60}$Co which originates from $^{59}$Co contamination and has a half-life of 5.26 years. The measurement from Ref.~\cite{Baudis:2015kqa} was taken after a relatively long $t_{\rm exp} = 345$ days compared to the much shorter $t_{\rm cool} = 14.8$ days. The saturation activity for $^{60}$Co was 340$\pm{^{82}_{68}}$ $\upmu$Bq~kg$^{-1}$. Taking the upper end of this measurement, we estimate that the rate of gamma-induced electron recoils for a 5~cm thick copper vessel will be $< 9.8\pm0.6\times10^4$~keV$^{-1}$~yr$^{-1}$ in the $1$--$10$~keV energy region. This rate is $\sim$6 times larger than the recoil rate due to $^{238}$U, $^{232}$Th, and $^{40}$K radioactivity in the same thickness of copper (see Sec.~\ref{subsec:vessel gammas}), assuming the upper limits are the actual radiopurity values.

Since the measurement and simulation of the cosmogenic activation in copper have shown good agreement~\cite{Baudis:2015kqa}, for a more detailed study we can make use of the simulation tool ACTIVIA~\cite{Back:2007kk}. This enables us to predict the minimum amount of time a material should spend above ground and cooling off underground in order to limit activation.

For SF$_{6}$, the longest-lived isotope of fluorine, $^{18}$F (with a half-life of 109.77 minutes), is still too short-lived to produce a lasting background contribution. Nearly all isotopes of sulfur have half-lives on the scale of seconds to minutes with the exception of $^{35}$S, which has a half-life of 87.5 days. We simulate 90 days of surface time with ACTIVIA, during which the gas is exposed to cosmic rays sampled from an energy spectrum ranging between $10$--$10000$~MeV. After this we assume a 180 day cooling-off period underground, during which time the gas is not exposed to cosmic rays. This resulted in a $^{35}$S production rate of $R = 0.021$ kg$^{-1}$ day$^{-1}$, giving a decay rate of 2.61 mBq/kg. A 1000 m$^{3}$ vessel filled with SF$_{6}$ at 20 Torr will hold $\sim160$~kg of gas, of which $\sim$35 kg is sulfur, so the total decay rate for the detector is 91.35 mBq. We simulate $10^5$ $^{35}$S decays in Geant4, originating from the center of the 20 Torr SF$_{6}$ gas volume. Only three gamma induced electron recoil events between $1$--$10$~keV were observed in the simulation time of $\sim12.5$~days. We estimate therefore a rate of $90\pm50$~yr$^{-1}$, too small to be a significant electron recoil background. 

In general the amount of cosmogenic activation is reduced by limiting $t_{\rm exp}$ and extending $t_{\rm cool}$. In addition one should limit the amount of time the material spends as air-shipment and therefore subject to a higher cosmic ray flux. One might also want to electroform the metallic components underground, at the construction site, where possible.

\subsection{Backgrounds at atmospheric pressure \label{atm_backgrounds}}

A 1000~m$^{3}$ vessel filled with low-pressure target gas would need to withstand outside pressures of $\sim$760~Torr, requiring larger amounts of structural support than a vessel filled with atmospheric pressure gas. Sections~\ref{subsec:Vessel Neutrons} and~\ref{subsec:vessel gammas} found that a steel or titanium support structure would increase the total vessel background. Therefore, running at atmospheric pressure rather than at 20 Torr SF$_{6}$ would be advantageous. An addition of 740 Torr He allows for TPC operations at atmospheric pressure, while increasing the density only by a factor of $\sim$2 when compared to 20 Torr SF$_{6}$ alone (see Table~\ref{tab:gas_specs}). Adapting our previous simulation to this 740:20 Torr mixture, we find that an increase in density corresponds to an increase in both the gamma and neutron background by around the same factor of $\sim$2. The results presented thus far throughout Sec.~\ref{sec:backgrounds} can scale straightforwardly between low pressure and atmospheric pressure by this factor. In addition, the results can be scaled to give an approximation of the background rates at other ratios of He:SF$_6$. For example, a ratio of 755:5 Torr would incur a density increase and, therefore, a background rate increase of around 25 percent compared to pure SF$_6$ at 20 Torr.

A major advantage of atmospheric pressure operation is not only the reduction in vessel support needed, but also that the vessel itself could be built with a lower average thickness. A 1000 m$^{3}$ \Cygnus vessel operating at 760 Torr would need an average thickness of around 30, 12, 10 or 8~mm of acrylic, copper, steel or titanium respectively~\cite{Gamble:2018}. Due to the strength of titanium it may even be possible to construct a vessel with as little as 6 mm average thickness. The acrylic estimate does not include additional structural support which would also need to be taken into consideration, but the copper, steel and titanium thicknesses are averaged over both the vessel walls and structural support. We employ the simulation described in Secs.~\ref{subsec:Vessel Neutrons} and \ref{subsec:vessel gammas} to re-estimate the neutron and gamma recoil rate due to vessel materials for the new 740:20~Torr \hesfsix gas mixture. Table~\ref{tab:vessel_gammas_atm} lists these estimates. Note that the rock and water shielding are not included now, but expected background changes due to these will be commented on below.

\begin{table}[ht]
\begin{centering}
\begin{tabular}{cccc}
\toprule
Material & Thickness & Gamma recoil rate & Neutron recoil rate\tabularnewline
 & (mm) & (yr$^{-1}$) & (yr$^{-1}$) \tabularnewline
\midrule
Acrylic & 30 & 3.57$\pm$0.03$\times{10^{5}}$ & 0.40$\pm$0.02 \tabularnewline
Copper & 12 & < 1.50$\pm$0.02$\times{10^{4}}$ & < 0.12$\pm$0.02 \tabularnewline
Steel & 10 & 5.75$\pm$0.08$\times{10^{5}}$ & 5.6$\pm$1.0 \tabularnewline
Titanium & 8 & < 2.21$\pm$0.02$\times{10^{5}}$ & < 4.0$\pm$0.7 \tabularnewline
Titanium & 6 & <1.88$\pm$0.02$\times{10^{5}}$ & < 2.4$\pm$0.4 \tabularnewline
\bottomrule
\end{tabular}
\par\end{centering}
\caption{Vessel backgrounds for different vessel materials at reduced thickness, for an atmospheric-pressure 740:20 Torr \hesfsix gas mixture scenario.}\label{tab:vessel_gammas_atm}
\end{table}

We can compare the 740:20 Torr mixture results of Table \ref{tab:vessel_gammas_atm} with the 20 Torr SF$_6$ results at the thinnest previously simulated vessel thickness, 5 cm, shown in Table \ref{tab:vessel_gammas}. While the gamma recoil rates are similar, the neutron rates from the metallic vessel walls are all lower with the atmospheric pressure. However our general conclusion from before---that the steel and titanium vessel neutron background exceed the design target---still persists at atmospheric pressure. 

It seems that the increase in background due to the increased gas density is approximately compensated by a decrease in background due to the reduction in required vessel material. So at the very least we have shown that the vessel background does not {\it increase} for the atmospheric gas mixture. However the rock-gamma shielding efficiency would suffer with a thinner vessel. If an additional copper shield were to be implemented then this too would contribute to the total background, requiring potentially a thickened water shielding to counterbalance and keep the total background acceptable.

We have seen that for a fixed detector, background rates are proportional to gas density. Hence can already conclude that a 755:5 Torr \hesfsix target gas, which has only 25\% higher mass density than 20 Torr SF$_6$, is already close to acceptable from a background point of view, provided we do not change the vessel design. It seems quite likely, that with thinner vessel material that can be considered for atmospheric pressure, backgrounds can be reduced further, probably below the 20 Torr SF$_6$ scenario that has been studied in detail here. We therefore are cautiously optimistic, but also note that detailed, dedicated simulations of complete atmospheric pressure scenarios should be carried out in future work with high priority.



\subsection{Conclusions \label{background conclusion}}
In this section we have discussed in detail the intrinsic backgrounds for a 1000 m$^3$ \Cygnus TPC. We have taken into account neutron and gamma backgrounds originating radioisotopes in the rock, vessel and detector, as well those originating from cosmic ray muons, radon emissions and cosmogenic activation. Following Sec.~\ref{subsec:Analysis}, we have aimed to determine the feasibility of limiting electron recoils to a rate of 10$^{4}$~recoils~keV$^{-1}$~yr$^{-1}$ in the energy range 1--10 keV$_{ee}$, and nuclear recoils to a rate of $<1$~yr$^{-1}$ between 1--100 keV. These numbers are compatible with zero electron background after offline electron rejection and limit nuclear recoils from neutrons to less than a sixth of the solar neutrino signal. The feasibility of achieving the electron background limit is essential as it will ultimately control the threshold of the experiment and the sensitivity to low WIMP masses, whereas meeting the nuclear recoil rate target is important for optimizing sensitivity to WIMPs and solar neutrinos.

While we have assumed a fill gas of SF$_{6}$ at 20~Torr in much of our discussion, we have also shown how the results scale to atmospheric pressure He:SF$_{6}$ mixtures, which appear most desirable for \Cygnus. From a background perspective, SF$_6$ is advantageous due to its minority carriers which enable $z$-fiducialization. A vessel running at atmospheric pressure also requires less structural support and thinner walls than a vessel running at near-vacuum pressure and therefore produces less intrinsic background. 

Summarizing the results we can state first that to shield rock neutrons the vessel would need to be surrounded by $75$~cm of water shielding to bring the nuclear recoil rate, above 1 keV$_{r}$, down to less than one event per year. Next, for the vessel material itself, we determined that 20--30~cm of copper is optimum. The electron recoils from the rock gamma background are already shielded to within our limit using this thickness, whereas the neutron and gamma backgrounds from the vessel fall only slightly above it. If it can be demonstrated in further screening tests that copper's $^{238}$U radioactivity is around half the upper limit used here ($<0.012$~mBq~kg$^{-1}$~\cite{Alvarez:2012as}) then it will be an attractive candidate for the \Cygnus TPC. Although such a vessel would likely need a steel support frame---introducing a further background---conceptual designs are feasible in which the frame is embedded in the copper, thereby providing the necessary additional shielding. Ideally this vessel would spend only a short time above and below ground, by limiting air-shipment of materials, procuring the materials and, if possible, electroforming, as close to the site as possible; all in the aim of reducing further cosmogenic activation.

In simulating the various readout technologies and amplification stages we conclude that while some readout-generated backgrounds can be kept within the limit, further R\&D is clearly still needed to establish the optimum choice for \Cygnus from a background standpoint. A few issues that would need to be addressed include: (1) for strip micromegas, GEMs and \mupic, a better understanding of the radiopurity, as well as an investigation into the use of lower background material alternatives such as acrylic; (2) for the pixel chip readout, measurements of the radioactivity levels of silicon, combined with an investigation of charge focusing to reduce the required number of chips; and (3) an improved field cage design using a smaller number of resistors, which ideally would contain lower levels of $^{238}$U than the ceramic resistors we assumed here. Resistive sheets \cite{Miuchi:2019qdp} should be evaluated as a potential resistor replacement.

The issue of the readout background may also be partially addressed on the analysis side. For instance, improving the electron/nuclear recoil discrimination by a factor of 10-100 would both widen the available choices of readout material, while also permitting background-free operation with a lower energy threshold. This may be possible by using a larger set of recoil discrimination parameters than already discussed in Sec.~\ref{subsec:Analysis}, and/or by decreasing the drift distance to limit diffusion. 


As long as the total neutron recoil background remains subdominant to solar neutrinos, then the WIMP cross section sensitivities from Sec.~\ref{sec:wimp_reach} would be largely unaffected. We reiterate that the most important impact on our final physics reach at this stage comes from the electron discrimination threshold. Hence the electron background and electron discrimination should take priority in future work. Ultimately, while we conclude that while there is more work still needed to prove that \Cygnus can sustain completely background-free operation, we do not yet consider there to be any immediate showstoppers, and have identified steps that should be taken to address our unresolved issues.

\section{Underground sites and engineering}
\label{sec:sites}



\begin{figure*}
\begin{centering}
\includegraphics[width=\textwidth]{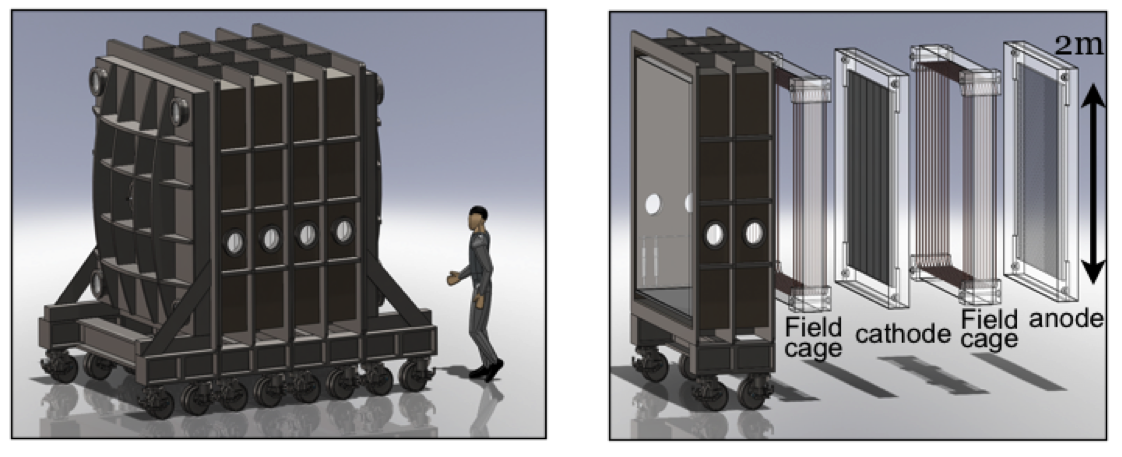}
\par\end{centering}
\caption{Design for a first stage 8 m$^{3}$ steel vessel (courtesy of E. Lee, University of New Mexico).\label{fig:8m3}}
\end{figure*} 

\begin{figure}
\begin{centering}
\includegraphics[width=\columnwidth]{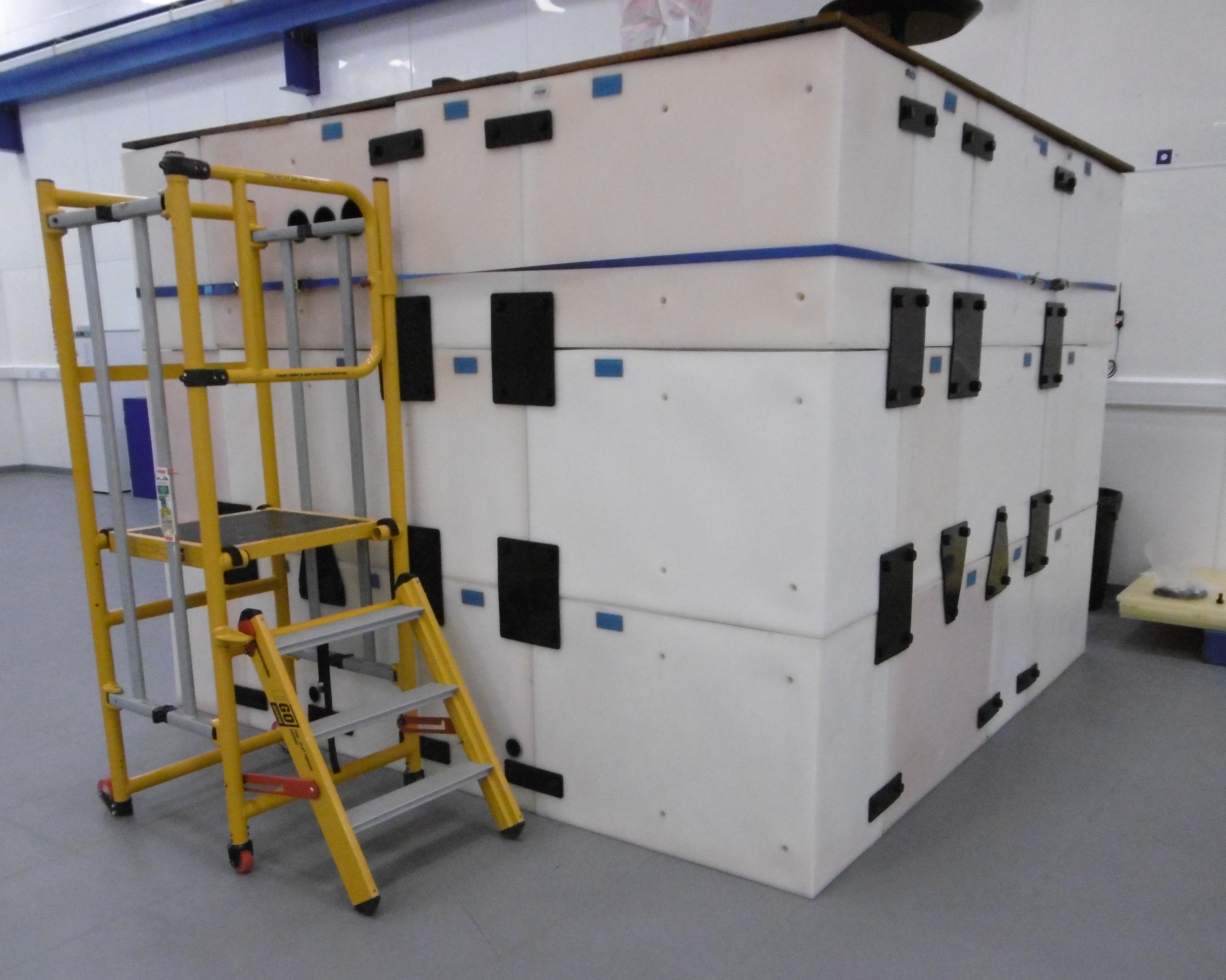}
\par\end{centering}
\caption{DRIFT-IId interlocking acrylic block shielding surrounding the detector and containing water, located at Boulby underground laboratory.\label{fig:DRIFT_shield}}
\end{figure}

We displayed the locations of several candidate underground sites that are currently under consideration for \Cygnus in Fig.~\ref{fig:map}. This section provides a summary of the significant logistical and engineering requirements that must be considered for these sites to host directional dark matter detectors. Principal of these for gas TPCs are space requirements and the need to provide handling of the gas target underground. Both of these factors depend on the choice of gas and operating pressure, so we center the discussion around a first stage \Cygnus TPC of order 10 m$^3$ operating at one atmosphere. 

\subsection{Site requirements}
While a TPC with 1000~m$^3$ fiducial volume is rather large in total volume, a corresponding advantage of the concept is that there is no particular restriction on the shape of that volume. For instance, one could consider an elongated `worm-like' sectional design where one dimension is substantially longer than the other two. This is not only feasible but actually has certain advantages for operations and maintenance. \Cygnus is therefore well-suited to sites where long tunnels and restrictions on available height are normal, \eg{}~mine sites like Stawell, Australia or Boulby, UK. The relative simplicity of the services needed---in particular that there are no cryogenics involved---is also is an advantage here. Additionally, the experiment can be built as separate detectors, either in one site, or in multiple sites. This also allows cross-correlation of data at different latitudes which could aid in the control of systematics. Such geometric features will not necessarily be required for a first 10 m$^{3}$ stage but will become important in 1000 m$^{3}$ designs aiming to go below the neutrino floor.

Due to finite door, elevator, and cavern sizes, certain sites will have more stringent constraints than others on the maximum size of an object able to be brought in, \eg{}~$2 \times 2$~m at Boulby. This would need to be accounted for in the vessel design. A baseline single $10\times10\times10$~m$^3$ detector could be engineered in sites such as Gran Sasso, where there is available head-room and good access. A different, segment-based design could be used for a detector of the same total volume distributed in different sites, as well as for an elongated version of, say, $5\times5\times40$~m$^3$ that would conceivably fit in an existing facility at Boulby. 

The site overburden requirement is not foreseen to be as important a consideration here since the candidate sites are all $>1$~km in depth. The muon-induced neutron event rate at this depth (calculated in Sec.~\ref{subsec:Muon-induced-neutrons}) can be rejected with a combination of an active external muon veto and the inherent charged particle tracking capability of TPCs, depending on the adopted TPC readout technology. The need for an external veto would have only a modest impact on the overall dimensions of the experiment, probably a $<1$~m increase in linear dimension for the case of plastic scintillator with support structure.

\subsection{Engineering requirements}
\begin{figure}
\begin{centering}
\includegraphics[width=\columnwidth]{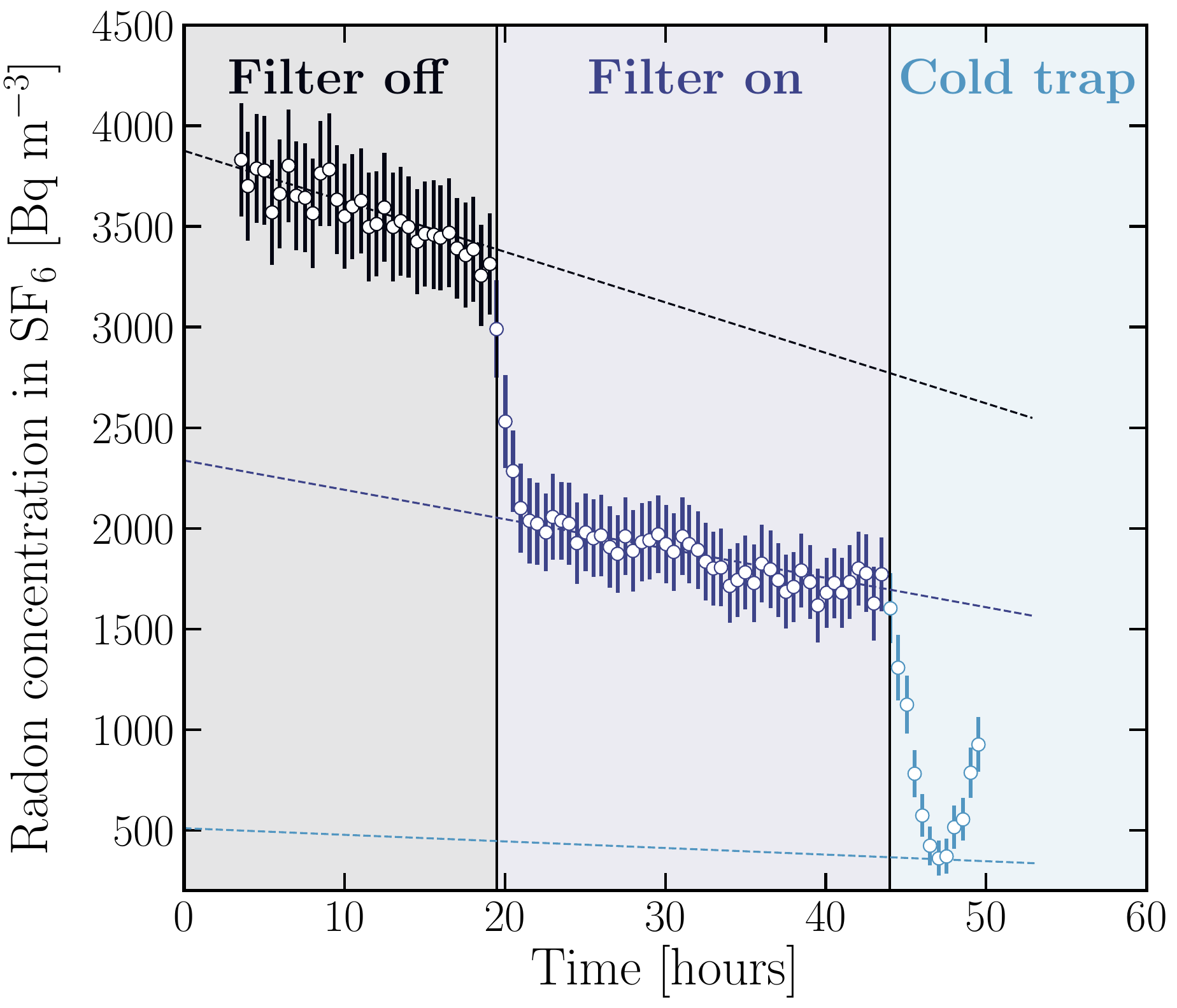}
\par\end{centering}
\caption{Experimental results for the radon in SF$_{6}$ as a function of time when using a 5~\si{\angstrom} molecular sieve. The black points show the radon concentration prior to being diverted through the sieve. Then dark blue points show the reduction when the filter is turned on. The filter is then cooled with dry ice resulting in a further reduction, shown in light blue.\label{fig:radon_sieve}}
\end{figure}
One of the most important engineering constraints for gas TPCs is that they typically need to be operated at low pressure, requiring a vacuum vessel. We have proposed here however that the addition of helium---such that the experiment operates at atmospheric pressure---alleviates this constraint whilst simultaneously gaining the experiment access to low WIMP masses and a larger rate of solar and reactor neutrinos. The impact on directional readout sensitivity and the intrinsic background were discussed in Secs.~\ref{sec:technology_comparison} and~\ref{sec:backgrounds} respectively, and the results were promising. From an engineering perspective however, it is possible that the TPC would still need significant certification as a pressure vessel if contamination is an issue. Purification is conventionally achieved via outgassing under vacuum, so it would need to be proven that flushing gas through the vessel prior to operation can sufficiently reduce contamination. If possible this would obviate the need for the vessel to be of vacuum standard, greatly reducing the mass of steel, titanium or copper needed. This option is under investigation. We note that the NID gas CS$_{2}$ is known to be highly tolerant of impurities so that the DRIFT experiment operates well with up to 1\% gas impurity, but SF$_6$ needs more study in this regard.

Figure~\ref{fig:8m3} shows an example design for an 8~m$^{3}$ \Cygnus vessel based on a steel construction. An alternative is to use a cylindrical design to reduce the mass by a factor of about two, but at the cost of additional complexity in the detector and shielding. 

For the design of Fig.~\ref{fig:8m3}, it is feasible to obtain the required $\sim$4 tons of steel with sufficient radiopurity to reach the goal of $< 1$ background neutron per year, however the intrinsic gamma background would still be too high (see Table~\ref{tab:vessel_gammas}). An alternative would be copper, as proposed for NEXT-100~\cite{NEXTHector} whose screening process proved ultra low-background levels of $<0.012$~mBq/kg U, $<0.0041$~mBq/kg Th, and $0.061$~mBq/kg K. Although it may not be possible to build a large-scale pure copper vessel, a hybrid design in which copper is used for the main vessel panels with steel or titanium support structures could be engineered instead. Ultimately, we envisage that the scale and engineering requirements for a 10~m$^{3}$ \Cygnus vessel are unlikely to be challenging for any of the possible underground sites.

In addition to the $\mathcal{O}(10\,{\rm m}^3)$ vessel, external neutron and gamma shielding would likely contribute an additional $\sim$1~m surrounding the detector. A variety of designs are possible for this. The use of interlocking plastic water containers is one possibility, as adopted for the DRIFT upgrade shielding shown in Fig.~\ref{fig:DRIFT_shield}. Another is to use a purpose-built water container. Though more expensive, it would have a potential advantage in that it could be instrumented with photomultipliers to provide a Cherenkov muon veto as well.

While there is no need for complex cryogenics engineering in \Cygnus, consideration is still needed with regards to the engineering aspects of the gas supply. In most current generation directional detectors the target gas is flowed and disposed of through filters to the atmosphere. However, for \Cygnus, recirculation will be needed, both for cost and environmental reasons: SF$_{6}$ is a powerful greenhouse gas. Recirculation also provides a potential means for the reduction of radon and water from the target. Purification of SF$_{6}$ is well known in industry~\cite{Bessede2015} so therefore is not seen as a major impediment here. Meanwhile, recent experiments by the Sheffield group have investigated active radon removal in SF$_{6}$~\cite{Ezeribe:2017phs}. Figure~\ref{fig:radon_sieve} shows the results from an experiment in which SF$_{6}$ was circulated through a vessel with a known level of radon added via a sealed radon source. When the gas is diverted through a 5\si{\angstrom} molecular sieve, the radon is seen to reduce. When the filter is cooled with dry ice a further reduction is seen. The final up-turn in this data is believed to be due to the gradual temperature increase of the dry ice over time. An earlier test also checked that the SF$_{6}$ itself was not absorbed by the filter. These experiments show for the first time that active radon removal in SF$_6$ is possible.

Summarizing this brief overview of the \Cygnus site and engineering issues we conclude that, while the required vessels do present challenges, they are surmountable. One of the major benefits of the \Cygnus experiment is the significant flexibility in the approach to its shape and modularization. This opens up prospects for construction in sections underground and at multiple locations so that various constraints imposed by each site can be met. It is clear that building \Cygnus at the sufficient scale to reach below the neutrino floor is a challenge. Nevertheless, it is worth noting that a 1000~m$^3$ vessel represents approximately 1/50th of the {\it cryogenic} internal vessel volume proposed for DUNE~\cite{Abi:2018dnh} at Sandford Underground Laboratory, for which rock excavation is now underway.

\section{Summary and recommendations}
\label{sec:conclusion}


\begin{figure*}[hbt]
\includegraphics[width=\textwidth]{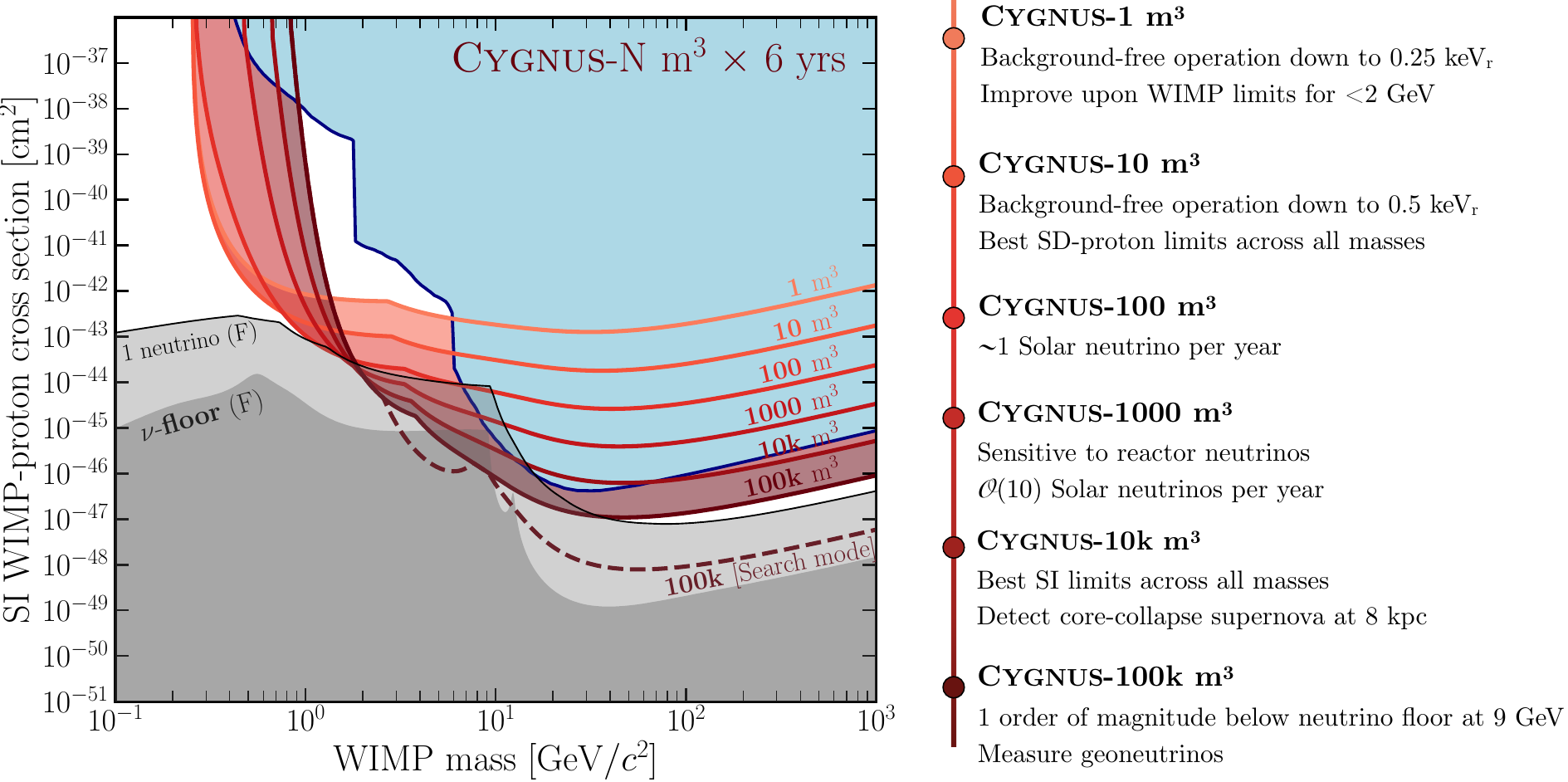}
\caption{Summary of the projected SI WIMP 90\% CL exclusion limits as a function of the total fiducial volume of the network of detectors comprising the \Cygnus experiment. All experimental exposures are multiplied by a running time of 6 years such that the listed benchmarks are all approximately multiples of 1 ton year (1 ton year corresponds to 1000 m$^3$). We shade in blue the currently excluded limits on the SI WIMP-nucleon cross section as of 2020. The thresholds are increased evenly between 0.25 and 8~\kevr and for each increasing volume to illustrate the range of possible thresholds envisaged for the final experiment. Below the final volume we also extend the reach by including the possibility of a ``search mode'' experiment which would have 1520 Torr of SF$_6$ (as opposed to 5 Torr), but would have no directional sensitivity.}\label{fig:timeline}
\end{figure*}

We have outlined the physics case, technology choices, and design of the first large-scale directional nuclear recoil observatory. We name the project \Cygnus after the constellation from which the wind of dark matter originates.

 
While previous work on the subject of directional detection (summarized in Ref.~\cite{Mayet:2016zxu}) was often based on idealized theoretical concepts, here we have calculated the sensitivity of a directional detector, incorporating experimental realities. These include energy-dependent performance parameters such as detection efficiency, energy resolution, head/tail recognition, and angular resolution; as well as background considerations such as radioactive impurities in the vessel, readout and environment, and the ability to discriminate between nuclear and electron recoils. This article is therefore the first to bridge the gap between the theoretical and experimental literature. Our principal conclusion is that, despite some remaining technical challenges, a \Cygnus nuclear recoil observatory is indeed feasible in line with the sensitivities displayed in Fig.~\ref{fig:limits}.

A ton-scale `\Cygnus-1000' detector, with a fiducial target volume of 1000~m$^3$, filled with a \hesfsix gas mixture at room temperature and atmospheric pressure, and with 1--3~\kevr~event detection thresholds, would have a non-directional sensitivity to WIMP-nucleon cross sections extending significantly beyond existing limits. For SI interactions this sensitivity could extend into presently unexplored sub-10~\gevcc parameter space, whereas for SD interactions, even a 10 m$^3$-scale experiment would be sufficient to compete with generation-two (G2) detectors currently under construction, such as LZ and SuperCDMS. 

As we discussed in Sec.~\ref{sec:science_case}, the physics motivation for \Cygnus is substantial, prior to, or even in the absence of, a positive detection of DM. Even for the worst-case scenario of an 8~\kevr nuclear recoil threshold, \Cygnus-1000 would observe around 13 \cevns events over six years from $^8$B and $hep$ solar neutrinos. This would be a significant achievement, given that \cevns will not become an appreciable signal in conventional direct detection experiments until LZ or XenonNT have taken data. For a threshold of 1~\kevr this number increases to 37 which would already be enough to begin to characterize the neutrino spectrum. 

In addition to solar neutrinos, supernovae occuring at distances closer than 3 kpc would be sufficient to produce a measurable number of highly energetic nuclear recoil events in \Cygnus-1000.  For even larger volumes, the possibilities open up for the detection of geological and reactor neutrinos. Another interesting signal, but one that is not explored in detail here, are the signals from neutrino electron scattering events (see Fig.~\ref{fig:NuRates}). A large number of high energy electron recoils would be expected in \Cygnus-1000, primarily from the lower energy components of the flux like $pp$ or $^7$Be neutrinos. \Cygnus could have a lower threshold for electron events than even Borexino, so the prospects to contribute to the physics of solar neutrinos may even extend beyond what we have discussed here. Just as the excellent nuclear/electron recoil discrimination permits the identification of a WIMP signal, this same capability also implies excellent potential background rejection if solar neutrino-electron recoils are the signal of interest. An investigation of both solar and geoneutrino-electron recoils in directional experiments is therefore worthy of a detailed follow-up study.

We expect \Cygnus to be capable of full 3d event reconstruction, good directional sensitivity and excellent electron/nuclear recoil discrimination even at energies well below 10~\kevr. This will require a highly segmented charge readout and the limiting of the diffusion of ionization in the target gas. High electron rejection will enable the detector to operate free of internal background, with the capability of distinguishing WIMP-nucleus scattering from \cevns or unexpected nuclear recoil backgrounds. For example, with a high-resolution pixel or strip readout, on average 4-5 detected 100-\gevcc WIMP-fluorine recoils above 50 \kevr are sufficient to rule out an isotropic recoil distribution at 90\% C.L. For a 10-\gevcc WIMP, less than 20 detected helium-recoil events above 6~\kevr~or 3-4 events above 20~\kevr~would suffice to rule out a neutrino-induced nuclear recoil distribution at 90\% C.L. This would constitute the first step in firmly establishing the galactic origin of a tentative dark matter signal.


We estimate that the most cost-effective way to achieve the desired high readout segmentation, low diffusion, and good directional performance at low recoil energies, is via strip readout technologies and negative ion drift. \Cygnus-1000 would then require 2000~\si{m^2} of strip readout planes. Thanks to LHC technology development, large strip micromegas planes from CERN that would meet our technical requirements are already available at a cost of order \$12,500 /m$^2$. If a radiopure version of these detector as well as preamps with integration time appropriate for NID are developed, then \Cygnus-1000 could be constructed relatively soon and at quite reasonable cost. Assuming 20 million readout channels at an electronics cost of US \$1/channel for mass production, the total charge readout cost of \Cygnus-1000 would then amount to US \$45 million. Downstream DAQ, gas vessels and shielding would add to the cost, but due to the ability of \Cygnus to operate with low noise at room temperature and atmospheric pressure, these costs will be kept reasonable.
	
We do not anticipate \Cygnus to be a monolithic experiment like most detectors currently in operation. The \SI{1000}{m^3} target volume would be best achieved using multiple smaller detectors (see Fig.~\ref{fig:TPCdiagram}), say of $5 \times 5 \times 10$~m$^3$, potentially located at multiple underground sites (see Fig.~\ref{fig:map}). As well as simply facilitating our the sensitivity goals while maintaining low pressure operation, the modularity and distribution of the experiment would allow for the control of systematics by comparison between detectors, and importantly, leaves room for future expansion at each site.  Utilizing multiple detectors would allow also the use of multiple target gases and pressures, but the currently favored scenario is to include a negative-ion drift gas, preferably SF$_{6}$ mixed with helium, to arrive at atmospheric total pressure. This mixture has multiple advantages: it improves the directionality of all recoil species, permits $z$-fiducialization via minority carriers, and extends the sensitivity to both low WIMP masses and neutrinos. Atmospheric pressure also avoids the need for a vacuum vessel.  

While a low-density gas such as 755:5 torr \hesfsix is essential to maintain low-mass WIMP and neutrino sensitivity with directionality, the planned segmentation of \Cygnus naturally enables operation of parts of the detector with a higher-density ``search mode" gas. If we choose a vacuum-capable gas vessel design, then this would be capable of withstanding a 1 atmosphere pressure differential. In that case the search mode could utilize 1520 torr of SF$_6$ for a factor 300 boost in exposure, and around a factor of $\sim$17 boost in sensitivity at high masses . The beauty of \Cygnus is that the exact partitioning of the target volume into low-density and search mode running can be optimized and varied even after construction, and be responsive to new developments in the field. This flexibility may prove particularly important for larger volume detectors, \eg ~a \Cygnus-100k, which will required a substantial investment of time and funding.



The new simulations presented here, validated in part by experiment, suggest that electron rejection is feasible down to 1~\kevee~in the atmospheric pressure \hesfsix mixture.  The discrimination power depends on the sophistication of the readout but rises exponentially with energy. With a simple discriminant the expected electron rejection factor exceeds $10^{6}$ at 5-8~\kevee for fluorine, and 10~\kevee for helium. New simulation studies with more sophisticated events shape variables suggest the electron rejection factor at a given recoil energy can be improved by another two orders of magnitude \cite{vahsenPrivate}. Efforts by the MIMAC collaboration~\cite{Riffard:2016mgw} and first CYGNUS investigations with deep learning neutral networks \cite{PeterSadowski} suggest that combining discriminants with machine learning can improve this even further, perhaps significantly so. The full optimization of electron rejection, and finding the optimal tradeoff of intrinsic background versus the sophistication of readout is the subject of future work.

To limit the background from the \Cygnus vessel, we envisage a skeletal design based on titanium, copper and acrylic, surrounded by an inner layered structure of gamma shielding, incorporating copper and steel.  Shielding from neutron backgrounds can be achieved by using water blocks outside the inner shield (see Fig. \ref{fig:DRIFT_shield}), and a conventional muon veto depending on site depth.  Some improvement in copper radiopurity will be needed for the shielding but a back-up to this would be to use entirely water shielding.

\subsection*{Final recommendations}
To clarify the route forward, we have five main recommendations: First, the strawman design discussed here should be advanced to a properly optimized technical design. We expect that the low-mass reach of a fully optimized strip readout detector will exceed the physics reach estimated here, as we found that our strawman design is diffusion-limited. Second, all energy-dependent performance metrics introduced here, such as the recoil efficiency, angular resolution, head-tail efficiency, and electron rejection, must be demonstrated experimentally in a small prototype with full drift length and high readout resolution: a `\Cygnus HD1 Demonstrator'. Third, the intrinsic radioactivities of the components of the strip readout must be reduced to where they are consistent with the measured electron rejection capabilities. Fourth, inexpensive, scaleable readout electronics for large-area strip readout planes should be designed. Lastly, the expected physics sensitivity of \Cygnus should be more broadly investigated. For example, we expect that the 3d ionization measurements should be uniquely capable of detecting and identifying non-conventional DM-nucleon scattering final states; and as mentioned above, the prospects for novel physics based on \emph{electron} recoils should also be investigated, such as in the context of neutrinos or axion-like particles.

Independently of this main development path, we also recommend the pursuit of alternative design approaches, based on electron drift. Electron drift gases will allow much higher avalanche gains than NID gases, but at the cost of increased diffusion. For electron drift, fiducialization would be performed via measurements of diffusion, rather than via observation of minority carriers. The higher gain could be a good match for optical readout, and hence the CYGNO experiment in Italy~\cite{CYGNO:2019aqp} will pursue that option. Another interesting configuration is a pixel or strip readout detector with electron drift and low threshold. This is being pursued by US CYGNUS collaborators. We leave the evaluation of these options for future work.

\section {Conclusion}
An exciting physics program will be possible with the anticipated network of \Cygnus detectors, as illustrated in Fig.~\ref{fig:timeline}. The initial 1~m$^3$ \Cygnus HD1 demonstrator, or a slight scale-up, could be used to demonstrate directional sensitivity to \cevns with reactor or spallation sources. This would guarantee sensitivity to a directional solar neutrino signal in the subsequent detectors. Then, for every factor ten increase in exposure, interesting new measurements are possible. \Cygnus-1000 would detect between 13 and 37 \cevns events over six years, depending on the exact energy threshold for background free operation. An ambitious \Cygnus-100k detector, with volume similar to that of DUNE~\cite{Abi:2018dnh}, would have non-directional WIMP sensitivity in excess of any proposed experiment, and would, in addition, allow us to utilize directionality to penetrate deep into the neutrino floor. Finally, if a dark matter signal is observed, this would mark the beginning of a new era in physics. The community would seek to firmly establish the galactic origin of the signal, which is only possible by observing some form of direction-dependence. After that, a worldwide effort would ensure to map the local velocity distribution and explore the particle phenomenology of dark matter. A large directional detector such as \Cygnus-100k would be essential to enable this.\\

\acknowledgments
SEV acknowledges support from the U.S.~Department of Energy (DOE) via Award Numbers DE-SC0018034 and DE-SC0010504. 
CAJO is supported by the University of Sydney and the Australian Research Council.
EB is supported by the European Research Council (ERC) under the European Union Horizon 2020 programme (grant agreement No.~818744).
DL acknowledges support from the U.S.~Department of Energy via Award Number DE-SC0019132 and the National Science Foundation via Award Numbers 1506329 and 1407773.
KS is supported by the U.S.~Department of Energy and the National Science Foundation.  
WL, NJCS, CE, ACE and FMM would like to thank STFC for continued support under grant ST/S000747/1.
KM is supported by the Japanese Ministry of Education, Culture, Sports, Science and Technology, a Grant-in-Aid for Scientific Research, ICRR Joint-Usage, Japan Society for the Promotion of Science (JSPS) KAKENHI Grant Numbers 16H02189, 26104005, 26104009, 19H05806 and the JSPS Bilateral Collaborations (Joint Research Projects and Seminars) program and Program for Advancing Strategic International Networks to Accelerate the Circulation of Talented Researchers, JSPS, Japan (R2607). 
Computing resources were provided by the University of Chicago Research Computing Center. 

\bibliographystyle{apsrev4-1}
\bibliography{DDR}

\end{document}